\pdfoutput=1

\documentclass[pdftex,a4paper,12pt,english]{MGthesis}
\usepackage{a4,amsmath,amssymb,fancyhdr,latexsym}
\usepackage[english]{babel}
\usepackage[pdftex]{graphicx}
\usepackage{fullpage}
\usepackage{booktabs}
\usepackage{caption}
\usepackage{appendix}
\usepackage{slashbox}
\usepackage{setspace}

\newcommand{\vel}{\,{\rm km\,s^{-1}}}
\newcommand{\mincir}{\raise
  -2.truept\hbox{\rlap{\hbox{$\sim$}}\raise5.truept \hbox{$<$}\ }}
\newcommand{\magcir}{\raise
  -2.truept\hbox{\rlap{\hbox{$\sim$}}\raise5.truept \hbox{$>$}\ }}
\newcommand{\siml}{\raise
  -2.truept\hbox{\rlap{\hbox{$\sim$}}\raise5.truept \hbox{$<$}\ }}
\newcommand{\simg}{\raise
  -2.truept\hbox{\rlap{\hbox{$\sim$}}\raise5.truept \hbox{$>$}\ }}
\newcommand{\msun}{$h^{-1}{\rm M}_\odot$}
\newcommand{\pc}{pc}

\newcommand{\circa}{$\sim$}
\newcommand{\K}{K}

\newcommand{\yr}{yr}

\newcommand{\cmt}{cm$^{-3}$}

\newcommand{\msunyr}{M$_\odot$ yr$^{-1}$}

\newcommand{\dens}{M$_\odot$ pc$^{-3}$}
\newcommand{\surf}{M$_\odot$ pc$^{-2}$}

\newcommand{\cl}{_{\rm cl}}

\newcommand{\HRule}{\rule{\linewidth}{0.5mm}}
\newcommand{\be}{\begin{equation}}
\newcommand{\ee}{\end{equation}}
\newcommand{\lum}{\,{\rm erg\,s^{-1}}}
\begin{document}
\thispagestyle{empty}
\begin{titlepage}


\begin{center}
 
\textsc{\LARGE University of Torino}\\[1cm]
\textsc{Department of General Physics}\\[0.5cm]
\textsc{\Large Ph.D. Thesis in Astrophysics}\\[0.5cm]
\HRule \\
{ \huge \bfseries Numerical approaches\\
to star formation and SuperNovae energy feedback\\[5mm]
 in simulations of galaxy clusters}\\
 \HRule \\[3cm]
\begin{minipage}{0.8\textwidth}
\begin{flushleft} \Large
\emph{Author:}\\
Martina \textsc{Giovalli}
\end{flushleft}
\end{minipage}
\begin{minipage}{0.8\textwidth}
\begin{flushright} \Large
\emph{Supervisors:} \\
Prof. Antonaldo \textsc{Diaferio}\\
Dr. Giuseppe \textsc{Murante}
\end{flushright}
\end{minipage}
\vfill
 
\end{center} 
\end{titlepage}

%
 


\newpage
\thispagestyle{empty}
\vspace*{8cm}
\begin{minipage}{0.8\textwidth}
\begin{flushright} 
``\textsl{Science is built up with facts, as a house is with stones. \\
But a collection of facts is no more a science than a heap of stones is a
house.}''\\ 
\textit{Jules Henri Poincare}
\end{flushright}
\vfill
\vspace*{2cm}
\begin{flushright}
``\textsl{Do you want to stride into the infinite?\\
 Then explore the finite in all directions.}''\\
\textit{Johann Wolfgang Goethe}
\end{flushright}
\end{minipage}
\clearpage

\oddsidemargin 0.0in 
\onehalfspacing
\setlength{\textwidth}{6.5in}
\newpage

\pagestyle{empty} 
\tableofcontents 
\cleardoublepage 
\pagestyle{plain} 

\pagenumbering{roman}
\addcontentsline{toc}{chapter}{INTRODUCTION}{\protect\thispagestyle{empty}}
\chapter*{Introduction}
\vspace*{1cm}
\textsf{If a century ago, Astronomy concerned almost completely the
description of the dynamics of objects that we could see, the situation
has now reversed, with most of the astronomers more interested in
how a certain observed situation might have come to be. To this
end, numerical techniques have become standard tools for studying
a wide range of cosmological and astrophysical problems.}

\textsf{Among the many open issues in modern Astronomy, galaxy formation is
certainly one 
of the most important and studied.  The physics of gravity only, which rules
the formation and evolution of the large scale structure of the universe in a
Cold Dark Matter cosmology, is sufficient to explain a number of relevant
observations, ranging from the Cosmic Microwave Background Radiation power
spectrum to the statistical properties of the distribution of galaxy and
galaxy cluster. On the other hand, observed properties of galaxies
(e.g. morphologies, luminosities, colours, stellar ages, Tully-Fisher
relation), both in clusters and isolated galaxies, are not satisfactorily accounted for
without considering a number of other astrophysical processes, in addition to
the already complex interaction of nonlinear gravitational evolution and
dissipative gas dynamics.}

\textsf{Observed cluster of galaxies are, in fact, composed by three
main distinct 
components, dark matter, diffuse gas and stars, which have a completely
different physics behind. Numerical codes following both dark matter and
baryonic particles are now commonly used, but they still have shortcomings,
mainly for two reasons.  The first one is the enormous resolution needed, for
example, to simultaneously follow the birth of stars and the Inter-Stellar
Medium physics, and the large scale physics responsible for structure
formation.  The second one deals with the complexity of the involved
physics.  For example, the physics of the Inter-Stellar Medium (ISM) and the
related star formation and energy feedback processes are currently not
understood in full detail.}

\textsf{To deal with such problems, numerical simulation codes often resort to
simplified, {\it sub-grid} models of the complex hydrodynamical and
astrophysical processes working at scales where star formation takes
place. Even then, detailed and satisfactory numerical models of galaxy
formation and evolution, which starts from the formation of cosmic structures
and self-consistently includes gas dynamics, star formation,
SuperNovae (SNe) energy
feedback and all the pertinent processes are still lacking.\\
Therefore, it is of paramount importance to improve the sub-grid treatment of
the ISM physics in numerical simulation, paying particular attention to the
feedback processes that arise from the energetic activity of massive dying
stars, and taking place through winds, ionising photons and SNe explosions
followed by the creation and propagation of hot expanding pressure fronts (SNe
super-bubbles). This energy input is, in fact, believed to be the fundamental
mechanism which shapes and sustains the Inter-Galactic Medium, thus preventing
the ``cooling catastrophe'' typically found in Cold Dark Matter cosmologies
and producing global galactic winds and ``fountains'' which can be observed in
the real Universe. Finally, such an energy provides a source of heating for
the Intra Cluster Medium, especially at the centre of established cooling
flows, whose importance in the global galaxy cluster energy budget has to be
carefully estimated.\\}

\textsf{The aim of this Thesis is thus to investigate different
numerical approaches 
and to introduce a new, physically-based sub-grid model for the ISM physics,
including a treatment of star formation and Type II supernovae energy feedback
(MUPPI, MUlti-Phase Particle Integrator). Our model follows the ISM physics
using a system of ordinary differential equations, describing mass and energy
flows among the different gas phases in the ISM inside each gas particle. The
model also includes the treatment of SNe energy transfer from star-forming
particles to their neighbours. We will show in this Thesis how this model is
able to reproduce observed ISM properties, while also providing an effective
thermal energy feedback and responding to variations
in the local hydrodynamical properties of the gas, e.g. crossing of a
spiral density wave in a galaxy disk.}

\textsf{This thesis work is organised as follows:
\begin{description} 
\item [Chapter 1] we provide the basics of the
cosmological framework upon which this thesis is based. We begin by
introducing the Hot Big-Bang theory and the theory of structure
formation. We then pose particular attention on describing some of the
observed physical properties of galaxy clusters, among which the
diffuse stellar light. We finally introduce some open questions (and
some possible answers) which are still unresolved in this field of Astronomy,
focusing our attention on the role covered by supernovae feedback.
\item [Chapter 2] we provide a review on existing star
formation and supernovae feedback models. We first describe
the numerical code GADGET 2, in which we implemented our novel algorithm
MUPPI (Chapter 4). After reviewing the more relevant star formation
models existing in 
literature, we describe the analytical model for the ISM by Monaco 2004,
upon which MUPPI is based.  
\item [Chapter 3] we present our published results on the study of the
formation mechanism of the diffuse stellar component in a
cosmological hydrodynamical simulation. After describing the numerical simulations and the techniques for
galaxy identification we use, we present the link between galaxy histories and the
formation of the diffuse light. Finally, we
discuss the dynamical mechanisms that unbind stars from galaxies in clusters
and compare with the statistical analysis of the cosmological simulation
we performed. We performed such work using the effective model for star
formation and feedback (Springel \& Hernquist 2003).
\item [Chapter 4] we present the novel algorithm for the Interstellar
Medium evolution, MUPPI (MUltiPhase Particle
Integrator), whose implementation in GADGET2 has been the principal
work of the present PhD thesis. We first describe how the model is
initialised; then we give details on the model core and on all the processes
regulating the evolution of the Interstellar Medium; finally, we
account for the supernova energy redistribution from star-forming gas
particles to their neighbours. In two appendices, we show the flow charts of
MUPPI model and of GADGET-2 code.
\item [Chapter 5] we present and discuss the results we obtained by
using MUPPI with our reference set of parameters in various
initial physical conditions, i.e. an isolated  model 
of the Milky-Way, a model of a typical dwarf galaxy and two isolated non-rotating
cooling-flow halos,  equivalent to those used for the Milky-Way and the dwarf-like
galaxy, but not containing any galaxy. Finally we show our test on the
behaviour of MUPPI when we change numerical resolution and the most important model parameters.  
\end{description}
}

\setcounter{page}{1} 

\pagenumbering{arabic}

\chapter{Basics of the cosmological framework}
\begin{quote}
\textit{The evolution of the world can be compared to a display of fireworks that has just ended;
 some few red wisps, ashes and smoke. Standing on a cooled cinder, we see the slow fading of the suns, and we try to recall the vanishing brilliance of the origin of the worlds.'} Lema\^itre.
\end{quote}
During last century, two observative discoveries have revolutionised our
view of the Universe: \textit{the Expansion of the Universe} in 1929 
by Edwin Hubble and \textit{the Cosmic Microwave background} in 1965 by
Arno Penzias and Robert Wilson. \\
A relevant implication of the Hubble discovery is the formulation of the
 \textit{Cosmological Principle}, which states that
``\textit{On sufficiently large scales the Universe is both homogeneous
   and isotropic}'', namely there are no special directions and no special
 places in the Universe. Since little was known empirically about the
 distribution of matter in the Universe, Einstein thought that the only way to
put theoretical cosmology on a firm footing was to assume there was
a basic simplicity to the global structure of the Universe enabling a
similar simplicity in the local behaviour of matter. So first cosmologists had
 to content themselves with the construction of simplified models
 based on the guiding Cosmological Principle, with the hope of
 describing some general aspects of the Universe.\\
Since the Hubble discovery, many cosmological models have been proposed to
describe the birth and evolution of the Universe. In the early 1960,
there were mainly two rival theories of cosmogony: the Big Bang
theory, which proposes that the universe was created in a giant
explosion, and the Steady-state theory, which denies any beginning or
end, being the Universe infinite and matter within it continuously
created.
The debate between these two philosophically opposite ideas, was over
in 1965 with the detection of the Cosmic Microwave Background (CAB)
radiation. The Big Bang theory was the clear winner for the simple
reason that the steady-state model did not predict and could not
reasonably account for the presence of the cosmic background
radiation. On the other side, the Big Bang theory not only predicted
the background radiation but required it. \\
Thanks to the remarkable progresses achieved by cosmological
observations, the formulation of cosmological models became much more
reliable. Our current knowledge of the birth and evolution of the
Universe and of the objects it contains is based on \textit{the Hot Big-Bang
theory}, or the \textit{standard cosmological model}. \\
Even though the Hot Big bang Theory has been succesfull in predicting
and explaining a high number of observed phenomena, there is still a
number of unresolved questions of a rather fundamental nature.\\
 First there is the issue of \textit{Dark Matter}. \\
About 80 years ago, Fritz Zwicky by studying the motion of galaxies in
the Coma cluster found that such cluster must contain
much more mass than can be seen in its
galaxies. This was the first indication of the existence of some form
of invisible matter. In fact, the mass estimated by the number of
stars belonging to the cluster was lower than that needed for
reproducing the observed velocities. But was just in the 70s that the
existence of the dark matter was posited. At that time, astronomers
demonstrated that the outer parts of spiral galaxies rotate much
faster than theory would predict. If galaxies consisted only of
luminous matter, they would quickly fly apart. 
The only plausible explanation for such behaviour is that
galaxies and clusters contain a healthy dose of dark matter that
provides the gravitational glue needed to hold them together.
It was just early in the 1980s that most astronomers became
convinced by the presence of the dark matter.\\
 Anyway, its nature still remains uncertain. Some observational
 evidences outlined the following properties of the Dark Matter:
\begin{description}
\item [Collisionless] the interaction cross-section between dark
  matter particles (and between dark matter and ordinary matter) is
  so small as to be negligible for densities found in dark matter
  halos (Ostriker \& Steinhardt 2003\nocite{OstSte03}).
\item [Cold] dark matter particles have low velocity dispersions
  (assumed to contain no internal thermal motions, i.e. they are
  cold); this could be due to the fact that DM decoupled from
  ordinary matter in the early Universe when was already non
  relativistic or that DM has never been in thermal equilibrium with
  the other components. Cold Dark Matter made small
  fluctuations in the density field to grow for a long time before the
  decoupling of matter from radiation occurred. When \textit{the
  matter era} begun, the ordinary matter has been rapidly drawn to the
  dense clumps of dark matter and then formed the observed structures
  (Ostriker \& Steinhardt 2003).
\item [Non-baryonic] The COBE satellite detected non-vanishing temperature
  anisotropies in the CMB angular distribution (Bennet et
  al. 96). These anisotropies generated in the presence of small
  primordial density fluctuations at the recombination
  epoch. According to the current and most reliable theory of
  structure formation, these 
  fluctuations serve as the seeds of all current structures in the Universe.
 Moreover, to form the observed large-scale structure
  through purely 
  gravitational processes, the amplitude of the fluctuations in the
  matter density at decoupling must have exceeded a minimum 
  value. It is demonstrable that quantum fluctuations of solely baryonic
  matter could not generate the observed structures without living a
  different imprint in the CMB. Furthermore, the abundances of
  primordial chemical elements show that baryons can not constitute
  more than some percent of the critical density. Otherwise, the
  predicted abundances would not agree with observations.
\end{description} 
These properties define the Cold Dark Matter (CDM) model. The presence
of such a large amount of unobserved matter forms a major challenge
for present day cosmology.\\

A further enigma is presented by the presence of a mysterious and
elusive all-pervading \textit{dark energy}, which is thought to
account for 73$\%$ of the total energy content of the universe (see
below). 
Even though speculations about its nature are plenty, it basically remains
a mystery. In fact, dark energy does not absorb nor emits any light and remains
early uniformly spread throughout the space, on the contrary with the
dark matter which instead collapses with ordinary matter during the
process of galaxy formation.\\

At present the universe contains a wealth of structures on all
  scales. Examples include our planet, the Earth, the stars, the Milky
  Way as well as other galaxies. On a Megaparsec scale we find the
  largest structures presently known to us: the \textit{galaxy
  clusters}. Here galaxies are grouped into huge and nearly spherical
  concentrations which may contain up to thousands of galaxies. These
  dense galaxy clusters are inter-conned by highly anisotropic filamentary and
  wall-like structures. These wide structures may extend over more
  than a hundred Megaparsec and are called \textit{super-clusters}. In
  between clusters and super-clusters of galaxies there are large
  regions almost devoid of galaxies. These are usually called
  \textit{voids}. The emergence of these structures in a otherwise perfectly
  isotropic and homogeneous universe is explained by postulating that
  in the early ages of the universe very small density fluctuation
  were present. The most accepted structure formation theory states
  that the continued action of gravity made these small
  fluctuations to grow, giving rise to the structures we presently
  observe. This gravitational instability, known as the \textit{Jeans
  instability} (Sec.~\ref{struform}), is now the cornerstone of the
  standard cosmological model for the origin and evolution of galaxies
  and large scale structures. \\ 
In the following sections, we briefly review some of the necessary
  astronomical background for the questions addresses in this PhD
  Thesis. 
 
\section{The \textit{Standard} Cosmology or the \textit{Hot Big Bang} theory}
When Albert Einstein in 1915 proposed his theory of gravitation (i.e. the
general relativity) and the equations describing the
dynamics of the 
universe, it was still believed that the universe was static and that
the Milky Way was the entire universe. Thus Einstein could not explain why
resolving his equations he found that the universe should be
expanding or contracting, something entirely incompatible with the
current notion of static universe. It was just after Edwin
Hubble's brilliant observations (1922, 1929) that the modern science
of cosmology was born. \\ 
Since on large scales the strongest force of Nature is gravity, the most
important part of any physical descriptions of the Universe is a
theory of gravitation.  Our best current theory of gravitation is
Einstein's General Theory of Relativity (1915). All modern cosmological
models are based on Einstein equations.\\
General relativity (GR, hereafter) is a metric theory of
gravity. Since gravitation in GR is transformed from being a  
force to being a property of space-time (i.e. gravity is a
manifestation of the local curvature of space), all modern cosmological models
also require a description of the space-time geometry.

The \textit{Einstein field equations} relate the curvature of the space
to matter and energy and are given by 
\begin{equation}
R_{ij} - \frac{1}{2}g_{ij}R = \frac{8\pi G}{c^{4}}T_{ij} +
\frac{\Lambda}{c^{2}}g_{ij} 
\label{eins}
\end{equation}
where $R_{ij}$ and $R$ are the Ricci tensor and Ricci scalar
respectively, $T_{ij}$ is the energy-momentum 
tensor and $\Lambda$ is the \textit{cosmological constant}. The
left-hand term (the Einstein tensor) holds all the 
necessary informations for describing a non-Euclidean space. In
cosmology, the tensor describing the matter distribution which is of greatest
relevance is that of a perfect fluid: 
\begin{equation}
T_{ij} = (p + \rho c^{2})U_{i}U_{j} - pg_{ij}
\end{equation}
where $p$ is the pressure, $\rho c^{2}$ is the energy density
(including the rest-mass energy), and $U_{k}$ is the fluid
four-velocity.\\
Substituting the $RW$ form of metric and the perfect-fluid tensor into
the field equations yields to the \textit{Friedmann Cosmological
  Equations} (1922)


\begin{equation}
\frac{\ddot{a}}{a} = -\frac{4\pi G}{3}\Big(\rho + \frac{3p}{c^{2}}\Big) +
\frac{\Lambda}{3}
\label{frie1}
\end{equation}
for the time-time component and
\begin{equation}
\frac{\dot{a}^{2}}{a^2} = \frac{8\pi G\rho}{3} - \frac{kc^{2}}{a^{2}} +
\frac{\Lambda}{3} 
\end{equation}
for the space-space component. Here $\rho$ is the mass density and $p$
the pressure. Given an equation of state, we can
solve the above equations. Since the Universe is approximated to be an
ideal perfect fluid, the equation of state is given by:
\begin{equation}
p = \omega \rho c^{2}
\end{equation}
where the parameter $\omega$ is a constant which lies in the range $0
\leq \omega \leq 1$. There are three main cases:
\begin{itemize}
\item $\omega = 0$ $\rightarrow$ $p = 0$   dust universe,
  \textit{matter} dominated
\item $\omega = \frac{1}{3}$ $\rightarrow$ $p =\frac{1}{3} \rho c^{2}$
  radiative universe, \textit{radiation} dominated
\item $\omega = -1$ $\rightarrow$ $p = -\rho c^{2}$ De Sitter
  universe, \textit{vacuum} dominated  
\end{itemize}
The Belgian priest Georges Lema\^itre, independently from
Friedmann, discovered solutions to Einstein equations (Eq.~\ref{eins})
and also presented the new idea of an expanding Universe.
Moreover, extrapolating backwards
in time, he saw that an expanding universe should have had a beginning
in an extremely hot and dense phase: the first hint at the Big Bang in
the history of Science. Soon after, in 1929,  Lema\^itre's theoretical
ideas were confirmed when Edwin Hubble discovered that galaxies recede
from us with a velocity which increases with increasing
distance (the \textit{Hubble flow}, see next paragraph).\\
 Without doubt, this discovery was one of the greatest scientific
 revolutions in human history.\\ 

Under the assumption the universe is homogeneous and isotropic (the
Cosmological principle), the field equations admit the \textit{Robertson-Walker metric}
\begin{equation}
ds^{2} = cdt^{2} - a(t)^{2}[\frac{dr^{2}}{1-Kr^{2}} +
 r^{2}(d\theta^{2} + \sin^{2} \theta d\varphi^{2})]
\label{RW} 
\end{equation}
where we have used spherical polar coordinates: $r$, $\theta$ and
$\phi$ are the comoving coordinates; $K$ is the curvature
parameter; $t$ is the proper time; $a(t)$, the \textit{expansion
 factor}, is a function to be determined in the following
which as the dimensions of a length and defines the spatial extent of
the expanding universe. The coordinates are comoving with the
expanding background, which means that the position \textbf{r} of a
point can be written as \textbf{r} = $a(t)\textbf{x}$, where \textbf{x}
is called the \textit{comoving position}. The expansion factor at the 
present time has been normalised to one ($a(t_{0}) = 1$) for
convenience. The curvature parameter $k$ parametrises the global
geometry of the universe, which thus can be \textit{closed} ($k >$ 0), 
\textit{flat} ($k$ = 0) or \textit{open} ($k <$ 0).\\  


The expansion of the Universe, modifies the proper
distance between galaxies. For this reason, the velocity of objects in
the Universe is given by two different components:
\begin{equation}
\boldsymbol{v} = v_{H} + \boldsymbol{v}_{p} \equiv Hd + \frac{dx(t)}{dt}\cdot a(t)
\end{equation}
where
\begin{itemize}
\item $v_{p}$ is the peculiar velocity, the velocity component due to
  the gravitational attraction of other objects;
\item $v_{H}$ is the Hubble flow, i.e. the recessional velocity from
  the observer given by the expansion factor $a(t)$:
  $H(t)$=$\frac{\dot{a}(t)}{a(t)}$ is called the \textit{Hubble parameter}
\end{itemize}
There has always been some uncertainty in the value of the Hubble
constant $H_{0}$, i.e. the measured present expansion rate of the
universe, with the result that usually cosmologist usually still
parametrise it in terms of a dimensionless number $h$, where:
\begin{equation}
h = \frac{H_{0}}{100 Km s^{-1} Mpc^{-1}}
\end{equation}
The same expansion provokes the spectral redshift of both baryons and photons,
which in turn lower their frequencies and lose energy as they
propagates through the space-time.\\
Plugging $ds^{2}=0$ (null geodesic for a light ray) in the $RW$ metric
(Eq.~\ref{RW}) gives the \textit{cosmological} redshift
\begin{equation}
z = \frac{a_{0}}{a} -1
\end{equation}\\
 The evolution of the Universe depends not only on the total density
 $\rho$ but also on the individual contributions from the various
 components present. We denote the contribution due to the
 \textit{i}th component to the total density as
\begin{equation}
\Omega_{i} = \frac{\rho_{i}}{\rho_{c}}
\end{equation}
where
\begin{equation}
\rho_{c} = \frac{3H^{2}_{0}}{8\pi G}
\end{equation}
 is the \textit{critical density}, the density sufficient to halt the
 expansion at t = $\infty$. 

Different cosmological models are selected depending on the value
  of a set of \textit{cosmological parameters}. Constraints of
  cosmological parameters have been placed so far using data coming from
  different types of observations. The basic set of cosmological parameters is
  shown in Tab.~\ref{tab1.1}.
\begin{table}
\centering
\begin{tabular}{c c c }
\hline 
\hline
Parameter & Symbol & Value \\ 
\hline
Hubble constant& $h$ & 0.73 $\pm$ 0.03 \\
Total matter density  & $\Omega_{m}$ & $\Omega_{m} h^{2}$ =
0.134 $\pm$ 0.006 \\ 
Baryon density & $\Omega_{b}$ & $\Omega_{b}h^{2}$ = 0.023 $\pm$
0.001 \\
Cosmological constant & $\Omega_{\Lambda}$ & $\Omega_{\Lambda}$ = 0.7
\\
Power spectrum normalisation & $\sigma_{8}$ & 0.9\\
Spectral index & n & 1.0\\
\hline
\hline
\end{tabular}
\caption{Key cosmological parameters, as deduced by data of different nature}
\label{tab1.1}
\end{table}
The cosmological model defined by the current set of cosmological parameters is
called {\it{standard model}}. Its main aim is to provide the more
accurate description of the actual and high-redshift Universe,
as indicated by various observational data.\\
The cosmological constant $\Omega_{\Lambda}$ plays a fundamental role
in modern cosmology, as well as the density parameters
$\Omega_{m_{i}}$= $\frac{\rho_{m_{i}}}{\rho_{c}}$, where $\rho_{m_{i}}$ is the density of the various
components of the mass-energy: baryons, radiation, dark matter and neutrinos.
There has been a conspicuous number of observational studies devoted
to the calculation of the density parameter $\Omega_{m}$ =
$\sum_{i}\Omega_{m_{i}}$ of the Universe and all of them indicate the
presence of a remarkable quantitative of dark matter
(Sakharov \& Hofer 2003\nocite{2003astro.ph..9326S}).\\
Since the end of nineties (Perlmutter et al. 1999\nocite{1999PhRvL..83..670P},
Mullis et al. 2003\nocite{2003ApJ...594....1T}), type Ia Supernovae
(among the most important 
cosmological distance indicators, resulting from the
violent explosion of a white dwarf star) have been used for precision
estimation of cosmological parameters: in fact, the relation existing
between the observed flux and their intrinsic luminosity depends on
the luminosity distance $d_{L}$, which in turn depends on the
cosmological parameters $\Omega_{m}$ and $\Omega_{\Lambda}$. The SNe
Ia data alone can only constrain a combination of $\Omega_{m}$ and
$\Omega_{\Lambda}$. The CMB data indicates
flatness, \textit{i.e.} $K=0$ $\rightarrow$ $\Omega_{m}$ +
$\Omega_{\Lambda}$ $\sim$ 1); estimates of the energy density
resulting from the distribution of matter, $\Omega_{m}$, sum of the
contributions of baryons and dark matter, provide $\Omega_{m}
\sim$0.3. When these data are analysed together, a component
contributing for the 70\% to $\Omega$ ($\Omega_{\Lambda}\sim$0.7) is found: a
 force of unknown nature, the \textit{dark energy}. From the FLRW
 equations it can be seen that such a component acts as a
 repulsive gravity at large scales, whose characteristics are very
 similar to a fluid with negative pressure. The cosmological constant
 is its simpler example.\\
Dark \textit{energy} does not absorb neither emits any light and remains
nearly uniformly spread throughout the space, on the contrary with the
dark \textit{matter} which instead collapses with ordinary matter
during the process of galaxy formation. The net effect of the presence
of such a component is to cause an accelerated expansion of the
expansion factor $a(t)$ since the moment it begins to dominate over
the others.\\
The theoretical and observational framework briefly reviewed here
provides the so-called \textit{Concordance Model}, whose parameters
are summarised in Tab.~\ref{tab1.1}. 
 
\subsubsection{A brief history of the Universe}
By inventing general relativity, Einstein introduced not only the
possibility that spacetime might be bent, but also the possibility
that it might be punctured (before his theory spacetime could be
treated just as a continuum). Anyhow, within the context of a
classical theory of gravitation (i.e. GR), it is not possible to
discuss the history of the universe meaningfully for instants less
than the \textit{Planck time} i.e. the time it would take a photon
travelling at the speed of light in vacuum to cross a distance equal
to the Planck length which is defined as the scale at which quantum
effects (estimated by the Heinsenberg uncertainty principle) have the
same order of magnitude of GR effects (estimated using the
Schwarzschild radius):
\begin{equation}
t_{P} = (Gh/c^{5})^{1/2} = 1.35 \cdot 10^{-43} sec.
\end{equation}
 Thus we can describe the Universe since the Planck Era ($10^{49}$ Gev,
$t_{P}$, T $\geq 10^{12}$K) from a primordial quantum fluctuation in
the density field. Between the epochs when the Universe was $10^{-35}$ and
$10^{-33}$ seconds old, there has been a post Big-Bang phase of \textit{cosmic
  inflation}, during which the vacuum energy density was dominating over the
energy density of the Universe. In this phase the expansion factor
$a(t)$ increased exponentially, allowing a small causally connected
region (with dimension $\sim H(t)^{-1}$) to inflate to a large region
containing our universe (with dimension $\approx
cH_{0}(t)^{-1}$). Quantum primordial fluctuations have been
consequently magnified to cosmic size giving rise to a primordial
spectrum of cosmological fluctuations whose  characteristics depend upon
the parameters of the model employed. These tiny perturbations seeded the
formation of structure in the later universe. Inflation is at the
moment more a paradigm than a precise theory; many different
inflation models are known which produce reasonable primordial
fluctuation spectra.\\ 
In the Standard Cosmological Model, the nucleosynthesis of light
elements begin at the start of the \textit{radiative era} ($t\sim$ 1 min),
at a temperature $T=10^{9}K$, when positron-electron annihilation
transferred a lot of energy to photons. During this epoch, the
temperature of the universe falls to the point where atomic nuclei can
begin to form.  \\
At $t_{eq}\sim 10^{5}$ years (\textit{equivalence era}),
 matter and radiation reach the same density: the Universe is a plasma
 of protons, electrons and photons.\\
Some time later, at $t_{rec}\sim3\cdot10^{5}$ years
 (\textit{recombination}),  electrons start to combine with nuclei:
 hydrogen and helium atoms begin to form  and the density of the
 universe falls. The universe, lacking free  electrons to scatter the
 photons, suddenly became transparent to  radiation. The liberated
 photons started streaming freely in all  directions: the Cosmic
 Microwave Background (CMB) was born.  

According to the most reliable large scale structure formation
theory, between the ephocs when the Universe was 1 Gyr and 10 Gyr old,
small inhomogeneities created during inflation grow and through
processes of gravitational accretion gave rise to galaxies, cluster
and super-cluster of galaxies. 
In the next section, we review the basics of the \textit{theory of
  structure formation}. \\ 
 
\section{Theory of structure formation}
\label{struform}
According to the Hot Big Bang model, collisionless dark matter
particles dominates the mass density of the Universe. In this picture,
dark matter clumps grow by gravitational collapse and hierarchical
aggregation of even less massive 
systems. Galaxies form at the centres of this haloes, where gas
condenses, cools and forms stars once it becomes sufficiently
dense. Groups and clusters of galaxies form as haloes aggregate into
larger systems. This process, resulting in a
gradual building-up of successively larger structures by the clumping
and merging of smaller-scale structures, is called
\textit{hierarchical structure formation}.\\
Although at sufficiently large scales the universe may be considered
homogeneous and isotropic, at smaller scales it contains all
kinds of structures. How could such structures emerge in a universe
if, according to the Cosmological principle, it is perfectly
homogeneous and isotropic? The more accreditated answer is given by
the \textit{gravitational instability theory} or \textit{Jeans
  theory}: at very early stages the universe was not perfectly
homogeneous, but instead small fluctuations were about.
The Jeans theory describes how these small density fluctuations have grown
with respect to the global cosmic background into the wealth of
structures observed today. 
Jeans (1902) demonstrated that, starting from an homogeneous and
isotropic ``mean'' fluid, small fluctuations in the
density,$\delta$$\rho$, and in the velocity, $\delta v$, could evolve
with time. These fluctuations grow if the stabilizing effect of
pressure is much smaller then the tendency of the self-gravity to
induce collapse.\\
 The basilar equations governing the evolution of a
nonrelativistic, self-gravitating fluid, in its proper frame of
reference, are:\\ 
\begin{equation}
\frac{\partial \rho}{\partial t} + \boldsymbol{ \nabla }\cdot 
(\rho \boldsymbol{u}) = 0 \qquad 
\textrm{The continuity equation} \label{cont}
\end{equation} 
which gives the mass conservation,
\begin{equation}
\frac{\partial \boldsymbol{u}}{\partial t} + ( \boldsymbol{u} \cdot
\boldsymbol{ \nabla }u) = - \frac{1}{\rho}\boldsymbol{ \nabla }p -
\boldsymbol{ \nabla }\boldsymbol{ \Phi } \qquad
\textrm{The Euler equation}  \label{euler}
\end{equation}
the equation of motion, which gives the relation between the
acceleration of the fluid element and the gravitational force, and 
\begin{equation}
\nabla^{2}\Phi = 4\pi G\rho \qquad  \label{eq:poi}
\textrm{The Poisson equation}
\end{equation}
which relates the matter distribution to the gravitational field.\\
The most convenient coordinates to use for cosmology are the
comoving coordinates
\begin{equation}
\textbf{r} = a(t)\textbf{x}
\end{equation}
and the peculiar velocity is:
\begin{equation}
\boldsymbol{v} = a(t)\boldsymbol{x}
\end{equation}
In terms of these coordinates, density fluctuations
  $\delta$(t,$\textbf{x}$) are expressed as,  
\begin{equation}
\delta(t,\boldsymbol{x}) = \frac{\rho(t,\boldsymbol{x})}{\bar{\rho}(t)}
-1  \label{dens_contr}
\end{equation}
where $\bar{\rho}(t)$ is the average value of the density field
$\rho(t,\boldsymbol{x}$. The peculiar gravitational potential
$\phi$(t,$\textbf{x}$) is 
defined as:
\begin{equation}
\phi(t,\boldsymbol{x}) = \boldsymbol{\Phi} + \frac{1}{2}a\ddot{a}\mid\boldsymbol{x}\mid^{2}
\end{equation}
In comoving coordinates, equations ~\ref{cont} -~\ref{poi}
simplify to the following:
\begin{equation}
\dot{\delta} + \frac{1}{a}\boldsymbol{ \nabla } \cdot [(1 +
  \delta)\boldsymbol{v}] = 0 \label{cont_com} 
\end{equation}
\begin{equation}
\dot{\boldsymbol{v}} + \frac{1}{a}(\boldsymbol{v} \cdot \boldsymbol{
  \nabla })\boldsymbol{v} + \frac{\dot{a}}{a}\boldsymbol{v} = -
  \frac{1}{\rho a}\boldsymbol{ \nabla }p - \frac{1}{a}\boldsymbol{ \nabla }\phi   \label{e} 
\end{equation}
\begin{equation}
\nabla^{2}\phi = 4\pi G\bar{\rho}a^{2} \delta  \label{poi_com} 
\end{equation} 
The hierarchical buildup of cosmic structures follows the evolution of
tiny primordial fluctuations in the density field according to
Eq.~\ref{cont_com} -~\ref{poi_com}.\\
During its earliest phases and at large scales, the Universe is
usually assumed to be homogeneous. When the fluctuations are still
small ($\delta << 1$, \textit{linear regime}),
equations~\ref{cont_com}-~\ref{poi_com} can be 
linearised by neglecting all terms which are of second order in the
fields $\delta$ and \textbf{v} as:
\begin{equation}
\dot{\delta} + \frac{1}{a}\boldsymbol{ \nabla } \cdot \boldsymbol{v}
  = 0 \label{g}
\end{equation}
\begin{equation}
\dot{\boldsymbol{v}} + \frac{\dot{a}}{a}\boldsymbol{v} = -
\frac{c^{2}_{s}}{a}\boldsymbol{ \nabla }\delta  -
\frac{1}{a}\boldsymbol{ \nabla }\phi   \label{h} 
\end{equation}
\begin{equation}
\nabla^{2}\phi = 4\pi G\bar{\rho}a^{2} \delta  \label{i} 
\end{equation}
where, in Eq.~\ref{h}, $c^{2}_{s} \equiv (\partial p/\partial \rho)$
is the sound speed.  \\
Transforming Eq.~\ref{g} -~\ref{i} in the \textbf{k} space, using the
Fourier transforms, equation~\ref{g} becomes:
\begin{equation}
\ddot{\delta}_{\boldsymbol{K}} +
2\frac{\dot{a}}{a}\dot{\delta}_{\boldsymbol{K}} + (
\frac{c^{2}_{s}k^{2}}{a^{2}}- 4\pi G\bar{\rho})
\delta_{\boldsymbol{K}} = 0 \label{l}
\end{equation}
If the sign of the third term is positive, $\delta_{\boldsymbol{K}}$
represents a monotone growing solution. This condition is equivalent
to the \textit{Jeans Gravitational Instability Criterion}: if the
wavelength of the fluctuation $\lambda$= 2$\pi/k$ is greater then the
Jeans length, defined as
\begin{equation}
\lambda_{J} \equiv c_{s}\sqrt{\frac{\pi}{G\bar{\rho}}} 
\end{equation}
where $\bar{\rho}$ is the enclosed mass density, then the effect of
self-gravity is stronger then the stabilising effect of 
pressure. The Jeans length characterises the maximum scale a sound
wave can reach by propagating throughout the medium, within a
dynamical time. \\
Being our Universe dominated by collisionless dark matter,
$\lambda_{J}$ can be neglected ($c_{s}\sim 0$). In this contest, the
evolution equation describing perturbations on scales of
cosmological relevance can be approximated as: 
\begin{equation}
\ddot{\delta}_{\boldsymbol{K}} +
2\frac{\dot{a}}{a}\dot{\delta}_{\boldsymbol{K}} - 4\pi G\bar{\rho}
\delta_{\boldsymbol{K}} = 0 \label{m}
\end{equation}
where the \textit{Hubble drag} term
$2\dot{a}/a \dot{\delta}_{\boldsymbol{K}}$ describes the
counter-action of the expanding background on the perturbation growth.\\
For a given set of cosmological parameters, the evolution of $a(t)$ is
completely specified and equation~\ref{m} can then be solved. Two
independent solutions are obtained, one growing and one decaying. The
decaying solution becomes negligible with time.\\
As long as the evolution is linear, a generic perturbation can be
represented as a superposition of plane waves (with wave vector
\textbf{k}) which evolve independently of each other starting from a
primordial density field assumed to be Gaussian. \\
Fundamental quantities of the linear perturbation theory are the
density contrast (Eq.~\ref{dens_contr}) and its Fourier
representation  
\begin{equation}
\delta_{\boldsymbol{K}} = \frac{1}{V}\int d\boldsymbol{x} \delta(\boldsymbol{x})e^{i\boldsymbol{k}\cdot\boldsymbol{x}}
\end{equation} 
Since $\delta_{\boldsymbol{K}}$ is a complex variable, we can rewrite it
as function of two real variables: the amplitude $D_{k}$ and the
phase $\phi_{k}$
\begin{equation}
\label{Fou_coeff}
\delta_{\boldsymbol{k}} \equiv D_{\boldsymbol{k}}e^{i\phi_{\boldsymbol{k}}} =
Re(\delta_{\boldsymbol{k}}) + iIm(\delta_{\boldsymbol{k}}) 
\end{equation}
Substituting the expression~\ref{Fou_coeff} for
$\delta_{\boldsymbol{k}}$ into Eq.~\ref{m} and following the time evolution
of the real and the imaginary part, the amplitude $D_{k}$ evolves as
the \textit{growing} solution coming from the linear theory (the decaying mode
rapidly goes to zero) while the phase $\phi_{k}$ rapidly converges to
a constant value.\\
If the probability distribution for the real and imaginary part of
$\delta_{\boldsymbol{k}}$ follows a Gaussian statistics, in a
homogeneous and isotropic Universe, then $\bar\delta_{k}=0$.\\
The Fourier phases $\phi_{\textbf{k}}$ have a random distribution
and, defining the \textit{power spectrum} of perturbations as $P(k)=
\langle \Arrowvert \delta_{\boldsymbol{K}} \Arrowvert^{2} \rangle$, the
variance of the fluctuation field $\sigma^{2}$ $\equiv$
$\langle$$\delta^{2}$$\rangle$ gives the measure of the
inhomogeneities, variance that with volume going to infinity is:
\begin{equation}
\sigma^{2} = \frac{1}{2\pi^{2}}\int_{0}^{\infty} P(k)k^{2}dk \label{o}
\end{equation}
It is usual to assume that the power spectrum $P(k)$, at least within
a certain interval in $k$, is given by a power law 
\begin{equation}
P(k) = AK^{n}
\label{eq1.1}
\end{equation}
where A is the normalisation and the exponent $n$ is the spectral
index, which can be determined from observations of the CMB (see
Table~\ref{tab1.1}). The convergence 
of the variance in~\ref{o} requires that $n > -3$ on large scales ($k
\rightarrow 0$) and $n < 3$ on small scales ($k \rightarrow \infty$).
Moreover, since the model of structure formation is
hierarchical, i.e. small structures form first then merge to form
larger object, it has to be $n > -3$.\\
At the end of inflation and after the entrance in the cosmological
horizon, the perturbation power spectrum $P(k)$ is modulated by the
physical processes characterising the evolution of the Universe.
The net effect of these processes is to change the shape of the
original power spectrum $P_{0}(k)$ in a manner described by a simple
function of the wave-number, the \textit{transfer function} $T(k)$:
\begin{equation}
P(k) = A[\frac{D(t)}{D_{0}(t)}]^{2}T^{2}(k,t_{f})P_{0}(k)
\label{ciao}
\end{equation}
where $D(t)$ describes the growing model of perturbations
according to the adopted cosmological framework. In the case of the
$\emph{concordance model}$, 
\begin{equation}
D(a) = \frac{5}{2}H_{0}^{2} \Omega_{m} \frac{\dot{a}}{a}\int \frac{da'}{\dot{a}'^{3}}
\label{evviva}
\end{equation}
where
\begin{equation}
\frac{\dot{a}}{a} = H_{0}[\frac{\Omega_{m}}{a^{3}} + \Omega_{\Lambda} + \frac{1-\Omega_{m} - \Omega_{\Lambda}}{a^{2}}]^{1/2}
\end{equation}
The shape of the power spectrum is essentially fixed once the matter
 $\Omega_{m}$ and baryon $\Omega_{bar}$ density parameters and the
 Hubble parameter $H_{0}$ are specified (Eisenstein \& Hu
 1999\nocite{1999ApJ...511....5E}).  
However, its normalisation A (defined in Eq.~\ref{eq1.1}) can only be
 fixed through a comparison of observational data of the large scale
 structure of the Universe or of the anisotropies of the CMB. One
 possible parametrisation of this normalisation is through the
 quantity $\sigma_{8}$, which is defined as the variance of the
 fluctuation field computed at the scale $R = 8h^{-1}Mpc$
\begin{equation}
\sigma_{8}^{2}=\frac{<\delta M^{2}>}{<M>^{2}} \equiv 0.9 
\label{sigma8}
\end{equation} 
where M is the average mass contained in a sphere having radius
$8h^{-1}Mpc$. The historical reason for this choice of the
normalisation scale is that the variance of the galaxy number counts,
within the first redshift surveys, was observed to be about unity
inside spheres of that radius (e.g. Davis \& Peebles
1983\nocite{1983ApJ...267..465D}). 
The value of $\sigma_{8}$ is obtained from mass and galaxy
distribution on galaxy cluster surveys. In this way, the power
spectrum is normalised over relatively small scales. The value of the
$\sigma_{8}$ is thus given by:
\begin{equation}
\sigma_{8}^{2}=\frac{1}{2\pi^{2}}\int_{0}^{\infty} P(k)W^{2}(kR_{8})k^{2}dk 
\label{eq1.2}  
\end{equation}
where $W^{2}(kR)$ is a \textit{filter function} which selects the
contributions to the Fourier amplitudes at the assigned scale
$R_{8}h^{-1}Mpc$. Substituting the power spectrum $P(k)$ given by
Eq.~\ref{eq1.1}, we obtain a value for $A$, being the other unknown
parameters fixed by the cosmological model. \\
As the dense regions become denser and the density contrast approaches
$\delta \sim 1$, the linear approximation begins to
break down  and a more detailed treatment, using the full theory of
gravity, becomes necessary. The complexity of the physical 
 behaviour of fluctuations in the non linear regime  
makes it impossible to study the details exactly using analytical
methods. For this task one must resort to \textit{numerical
  simulation} methods. In the next chapter (Ch. 2), we will
extensively review the major numerical techniques to compute the
nonlinear evolution of gravitational instabilities and the
hydrodynamical processes describing the gas dynamics, processes which
became necessary once one needs to properly follow the evolution of
luminous matter.\\
The nonlinear evolution of an inhomogeneity can be computed exactly without
resorting to numerical simulations just in the case it has some
particularly simple form, as we shall see in the following.\\
\subsubsection{The spherical top-hat collapse}
The first step in modelling the large-scale evolution of matter into non-linear
structure is a description of the 'microscopic' case: the
collapse of a single spherical region at constant density into a
self-gravitating halo, via a model known as spherical \textit{top hat}
collapse. Although the restrictive assumptions, this model serves
as a very useful guideline to describe the process of evolution and
formation of virialised DM halos. In the following, we just recall some
background about the spherical top-hat collapse model, for a detailed
review, see Peebles 1993\nocite{1993ppc..book.....P}, Peacock
1999\nocite{1999coph.book.....P}, Coles \& Lucchin
2002\nocite{2002coec.book.....C}, Borgani
2006\nocite{2006astro.ph..5575B} and references 
therein. \\ 
The \textit{halo} model describes the formation of a collapsed
object by solving for the evolution of a sphere of uniform over-density
$\delta$ in a smooth background of density $\bar{\rho}$.
The over-dense region evolves as a positively curved
Friedmann-Lema\^itre-Robertson-Walker (FLRW) universe 
whose expansion rate is initially matched to that of the
background. After reaching the maximum expansion, the perturbation
then evolves by detaching from the general Hubble expansion and then
re-collapses, reaching virial equilibrium supported by the velocity
dispersion of DM particles. This happens at the
virialisation time $t_{vir}$, at which the perturbation meets by
definition the virial condition $E = K + U = -K$, being $E$, $K$ and
$U$ the total, the kinetic and the potential energy respectively.
  In an Einstein-de Sitter (EdS) model ($\Omega_{m} = 1$)
 the over-density at $t_{vir}$ is 
\begin{equation}
\Delta_{\rm  vir} = \frac{\rho_{p}}{\rho} \sim 178
\end{equation} 
which is usually approximated to $\Delta_{\rm vir} = 200$ for a typical
 DM halo which has reached the condition of virial equilibrium. 
$\Delta_{\rm 200}$ is often used as a threshold parameter in
 the Press-Schechter (PS) theory (Press \& Schechter
 1974\nocite{1974ApJ...187..425P}, Sheth, Mo \& Tormen 2001\nocite{2001MNRAS.323....1S}) for predicting the shape and evolution of
 the mass function of bound objects.\\

\section{Physical properties of galaxy clusters}
Nearby galaxy clusters provides important
 constraints on the amplitude of the power spectrum at cluster scale
(e.g. Rosati et al. 2002\nocite{2002ARA&A..40..539R} and references therein), on the
 linear growth rate of density perturbations and thus on the matter
 and dark energy density parameters. 
Observations of galaxy clusters
 provide other fundamental constraints on the determination of
 cosmological parameters. For example, clustering properties
 (the correlation function and the power spectrum) of their
 distribution provide direct measurements on the
 shape and amplitude of the underlying DM distribution power spectrum;
 the estimation of the baryon fraction in nearby galaxy clusters
 provides constraints on the matter density parameter. That is why
  galaxy clusters are now the best astrophysical laboratories for
 investigating the correctness of the current cosmological model.

In the following section, we will briefly review
 the basic observational properties of galaxy clusters.

Clusters of galaxies are the largest virialised objects in the
Universe and, therefore, outline the network of the distribution of
visible matter in the Universe. Clusters were first identified as
large concentrations in the projected galaxy distribution 
(Abell 1958\nocite{1958ApJS....3..211A}, Zwicky et
al. 1966\nocite{1966cgcg.book.....Z} and Abell et
al. 1989\nocite{1989ApJS...70....1A}). They arise from the
gravitational collapse of the highest 
peaks (with size $\sim$ 10 Mpc) of the primordial matter density
field in the hierarchical scenario for the formation of structures
(e.g. Peeble 1993, Coles \& Lucchin 2002). The
scientific importance of studying galaxy clusters resides in their marking the
transition between two distinct regimes in the study of the formation
of cosmic structures: on one hand, the \textit{large} scale, driven by
gravitational instability of the DM particles; on the other hand,
 the \textit{galaxy} scale, ruled by the combined action of gravity and complex
 hydrodynamical and astrophysical processes (e.g. cooling, star
 formation, energy feedback by SNe).\\
Clusters of galaxies are used both as cosmological tools
and as astrophysical laboratories (see Rosati et al. 2002 and Borgani
\& Guzzo 2001 for a review).\nocite{2001Natur.409...39B}
While the evolution of the population of
galaxy clusters and their overall baryonic content provide powerful
constraints on cosmological parameters, the physical properties of the
intra-cluster medium (ICM) and their galaxy population give fundamental
insights on astrophysical-scale problems. \\ 

The first observations showed that galaxy clusters are associated with deep gravitational potential well
and comprised galaxies with velocity dispersion $\sigma_v$ = $10^{3}$ km $s^{-1}$. Defining the crossing time for a cluster of size
 $R$ as 
\be
t_{cr} = \frac{R}{\sigma_v} \sim 1 \big( \frac{R}{1 Mpc} \big) \big(\frac{\sigma_v}{10^3 km s^{-1}} \big ) Gyr
\ee
and, given the Hubble time $t_{H} \sim 10 h^{-1}$Gyr, one can easily deduce that such a system 
has enough time in its internal region ($\leq1$ Mpc) to dynamically relax, on the contrary with the surroundings ($\sim10$ Kpc).
Assuming virial equilibrium, from this type of analysis one typically obtains for the cluster mass
\be
M \cong \frac{R \sigma_v^2}{G} \cong \big ( \frac{R}{1 h^{-1} Mpc} \big ) \big ( \frac{\sigma_v}{10^3 km s^{-1}} \big)^2 10^{15} h^{-1} M_{\odot}
\ee
Smith (1936\nocite{1936ApJ....83...23S}) first discovered in his study of the Virgo cluster that the mass implied by cluster 
galaxy motions was largely exceeding that associated with the optical component. This was confirmed by Zwicky (1937\nocite{1937ApJ....86..217Z}) and was the first 
evidence of the presence of dark matter. \\
Clusters are now believed to be made out of three main ingredients: non-collisional dark matter ($\approx 80\%$),
hot and warm diffuse baryons ($\approx 15\%$) and cooled baryons (stars $\approx 5\%$). As cosmic baryons collapse following 
the dynamically dominant dark matter, a thin hot gas permeating the cluster gravitational potential well is formed,
 as a result of adiabatic compression and accretion shocks generated by supersonic motions during shell crossing and virialisation.
For a typical cluster mass of $10^{14}$-$10^{15}$ $M_{\odot}$ this gas reaches temperatures of several $10^7$ K, becomes fully ionised, and therefore emits 
via thermal bremsstrahlung in the X-ray band. Observations of clusters in the X-ray band have became a standard tool for studying the ICM 
properties and moreover provide an efficient and physically motivated method of identification.
\subsection{X-ray properties}
X-ray surveys with the ROSAT satellite, supplemented by follow-up studies with ASCA and Beppo-SAX, have allowed an assessment of the evolution
 of the space density of clusters out to z $\simeq 1$ and the evolution of the physical properties of the intra-cluster medium out to z $\simeq$ 0.5. In the following, we briefly outline the fundamental properties of galaxy clusters and the key X-ray observations about them.
\\
Observations of galaxy clusters in the X-ray
band have revealed a substantial fraction, 
$\sim\! 15\%$, of the cluster
mass to be in the form of hot diffuse gas, permeating its potential
well. If this gas shares the same dynamics as member galaxies, then it
is expected to have a typical temperature
\be
k_BT \simeq \mu m_p \sigma_v^2\simeq 6 \left({\sigma_v\over10^3\vel}\right)^2 {\rm keV}\,,
\label{eq:tsig}
\ee
where $m_p$ is the proton mass and $\mu$ is the mean molecular weight
($\mu=0.6$ for a primordial composition with a 76\% fraction
contributed by hydrogen). 
\nocite{1997ApJ...478..462C}
\nocite{2001ApJ...548...79G}
\nocite{1999ApJ...517..587L}
\nocite{2001ApJ...562..124J}
\nocite{1999AJ....117.2608G}
\nocite{2002ApJ...567..716R}

\begin{figure}
\centering{
\includegraphics[viewport= 15 120 650 650,clip,height= 6.5cm]{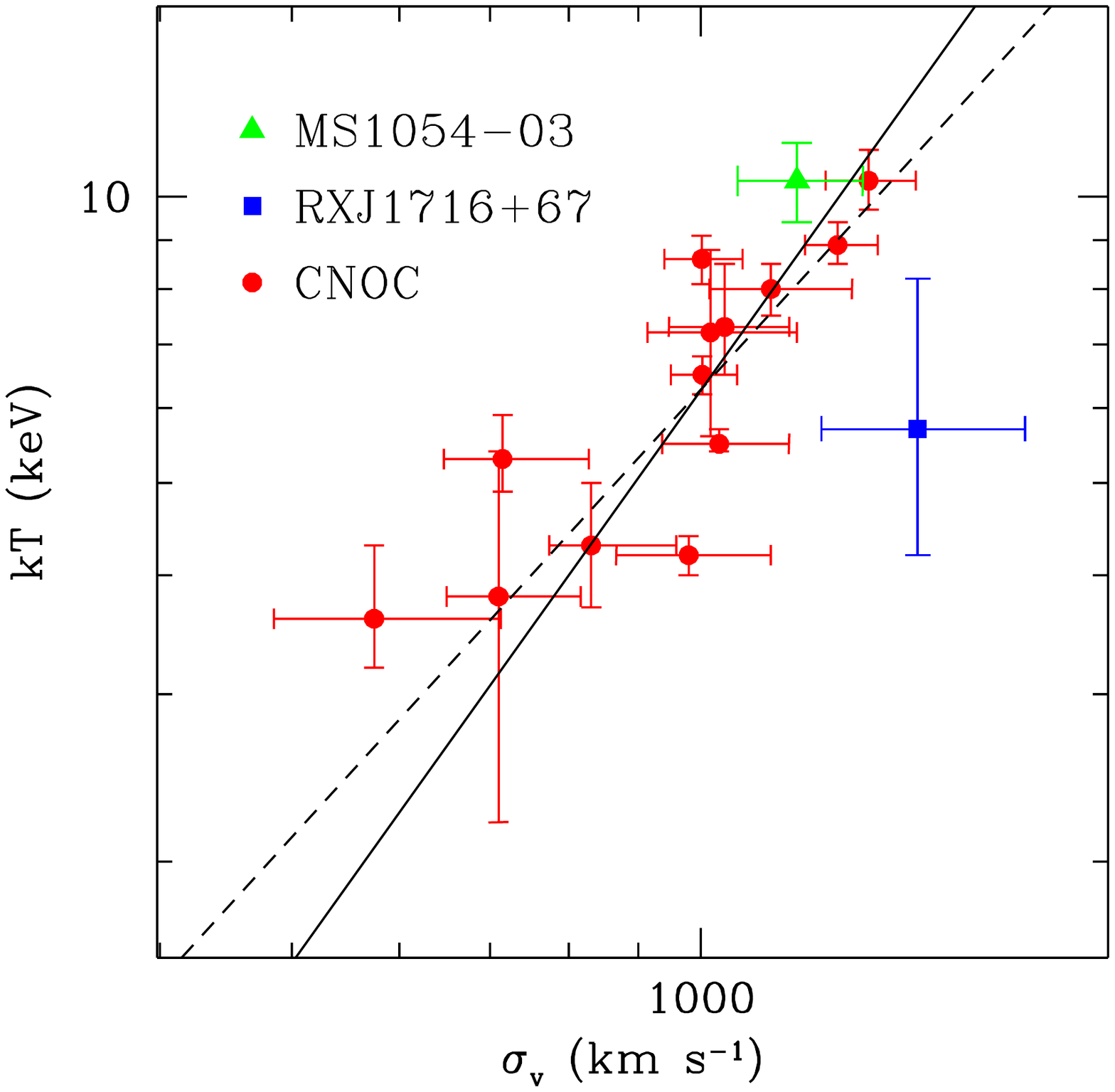} 
\includegraphics[viewport= 90 350 700 700,clip,width=7.3cm,height=6.3cm]{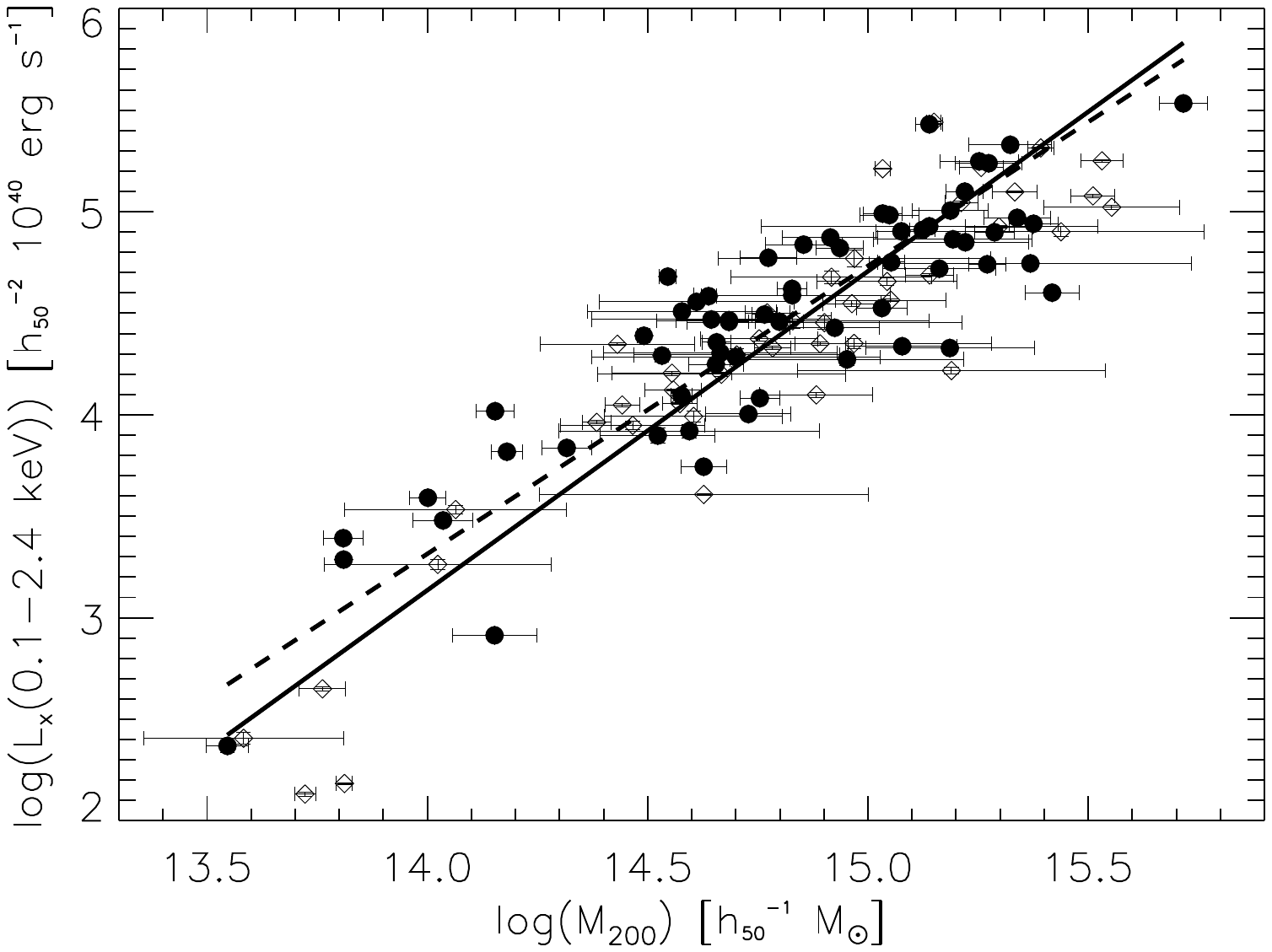} 
\captionsetup{font={normalsize,sf}, width= 0.9\textwidth}
\caption{ ({\it Left}) The relation between galaxy velocity dispersion,
$\sigma_v$, and ICM temperature, $T$, for distant ($z>0.15$) galaxy
clusters. Velocity dispersions are taken from Carlberg et
al. (1997a)) for CNOC clusters and from Girardi \& Mezzetti (2001)) 
for MS1054-03 and RXJ1716+67. Temperatures are taken from Lewis et
al. (1999) for CNOC 
clusters, from Jeltema et al. (2001) or MS1054-03 and from Gioia et
al. (1999) for RXJ1716+67. The solid line shows the relation $k_BT=\mu
m_p \sigma_v^2$, and the dashed line is the best--fit to the
low--$z$ $T$--$\sigma_v$ relation from Wu et al. (1999). ({\it Right})
The low-$z$ relation between X-ray luminosity and the mass
contained within the radius encompassing an average density
$200\rho_c$ (from Reiprich \& B\"ohringer 2002). The two lines are the
best log--log linear fit to two different data sets indicated with filled
and open circles.}}
\label{fi:sigv_tx}
\end{figure}

Observational data for nearby clusters (e.g. Wu et al. 1999\nocite{1999ApJ...524...22W})
and for distant clusters (see Figure~\ref{fi:sigv_tx}) roughly follow
the above relation for the gas temperature.
This correlation indicates that the assumption that clusters are
relaxed structures in which both gas and galaxies feel the same
dynamics is reasonable. Anyway, there are some exceptions
that reveal the presence of a more complex dynamics.

At high energies, the ICM behaves as a fully ionised plasma, whose
emissivity is dominated by thermal bremsstrahlung. 
The pure bremsstrahlung emissivity is a good approximation
for $T\gtrsim 3$ keV clusters, but a further contribution from metal
emission lines should be taken into account when considering cooler
systems (e.g. Raymond \& Smith 1977\nocite{1977ApJS...35..419R}).  By
integrating the above 
equation over the energy range of the X-ray emission and over the gas
distribution, one obtains X-ray luminosities $L_X \sim 10^{43}$-$10^{45}\lum$.
These powerful luminosities allow clusters to be identified as
extended sources out to large cosmological distances. 

The condition of hydrostatic equilibrium
connects the local gas pressure $p$ to its density $\rho_{\rm gas}$
according to

\be
{dp\over dR}\,=\,-{GM(<R)\rho_{\rm gas}(R)\over R^2}\,.
\label{eq:hy1}
\ee

By substituting the equation of state for a perfect gas, $p=\rho_{\rm
gas}k_BT/\mu m_p$ into the above equation, one can express the total gravitating mass within R as

\be 
M(<R)\,=\,-{k_BT R\over G\mu m_p}\,
\left({d\,\log\rho_{\rm gas}\over d\log R}+{d\,\log T\over d\log R}\right)\,.
\label{eq:hy2}
\ee

At redshift $z$, we have $M\propto R^3
\bar\rho_0(1+z)^3\Delta_{vir}(z)$, where $R$ is the virial radius, $\bar\rho_0$ is the mean
cosmic density at present time and $\Delta_{vir}(z)$ is the mean
overdensity within a virialized region. For an Einstein--de-Sitter cosmology, $\Delta_{vir}$ is constant
and therefore, for an isothermal gas distribution, Equation
(\ref{eq:hy2}) implies $T\propto M^{2/3}(1+z)$.
By measuring quantities such as $\rho_{\rm
gas}$ and $T$ from X-ray observations, one can easily derive the mass of the selected cluster. Thus, in addition to providing an
efficient method to detect clusters, X-ray studies of the ICM allow
one to quantify the total gravitating cluster mass, which is the
quantity predicted by theoretical models for cosmic structure
formation.

A popular description of the gas density profile is the
$\beta$--model,
\be
\rho_g(r)\,=\,\rho_{g,0}\,\left[1+\left({r\over
r_c}\right)^2\right]^{-3\beta/2}, 
\label{eq:betam}
\ee
which was introduced by Cavaliere \& Fusco--Femiano
(1976)\nocite{1976A&A....49..137C}; 
see also Sarazin 1988\nocite{SA88.1}, and references therein) to describe an
isothermal gas in hydrostatic equilibrium within the potential well
associated with a King dark-matter density profile. The parameter
$\beta$ is the ratio between kinetic dark-matter energy and
thermal gas energy.
This model is a
useful guideline for interpreting cluster emissivity, although over
limited dynamical ranges. Now, with the {\it Chandra} and {\it
Newton-XMM} satellites, the X-ray emissivity can be mapped with high
angular resolution and over larger scales. These new data have
shown that Equation \ref{eq:betam} with a unique $\beta$ value cannot
always describe the surface brightness profile of clusters
(e.g. Allen et al. 2001\nocite{2001MNRAS.328L..37A}). 

Kaiser (1986\nocite{1986MNRAS.222..323K}) described the thermodynamics
of the ICM by assuming it 
to be entirely determined by gravitational processes, such as
adiabatic compression during the collapse and shocks due to supersonic
accretion of the surrounding gas. As long as there are no preferred
scales both in the cosmological framework (i.e.  $\Omega_m=1$ and
power--law shape for the power spectrum at the cluster scales), and in
the physics (i.e. only gravity acting on the gas and pure
bremsstrahlung emission), then clusters of different masses are just a
scaled version of each other, because bremsstrahlung emissivity predicts
$L_X\propto M\rho_{\rm gas}T^{1/2}$,  $L_X\propto
T_X^2(1+z)^{3/2}$ or, equivalently $L_X\propto
M^{4/3}(1+z)^{7/2}$. Furthermore, if we define the gas entropy as
$S=T/n^{2/3}$, where $n$ is the gas density assumed fully ionized,
we obtain $S\propto T(1+z)^{-2}$.

It was soon recognized that X-ray clusters do not follow these scaling
relations. The observed luminosity--temperature relation for clusters
is $L_X\propto T^{\alpha}$, where $\alpha\sim 3$ for $T\gtrsim 2$ keV, and possibly even steeper
for $T\lesssim 1$ keV groups. This result is consistent with the finding that
$L_X\propto M^\alpha$ with $\alpha\simeq 1.8\pm 0.1$ for the observed
mass--luminosity relation (e.g. Reiprich \& B\"ohringer 2002; see
right panel of Figure~\ref{fi:sigv_tx}). Furthermore, the
low-temperature systems are observed to have shallower central
gas-density profiles than the hotter systems, which turns into an
excess of entropy in low--$T$ systems with respect to the $S\propto T$
predicted scaling (e.g. Ponman et
al. 1999\nocite{1999Natur.397..135P}, Lloyd--Davies et
al. 2000\nocite{2000MNRAS.315..689L}).  
\subsection{Breaking of the scaling relations: the importance of
  non-gravitational heating}
\label{cap1:overcool}
\begin{figure}
\centering{
\includegraphics[width=0.33\linewidth]{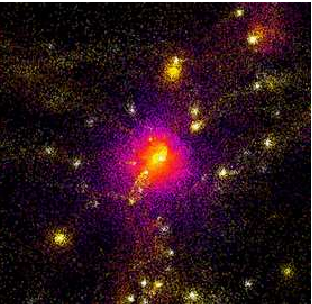} 
\includegraphics[width=0.33\linewidth]{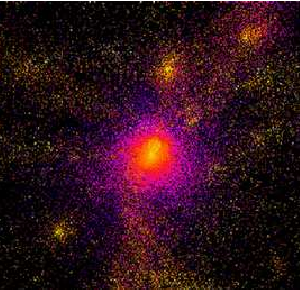} 
\captionsetup{font={normalsize,sf}, width= 0.9\textwidth}
\caption{Map of gas entropy from hydrodynamical simulations of a
galaxy cluster (from Borgani et al. 2001a). ({\it Left}): gravitational
heating only. ({\it Right}): entropy floor of 50 keV/cm$^2$ imposed at
$z=3$, corresponding to about 1 keV/part. Light colors correspond to
low entropy particles, and dark blue corresponds to
high--entropy gas. }}
\label{fi:entr_sim}
\end{figure}

A possible interpretation for the breaking of the scaling relations
assumes that the gas has been heated at some earlier epoch by feedback
from a non-gravitational astrophysical source (Evrard \& Henry
1991\nocite{1991ApJ...383...95E}).  
This heating would
increase the entropy of the ICM, place it on a higher adiabat, prevent it
from reaching a high central density during the cluster gravitational
collapse and, therefore, decrease the X-ray luminosity (e.g. Balogh et al.
1999\nocite{1999MNRAS.307..463B}, Tozzi \& Norman
2001\nocite{2001ApJ...546...63T}, and references therein). For a fixed
amount of extra energy per gas particle, this 
effect is more prominent for poorer clusters, i.e. for those objects
whose virial temperature is comparable with the extra--heating
temperature. As a result, the self--similar behavior of the ICM is
expected to be preserved in hot systems, whereas it is broken for
colder systems.  Both semi--analytical works (e.g. Cavaliere et al.
1998\nocite{1998ApJ...501..493C}, Balogh et al. 1999, Wu et
al. 2000\nocite{2000MNRAS.318..889W}; Tozzi et
al. 2000\nocite{2000ApJ...542..106T})  
and numerical simulations (e.g. Navarro et
al. 1995\nocite{1995MNRAS.275..720N}, Brighenti \& Mathews
2001\nocite{2001ApJ...553..103B}, Bialek et
al. 2001\nocite{2001ApJ...555..597B} 
Borgani et al. 2001a\nocite{2001ApJ...559L..71B}) converge to indicate that
$\sim 1$ keV per gas particle of extra energy is required.  A visual
illustration of the effect of pre--heating is reported in Figure
\ref{fi:entr_sim}, which shows the entropy map for a
high--resolution simulation of a system with mass comparable to that
of the Virgo cluster, for different heating schemes (Borgani et
al. 2001b\nocite{2001ApJ...561...13B}). The effect of extra energy
injection is to decrease the gas 
density in central cluster regions and to erase the small gas clumps
associated with accreting groups.

The gas-temperature distributions in the outer regions of clusters are
not affected by gas cooling. These temperature distributions have
been studied with the {\it ASCA} and {\it Beppo--SAX} satellites.
General agreement about the shape of the temperature profiles has
still to be reached (e.g. Markevitch et
al. 1998\nocite{1998ApJ...503...77M}, White
2000\nocite{2000MNRAS.312..663W}, Irwin \& Bregman
2000\nocite{2000ApJ...538..543I}). De Grandi \& Molendi
(2002)\nocite{2002ApJ...567..163D} analyzed a set of 21 
clusters with {\it Beppo--SAX} data and found the gas to be isothermal out
to $\sim 0.2 R_{\rm vir}$, with a significant temperature decline at
larger radii. Such results are not consistent with the temperature
profiles obtained from cluster hydrodynamical simulations (e.g.
Evrard et al. 1996\nocite{1996ApJ...469..494E}, Borgani et al. 2003), thus indicating that
some physical process is still lacking in current numerical
descriptions of the ICM. Deep observations with {\it Newton--XMM} and
{\it Chandra} will allow the determination of temperature profiles
over the whole cluster virialized region.

\subsubsection{Cooling in the Intra Cluster Medium}
\label{cool_ICM}
In order to characterize the role of cooling in the ICM, it is useful
to define the cooling time--scale, which for an emission process
characterized by a cooling function $\Lambda_c(T)$, is defined as
$t_{cool}= k_BT/(n\Lambda(T))$, $n$ being the number density of
gas particles. For a pure bremsstrahlung emission:
$
t_{cool}\simeq 8.5\times 10^{10}{\rm yr}\,(n/10^{-3}{\rm
cm}^{-3})^{-1} \,(T/10^8 K)^{1/2}\,.
$ (e.g. Sarazin 1988).
Therefore, the cooling time in central cluster regions can be shorter
than the age of the Universe. A substantial fraction of gas undergoes
cooling in these regions, and consequently drops out of the hot
diffuse, X-ray emitting phase. Studies with the {\it ROSAT} and {\it
ASCA} satellites indicate that the decrease of the ICM temperature in
central regions has been recognized as a widespread feature among
fairly relaxed clusters (see Fabian 1994\nocite{1994ARA&A..32..277F},
and references therein). The canonical picture of cooling flows
predicted that, as the
high--density gas in the cluster core cools down, the lack of pressure
support causes external gas to flow in, thus creating a superpositions
of many gas phases, each one characterized by a different
temperature. Our understanding of the ICM cooling structure is now
undergoing a revolution thanks to the much improved spatial and
spectral resolution provided by {\it Newton--XMM}. Recent observations
have shown the absence of metal lines associated with gas at
temperature $\lesssim 3$ keV (e.g. Peterson et
al. 2001\nocite{2001A&A...365L.104P}, Tamura et 
al. 2001\nocite{2001A&A...365L..87T}), in stark contrast with the
standard cooling flow prediction 
for the presence of low--temperature gas (e.g. B\"ohringer et
al. 2002a\nocite{2002ApJ...566...93B}, Fabian et
al. 2001a\nocite{2001MNRAS.321L..20F}). 

Radiative cooling has been also suggested as an alternative to extra
heating to explain the lack of ICM self--similarity (e.g. Bryan 2000,
Voit \& Bryan 2002). When the recently shocked gas residing in
external cluster regions leaves the hot phase and flows in, it
increases the central entropy level of the remaining gas. In the
meanwhile, gas ``cools off'' the hot phase: the
decreased amount of hot gas in the central regions causes a
suppression of the X-ray emission (Pearce et al. 2000, Muanwong et
al. 2001). This solution has a number of problems. Cooling in itself
is a runaway process, leading to a quite large fraction of gas leaving
the hot diffuse phase inside clusters. Analytical arguments and
numerical simulations have shown that this fraction can be as large as
$\sim 50\%$, whereas observational data indicates that only $\lesssim
10\%$ of the cluster baryons are locked into stars (e.g. Bower et
al. 2001, Balogh et al. 2001). This calls for the presence of a
feedback mechanism, such as supernova explosions (e.g. Menci \&
Cavaliere 2000, Finoguenov et al. 2000, Pipino et al. 2002; 
Kravtsov \& Yepes 2000) or Active Galactic Nuclei
(e.g. Valageas \& Silk 1999, Wu et al. 2000, Yamada \& Fujita 2001),
which, given reasonable efficiencies of coupling to the hot ICM, may
be able to provide an adequate amount of extra energy to balance
overcooling. We will summarize feedback mechanisms on
Sec.~\ref{SF:AGN_FB} and ~\ref{SF:SN_FB}.

\subsection{The diffuse light in galaxy clusters}

A significant stellar component of galaxy clusters is found outside of
the galaxies. The standard theory of cluster evolution is one of
hierarchical collapse, as time proceeds, clusters grow in mass through
the merging with other clusters and groups. These mergers as well as
interactions within groups and within clusters strip stars out of
their progenitor galaxies.\\
In this section, we review the basic properties and observed
characteristics of the optical diffuse light in clusters. In Chapter~\ref{ICL}
we will show how studying this intracluster component in cosmological
simulations of galaxy clusters can
inform hierarchical formation models as well as tell us something
about physical mechanisms involved in galaxy evolution within
clusters. \\

The first reference in the literature about the diffuse light in 
cluster of galaxies was given by Zwicky
(1951\nocite{1951PASP...63...61Z}): ``One of the most interesting 
discoveries made in the course of this investigation [in the Coma
  cluster] is the observation of an extended mass of luminous
intergalactic matter of very low surface brightness. The objects which
constitute this matter must be considered as the faintest 
individual members of the cluster. [We report] the discovery of
luminous intergalactic matter concentrated generally and
differentially around the center of the cluster 
and the brightest (most massive) galaxies, respectively''. This is a
perfect characterization of the optical diffuse light in cluster:
extended, low surface brightness and around the center of the cluster.

The characteristics of this diffuse stellar component published by
Zwicky (1951, 1957\nocite{1957moas.book.....Z},
1959\nocite{1959HDP....53..390Z}) were qualitative: it has an
extension of around 150 kpc, the color index is rather blue and
produces a local absorption of light of the order of six tenth of a magnitude. 

The first published attempt to obtain a value for the surface
brightness $\mu$ of the faint intergalactic matter in Coma corresponds to de
Vaucouleurs (1960\nocite{1960ApJ...131..585D}). He reported an upper limit of  
$\mu$ $>$ 29.5 mag\,arcsec$^{-2}$. With this value, de
Vaucouleurs reasons out that ``a stellar population composed
exclusively of extreme red dwarfs of mass M$< 0.1$ M$_{\odot}$ and absolute  
magnitudes $M(pg) > +15$ would, in principle, give a mass-to-light
ratio (M/L) ratio of the order measured in Coma. While such stars are 
known to exist in the neighborhood of the sun, it seems very difficult
to admit that they could populate intergalactic 
space with the required density and to the exclusion of all other
stars of slightly greater mass. Thus, de Vaucouleurs 
concludes that the mass of the intergalactic matter is not enough to
account for the mass value estimated through the virial theorem.

Before the CCD detectors were widely used, most of the observations
and study of the diffuse light in clusters was carried 
out in the Coma cluster (e.g. Abell
(1965\nocite{1965ARA&A...3....1A}); Gunn
(1969\nocite{1969BAAS....1R.191G})) and in other
rich clusters (e.g. the Virgo cluster: Holmberg
(1958\nocite{1958MeLu2.136....1H}); de Vaucouleurs
(1969\nocite{1969ApL.....4...17D})). 
The first accurate measurements of the diffuse light in Coma dated
back to the beginnig of the nineties thanks to the introduction of the CCD
photometry (Bernstein et al. 1995\nocite{1995AJ....110.1507B}).

Observing the Intra Cluster Light (ICL) is quite problematic, first of
all because it is expected to be extremely faint (about 25--26 mag
arcsec$^{-2}$). Than, there a number of problems associated to the use of
CCD's, among them we review the following:
\begin{itemize}
\item ICL is expected to be extremely faint, thus detections are
subjected to spurios effects, intrumental scattering and contamination
due to bright stars or faint galaxies;
\item if the cleanliness of the 
telescope optics is not correct enough, some of the results that we could ascribe to the intracluster light would be masked
or spoiled;
\end{itemize}

The most important characteristics associated with the diffuse light
in clusters of galaxies can be summarized as follows:
\begin{description}
\item{$\pmb{Luminosity}$} It shows a wide range. The intracluster
  light can represent between the 10\% and the 50\% 
of the total light of the region where it is detected. Schombert
  (1988) finds some correlation, but faint,  
between the luminosity of the cD envelope and that of the underlying
  galaxy. This correlation can hint that the process of formation of
  the brightest cluster galaxy (BCG) has some reflection in the origin
  of its envelope.  
 
\item{$\pmb{Color}$} Different authors have report various results. 
Valentijn (1983\nocite{1983A&A...118..123V}) in $B - V$ and Scheick \&
Kuhn (1994\nocite{1994ApJ...423..566S}) in $V - R$ find blueward
gradients that vary between 0.1 to 0.6 mag drop. 
Schombert (1988\nocite{1988ApJ...328..475S}) in $B - V$ doesn't find
any evidence of strong color gradients or blue envelopes
colors. Finally, Mackie (1992\nocite{1992ApJ...400...65M}) in $g - r$
reports a reddening at the   
end of the envelopes, in one case of the order of 0.15
mag. 

\item{$\pmb{Structure}$} Schombert (1988) and Mackie, Visvanathan \&
Carter (1990\nocite{1992ApJ...400...65M}) find a apparent break in the  
surface brightness profile of the underlying cD galaxies. According to
Schombert (1988), this break is found near the 
$24 V$ mag arcsec$^{-2}$ but there are no sharp changes in either
eccentricity or orientation between the galaxy and the 
envelope. However, Uson et al. (1991) and Scheick \& Kuhn (1994) don't
see such a break in their studies.   
Reinforcing the idea of common evolutive processes, Schombert (1988)
and Bernstein et al. (1995) find that the  
diffuse light, globular cluster density and galaxy density profiles
seem to have similar radial structure.
\end{description}


Basically, there are three processes that could be responsible for the
origin of the ICL. According to the first one, ICL originates from
stars lying in the outer envelopes of galaxies. Sometimes
  the extension of the diffuse light is so large (several core radius)
  that is hard to believe that these stars are gravitationally bound to any 
galaxy, and probably, they are stripped material after the interaction
  between galaxies. This could be  
the case in Cl 1613+31 (V\'{\i}lchez--G\'omez et
  al. 1994a\nocite{1994A&A...283...37V}). Also, it
  could be that the stars have born directly in the intergalactic
  medium from a cooling flow, for example (Prestwich \& Joy
  1991\nocite{1991ApJ...369L...1P}).\\
According to the second process, ICL is given by dwarf galaxies and
globular clusters. Part of the light 
  in the intergalactic medium in 
distant clusters, where it is not possible to resolve dwarf galaxies
  and globular clusters, can have this 
origin. Nevertheless, Bernstein et al. (1995) have measure in the Coma
  cluster a diffuse light apart from dwarf galaxies and globular
  clusters.\\
Finally ICL can originate from light scattered by intergalactic
dust. The existence of dust in rich clusters of galaxies 
as established by Zwicky (1959) or Hu (1992) would suggest the production of diffuse scattered light. 


There are al least three theories that try to elucidate what is the
origin and evolution of cD envelopes.
None of them offers a complete picture of the problem.
\begin{description}

\item [Stripping theory] This theory was initially proposed  by
  Gallagher \& Ostriker (1972\nocite{1972AJ.....77..288G}). According
  with this theory, the origin 
  of the envelope is on the debris due to tidal interactions between the
cluster galaxies. These stars and gas are then deposited 
in the potential well of the cluster where the BCG is located. This
  process begins after the 
cluster collapse and the envelope grows as the cluster evolves. 
The fact that different cD envelopes show different color gradients
  can be explained as the result of different  
tidal interaction histories: in some clusters the tidal interactions
  involve mainly spirals, but in others, early type galaxies are the
  source material (Schombert 1988). The main problem to this hypothesis is the
difficulty to explain the observed smoothness of the envelopes as the
  timescale to dissolve the clumps is on the order of 
the crossing time of the cluster (Scheick \& Kuhn 1994).

\item [Primordial origin theory]

This hypothesis, suggested by Merrit (1984), is similar to the previous one but, in this case, the process of removing
stars from the halos of the galaxies was carried out by the mean cluster tidal field and took place during the initial
collapse of the cluster. The BCG, due to its privileged position in relation with the potential well,
gets the residuals that make up its envelope. However, this picture is difficult to reconciliate with the fact that there
are cD's with significant peculiar velocities (Gebhardt \& Beers
1991\nocite{1991ApJ...383...72G}) as well as with the  
smoothness of the diffuse light 
either the envelope is affixed to the cD or fixed and the cD is moving through it. Moreover, if the origin of the 
diffuse light is primordial, how can we explain the observation of blue color gradients in some envelopes,
supposed little activity after virialization?

\item [Mergers] Villumsen (1982,
  1983\nocite{1983MNRAS.204..219V}\nocite{1982MNRAS.199..493V}) found that after a merger with the BCG, and under special conditions, it is possible to form
an halo similar to that present in cD galaxies since there is a
  transfer of energy to the outer part of the mergers resulting an extended envelope. Although this theory reproduces the profile observed for the envelopes, it is not
possible to accomplish for the luminosities and masses seen for the diffuse light. However, in poor clusters where 
there are cD-like galaxies without a clear envelope this mechanism can
play a more important role (Thuan \& Romanishin
  1981\nocite{1981ApJ...248..439T}; Schombert 1986). 
\end{description}

\newpage

\section{Open questions in galaxy formation}

After a decade of spectacular breakthroughs in physical cosmology, the
focus has now been shifted away from determining cosmological
parameters towards attacking the problem of galaxy formation. Consequently,
 the origin and evolution of galaxies are one of the current 
major outstanding questions of astrophysics.\\
Galaxy formation is driven by a complex set of physical processes with
very different spatial scales. Radiative cooling, star formation and
supernovae explosions act at scales less than 1 pc, but they affect
the formation of the whole galaxy (Dekel \& Silk, 1986). Active
Galactic Nuclei act on galaxy scale and thus are thought to be
play a fundamental role in regulating galaxy evolution. In addition,
large--scale cosmological processes, such as gas accretion through
cosmic filaments and galaxy mergers, control the general galaxy
assembly.  \\
Galaxies are, in their observable constituents, basically large bound
systems of stars and gas whose components interact continually with
each other by the exchange of matter and energy. The interactions that
occur between the stars and the gas, most fundamentally the 
continuing formation of new stars from the gas, cause the properties
of galaxies to evolve with time, and thus they determine many of the
properties that galaxies are presently observed to have. Star
formation cannot be understood simply in terms of the transformation
of the gas into stars in some predetermined way, however, since star
formation produces many feedback effects that control the properties
of the interstellar medium, and that thereby regulate the star 
formation process itself. A full understanding of the evolution of
galaxies therefore requires an understanding of these feedback effects
and ultimately of the dynamics of the entire galactic 
ecosystem, including the many cycles of transfer of matter and energy
that occur among the various components of the system and the
magnetohydrodynamical (MHD) instabilities that regulate
the cooling flow in hot galaxy clusters.\\
 A comprehensive review on star formation and related
 topics (molecular clouds, triggering mechanisms, energy injection by
 SNe) will be given along chapters 2 and 3 by means of numerical models.  
In the following we first give some clues on the role that
galactic magnetic field may have on the dynamics of galaxy formation
and finally summarize the most accreditated form of
energy feedback which are considered as an alternative to extra
heating in explaining the lack of ICM self--similarity (see
Sec.~\ref{cool_ICM}).
\subsection{Galactic magnetic fields}
 Magnetic fields may significantly
influence the structure and evolution of Inter Stellar Medium (ISM
hereafter). This has been proven extensively by local
magnetohydrodynamic (MHD) simulations of the ISM (Mac Low et al. 2005,
Balsara et al.2004, de Avillez \& 
Breitschwerdt 2005, Piontek \& Ostriker 2007, Hennebelle \& Inutsuka
2006) and by observations (e.g. Beck 2007, Crutcher
1999). On a larger scale, simulation are quite close to having the
resolution necessary to properly describe the magnetic field
components down to the observed scales. Cosmological MHD simulations
have been done in SPH (e.g. Dolag 1999, 2002, 2005) in the context of galaxy
cluster formation and in Eulerian-code simulations (e.g. Br\"uggen et
al. 2005, Li et al. 2006).\\ 
The main open question in galactic MHD concerns the origin and the
evolution of the magnetic field (MF) in galaxy clusters. Besides the
variety of the possible contributors, the corresponding generated MF
will be compressed and amplified by the process of structure
formation.  
There are basically three main
classes of models for the origin of cosmological MF:
\begin{itemize}
\item MF are produced ``locally'' at relatively low redshift ($z \sim
  $2--3) by galactic winds (e.g. V\"olk \& Atoyan 2000) or Active 
  Galactic Nuclei (e.g. Furlanetto \& Loeb 2001);    
\item MF seeds are produced at higher redshifts, before galaxy
  clusters form gravitationally bound systems; the origin could still
  be starburst
  galaxies and AGN but at earlier times ($z \sim$ 4--5) ot seeds
  may have a comological origin;
\item MF seeds are produced by the so--called Biermann battery effect
  (Kulsrud et al. 1997; Ryu, Kang \& Biermann 1998). In few words,
  merger shocks generated during the hierarchical structure formation
  process give rize to small thermionic electric currents which in
  turn may generate MFs. 
\end{itemize}
Supported by simulations of individual events/environments like shear
flows, shock/bubble interactions or turbolence/merging events, all
the above different models of seed MFs predict a super--adiabatic
amplification of the MF. Anyhow, none of the present simulations
include the creation of MF by all the feedback processes happening
with the Large Scale Structure (e.g. radio bubbles, AGNs, galactic
winds, etc.). Moreover, all the simulation done so far neglect
radiative losses and thus the corresponding increase in density in the
central parts of the clusters which would lead to a further MF
amplification.\\
Following the dynamics of galaxies in cosmological simulation is a
real challenge within LSS simulations. It is expected that once these
limitations will be overcame, the dynamical impact of the MF on
regions likes the cooling flows at the centre of galaxy clusters will
be significant and will eventually contribute to solve important
remarkable enigmas.  

\subsection{Active Galactic Nuclei feedback}
\label{SF:AGN_FB}
Many galaxies reveal an \textit{active nucleus}, a compact central
region from which one observes substantial radiation that is \textit{not} the
light of stars or emission from the gas heated by them. Active Nuclei
emit strongly over the whole electromagnetic spectrum, including the
radio, X--ray, and $\gamma$--ray regions where most galaxies hardly
radiate at all. The most powerful of them, the quasars, easily
outshine their host galaxies. Many have luminosities exceeding
$10^{12}$$L_{\odot}$ and are bright enough to be seen most of the way
across the observable Universe.\\
 In the standard model of AGN, cold
material close to the central Black Hole (BH) forms an accretion
disk. Dissipative processes during accretion transport matter inward
and angular momentum outward, while causing the accretion disk to
heat up. The radiation from the accretion disk excites cold atomic
material close to the BH and this radiates via emission lines. At
least some accretion disks produce \textit{jets}, twin highly
collimated and fast outflows that emerges form close to the disk.\\
The accretion process release huge amounts of energy to their
surroundings, in various forms:
\begin{itemize}
\item in \textbf{luminous AGN} (Seyfert nuclei and quasars) the output is
  mostly \textit{radiative}: this radiation can affect the environment
  through radiation pressure and radiative heating;
\item in \textbf{most accreting BH} the \textit{kinetic} energy output is as
  important as the radiative one, due to the presence of strong winds and jets.
\end{itemize}
In all cases, it is also present a significant output of energetic
particles (``cosmic rays'', relativistic neutrons and neutrinos). \\ 
In the following we recapitulate the various forms of energy injection
that we hinted above and their plausible effects on the gas surrounding
an AGN:
\begin{description}
\item[Radiation Pressure]: exerts a force on the gas via electron
  scattering, scattering and absorption on dust, photoionisation or
  scattering in atomic resonance lines;
\item[Radiative heating]: gas exposed to ionising radiation from AGN
  tends to undergo and abrupt transition from the typical CII region
  temperature $\sim 10^4$K, to a higher temperature and ionisation
  state when $P_{gas}$/$P_{rad}$ falls below some critical value. The
  main effect on the surrounding are mass evaporation from clouds,
  elimination of cool ISM phase, modification of ISM phase structure. 
\item[Kinetic energy]: when charged particles cross shock
  waves generated by jets in radio galaxies and
 by outflows in accreting BH, they are expected to lose
  considerable energy as they move out of the denser regions and of
  the radiation field. However, the exact mechanism of ``cosmic rays'' 
  acceleration in Ages is still not known.
\end{description}

As we reviewed in this section, AGN feedback effects (enormous on the
basis of energetic arguments) depend sensitively on both the form of
feedback and the detailed structure of the environment. While the
efficiency of feedback due to radiation is often small, the kinetic
energy injected by Ages tends to be trapped inside the ambient medium,
leading to a higher efficiency. For a deep description of AGN
feedback mechanisms we address the reader to Begelman 2001\nocite{begelman}.\\
Recent X-ray observations show that radio galaxies can blow long
lasting ``holes'' in the ICM, and may offset the effects of radiative
losses, providing a possible interpretation for the entropy
``excess'' with respect to the self-similar expectations and a
possible source of heating for reproducing the break at the bright end
of the luminosity function. As discussed in
Sec.~\ref{cap1:overcool}, it has been proposed that the entropy excess results
from some universal \textit{external} pre--heating process 
(AGN, population III stars, etc.) that occurred before most of the gas 
entered the dark halos. Alternatively, the hot gas in groups may be
heated \textit{internally} by Type II supernovae when the galactic
stars form. 
\subsection{Type II Supernovae energy feedback} 
\label{SF:SN_FB}
Stars more massive than
about nine times the mass of the Sun become  
internally unstable and violently explodes as they end their lives,
turning into a type II supernova or core-collapse supernova. By
dying, they create and disperse their stardust, including the elements
with masses near that of oxygen, and inject in the surrounding medium
approximately $10^{51}$ erg.
Therefore, the formation of massive stars leads to a number of
negative feedback effects including ionisation, stellar winds, and
supernovae explosion that reduce the efficiency of star formation by
destroying star--forming clouds and dispersing their gas before most of
it has been turned into stars. The physics behind these star formation
related processes is in general poorly understood. 
Feedback processes arguably have the
 largest impact on the form of the theoretical predictions for galaxy
 properties, while at the same time being among the most difficult and
 controversial phenomena to model.\\
The initial motivation for invoking SNe feedback was to reduce the
efficency of star formation in low mass haloes, in order to flatten
the slope of the faint end of the predicted galaxy luminosity
function and make it in line with observations (Cole 1991; White \&
Frenk 1991). In current simulations of hierarchical galaxy formation
models, there are two main 
physical mechanisms that can transfer energy from SNe to the
surrounding medium:   
\begin{itemize}
\item \textbf{Kinetic feedback}: the energy released by the supernova
is directly injected to surrounding gas
   via outwards velocity kicks. This causes cold gas to be
  ejected from the parent galaxy, mimicking the effect of a supernova driven
  wind (Larson 1974, Dekel \& Silk). Ejection of cold gas out of star
  forming regions has been proven to be very efficient in lowering the
  star formation rate;
\item \textbf{Thermal feedback}: the energy from the supernova heats
  the interstellar medium (katz 1992). This causes the ablation of
  cold clouds and a net reduction of the star formation efficiency. 
\end{itemize}
In the following chapters we will study in detail the type II SNe
feedback physics from the numerical point of view. The aim of the
present PhD Thesis, in fact, is to investigate different 
numerical approaches and to introduce a new, physically-based sub grid
model of star formation and SNII energy feedback. \\

\chapter{Numerical techniques for galaxy formation simulations}
\begin{quote}
\textit{An important issue in theories of galaxy formation is the relative
importance of purely gravitational processes (as N-Body effects,
clustering, etc..) and of gas-dynamical effects involving dissipation
and radiative cooling.}  White $\&$ Rees 1978.\\
\end{quote}
Structure formation refers to a fundamental problem in physical
Cosmology. When inhomogeneities in the matter field are still very
small, we can describe the evolution of perturbations using simple
linear differential equations. The complexity of physical behaviour of
fluctuations in the non-linear regime makes it impossible to study the
details exactly using analytical methods. For this reason, numerical
simulations and semi--analytical models have became standard tools for
studying galaxy formation.\\
A detailed understanding of galaxy formation in cold dark matter
scenarios remains a primary goal of modern astrophysics. While the
large scale range physics is sufficient to describe a number of
observations, using almost solely gravitational forces, the scales
relevant for galaxy formation 
requires many physical processes to be considered in addition to the
already complex interaction of nonlinear gravitational evolution and
dissipative gas dynamic. Observed cluster of galaxies are, in fact, composed by
three distinct components, dark matter, diffuse gas and stars, which
have a different physics 
behind. Codes following both DM and baryonic particles already exist but they
still have short comings, mainly for two reasons: the lack of resolution
and the complexity of the involved physics. In fact, if from one side we
need to account for star formation physics, acting on small
scales, from the other we need to consider
large--scale cosmological processes, such as
gas accretion through cosmic filaments and galaxy mergers, which
instead control the general galaxy assembly. For example, the
typical size of a cold gas cloud is about 10 -- 100 pc, while that
of a galaxy like the Milky Way is of order 20 kpc, and that of a galaxy
cluster is of 
order 1 Mpc. Following in details the whole dynamical range is too
computationally expensive, so usually one resort to simplified models of the
complex hydrodynamical and astrophysical processes working at the interstellar medium
scales (see Sec.~\ref{sf:models}). \\

\section{N-body and SPH codes}
\label{algorithm}
Given the initial conditions (which depend on the adopted cosmological
model), the purpose of any cosmological code is to follow the
evolution of density fluctuations from the linear regime (up to $z
\sim 50-100$) till the actual time ($z = 0$). It is possible to represent part of the expanding Universe as a
``box''containing a large number N of point masses interacting through
their mutual gravity. To start a numerical simulation one has first to
decide about the size of the box the evolution of which should be
simulated. For a given number of particles (which is limited 
by the power of the computer) this is always a compromise between
higher mass resolution 
(smaller boxes) and a representative volume of the Universe (larger
box). The Universe is
assumed to be homogeneous on scales larger than the box size by means
of periodical boundary conditions.\\
 Numerical N-body techniques offer a simple and effective tool for
 describing the dynamical evolution of self-gravitating systems and for
 investigating non-linear cosmological gravitational evolution. 
A number of numerical techniques are available at the present time;
they differ, for the most part, only in the way the forces on each
particle are computed. Whereas N-body simulations are applied to a
wide range of different astrophysical problems, the most appropriate
technique to use depends on the specific context. In the following we
briefly describe some of the most popular methods. For a comprehensive
review on simulation techniques for cosmological simulations see e.g Dolag
et al. 2008.
\begin{description}
\item [Direct sum (Particle-Particle, PP)] the gravitational interaction
  is computed by summing \textit{directly} the (pairwise) contributions of all
  the individual particles to calculate the Newtonian forces. Despite
  its accuracy, direct summation requires a prohibitive computational
  cost (i.e. computing time $t$ scales as $N^2$) and is thus
  of difficult application in large scale numerical simulations.
\item [Particle-Mesh (PM) methods] are the fastest scheme for
  computing the gravitational field. The forces are solved by assigning
  mass points to a regular grid and then solving Poisson's equation on
  it in the Fourier space
The use of a regular grid naturally provide periodic boundary
conditions  and allows one to use Fast Fourier Transform (FFT) methods
to recover the potential, which leads to a considerable increase in
speed ($t \sim N_g\log N_g$, where $N_g$ is the number of grid
points). This method is well suited for cosmological simulations since
allows the use of a large number of particles but the force resolution
is given by the finite spatial size of the mesh which can't be
  infinitely large. 
A substantial increase in spatial resolution can be achieved by using
a hybrid ``particle-particle-particle-mesh'' method, which solves the
the short range forces by direct summation (PP) but uses the mesh to
compute those of longer range (PM). This scheme is usually called
$P^3M$ ($PP + PM$) and achieves a good compromise between 
computational cost and accuracy of the solutions coming from PP
and PM methods. In high density regions, however, the PP method
dominates, degrading the code performance. If this is the case, the
\textit{adaptive} $P^3M$ method defines a number of sub-grids over
which the PM computation is repeated, thus limiting the use of PP summation.

\begin{figure}
\centering{
\includegraphics[width=0.5\linewidth]{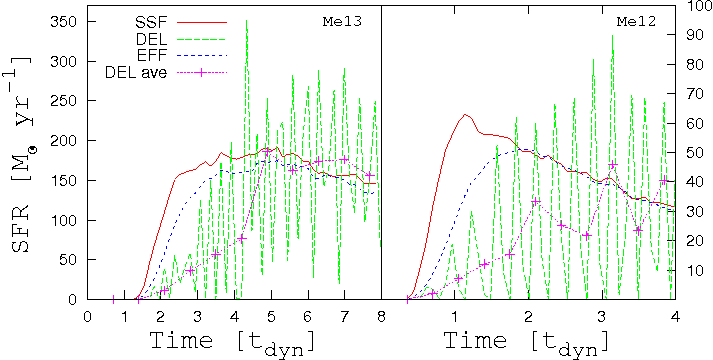}
\captionsetup{font={normalsize,sf}, width= 0.9\textwidth}
\caption{Schematic illustration of the Barnes \& Hut (1986)
oct-tree in two dimensions.  The particles are first enclosed in a square
(root node). This square is then iteratively subdivided into four
squares of half the size, until exactly one particle is left in
each final square (leaves of the tree). In the resulting tree
structure, each square can be the progenitor of up to four siblings.
Taken from Springel et al. 2001.}}
\label{fig:tree}
\end{figure}
\item [Tree codes] the computational domain is recursively partitioned
  into a hierarchy of cells containing one or more particles. Every
  cell which contains more than one particle is divided into $2^3$
  sub-cells. If any of the resulting sub-cells contains more than one
  particle, that cell is subdivided again (see Fig.~\ref{fig:tree}). 
  The essence of a tree code is the recognition that the gravitational
  potential of a distant group of particles can be well-approximated by
  a low-order multipole expansion. In a tree code, a set of particles
  is arranged thus in a hierarchical system of groups that form of a tree
  structure.
  Thus, when the force on a particular particle is computed, the force
  exerted by distant particles is treated using the  
  coarsely grained distribution contained in the higher level of the
  tree, the force being consequently approximated by their lowest multipole
  moments. In this way, the computational cost for a complete force
  evaluation can be reduced to ${\cal O}(N\log N)$ (Appel, 1985).
 In Sec.~\ref{GDT:gravity}, we will study in detail the tree
  algorithm implemented in the GADGET-2 code {Springel 2005}. The
  greatest problem with such codes is that, although they run quite
  quickly in comparison with particle-mesh methods with the same
  resolution, they do require considerable memory resources. Moreover
  implementing periodic boundary conditions is much more difficult
  than in PM codes. Periodic boundary conditions for tree codes are
  usually obtained by a numerical technique called ``Ewald summation''
  (Ewald, 1921).
\end{description}

Following the non-linear gravitational evolution of collisionless dark
matter however is not enough if one aims to obtaining precise
informations on the distribution of luminous matter. In this case,
collisionless 
dynamics must be coupled to gas dynamics. \\
The fundamental equations regulating the evolution of a collisional
fluid (gas) are, in comoving coordinates:
\begin{equation}
\label{hydr:masscont}
\frac{\partial}{\partial t} (\frac{\rho_{b}}{\bar{\rho_{b}}}) + \frac{1}{a}
\boldsymbol{ \nabla } \cdot \boldsymbol{v_{b}} = 0  
\end{equation}
\begin{equation}
\frac{\partial \boldsymbol{v_{b}}}{\partial t} + \frac{1}{a}\boldsymbol{v_{b}}
 \cdot \boldsymbol{ \nabla}\boldsymbol{v_{b}} + H\boldsymbol{v_{b}} =
    -\frac{1}{a\rho_{b}}\boldsymbol{ \nabla}p + \boldsymbol{g} \label{CIAO2}
\end{equation}
where $\rho_{b}$ is the density of baryons, $\bar{\rho_{b}}$ is
the average density of baryons, $\boldsymbol{v_{b}}$ is the
peculiar velocity, $p$ the pressure, $a$ the scale factor e
$\boldsymbol{g}$ the gravitational field.\\
An equation for energy or entropy has to be added to the above
equations. Outside shocks, the equations for thermal energy  and entropy
evolution can be written respectively as:
\begin{equation}  
\frac{\partial U}{\partial t} + \frac{1}{a}\boldsymbol{v_{b}} \cdot
\boldsymbol{ \nabla}U = \frac{p}{a\rho_{b}}\boldsymbol{ \nabla} \cdot
\boldsymbol{v_{b}} + \frac{1}{\rho_{b}}(\Gamma - \Lambda) 
\end{equation}
\begin{equation}
\frac{\partial S}{\partial t} + \frac{1}{a}\boldsymbol{v_{b}} \cdot
\boldsymbol{ \nabla}S = \frac{1}{p}(\Gamma - \Lambda)  
\end{equation}
where $\Gamma$ and $\Lambda$ are respectively the heating and the
cooling rates. \\
For a perfect gas with specific heat $c_{v}$
\begin{equation}
U = \frac{p}{[(c_{v} -1)\rho_{b}]} \qquad \textrm{thermal energy per
  unit mass}
\end{equation}
\begin{equation}
S = (c_{v} -1)^{-1}ln(p\rho_{b}^{-c_{v}})\qquad \textrm{entropy per
  unit mass}
\end{equation}
To close the system, an equation of state relating the thermodynamical
variables needs to be specified. This is:
\begin{equation}
P = (\gamma - 1 )\rho U
\end{equation}
where $\gamma = \frac{5}{3}$ is the adiabatic exponent for a
monoatomic ideal gas.\\
Due to the non-linear nature of the equations of hydrodynamics, exact
solutions of these equations are rarely found. \\
There are two principle algorithms in common use to follow the 
hydrodynamics of gas in an expanding universe: particle based, Lagrangian 
schemes, which employ a technique called smoothed particle hydrodynamics 
(SPH, Lucy, Gingol \& Monaghan 1977, Monaghan 1992; Couchman, Thomas
\& Pearce 1995;  
Gnedin 1995; Springel \& Hernquist 2003; 
Wadsley, Stadel \& Quinn 2004) and grid based, 
Eulerian schemes (e.g. Ryu et~al. 1993; Cen \& Ostriker 1999). We here
focus on Lagrangian scheme, for a description of grid-based methods
see e.g. Dolag et al. 2008.\\
In \textit{smoothed particle hydrodynamics} (SPH) one typically
represents the fluid as a set of particles in the same way as in the
N-body gravitational case, described before in this
section. SPH is thus an extension of N-body techniques, making simple
its integration in pre-existing cosmological codes.\\
 The basic idea characterising SPH is to discretise
the fluid by mass elements: fluid variables, such as
baryonic density, temperature and velocity, are evolved using
particle with constant mass. Since SPH is a Lagrangian method,
equation~\ref{hydr:masscont} (the mass continuity equation) can be
neglected. The thermodynamics variables $q$ are estimated by a
\textit{smoothing kernel} $W$ from which we obtain:
\begin{equation}
q(\boldsymbol{x}) = \sum_{i=1}^{N} q_{i}W(\boldsymbol{x}-
\boldsymbol{x}_{i}, h) \label{r} 
\end{equation}
where $q_i$ is the studied quantity and ${\vec x_{i}}$ its position
while $h$ is the smoothing length. This method consists in assuming
that the quantities are ``smoothed'' into a region of finite dimension,
centred on each particle, rather than being punctiform. The summation
~\ref{r} extends only over particles lying within a cutoff radius
proportional to $h$. Usually $h$ is taken to be proportional to
$\rho_b^{-1/3}$: in this way, we are sure to include in the above
summation at least $N_{SPH}$= 30-40 particles. In this case, the \textit{number}
of particles inside smoothing radius $h$ is nearly equal to $N_{SPH}$.
 Alternatively, adaptive smoothing
lengths $h_i$ of each particle are defined such that their kernel
volumes contain a constant mass for the estimated density,
i.e. smoothing lengths and densities obey the equation:
\begin{equation}
\frac{4\pi}{3} h_i^3\rho_i = N_{\small SPH} \bar m
\end{equation}
where $N_{\small SPH}$ is the typical number of smoothing neighbours
and $\bar m$ is an average particle mass. \\
SPH has some special advantages over the traditional grid-based
methods, the most important among which is its \textit{adaptive}
Lagrangian nature: SPH follows the motion of the fluid and thus is not
affected by the arbitrariness of the particle distribution. In
practical terms, this means that most of the computing effort is
directed towards places where most of the particles are. An important
disadvantage of SPH is that it has to rely on an
\textit{artificial} viscosity for supplying the necessary entropy
injections in shocks (see Sec.~\ref{GDT:hydro}). On the contrary,
Mesh-codes offer superior resolving power for hydrodynamical
shocks. However, static meshes are only poorly suited for the high
dynamic range encountered in cosmology.\\
\section{GADGET-2}
\label{GDT}
GADGET (\textbf{GA}laxies with \textbf{D}ark matter and \textbf{G}as
int\textbf{E}rac\textbf{T}, Springel et al. 2001) is a freely
available$\footnote{http://www.mpa-garching.mpg.de/gadget/}$, parallel
code which uses the standardised MPI communication interface. \\
GADGET computes gravitational forces with a hierarchical tree
algorithm (optionally in combination with a particle-mesh scheme for
long-range gravitational forces) and represents fluids by means of
smoothed particle hydrodynamics (SPH). The code can be used both for
studies of isolated self-gravitating systems, and for cosmological
simulations. Both the force computation and the 
time stepping of GADGET are fully adaptive.\\
A complete review of its features is given in a number of articles by
its author, Volker Springel, so we will refer to them for all the
details about the code. In the following, we will give a brief summary
of the physics implemented in GADGET.
\subsection{Collisionless dynamics}
\label{GDT:gravity}
In an expanding background Universe, DM and stars are commonly treated
as collisionless fluids and are described by the
collisionless Boltzmann (or Vlasov) equation (CBE), which in comoving
coordinates becomes: 
\begin{equation}
   \frac{\partial f}{\partial t} + \frac{{\vec p}}{m a^2} {\vec \nabla} f
 - m {\vec \nabla}\Phi \frac{\partial f}{\partial {\vec p}} = 0 \qquad
\qquad \textrm{CBE}
\end{equation}
where $a$ is the \textit{scale} factor describing the spatial extent of
the Universe, $f({\vec x}, {\vec p}, t)$ is the phase--space
distribution function of the fluid and ${\vec \nabla}\Phi$ is the
solution to the Poisson equation (Eq.~\ref{eq:poi}). The fluid is thus
represented in the phase space by a finite number N of tracer
particles which are integrated along the characteristic curves of the
CBE.\\
The gravitational attraction between particles is expressed in terms
of the Newton's equations of motion. In order to suppress large-angle
scattering in two-body collisions, GADGET introduces a softening into
the gravitational potential at small separations, which effectively
introduces a lower spatial resolution cut-off.\\
As we mentioned before, the gravitational force is described by a tree
code (see Sec.~\ref{algorithm}), where particles are arranged in a
hierarchy of groups. The authors employed the Barnes and Hut (BH,1986)
tree construction: the computational domain is hierarchically
partitioned into a sequence of cubes, where each cube contains eight
siblings, each with half the side-length of the parent cube. The tree
is constructed such that each node (cube) contains one particle or is
the progenitor to further nodes (see Fig.~\ref{fig:tree}), in which
case the node carries the monopole and quadrupole moments of all the
particles that lie inside the cube.\\
The force computation then proceeds by ``walking'' the tree and
summing up the appropriate force contributions form tree nodes. In the
standard BH tree walk, the multipole expansion of a node of size $l$
is carried out if the distance $r$ of the point of reference to the
centre-of-mass of the cell satisfies:
\be r>\frac{l}{\theta} \, ,
\label{eqopen}
\ee 
where $\theta$ is an accuracy
parameter which plays the role of a angle of view. If a node fulfils
the criterion (\ref{eqopen}), the tree walk along this branch can be
terminated, otherwise it is `opened', and the walk is continued with
all its siblings. For smaller values of 
the opening angle, the forces will in general become more accurate, 
but also more costly to compute. The geometric
criterion ~\ref{eqopen} can therefore cause high errors in the force
computation and does not guarantee a sufficient accuracy, so that
other conditions must be fulfilled. We leave a complete discussion to
Springel 2001, 2005.
\subsection{Hydrodynamics}
\label{GDT:hydro}
The gas, defined as perfect and non-viscous, is described by the SPH
method. \\
The computation of the hydrodynamic force and the rate of change of
internal energy proceeds in two phases.  In the first phase, new
smoothing lengths $h_i$ are determined for the {\em active} particles
(these are the ones that need a force update at the current time-step),
 and for each of them, the neighbouring particles $j$ inside
their respective smoothing radii are found. The Lagrangian nature of
SPH arises when this number of neighbours is kept either exactly, or
at least roughly, constant.  This is achieved by varying the smoothing
length $h_i$ of each particle accordingly. \\
Having found the neighbours, the SPH density is estimated as
\begin{equation}
\rho_i = \sum_{j=1}^{N}m_{j} W_{ij} (|r_i - r_j|, h_i)
\end{equation}
where $W_{ij}$ is the SPH smoothing kernel (see below).\\
In the second phase, the
actual forces are computed.
\subsubsection{Entropic formulation of SPH}
\label{GDTentro}
 In the conventional implementation of SPH (Monaghan 1992), the
 hydrodynamical equations are implemented using the thermal
 energy as an independent
variable: while total energy is manifestly conserved, the same is not
 true for the entropy. Moreover the thermal energy version of SPH
 leads to substantial overcooling in halos that are resolved with up
 to a few thousand particles. Under low resolution conditions, the
 cooling rates of the gas are substantially overestimated, behaviour
 that may in part explain why cosmological SPH simulations predict
 more gas cooling than is expected based on simple analytic models
 (Pearce et al. 2001; Benson et al. 2001). Moreover, in hierarchical
 scenarios of structure formation, larger systems will inherit this
 problem if their progenitors suffered from overcooling. A further
 problem of the standard implementation of SPH is that point-like energy
 injection in the ISM, as those due to SNe feedback, can lead to
 unphysical negative temperatures. Hernquist (1993) has shown that,
 using the ``energy'' implementation of SPH, it is not possible to conserve
 simultaneously energy \textit{and} entropy.\\
GADGET employs a new formulation of SPH which
conserves energy and entropy (when appropriate) by construction, in
which the dynamical equations are given as a function of entropy
(Springel \& Herquist, 2002). In this new implementation, the
equations describing the evolution of a fluid are expressed as:
\begin{description}
\item [Discretisation of the equation for entropy evolution]
\begin{equation}
 \frac{dA_{i}}{dt} = -\frac{\gamma
   -1}{\rho_{i}^{\gamma}}L(\rho_{i},u_{i}) + \frac{1}{2}\frac{\gamma
   -1}{\rho_{i}^{\gamma-1}}\sum_{j=1}^{N}m_{j}\Pi_{ij}
 \boldsymbol{v_{ij}} \nabla_{i}\bar{W_{ij}}  \label{u} 
\end{equation}
where $A(s) = u \frac{\gamma - 1}{\rho^{\gamma - 1}}$ is the entropic
function, $\gamma$ is the adiabatic 
index, $L(\rho_{i},u_{i})$ is the emissivity per unit volume
(introduced to describe external sinks or sources of 
energy due to radiative cooling or heating), $\Pi_{ij}$  denotes the
artificial viscosity, $W_{ij}$ is the smoothing kernel (see below) and
$v_{ij}$  is the relative velocity between fluid elements $i$ and $j$.
The artificial viscosity term (Steinmetz, 1996) is introduced to take
into account numerical effects due to gas shocks.
\item [Equation of motion] 
\begin{equation}  
\frac{d\boldsymbol{v_{i}}}{dt} = -\sum_{j=1}^{N}m_{j}[ f_{i} \frac{P_{i}}{\rho_{i}^{2}}\nabla_{i} W_{ij}(h_{i}) + f_{i}\frac{P_{j}}{\rho_{j}^{2}}\nabla_{i} W_{ij}(h_{j})]  \label{t}
\end{equation}
where the $f_{i}$ are defined by:  
\begin{equation}
f_{i} = (1 + \frac{h_{i}}{3\rho_{i}}\frac{\partial \rho_{i}}{\partial h_{i}})  
\end{equation} 
As mentioned before, an artificial viscosity term is incorporated for
handling of shocks. The evolution of the corresponding ``artificial''
viscous force is given by:
\begin{equation}
\frac{d\boldsymbol{v_{i}}}{dt}\mid_{visc.} =
-\sum_{j=1}^{N}m_{j}\Pi_{ij} \nabla_{i}\bar{W_{ij}}   
\end{equation}
This term added to acceleration ~\ref{t}, allows to completely
describe the shocks. The resulting
dissipation of kinetic energy is exactly balanced by a corresponding
increase in thermal energy if the entropy is evolved according to
equation (\ref{u}).
\end{description}   
The smoothing kernel used in GADGET is of the form:
\begin{equation}
W(r, h)= \frac{8}{\pi h^{3}}\left\{ \begin{array}{ll}
1 - 6(\frac{r}{h})^{2} + 6 (\frac{r}{h})^{3}, & 0\leqslant\frac{r}{h}
\leqslant \frac{1}{2} \\
2( 1 - \frac{r}{h})^{3}, & \frac{1}{2}< \frac{r}{h} \leqslant 1,\\
0, & \frac{r}{h} > 1 .
\end{array} \right. 
\label{kernel} 
\end{equation}
The entropic formulation of SPH solves satisfactorily (but not
completely) some problems of the standard implementation. It provides
one of best technique for modelling the process of point-like energy
injection, which is relevant for certain feedback algorithms. It also
reduces artificial overcooling in poorly resolved halos by
a factor of $\sim$ 2 with respect to the standard implementation. \\
A more complete physical treatment of radiative heating/cooling
and of energy feedback in GADGET is given in the following section.

\subsection{Cooling and star formation}
\label{GDT:coolSF}

A proper modelling of the formation and evolution of the luminous
component of galaxies is known to be an hard task. In order to
properly describe the large number of physical processes involved in
the \textit{interstellar medium} (ISM) dynamics, we
need to couple the already complex interaction of non-linear
gravitational evolution and dissipative gas dynamics to a treatment
for radiative heating and cooling  and for the coagulation and
fragmentation of molecular cold clouds.\\
\begin{description}
\item [Radiative cooling]
On the contrary with dark matter, gas can cool via a number of
mechanisms (see, for example, the discussion  
in Kauffmann \& White 1994). The relative importance of the various 
mechanisms depends upon the conditions in the universe at the time 
the gas is cooling and the temperature of the gas. The principal
cooling channels are:   
(i) Inverse Compton scattering of CMB photons by electrons in the hot halo gas
(independent from density and temperature); (ii)Bremsstrahlung
radiation as electrons are accelerated in an ionised  
plasma; (iii) decay/excitation of rotational or vibrational energy levels in 
molecular hydrogen through collisions.\\
Gas undergoing radiative cooling plays a key role in
the process of star formation, since it rules the collapse of gas in
the dark matter potential wells.\\
In GADGET, the equation regulating the energy loss per unit mass
due to radiative cooling is given by:
\begin{equation}  
(\frac{du}{dt})_{cool} = -\frac{\Lambda_{net}(\rho,T)}{\rho}
\end{equation}
The authors have computed the cooling function $\Lambda_{net}(\rho,T)$
from the radiative processes appropriate for a primordial plasma of
hydrogen and helium, neglecting variations in the metalicity
of the gas as described in Katz et al. (1996). The suffix
``net'' denotes 
that the cooling function accounts for the presence of an
external UV background field, which the authors take to be a modified
Haardt \& Madau (1996) spectrum.  In high density regions, where
radiative cooling is strong, the cooling times can become shorter
than the dynamical free-fall timescale. In this case, the gas cools so
quickly that dynamical processes are unable to adjust the pressure
distribution: pressure support is lost and the gas undergoes a rapid
collapse, causing the so-called ``cooling
catastrophe'' and the subsequent formation of cold dense knots of
gas. 
\end{description}
White \& Rees (1978) proposed a solution to this problem: energy
released from stars in the course of their evolution would act as a
 negative feedback on the gas, thus limiting its cooling rate and
 associated star formation.\\
In the ISM, the gas exists in a wide range of density and
temperature states (e.g. McKee \& Ostriker 1977) and may thus be
considered as a \textit{multiphase} system resulting from the
interplay of processes such as gravity, hydrodynamics, star formation,
shocking by SNe and stellar winds, magnetic fields , cosmic rays, 
chemical enrichment and dust formation (Field 1965; Ferrara et
al. 1995; Efstathiou 2000). Each process introduces its own length and
time scales which often differs by orders of magnitude form those of
the galaxy as a whole. As a result, a realistic description of the
galactic environment is one of the most outstanding challenges in
modern theoretical physics.

\subsubsection{Effective hybrid multiphase Star Formation model}
\label{GDT:sf_eff}
Springel \& Hernquist (2003, SH03) named their treatment of star formation and
feedback a ``hybrid'' method because it does not attempt to explicity
resolve the spatial multiphase structure of the ISM on small scales,
but rather makes the assumption that important aspects of the global
dynamical behaviour of the ISM can be characterised by an effective
``sub-resolution'' model that uses only spatially averaged properties
to describe the medium. The ISM is depicted as a fluid containing
condensed cold clouds in pressure
equilibrium with an ambient hot gas. The
basic processes driving mass exchanges between the hot and the cold
phase are: star formation, cloud evaporation due to SNe, cloud growth
due to cooling and finally processes leading to the development of
galactic winds.\\ 

There are two variants of the SH03 multi-phase model: one ``explicit''
where the 
treatment of mass and energy exchange among the different gas phases
is followed explicitly and  one
``effective'', where the ISM sub-grid
properties are derived each time-step from equilibrium solutions of equations
describing the self-regulated SF. In the following we describe the
effective SH03 model, being the one used in this PhD work.
Within the effective star formation model, SH03
assumes that conditions for 
self-regulated star formation always hold without following the
response of the star formation and feedback physics to variations of
the thermodynamics state of the gas particle from one timestep to the
following. The star forming particle thus passes from an equilibrium
solution to the next one in a very short time, given by an external decay time
parameter. The model
provides the mass fraction in cold clouds 
$\emph{x}$ = ${\rho_{c}}/{\rho}$ and the density threshold $\rho_{\rm thr}$
for the onset of star formation.  A detailed account of the star formation
algorithm is presented in SH03, below we summarise the main
features of the model. In the following description,  $\rho_h$ denotes the local density of the hot
ambient gas, $\rho_c$ is the density of cold clouds, $\rho_\star$ 
the density of stars, and $\rho=\rho_h+\rho_c$ is the total gas
density.  The average thermal energy per unit
volume of the gas is then written as $\epsilon = \rho_h u_h +
\rho_c u_c$, where $u_h$ and $u_c$ are the energy per unit mass of the
hot and cold components, respectively. 

The density threshold is fixed by requiring that the specific internal energy
of a gas particle having a temperature of $10^4$ K is equal to the effective
specific internal energy given by the model. In our runs, this corresponds to
a gas numerical density $n_{\rm thr} \sim 0.25$ cm$^{-3}$.  Once a gas
particle density is above the star formation threshold it becomes eligible to
form stars. In this PhD thesis work, we used the values
  suggested by SH03 for all parameters.

Star formation converts cold clouds into stars on a characteristic timescale
$t_{\star}$. A mass fraction $\beta$=0.1 of these stars are short-lived and
instantly die as SNe. Calling $\rho_c$ the average density of the cold gas,
this process is described by the equation:

\be
\frac{d\rho_{\star}}{dt} = (1 - \beta)\frac{\rho_{c}}{t_{\star}}\qquad \qquad
\textrm{\small{STAR FORMATION}}
\ee
The star formation rate (SFR) of a gas particle of mass $\emph{m}$ is given by: 

\be
\label{sfr_eff}
\dot{M_{\star}} = (1 - \beta) x \frac{m}{t_{\star}}
\ee 
$x$ being the fraction of gas in cold clouds.

In addition to returning gas to the hot phase of the ISM, SNe also release
energy $\epsilon_{\rm SN}$. Given the adopted Salpeter IMF, EFF expects an
average return of energy per unit mass in formed stars of $\epsilon_{\rm SN} =
4\cdot10^{48}$ erg $M_{\odot}^{-1}$.  This energy is injected into the hot
phase, whose density and internal energy are $\rho_h$ and $u_h$. The heating
rate due to SNe is then:

\be
\label{SNe_rate}
\frac{d}{dt}(\rho_{h}u_{h})\mid_{\rm SN} = \beta u_{\rm SN}
\frac{\rho_{c}}{t_{\star}}
\qquad \qquad
\textrm{\small{SNe ENERGY FEEDBACK}}
\ee
where $u_{\rm SN}\equiv(1 - \beta)\beta^{-1}\epsilon_{\rm SN}$ is the specific
energy released by one SN.\\

Exploding
supernovae, besides directly heating the ambient hot phase, evaporate
the cold clouds residing inside the supernova-generated bubbles ,
essentially by thermal conduction. This can be described by:  
\begin{equation}
\frac{d\rho_{c}}{dt}\mid_{EV} = A\beta\frac{\rho_{c}}{t_{*}}
\qquad \qquad
\textrm{\small{EVAPORATION}}
\end{equation}
The efficiency
$A$ of the evaporation process is expected to be a function of the
local environment. In this work, $A$ is et to 1000.\\
The process by which cold clouds come into existence and grow is
driven by radiative cooling (described above in this section). SH03
assume that a thermal instability operates in the region of
coexistence between the cold and the hot phases, leading to mass
exchanges among them. This mass flux is described by:
\begin{equation}
\frac{d\rho_{c}}{dt}\mid_{TI} = -\frac{\rho_{h}}{dt}\mid_{TI} =
\frac{1}{u_{h}-u_{c}}\Lambda_{net}(\rho_{h},u_{h})\qquad
\textrm{\small{CLOUD GROWTH due to COOLING}}
\end{equation}
Temperatures of and total volumes occupied by the hot and cold phases
are assumed to remain constant during cloud growth.\\

Quantitatively, the evolving rates of hot and cold masses can be written as:
\begin{equation}
\frac{d\rho_{c}}{dt} = -\frac{\rho_{c}}{t_{*}} - A\beta\frac{\rho_{c}}{t_{*}} + \frac{1 - f}{u_{h}-u_{c}}\Lambda_{net}(\rho_{h},u_{h}) \label{v}
\end{equation}
\begin{equation}
\frac{d\rho_{h}}{dt} = \beta\frac{\rho_{c}}{t_{*}} + A\beta\frac{\rho_{c}}{t_{*}} - \frac{1 - f}{u_{h}-u_{c}}\Lambda_{net}(\rho_{h},u_{h})\label{z}
\end{equation}
The energy budget of the gas is thus given by:
\begin{equation}
\frac{d}{dt}(\rho_{h}u_{h} + \rho_{c}u_{c}) = - \Lambda_{net}(\rho_{h},u_{h}) + \beta\frac{\rho_{c}}{t_{*}}u_{SN} - (1 - \beta)\frac{\rho_{c}}{t_{*}}u_{c} \label{w} 
\end{equation}
where $u_{SN} = (1 - \beta)/ \beta \epsilon_{SN}$, otherwise defined
as a ``supernova temperature" $T_{\rm SN} = 2\mu u_{\rm SN}/(3K) \sim
10^8 {\rm K}$.\\
The above equation can be splitted into two separate relations for the
energy of the hot and cold phase:
\begin{equation}
\frac{d}{dt}(\rho_{c}u_{c}) = - \frac{\rho_{c}}{t_{*}}u_{c} - A\beta\frac{\rho_{c}}{t_{*}}u_{c} + \frac{(1 - f)u_{c}}{u_{h}-u_{c}}\Lambda_{net}
\end{equation}
\begin{equation}
\frac{d}{dt}(\rho_{h}u_{h}) =  \beta\frac{\rho_{c}}{t_{*}}(u_{SN} + u_{c}) + A\beta\frac{\rho_{c}}{t_{*}}u_{c} - \frac{u_{h}- fu_{c}}{u_{h}-u_{c}} \Lambda_{net} 
\end{equation}
The equation for the energy of the hot phase after some calculus
becomes:
 \begin{equation}
\rho_{h}\frac{d}{dt}(u_{h}) =  \beta\frac{\rho_{c}}{t_{*}}(u_{SN} + u_{c} - u_{h}) - A\beta\frac{\rho_{c}}{t_{*}}(u_{h} - u_{c}) - f\Lambda_{net} 
\end{equation} 
which gives the evolving equation for the temperature of the hot phase.
It is easy to show from the previous equation that the internal
energy of the hot phase is set by: 
\be
\label{EFF_uh}
u_{h} = \frac{u_{\rm SN}}{1+A} + u_{c}
\ee
where $u_{c} = 2\cdot10^{44}$ erg $M_{\odot}^{-1}$ is the internal energy of
the cold phase (assumed to be at 1000 K) and $A = (\rho/\rho_{\rm thr})^{-0.8}
\cdot \beta$ is the efficiency of evaporation. Deviations from this
temperature decay on a timescale $\tau_h = t_{\star}\rho_h/(\beta(A +
1)\rho_c)$, therefore, provided SF is rapid compared to adiabatic heating and
radiative cooling, the temperature of the hot phase will be
maintained at the value set by Eq.~\ref{EFF_uh},
\textit{independent} of $t_{\star}$ and on the thermodynamics of the
gas as given by its SPH evolution.\\
A further interesting features of SH03 model is that it leads to a
self-regulated cycle of SF, where the growth of clouds is balanced by
their evaporation owing to SNe feedback.

In order to help simplify the model, in the effective version of the
SH03 model, the build-up of the 
stellar component is not described `smoothly', as in the explicit one, but
probabilistically.  Given a time-step $\Delta t$, a new star particle
of mass $m_{\star}$ = $m_{0}/N_{G}$ is spawned once a random number drawn
uniformly from the interval [0,1] falls below:
\be
\label{prob_EFF}
p_{\star} = \frac{m}{m_{\star}} \bigg\{1- exp\bigg[-\frac{(1-\beta) x
	\delta T}{t_{\star}}\bigg] \bigg\}
\ee
Here $m_{0}$ is the initial gas mass of the SPH particle and $N_{G}$ (set to
$N_G$=4 in our runs) gives the number of ``generations'' of stars that each
particle may form. Each star particle always gets a mass
$m_0/N_G$. This approach is quite mandatory  to avoid both an uncontrolled
multiplication of the number of particles and an artificial dynamical
coupling of the gas with the stars. \\

\subsubsection{Galactic winds}
\label{GDT:wind}
Galactic winds and outflows may be a fundamental mechanism in
regulating star formation on galactic scales (Scannapieco, Ferrara \&
Broadhurst 2000). In fact, winds can reheat and trasport collapsed
material from the center of a galaxy back to its extended
dark matter halo and therefore can help to reduce the overall
cosmic star formation rate to a level consistent with observational
constraints. Moreover, galactic winds may help in solving the
``overcooling'' problem, provided that they can expel sufficient
quantities of gas from low-mass galaxies.
For these reasons, the SH03 model has been extended
to account for galactic winds driven by star formation. \\
The \textit{wind} model can be summarized as follows. First, the disk
mass-loss rate that goes into wind, $\dot M_w$ is proportional to the
star formation rate itself $\dot M_{\rm w}=\eta \dot M_\star$, where
$\eta$ is the wind efficiency. This assumption is based on
observational evidences (Martin 1999) and tells nothing about the
ability of this gas to escape from the potential well. Second, it is
assumed that the wind carries a fixed fraction $\chi$ of the supernova
energy. The wind velocity is thus obtained by equating the kinetic
energy in the wind with the energy input by supernovae,
 \begin{equation} 
\frac{1}{2}\dot{M_{w}}v_{w}^{2} = \chi\epsilon_{SN}\dot{M_{*}} 
\end{equation}
and it is equal to:
\begin{equation} 
v_{w} = \sqrt \frac{2\beta\chi u_{SN}}{\eta(1-\beta)}
\label{GDT:eq:wind}
\end{equation}
All the star forming gas particles can enter the wind if chosen by the
probabilistic criterion given by Eq.~\ref{prob_EFF}. The net effect of the wind
is to remove cold gas from the potential well, thus halting star formation.\\
When wind particles depart from the inner parts of the star-forming
regions, their kinetic energy is thermalized inside the
region itself due to hydrodynamical interactions. To let the wind
particles to freely escape from star-forming dense regions, a
``decoupling'' mechanism of the wind particles from the hydrodynamical
interactions is provided (see Sec.~\ref{DVS08} for the Della Vecchia
\& Schaye 2008 ``coupled'' wind model).

\section{Modelling star formation and SNe energy feedback}
\label{sfsimu}

 Including astrophysical processes in
simulations such as radiative cooling of 
the gas, star formation and energy feedback from SuperNovae (SNe) (see
Dolag et al. 2008\nocite{dolag2008}), is a hard task for several
reasons. The 
physics of star formation is complex and currently not understood in full
detail. Moreover, the ISM besides being multi-phase is also multi-scale; the
dynamical range needed to simultaneously resolve the 
formation of cosmic structures and the formation of stars is huge,
since the former process happens on Mpc scales at densities of 10$^{-7}$
atoms ${\rm cm^{-3}}$ and the latter on parsec 
scales at typical densities greater than 100 atoms ${\rm
  cm^{-3}}$. This calls for resolving seven orders of magnitude in 
length  
scale, and about ten orders of magnitude in density. This can only get
worse if we aim at following directly the process of star formation.
As a consequence, numerical simulations commonly use simple ``sub-grid''
prescriptions (Yepes et al. 1997\nocite{1997MNRAS.284..235Y}, SH03,
Marri \& White 2003\nocite{2003MNRAS.345..561M}, Scannapieco et 
al. 2006\nocite{2006MNRAS.371.1125S}, Booth et
al. 2006\nocite{2007MNRAS.376.1588B}) for example to 
reduce the required dynamical range and account for star formation and SN
energy feedback, thus hiding the complexity of the star
formation process. Early attempts to introduce stellar feedback into
simulations (Baron \& White 1987\nocite{1987ApJ...322..585B}, Cen \&
Ostriker 1992\nocite{1992ApJ...393...22C}, Katz
1992\nocite{Katz92}, Navarro \& White
1993\nocite{1993MNRAS.265..271N}) showed that if SN energy is 
deposited as thermal energy onto the star-forming gas particle, it is
quickly 
dissipated through radiative cooling before it has any relevant effects. In
fact, the characteristic timescale of radiative cooling at the typical
density 
of star-forming regions is far shorter then the free-fall gravitational
timescale (Katz 1992).  Several solutions have been proposed to solve this
over-cooling problem.  One possibility consists in depositing SNe energy in
the form of kinetic energy instead of thermal energy (see
e.g.Navarro \& White 1993 and SH03).  A different 
solution simply consists in turning off radiative
cooling of the star-forming gas particle when SN energy is given to it
(Gerritsen \& Icke 1997\nocite{Ger97}, Thacker \& Couchman
2000\nocite{ThacCouch00}, Governato et al. 2007\nocite{Gov07}).
 
Such prescriptions are motivated on physical grounds, and usually tested
against observations of local spiral galaxies; for instance, the
Kennicutt law  $\rho_{\star}
\propto \rho^n$ 
(Kennicutt et al. 1998\nocite{1998ApJ...498..541K}), which relates the
surface densities of gas 
and star formation rate (SFR), must be recovered in simulations with star
formation prescriptions. Since the '50, ${\rm n} = 1.5$ has been
widely used as a gross estimate of the rate of star formation in very
different environments.\\

\label{sf:models}
\subsection{Simple Star Formation Models}
\label{sf:models:simple}
The advantage of self-consistent numerical simulations over
semi-analytic models is that of being able to 
provide a consistent description of the evolution of the structures
in the non-linear regime. As a consequence, physical processes related to
the evolution of the dissipative component can be included and modelled
upon a more physical basis.\\
Since it is difficult to arrive at a realistic star formation algorithm,
 several authors include simple schemes to transform the cold dense
gas into stars. Among them, the most widely known is Katz (1992,1996) star
formation and feedback algorithm, which we briefly summarise in the
following paragraph.

\subsubsection{A phenomenological conversion of gas into stars (Katz
  1992,1996)}
\label{katz} 
Katz (1996, KA96)  use an easily parameterised scheme that
incorporates most of the known gross properties of star formation
without invoking a detailed mechanism and thus ignoring 
the multiphase physics of the ISM.\\ 
In this scheme the criteria for a gas particle to
become eligible for star formation are:
\begin{itemize}
\item ({\rm i}) It is part of a convergent flow, i.e. the gas forming
  a star should be in a 
in regions that are in a state of collapse.
\item ({\rm ii}) It is Jeans-unstable, i.e. $\frac{h_i}{c_i} >
  \frac{1}{\sqrt{4 \pi G \rho_i}}$ where $h_i$ is the particle
  smoothing length and $\rho_i$ is the local particle density.
\item ({\rm iii}) Its density is greater than $n_{min} = 0.1$ cm$^{-3}$
\item ({\rm iv}) Its temperature is colder than 3$\cdot 10^4$ $\rm K$ 
\end{itemize} 
Once all the above conditions are satisfied, the rate at which gas is
converted into stars (i.e. the SFR) is given by:
\be
\label{sfr_eff_ssf}
\frac{d{\rho_{\star}}}{dt} =  c_{\star} \frac{\rho}{t_{\rm dyn}}
\ee
where $c_{*}$ is a constant star formation efficiency parameter
and $t_{\rm dyn}$ is the dynamical time of the particle: 
\be
\label{tdyn}
t_{\rm dyn} = \bigg(\frac{1}{4\pi G \rho} \bigg)^{1/2} 
\ee 
Note that the star formation efficiency simply depends on the gas
density.\\
As in the SH03 model, also Katz (1996) uses a probabilistic method for
forming new star particles (see Eq.~\ref{prob_EFF}).
The masses of newly created star particle and its parent gas particle
are further adjusted for stellar mass loss form SNe. The author, in
fact, supposes that a mass fraction
$\beta$ of these stars are short-lived and instantly die as SNe. In addition to
adjusting the mass of the particles, SNe also add heat: their energy
is directly injected in the 
star-forming gas particle.  As soon as a gas particle enters the star
formation regime, its energy is updated taking into account the incremental
amount of energy, $\Delta E$, introduced by the SNe in the time-step $dt$:

\be
\Delta E =  \epsilon_{\rm SN} \beta  \dot{M_{\star}} dt
\ee
where $\epsilon_{\rm SN}$ is the SN energy per solar mass returned to
the gas.\\

\subsection{Multiphase Star Formation Models}
\label{sf:models:multi}
As we hinted at along previous sections, the problem of star formation
can be dealt with algorithms which explicity modelize the multi-phase
structure of the ISM. In the following,
we summarise some multi-phase SF models existing in the literature which are
particularly relevant to the star formation and feedback model
 implementation (${\rm MUPPI}$) which we will present in
 Chapter~\ref{MUPPI_chap} (Giovalli et. al 2008, in preparation).

\subsubsection{The \textit{promotion} scheme for SN energy feedback
  (Scannapieco et al. 2006)}
Scannapieco et al., 2006 (SC06), have implemented a new scheme for chemical
enrichment and energy feedback by SNe in the GADGET-2 code (see
Sec.~\ref{GDT} for a detailed description) but they do not use its original
effective model for star formation and feedback (SH03, described in
Sec.~\ref{GDT:sf_eff}). In brief, gas particles become eligible to
form stars if they are denser then a physical threshold density
$\rho_{th} = 7\cdot 10^{-26}$$g$ $cm^{-3}$ and lie in a convergent flow
($\nabla \vec{v} <  0$). For these particles, they assume a star formation
rate per unit volume equal to:
\be
\dot{\rho_{\star}} = c \frac{\rho}{t_{dyn}}
\ee
where $c=0.1$ is the star formation efficiency and $t_{dyn} =
1/\sqrt{4 \pi G \rho}$ is the dynamical time. New stellar particles
are created according to the stochastic approach of SH03. 
It is beyond the scope of this
work to go into detail on the chemical enrichment model (see
\cite{2005MNRAS.364..552S}, SC05). We are indeed interested on the
treatment 
of the multiphase structure of gas particles and on the
\textit{promotion} scheme for depositing the energy
feedback by SNe. \\

One of the main innovation brought by SC06 SPH multiphase scheme
(similar to that presented by \cite{2003MNRAS.345..561M}), relates to
the selection of neighbours. Here gas particles with very different
thermodynamic variables do not see each other as neighbours,
i.e. the model decouples phases with very different entropies.
This allow hot, diffuse gas to coexist with cold, dense gas without
introducing any ad-hoc characteristic scales, thus leading to the
natural formation of a multiphase structure in the gas composition. \\
For what concerns the feedback model, the SC06 scheme  
 resorts to a an explicit segregation of the
gas surronding a star particle with exploding SNe into a
cold dense phase and a diffuse phase. The
\textit{cold} phase consists of gas with T $< 8 \cdot 10^4 K$ and
$\rho > 0.1\rho_{th}$, while the rest of the gas is considered to
belong to the \textit{hot} phase, even if much of it may be
cold.\\
The amount of energy injected by each newly formed solar mass
of stars is supposed to come both from SNII and SNIa
explosions. At each timestep, the number of exploding SNII is
calculated by adopting an IMF and by assuming that stars with mass
greater than 8$M_{\odot}$ end their lives as SNII after $\sim
10^{6}$yr. For estimating the SNIa number, the authors adopted an
observationally motivated relative rate with respect to SNII (see
SC05). Each SN explosion is assumed to release $10^{51}$erg in the
ISM.\\
The SN energy produced by a single star
particle is then distributed to neighbouring gas particle taking into
explicit account if they belong to the
hot or to the cold phases. Neighbours residing in different phases  
receive a different fraction of the SN energy: a fraction $\epsilon_h$ is
instantaneously thermalized in the  
hot phase, a fraction $\epsilon_r$ is immediately radiated away by the
cold phase while the remaining fraction of the SN energy $\epsilon_c$
is accumulated by the cold phase in a \textit{reservoir}
$E_{res}$. The value of $\epsilon_c$ has been adjusted to $0.5$ so as to
reproduce the observations of star-forming systems.  
Once the accumulated energy becomes high enough to modify the thermodynamic
properties of a cold particle in such a way that its new properties
will resemble that of the local hot environment, they \textit{promote}
the cold particle, 
dumping its energy reservoir into its internal energy. In
practise, the authors require the promoted particle to have an
entropic function at least as high as the mean of those of its hot
neighbours ($A_{Avg}^{hot}$), i.e. in terms of specific energy:
\be
E_{res} > E_{prom} = \frac{\gamma}{\gamma - 1}m_i \Big [
  A^{hot}_{Avg}(\rho^{hot}_{Avg})^{\gamma - 1} - A_{i}\rho_i^{\gamma -
  1} \Big ]
\ee
This leads to
\be
A_{new} > A_{Avg}^{hot}
\ee
where the new value for entropy $A_{new}$ is calculated assuming that
the energy of the particle after promotion will be its actual energy
plus the reservoir $E_{res}$. 
The promotion scheme for distributing energy feedback by SNe ensures
that the receiving gas particles will remain hot at least as long as
nearby material, since, after being promoted, gas particles will have
thermodynamic properties matching those of its local hot
environment.\\
The authors tested their SNe feedback model on a set of idealized
simulation of the formation of disc galaxies. They found their scheme
to be efficient in regulating star formation by reheating cold gas and
generating winds. Furthermore, their scheme can reproduce the Kennicut
relation if the star formation efficiency parameter is fixed to $c = 0.01$.
\subsubsection{Galactic outflows with kinetic supenovae feedback
  (Dalla Vecchia $\&$ Schaye 2008)}
\label{DVS08}
\nocite{2008MNRAS.387.1431D} 
In Sec.~\ref{GDT:wind} we described the recipe for galactic winds
implemented in the 
kinetic feedback model by SH03. Here wind particles are
stochastically selected from \textit{all} the star forming particles in the
simulation and thus are not constrained to be neighbours of
newly-formed stars, i.e. the feedback is not local. Another important
aspect of the SH03 recipe is that 
hydrodynamical forces on the wind particles are temporarly switched
off (for 50 Myr), so that wind particles can ``freely'' travel outside the disc
without being influenced by the pressure forces exterted by and on the
surronding gas. Dalla Vecchia \& Schaye 2008 (DS08 hereafter) 
have implemented a variation of the SH03 recipe in which winds are local
and not decoupled hydrodynamically. These authors showed that (ram) pressure
forces exterted by expanding SN bubbles have indeed a quite dramatic
effect on the ISM structure and on 
the galactic winds themselves. 
In the following, we give a brief
summary on the novelty brought by DS08 to the SH03 kinetic feedback 
model.\\ 
As in Aguirre et al. (2001) and SH03, the kinetic feedback is entirely
specified by the initial mass loading $\dot{M_w} = \eta \dot{M_{*}}$
(with $\eta =2$) and the wind velocity $v_w = 600$ $km$
$s^{-1}$. The wind carries a fixed fraction of the SN energy, $f_{w} =
\eta v^{2}_w/2\epsilon_{SN}$. Using the ``top-heavy'' Chabrier (2003)
initial mass function (IMF), the value of 
$\epsilon_{SN}$ is $  \sim 1.8 \cdot 10^{49}$ $\rm erg M_{\odot}^{-1}$ 
    i.e. the wind carries about the forty per cent
of the overall supernova energy input while the remainder is assumed
to be lost radiatively. 
Once a star particle reaches an age $t_{SN}$=$3\times 10^{7}$ yr, it
is allowed to inject kinetic energy into its surrounding by kicking one
or more neighbours.  
On the contrary with the non-locality of SH03 stochastic approach, DS08 select new wind particles
still stochastically but considering only the neighbouring gas particles of each newly spawned star particle.
Thus, a probability of becoming a wind particle
\be
P_w = \eta \frac{m_{*}}{\sum_{i=1}^{N_{ngb}}m_{g,i}}
\ee 
is associated with each neighbour, where $m_{*}$ is the mass of the star particle, $m_{g,i}$ the mass of gas particle $i$, $N_{ngb}$ the number of neighbours ($N_{ngb}$= 48 in their runs), and the sum is over all gas particle that are not already wind particles.\\
The authors tested their kinetic feedback scheme on simulations of isolated disc galaxies of masses $10^{10}$ (dwarf) and $10^{12}$ (massive) $M_{\odot}h^{-1}$.  They found that their prescription causes a strong reduction of the SFR and has a dramatic impact on the morphology of the galaxies. The differences between the predictions for the DS08 scheme (\textit{coupled} wind particles) and the SH03 model (\textit{decoupled} wind particles) are remarkable:
\begin{itemize}
 \item Decoupled wind has almost no effect on the morphology of the disc;
 \item Coupled wind model slightly increases the size of the gas disk, while the decoupled one continuously shrinks the disk. 
  \item Coupled winds generate a large bipolar outflow from the dwarf galaxy and a galactic fountain in the massive galaxy while the decoupled wind produces isotropic outflows in both cases.
\end{itemize}

\subsubsection{Molecular cloud regulated star formation: a model for
  cloud formation using \textit{sticky} particles (Booth et al. 2006)} 
   
Motivated by the fact it is not (yet) possible to reasonably resolve the Jeans scale for molecular clouds in galaxy simulations and that the formation of cold clouds is crucial for star formation, Booth et al. 2006 (BTO06) introduced a new prescription for SF and feedback. BTO06 scheme aims to mimic the interstellar multiphase medium using a different approach than the ones we describe before. Instead of adopting implicit differential equations for regulating the interactions between phases (SH03, SC06, DV08) or explicity decoupling SN heated gas from the surrounding (Thacker \& Couchman, 2000),
BTO06 decouples the cold molecular phase from the hot phase. This is achieved by following the ambient phase with a hydrodynamical simulation, whereas the cold phase uses a statistical model that encapsulates the physics relevant to the formation and evolution of cold clouds. The model works as follows. When a thermally instable gas particle is identified i.e. when $\rho > \rho_{th}\equiv$ 1 $cm^{-3}$ and $T < T_{h} \equiv 10^{5}$K, the gas particle begins to be converted to molecular clouds. The authors used a temperature threshold to prevent gas heated by SNe in dense regions from collapsing straight to the cold phase. If the gas particle does not fulfil the threshold for entering the \textit{cold} state, than it goes to ordinary radiative cooling. When the amount of mass in the molecular phase in a particle reaches the resolution limit of the simulation a separate \textit{sticky} particle is created, representing a Giant Molecular Cloud (GMC) containing many sub-resolution clouds.\\
Molecular clouds interact only gravitationally with the other phases in the simulation and are governed by a different set of rules than the ambient medium. Clouds are assumed to be approximately spherical objects that grow as mass is added to them as
\be
r_c = r_{ref}(\frac{M_c}{M_{ref}})^{\alpha_c} = 36(\frac{M_c}{10^5M_{\odot}})^{\alpha_c}
\ee
where $M_{ref}$, $r_{ref}$ and $\alpha_c$ are set by comparison with
observations of molecular clouds in nearby galaxy M33 (Wilson \&
Scoville 1990\nocite{1990ApJ...363..435W}). Lower and upper mass
bounds are respectively 100 $M_{\odot}$ (Monaco 2004\nocite{MO04}) and
$10^{6}M_{\odot}$ (i.e $M_{sf}$, beyond which clouds are converted
into stars). Thus, each cloud contains an entire mass spectrum of
``multiple``clouds statistically inside, where the evolution of the
mass function is based on the Smoluchowsky
(1916\nocite{1916ZPhy...17..557S}) equation of kinetic aggregation. This coagulation is driven by the \textit{coagulation kernel} $k(m_1, m_2)$ that represents the formation rate of clouds with mass m = $m_1$ + $m_2$
\be
K = < v_{app} \Sigma >_{v}
\ee
where $v_{app}$ is the relative velocity of the clouds, $\Sigma$ is the collision cross section. The product of the approach velocity and the collision cross section is averaged over the distribution of relative velocities.  To model the cooling of sub-resolution molecular molecular clouds the authors assumed that if:
\begin{description}
 \item[$v_{app}$ $>$ $v_{stick}$] the clouds merge
 \item[$v_{app}$ $<$ $v_{stick}$] the clouds collide, bouncing back with relative velocity a fraction $\eta$ of the initial approach velocity
 \end{description}
The free parameter $v_{stick}$ represents the maximum relative cloud velocity for mergers, which has been calibrated to 7 $kms^{-1}$ in order to reproduce the observed Schmidt law.\\ 
As soon as a cloud with mass $m$ $>$ $M_{sf}$ forms, it is assumed to
collapse on a very short timescale (10 Myr, see B{\"o}hringer et al. 2002\nocite{2002ApJ...566..302M}). A fraction $\epsilon_* $ of its mass is converted into stars imposing a star formation rate given by the Schmidt law while the rest is in part disrupted by SN feedback, photodisintegration and winds and part fragmented into the smallest allowed clouds. Only energy feedback form type II SNe is included and the mechanism by which it is implemented is briefly explained as follows. Each star of mass $M > 8 M_{\odot}$ releases $10^{51}$ ergs in thermal energy when undergoes a SN event. 
The SN explosion has been modelled using the Sedov solution for blast
waves (Sedov 1959\nocite{1959sdmm.book.....S}). According to this solution, if at time $t =0$ we release an amount of energy $E_{b}$, then after a time $t$ the blast wave will have a radius
\be
 r_b = \frac{E_b}{\rho_h}^{1/5} t ^{2/5}
\ee
The gas particle that inherits the SN energy $E_b$ is chosen randomly among the neighbours, thus each SN explosion can be approximated as an injection of energy at a single point in space. Moreover, the BSO06 scheme does not transfer all the SN energy to gas particles at each timestep. For each ambient gas particle, the authors calculate the \textit{porosity}
\be
Q = \frac{V_B}{V_A}
\ee
where $V_B = 4\pi/3$ $ \Sigma (r_b)^3$ is the total volume of SN bubbles in this particle and $V_A = 4\pi/3$ $h ^3$ is the volume associated with gas particles. If $Q$ is above a given limit, the ambient phase is heated, otherwise the available SN energy is carried over the next timestep. This procedure is to ensure that the ambient phase is heated \textit{only} when hot SN bubbles make up a significant fraction of the volume.\\
 Another important ingredient of any model which aims to describe the ISM structure is the thermal conduction, which has two primary effects on the surroundings:
\begin{itemize}
 \item smooth out the temperature and the density profiles inside SN remnants
  \item evaporate the cold clouds   
\end{itemize}
In BSO06 feedback model, those effects are implemented as follows. In order to smooth the temperature after a SN explosion, the temperature of the interior of the blast wave $T_b$ is assumed to be constant and equal to the mean temperature of the blast
\be
T_b \rtimes (r_b)^{-3}(n_b)^{-1}(E_{b})
\ee
where $n_b$ is the mean density inside the blast. The authors treated the evaporation of clouds taking into specific account that the evaporation is different if the clouds reside in the ambient medium or inside a SN bubble. Then, for a cloud of mass $M_{cloud}$ residing in a porous medium, the cloud mass loss rate is described by:
\be
\dot{M_{cloud}} = -Q\dot{M}_{bubble} - (1 - Q)\dot{M}_{ambient} 
\ee
where $\dot{M}_{bubble}$ and $\dot{M}_{ambient}$ give respectively the rate of mass loss for a cloud inside a SN bubble and situated in the ambient medium. 
 With the explicit assumption that the temperature $T_{b}$ remains constant over a timestep, BSO06 derived the following expression 
for the evaporation of a cloud with mass $M_i$ to the mass $M_f$, over a time $\Delta t$:
\be
M_f = (M_{i}^{1 - \alpha_c} - \lambda \Delta t)^{1/(1 - \alpha_c)}
\ee
The authors tested the effectiveness of this thermal conduction
implementation and proved that their model efficiently destroys smaller
clouds, even if its effects on larger cloud is far less dramatic.\\ 
 The authors tested the sticky particle star formation model to three
 different types of simulation: a simple one zone model (i.e. a static
 periodic box with no mass outflows) for calibrating the free
 parameters, the rotating collapse of a gas and dark matter sphere,
 and a model of an isolated Milky Way-like galaxy. They found that
 many observed properties of disk galaxies are reproduced well,
 including the molecular cloud mass spectrum., the Schmidt law, the
 molecular fraction as a function of radius and finally the appearance
 of a galactic fountain. 
\subsubsection{The Stinson 2006 star formation scheme with delayed
  radiative cooling }
\label{ST06}
While in SH03 SN energy sets the internal energy of the
star-forming gas particle (see Eq.~\ref{EFF_uh}), in KA96 SN feedback energy is
directly added to the particle internal energy. Since star formation occurs
in dense regions, where typical timescales for radiative cooling are very
short, this injected energy tends to be quickly radiated away.  As a
consequence, thermal feedback does not have a large effect on the large-scale
hydrodynamics.

As mentioned before, one possible method to alleviate the
overcooling problem is to turn off radiative cooling for some time when the SN
energy is deposited (Gerritsen \& Icke 1997, Thacker \& Couchman 2000,
Stinson et al. 2006). The motivation behind this approach is to mimic the
adiabatic expansion phase of SN remnants, maintaining high temperature and
pressure in the gas surrounding the explosion so as to allow it to expand and
sweep the surrounding cold gas.

This has two main effects on the gas particles surrounding the SN explosion:
first, besides cooling star formation is also quenched; second, the gas
absorbs energy coming from SNe and adiabatically expands, flowing toward less
dense regions; this further delays star formation even when the quenching
phase is over.

Stinson et al. 2006 (ST06) have implemented a revised version of the delayed
radiative cooling scheme on top of the KA96 model. Here after a gas
particle spawns a star particle, ST06 first  
estimate the available SN energy as: 
\be \Delta E_{\rm SN} = \epsilon_{\rm SN}\cdot \beta M_{*} 
\ee 
where $M_{*}$ is the mass of the star particle that
has just been spawned, $\epsilon_{\rm SN}$ is the specific
energy released by one SN and $\beta$ is the fraction of stars which
instantaneously die as SNe. ST06 then estimate the radius
$r_{\rm lim}$ of the current blast-wave as the radius of the sphere containing
a factor $\eta$=0.3 of the mass of stars that go SNe in one time-step ($M_{\rm
SN}$) and having an average density $\rho$

\be r_{\rm lim} = \left(\frac{3\eta M_{\rm SN}}{4\pi
  \rho}\right)^{1/3} \; . 
\ee 
The gas particles within the sphere of radius $r_{\rm lim}$ from the
star-forming particle receive a fraction $\Delta E_{\rm SN,i}$ of the total
feedback energy $\Delta E_{\rm SN}$.  This fraction is computed as

\be
\Delta E_{\rm SN,i} = \frac{m_i\cdot W(|r_i - r_*|,r_{\rm lim})\Delta
  E_{\rm SN}}{\rho_i} 
\ee 
where $W$ is the SPH kernel. Note that the authors use the same SPH
kernel used in the hydrodynamics \nocite{2002MNRAS.333..649S}, so farther
particles get an energy 
fraction which is significantly lower than the ones nearer to the newly formed
star. Immediately after deposition, ST06 disable cooling for the selected gas
particles for a time $\tau$=30 Myr..\\
In Governato et al. 2007, the ST06 "adiabatic" feedback has been applied to
cosmological simulations and has been proved to be even more effective
than kinetic feedback in producing an extended disk component.

\subsubsection{The role of runaway stars in modelling stellar feedback
  (Ceverino \& Klypin)}
As a last example, we describe the Ceverino \& Klypin (arXiv 2007)
approach for dealing with supernovae feedback. Instead of artificially
stopping cooling when SN energy is deposited (see previous section) or
using a sub-resolution model of the multiphase ISM (see Sec.~\ref{sf:models:multi}),
 these authors 
\textit{resolve} that multiphase ISM and, moreover, add to its dynamics the spreading effect of runaway stars.\\
According to Ceverino \& Klypin 2007 (hereafter CK07), the
main problems of current simulations of galaxy 
formation derive from the lack of the necessary resolution and the use
of too simplified models of the complex hydrodynamic processes in the multiphase ISM.
Therefore, they developed a model for SNe
explosions and stellar winds without the
ad-hoc assumptions typically used on stellar feedback and run it on
parsec scale (35 {\rm pc}) simulations of a piece of a galactic
disk. In the following we outline the basic
ingredients of CK07 model.\\
The thermodynamical state of the gas depends on two competing
processes: heating from stellar feedback and cooling from radiative
processes. In the first law of thermodynamics, these two competing
processes appear as source and sink terms,  

\begin{equation}
\frac{du}{dt} + p \nabla \cdot \mathbf{v} =   \Gamma - \Lambda
\label{eq:1}
\end{equation}

CK07 assume that the feedback heating has an effect on the ISM
\textit{only} when it dominates over radiative cooling, i.e. when
$\Gamma \geq \Lambda$. \\
The heating rate from stellar feedback in a volume element $V$ is
modelled as the rate of energy losses from a set of single stellar
populations present in that volume:
\begin{equation}
\Gamma  =  \frac{1}{V} \sum_i M_i \Gamma'(t_i),
\end{equation}
where $M_i$ and $t_i$ are the mass and the age of each single stellar
population. \\
The radiative cooling is followed with the model described in Kravtsov
(2003), which accounts for temperatures in the range
$10^2$ K $< T < 10^9$ K. The authors point out how a model of cooling
below $10^4$ ${\rm K}$ (given by molecular and atomic transitions, and
metallicity dependent) is fundamental for the efficiency of the
stellar feedback.\\ 
The biggest novelty brought by CK07 is anyway the treatment of runaway
stars.  These authors believe that is crucial to understand
where and how the energy from massive stars is released back to the
ISM. 10-30\% of massive stars is in fact found in the field, away
from typical star-forming regions (Gies 1987; Stone 1991). Their
velocity dispersion is about 30~km~s$^{-1}$, but can reach peculiar
velocity as large as 200~km~s$^{-1}$ (Hoogerwerf et al. 2000).
The importance of runaway stars is that they may help to spread the effect of 
stellar feedback, giving the fact that they usually explode as
supernovae in low-density regions. This is an
effect found in nature (Stone 1991), which enhances the feedback.\\
Once these effects are implemented into cosmological
simulations, galaxy formation proceeds more realistically. For
example, CK07 do not have the overcooling problem and produce
cold clouds, hot super-bubble and galactic chimneys. However, this
picture is only reproduced if the resolution is high enough to
resolve the physical conditions of densities and temperatures of
molecular clouds: CK07 cosmological simulations reach a resolution of
35 pc, which is 10 times better than the typical resolution found in
previous cosmological simulations (Sommer-Larsen et al. 2003; Abadi
et al. 2003; Robertson et al. 2004; Brook et al. 2004; Okamoto et
al. 2005; Governato et al. 2007). We note however that such an
extreme resolution is obtained at the cost of using a small (10 Mpc)
cosmological volume. Most important, from the published preprint, it
is not clear if they succeeded in bringing the simulation up to
redshift $z = 0$, since the lowest reported results refer to $z = 3.4$.

\section{The Monaco 2004 model: an example of semi-analytical approach}
\label{MO04}


As described in the previous sections, different approaches have been
developed over the years in order to link the observed properties of
the luminous component of galaxies to those of the dark matter haloes
in which they reside. Among these, semi-analytical models (SAMs) of galaxy
formation have developed into a flexible and widely used tool that
allows a fast exploration of the parameter space and an efficient
investigation of the impact of varying the physical assumptions.\\
While the processes dominant in the large-scale range are succefully
addressed in numerical simulations of whole galaxies, the intermediate
and small scales (the ones where star formation and feedback act) are
"sub-grid'' physics and are treated with simple heuristic models (see
previous sections). In this framework (semi-)analytic work can give a
very useful contribution in selecting the physical processes that are
most likely to contribute to feedback.\\

In what follows, we present the (semi-)analytic ISM, SF and SNe
feedback model by P. Monaco 2004 (MO04 hereafter). For a detailed description of the
physics implemented see MO04.
 We believe that this model provides a starting point for constructing
 a realistic grid of feedback solutions to be used in 
galaxy formation codes. \\
\subsection{Star Formation and feedback by steps}
The ISM is depicted as a two phase medium in 
pressure equilibrium and it is described by four components: cold and hot
phases, stars and the external halo. 
Star formation related events are assumed to take place through a
chain of processes.
In the following we describe these steps.  All distances are
given in \pc, masses in \msun, times in \yr, temperatures in \K, gas
densities in \cmt, average densities in \dens, surface densities in
\surf, energies in $10^{51}$ erg, mechanical luminosities in
$10^{38}$ erg s$^{-1}$, mass flows in \msunyr, energy flows in
$10^{51}$ erg/yr.  Pressures are divided by the Boltzmann constant
$k$ and given in \K\ \cmt. The subfix $\rm c$ denotes \textit{cold}
phase quantities, while subfix $\rm h$ denotes \textit{hot}
phase ones.
\subsubsection{\small{PRESSURE EQUILIBRIUM}}
The densities $n$ and filling factors $f$ of the
two phases are determined by pressure equilibrium, i.e. 
\begin{equation} 
n_h T_h = n_c T_c
\label{eq:presseq}
\end{equation}
where $T$ is the temperature, 
from which we obtain the filling factor of the cold phase as:
\begin{equation} 
f_c = \frac{1}{1+\frac{F_{\rm h}}{1-F_{\rm h}}\frac{\mu_c}{\mu_h}
\frac{T_h}{T_c}}\, , \label{eq:filcold}
\end{equation}
where $\mu$ is the mean molecular weight and $F_h = M_h/(M_c + M_h)$
is the fraction of hot gas.
\subsubsection{\small{FRAGMENTATION OF THE COLD PHASE}}
 It is assumed that the self-gravitating
clouds are reasonably stable (in the sense that they are not
significantly reshuffled by turbulence) within one or two dynamical
times.\\ 
The cooled or infalled gas fragments into clouds (subfix $\rm cl$) with a mass spectrum
assumed to be a power law: 
\begin{equation}
N_{cl}(m_{cl})dm_{cl} = N_0 (m_{cl}/1{\rm M}_\odot)^{-\alpha_{cl}} dm_{cl} 
\label{eq:cldistr} 
\end{equation}
where $N_0$ is a normalisation constant and $\alpha_{cl}$ is the
typical radius of the cloud in {\rm pc}.\\
The mass function of clouds is truncated both at low ($m_l$) and high ($m_u$)
masses. At the high mass end, the mass function is truncated by
gravitational collapse, because clouds that form stars are quickly
destroyed. At low masses, clouds are easily destroyed, for example by
photo-evaporation.   

\subsubsection{\small{CRITICAL MASS FOR CLOUDS}}
The collapse is triggered in clouds larger than the Jeans mass. In the
absence of turbulence and magnetic fields, the $M_{thre}$ for collapse
is fixed by the Bonnor-Ebert criterion (Ebert 1955; Bonnor 1956) and
depends on external pressure $P_{ext}$. If the external pressure is
fixed to the thermal one (no kinetic pressure support), the criterion
reduces to the classical Jeans mass:
\begin{eqnarray} m_{\rm J} &\simeq& 1.18 \frac{c_{\rm
      s,c}^4}{\sqrt{G^3\mu_{shape}^3 P_{\rm ext}}} 
\label{eq:lombert} 
\\ &\simeq& 20.3\, T_c^{3/2} n_c^{-1/2} \mu_c^{-2} \mu_{shape}^{-3/2}
\ {\rm M}_\odot 
\nonumber 
\end{eqnarray}
The parameter $\mu_{shape}$ is a free parameter and takes into account
non-spherical collapsing cloud. In the following we adopt
$\mu_{shape}$ = 1.
\subsubsection{\small{COAGULATION OF COLD CLOUDS}}
\label{MO:cloucoag}
Clouds larger than the Jeans mass are continually created by kinetic
aggregation (coagulation) of smaller clouds. The approach followed for
treating cloud coagulation is that of Cavaliere, Colafrancesco \& Menci
(1991, 1992). In brief, the coagulation of clouds is driven by a
kernel:
\begin{equation}
 K = \bar{\rho}_{\rm c} \left \langle\left \langle\Sigma_{\rm coag}
v_{\rm ap} \right\rangle_{\rm v} \right\rangle_{\rm m}
\label{eq:kernel}
\end{equation}
where $v_{\rm ap}$ is the approach velocity while $\Sigma_{\rm coag}$
 (Saslaw 1985) is the cross-section for the coagulation of two clouds
 (denoted by 1 and 2):
\begin{equation}
 \Sigma_{\rm coag} = \pi (a_1+a_2)^2\left(1+2G\frac{(m_1+m_2)}
{a_1+a_2} \frac{1}{v_{\rm ap}^2}\right)
\label{eq:crosssect} 
\end{equation}
where $a$ is the radius of the cloud.
The first term on the right-hand side accounts for
geometric interaction while the second one for resonant
interactions. In the following we will neglect the resonant interaction
contributions.\\
The time-scale for coagulation is set from the Smoluchowski equation
for kinetic aggregations (see Cavaliere et al. 1992):
\begin{equation}
t_{\rm coag} = \left(\frac{4\pi}{3}\right)^{2/3} \frac{1}{\pi}
\bar{\rho}_{\rm c}^{-1/3} (\frac{\rho_{\rm c}}{\bar{\rho}_{\rm c}})^{2/3}
\frac{m_{\rm J}^{1/3}}{\langle v_{\rm ap} \rangle
} \label{eq:coag1}
\end{equation} 
while the time at disposal for cloud accretion (i.e. time necessary to
a Jeans mass cloud to be destroyed) is related to the dynamical time:

\begin{equation}
t_{dyn} = \sqrt{\frac{3\pi}{32G\rho_{\rm c}}} \simeq 5.15 \times 10^7
(\mu_c n_c)^{-1/2} \ {\rm yr}
\label{eq:freefall}
\end{equation}
The author allows mass accretion to persist for 2 $t_{dyn}$ because soon
after star formation is triggered, early feedback from young stars
destroys the cloud in less than 1 dynamical time. Thus, the upper mass
cut-off is fixed as:
\begin{equation}
 m_{\rm u} = m_{\rm J} 
\left(1+\frac{2 t_{dyn}}{3t_{\rm coag}}\right)^3
\label{eq:upper}
\end{equation}

The mass of the typical collapsing cloud is then:

\begin{equation}
 M_{cc}= \frac{\int_{m_{\rm J}}^{m_{\rm u}} m_{cl} N_{cl}(m_{cl}) dm_{cl}}
{\int_{m_{\rm J}}^{m_{\rm u}} N_{cl}(m_{cl}) dm_{cl}}
\label{eq:mcc}
\end{equation}

and the fraction of cold gas presently available for star formation
is:
\begin{equation}
f_{coll} = \frac {\int_{m_{\rm J}}^{m_{\rm u}} m\cl N\cl(m\cl) dm\cl} 
{\bar{\rho}_{\rm c}}
\label{eq:fcoll} 
\end{equation}

The total number of collapsing clouds is:

\begin{equation}
n_{cc} =  f_{coll} \frac{M_{cold}}{M_{cc}} 
\label{eq:ncc} 
\end{equation}
Coagulation of small clouds is a physically motivated mechanism to
explain the growth of cold clouds. In fact, cooling alone is not going
to produce clouds larger than the Jeans mass.

\subsubsection{\small{STAR FORMATION AND EARLY FEEDBACK}}
Early feedback from massive stars can destroy the collapsed cloud
before the bulk of type II SNe has exploded. In fact, due to the
dense environment, the net effect
of the first SNe is that of collapsing again the diffuse material
heated up by SuperNova Remnants (SNRs). After a few SNe, most gas is
re-collapsed into cold clouds with a low filling factor: the diffuse
phase has such a low density that SNRs can now emerge from the cloud
before cooling.\\
 The author assumes that each SN releases
$10^{51}$${\rm erg}$ in the ISM and that all this energy is used for
 driving the SN Super Bubble (SB, see next paragraph). A fraction $f_{evap}$ is assumed to
evaporate to the temperature $T_{evap}$, while the rest (amounting
to a fraction $1- f_{\star}- f_{evap}$) is re-collapsed into cold clouds.
The reference values are $f_{evap}$= 0.1 and $T_{evap}$= $10^6\ K$.

At last, the contribution of a single collapsing cloud to the global
star formation rate is:
\begin{equation}
\dot{m}_{\rm sf} = f_{\star} \frac{M_{cc}}{t_{dyn}}
\label{eq:clsfr}
\end{equation}

\subsubsection{\small{SUPER BUBBLES}}
Given the fact that SNRs soon percolate into a single SB, it is
assumed that all the SNe exploding in a cloud will drive a single SB
into the ISM (see Mac Low \& McCray 1988).\\
Stars are formed with a given Initial Mass Function (hereafter IMF)
that must be specified.  For the model the only information needed is
the mass of stars formed for each supernova, $M_{\star.SN}$.  One
SN is associated to each $>8\ {\rm M}_\odot$ star.
 the number of SNe
that explode in a collapsing cloud and the resulting rate are:
\begin{equation}
N_{sn} = f_{\star} \frac{M_{cc}}{M_{\star,SN}}
\label{eq:nsn}
\end{equation}

\begin{equation}
R_{sn} = f_{\star}\frac{M_{cc}}{t_{life}M_{\star,SN}}
\label{eq:rsn}
\end{equation}
where $t_{life}$ is the difference between the lifetime of a
8$M_{\odot}$ star and that of the largest star, i.e. $t_{life}$ is the
lifetime of an 8-$M_{\odot}$ star.\\
The mechanical luminosity of the SB is then $L_{\rm mech} = L_{38} \times
10^{38}\ {\rm erg\ s}^{-1}$, where:

\begin{equation}
L_{38} = \frac{10^{13}R_{sn} E_{51}}{1\ {\rm yr}}
\label{eq:l38}
\end{equation}
In a two-phase medium, the SB expands into the more pervasive hot
phase. The cold clouds will pierce the blast, but this will reform
after the cold cloud has been overtaken (McKee \& Ostriker 1977; Mac
Low \& McCray 1988; Ostriker \& McKee 1988).\\
The evolution of the SB is described following the model of Weaver
et. al (1977).\\
There are mainly two stages:
\begin{description}
\item [1$^{st}$ STAGE: ADIABATIC] the hot phase is shock-heated by the
  blast. \\
This phase is called ``Sedov expansion''. Briefly, the gas begins to be
  swept up, while not cooling efficiently due to a too high
  temperature. As the SB expands, the temperature falls. Ions begins
  to acquire bound electrons and radiative cooling starts.\\
This stage ends when post shocked mass elements begin to cool, with
cooling time sets by: 
\begin{equation}
 t_{\rm cool}=3kT/n_h \Lambda(T)
\label{eq:tcool}
\end{equation}
\item [2$^{nd}$ STAGE: SNOWPLOUGH] the swept mass collapses into a
  thin cold shell.\\
The emission of the SNRs becomes dominated by line emission from the
outer part of the shell (rather than by the interior). This shell acts
like a snowplough, making the swept ISM collapse into it. The interior gas
is mainly cold and thus provides no pressure to drive the
expansion. Anyway, some of the hot gas will remain inside the bubble,
pushing the snowplough with its pressure. In the Pressure Driven
Snowplough (PDS) phase, the amount of ISM swept by the SB that is collapsed
into the shell is estimated as a fraction of the internal material for
which $t_{cool} < {\rm t}$.\\
The explosion of the last SN marks the exhaustion of energy injection
into the SB, so the evolution after this event should follow that of a
SNR. \\
There are some processes such the thermo-evaporation of the clouds that
are difficult to be modelled analytically and thus have been neglected. 
\end{description} 

\subsubsection{\small{THE FATE OF SBs}}
\label{fateSN}
SBs can end in two ways: ($\small{1}$) being confined by external
pressure; ($\small{2}$) blowing out of the system.\\
The \textbf{\small{CONFINEMENT}} case takes place at $t_{conf}$ when
the shock speed is equal to the external, thermal one. After
confinement, the blast dissolves in the hot phase or the shell
fragments because of instabilities. Then the hot phase mixes with the
interior hot gas. As long as $t_{conf}<\ t_{life}$ ( where $t_{life}$
is the lifetime of a 8$M_{\odot}$ star particle) many other SNe
explode after $t_{conf}$: this will allow the creation of secondary
bubbles.\\
The \textbf{\small{BLOWOUT}} case takes place when the SB overtakes
the vertical scale-height $H_{eff}$ of the system, defined as (MacLow
\& McCray 1988; Koo \& McKee 1992):
\begin{equation}
H_{eff} \equiv \frac{1}{\rho_0}\int_0^\infty\rho_{\rm h}(z)dz
\label{eq:heff}
\end{equation}
where $z$ is the vertical direction (that for which $H_{eff}$ is minimal)
and $\rho_0=\rho_{\rm h}(z=0)$.  
The SB does not stop immediately after blow-out, as the rarefaction
wave that follows blow-out must have time to reach the blast travelling
in the midplane. At blow-out, part of the hot interior gas of the SB
escapes to the halo while the swept gas receives momentum from the
blast in the radial direction. The blowing-out gas is that
contained in a double cone with opening angle $\theta$,
calculated from $cos \theta = H_{eff}/R_{fin}$ where $R_{fin}$ is the
final radius of the SB. Neglecting the gas residing outside the cone
(shaded areas in Fig. 2.2), the fraction
of gas that is blown out is 

\begin{equation}
f_{bo} =  \left\{\begin{array}{lll} \frac{1}{2}[H_{eff}/R_{fin} -
    (H_{eff}/R_{fin})^3] & {\rm if} 
& R_{fin} > H_{eff} \\ 0 & {\rm if} & R_{fin} < H_{eff} \end{array}\right.
\label{eq:fbo}
\end{equation}
Note that here
the gas remains in the halo, it is not expelled out of the
system. With this simple model, that contains no free parameters, 
the fraction $f_{bo}$ ranges from 0 to \circa0.2; this is roughly
consistent with Mac Low \& McCray (1988) and Mac Low, McCray \& Norman
(1989), who report that most of the internal hot gas remains in the
disc.

\begin{figure}
\centering{
\includegraphics[width=0.35\linewidth]{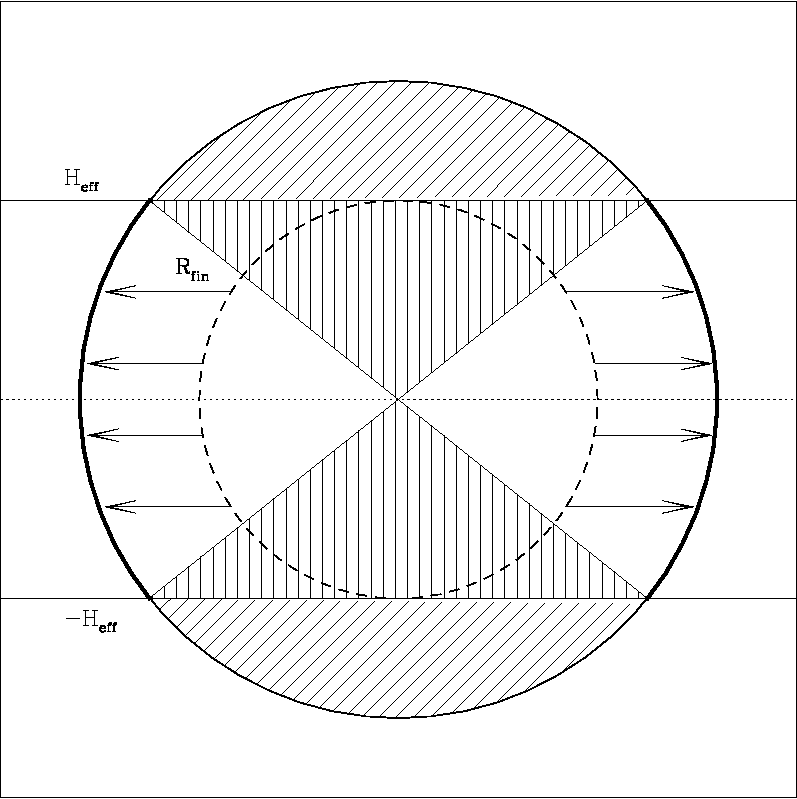}
\captionsetup{font={normalsize,sf}, width= 0.9\textwidth}
\caption{Geometrical model for blow-out.  The SB starts blowing out
when its radius is equal to $H_{eff}$, but continues to expand for one
sound crossing time, finally reaching the radius $R_{fin}$.  The two polar
cups (diagonal-shaded regions), defined by the intersection of the
final SB and the two planes at $H_{eff}$, are assumed to be devoid of
matter.  All the matter present in the double cone (vertical-shaded
regions) with an aperture equal to that of the polar cups receives a
radial momentum that allows it to blow-out into the halo. Taken from
MO04.}}
\label{fig:blowout}
\end{figure}

\begin{figure}
\centering{
\includegraphics[width=0.35\linewidth]{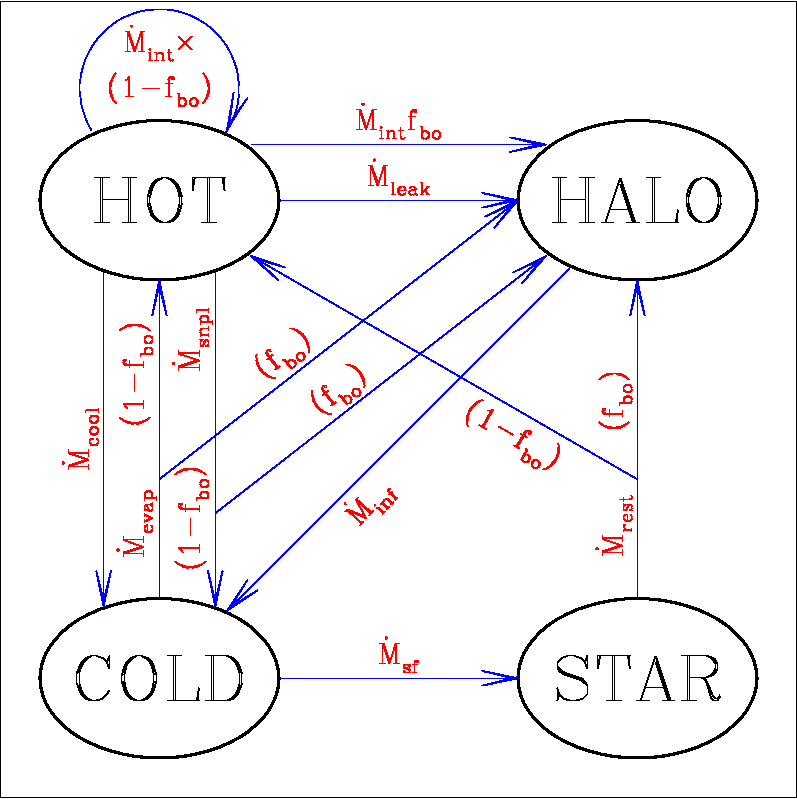}
\captionsetup{font={normalsize,sf}, width= 0.9\textwidth}
\caption{Mass flows between the four components described in the
model.  Arrows denote the flows connected to infall ($\dot M_{inf}$), star
formation ($\dot M_{sf}$), restoration ($\dot M_{rest}$), cooling ($\dot M_{cool}$),
evaporation ($\dot M_{evap}$), snowplows ($\dot M_{snpl}$), leak-out
($\dot M_{leak}$) and the 
rate at which the hot phase is engulfed by SBs ($\dot M_{int}$).  Blow-out
takes mass by a fraction $f_{bo}$ from the internal, evaporation,
snowplow and restoration mass flows. Taken from
MO04.}}
\label{fig:massflows}
\end{figure}
\newpage

\subsection{The system of equations} 

\subsubsection{\small{MASS FLOWS}}
Fig. 2.3 shows all the mass flows between the
different components that are taken into account. \\
Cold gas continually infalls from the halo:
\begin{equation}
\dot M_{inf} = \frac{\dot M_{halo}}{t_{inf}}\, , 
\label{eq:dminf}
\end{equation}
where $t_{inf}$ is the infall time-scale.\\
The hot phase cools at the rate $t_{cool}$, and thus the cooling mass
is modelled as:
\begin{equation}
\dot M_{cool} = f_{cool} \frac{M_{hot}}{t_{cool}}\, , 
\label{eq:dmcool}
\end{equation}
where $f_{cool}$ is a free parameter regulating the fraction of the
gas that is allowed to cool, while $t_{cool}$ is calculated from
Eq.~\ref{eq:tcool}.\\
While the cold phase is easily confined even by a modest gravitational
well, the hot phase is generally able to leak out of the volume $V$ to
the halo. This is described as:
\begin{equation}
\dot M_{leak} = \frac{M_{hot}}{t_{leak}}\, 
\label{eq:dmleak}
\end{equation} 
where $t_{leak}= \sqrt{3/d} \frac{H_{eff}}{c_{\rm s,h}}$ is the
timescale connected to this leak-out, with $d$ equal to one if the
leak-out is in one preferential direction, 3 if it is spherically symmetric.\\
The SFR is due to the contribution from the total number of collapsing
clouds. Therefore cold gas is converted into stars at the rate:
\begin{equation}
\dot M_{sf}= f_{star} f_{coll} \frac{M_{cold}}{t_{dyn}}\,
\end{equation} 

A fraction $f_{rest}$ is instantaneously restored to the hot phase:
\begin{equation}
\dot M_{rest} = f_{rest} \dot M_{sf}, . \label{eq:dmrest}
\end{equation}
This flux is responsible for chemical enrichment; this
equation implies instantaneous recycling.\\
The rate at which the mass of collapsing clouds is evaporated back to
the hot phase is:
\begin{equation}
\dot M_{evap} = f_{evap} \frac{\dot M_{sf}}{f_{\star}}
\end{equation}

At the final time $t_{fin}$ each SB has swept a mass $M_{\rm sw}(t_{fin})$,
of which a part $M_{\rm int}(t_{fin})$ is in hot internal gas and the
rest $M_{snpl}$ is in the
snowplow. The rate at which the hot phase becomes internal mass of a
SB is:
\begin{equation}
\dot M_{int} = N_{cc} \frac{M_{\rm int}(t_{fin})}{t_{dyn}}\, , \label{eq:dmint}
\end{equation}

while the rate at which it gets into a snowplow is:
\begin{equation}
\dot M_{snpl} = N_{cc} frac{M_{\rm sw}(t_{fin})-M_{\rm int}(t_{fin})}{t_{dyn}}\, .
\label{eq:dmsnpl}
\end{equation}

We recall that a fraction $f_{bo}$ (Eq.~\ref{eq:fbo}) of the swept
material and of the restored and evaporated mass
is blown-out to the halo.  If we define $\dot M_{bo} =
f_{bo}(\dot M_{evap}+\dot M_{rest}+\dot M_{int}+\dot M_{snpl})$, the
system of equations that 
describes the mass flows is:
\begin{equation}
\begin{array}{lcl}      
\dot M_{cold} &=& \dot M_{inf}+\dot M_{cool}-\dot M_{sf}-\dot
M_{evap}+(1-f_{bo})\dot M_{snpl} \\
\dot M_{hot}  &=& -\dot M_{cool} -\dot M_{snpl} -\dot M_{leak}
-f_{bo}\dot M_{int}+(1-f_{bo})(\dot M_{evap}+\dot M_{rest})\\
\dot M_{\star} &=& \dot M_{sf} - \dot M_{rest} \\
\dot M_{halo} &=& -\dot M_{inf} + \dot M_{leak} + \dot M_{bo} \\
\end{array} 
\label{eq:massfluxes}
\end{equation}
Mass conservation is guaranteed by the condition
$\dot M_{hot}+\dot M_{cold}+\dot M_{star}+\dot M_{halo}=0$.
\subsubsection{\small{ENERGY FLOWS}}
We here focus on the flows regulating the energy $E_{hot}$ of the hot
component, which moreover determines $T_{h}$. The total energy
released by SNe is:
\begin{equation} 
\dot E_{sn}= N_{cc} \frac{E_{51}N_{sn}}{t_{dyn}}
\label{eq:desn}
\end{equation}
Four different processes lead to energy losses:
\begin{equation} 
\dot E_{cool} = \frac{E_{hot}}{t_{cool}}
\qquad \qquad \qquad \qquad\textrm{\small{COOLING}}
\end{equation}

\begin{equation} 
\dot E_{snpl} = \dot M_{snpl} T_h \frac{3}{2}\frac{k}{\mu_h m_p}
\qquad \qquad \textrm{\small{SNOWPLOUGH}}
\end{equation}

\begin{equation} 
\dot E_{cool} = f_{bo}\dot M_{int} T_h\frac{3}{2}\frac{k}{\mu_h m_p}
\qquad \qquad \textrm{\small{BLOWOUT}}
\end{equation}

\begin{equation} 
\dot E_{cool} = \frac{E_{hot}}{t_{leak}}
\qquad \qquad  \qquad \qquad\textrm{\small{LEAK-OUT}}
\end{equation}

The energy budget of the SB $E_{sb}$ is in part given to the ISM as thermal
energy, and in part as kinetic energy, which is transformed into
turbulence and then partially thermalized. \\
The SNe feedback energy injected in the ISM is modelled according to
the physical stage of the SB.\\ 
In the \textit{blow-out} regime,
\begin{equation} 
\dot E_{sn} = (1-f_{bo})\left(N_{cc}\frac{E^{th}_{sb}+E^{kin}_{sb}}{t_{dyn}}
+\do M_{evap}T_{evap}\frac{3}{2}\frac{k}{\mu_h m_{\rm p}}\right)
\label{eq:defbbo} 
\end{equation}
where the first term at the right hand side models the energy
the ISM received by the SB, while the second describes the energy
connected to the evaporated mass.\\
In the \textit{adiabatic confinement} regime, all the energy is given
to the ISM:
\begin{equation} 
\dot E_{fb} = \dot E_{sn}
\end{equation}
In the \textit{PDS} regime,
\begin{equation} 
\dot E_{fb} = N_{cc}\frac{1}{t_{dyn}}(E^{th}_{sb}+E^{kin}_{sb} + f_{pds} E_{\rm rest})
\end{equation}
the ISM receives the thermal and kinetic energy of the SB.\\
Finally, the equation regulating the energy flows in the hot component
is :
\begin{equation}
\dot E_{hot} = -\dot E_{cool} -\dot E_{leak} +\dot E_{sb} -\dot E_{bo}
- \dot E_{snpl}
\label{eq:nrgflux}
\end{equation}

In this section, we have listed the mass and energy flows implemented in the
MO04 model. Describing the processes regulating the metal flows
is beyond the scope of this work, we address the reader to MO04
for further reading. 

In order to highlight its predictive power,
  MO04 ran the model in a Milky-Way-like system with the assumption
 that the SN bubbles are in the blow-out regime (see
  Sec.~\ref{fateSN}). In this case,
  the main characteristic of the ISM 
  of the Galaxy are broadly reproduced, but, with respect with
  previous models, here parameters such as the efficiency of feedback
  or the Schmidt-law are consistently \textit{predicted} by the model.
Moreover, the MO04 model distinguishes from the previous ones because
does not restrict to self-regulated, equilibrium 
solutions and neglects the global structure of the
galaxy. It presents a
rich variety of solutions with a relatively limited set of parameters.
Although the turbulent nature of the ISM is not explicitly taken into
account, the model is thought to give a good approximation to the
solution of the feedback problem.

The MO04 model can thus
construct a realistic grid of feedback
solutions to be used in galaxy formation codes, either semi-analytic or
numeric. For these reasons, we decided to implement the MO04 model
into the GADGET-2 code (see Chapter~\ref{MUPPI_chap}).

In Monaco et al. 2007, the MO04 multiphase analytic model for
supernovae feedback has been coupled to a model for the evolution of
the DM haloes (thus turning into a \textit{semi}-analytical model) and
has been proven to be in line with the predictions of the other
semi-analytical or N-body models.

\newpage

\section{Implementation and comparison of different ISM models in Gadget-2} 
Star formation prescriptions for numerical simulations (e.g. see
previous sections for some examples) are motivated on physical
grounds, and usually tested 
against observations of local spiral galaxies. In cosmological
simulations the physical state of 
the star-forming gas can be very different from that of
spiral galaxies: it can reside at the centre of cooling flows, forming
or not forming a thin disk 
depending on the angular momentum content of the gas itself.  The main
physical difference between these two cases is the way gas is supported: in
the disk case, gas is cold and dense, and its distribution is supported by its
angular momentum, while in the cooling flow case the gas is partially
pressure-supported.  In a cosmological simulation, of course, a range of
intermediate situation may and will happen.
Among the many different star formation and feedback prescriptions proposed in
the literature, we consider here three models that, in our opinion, give a
good sampling of the possible numerical choices, namely the multi-phase
effective model of SH03 (see Sec.~\ref{GDT:sf_eff}), an implementation
of the simple model by KA96 (see
Sec.~\ref{katz})\nocite{Katz96} 
and an implementation of the Thacker \& Couchman 2000
 prescription (TC00) that takes into account the improvement 
proposed by Stinson et al. 2006\nocite{Stinson06} (see Sec.~\ref{ST06}). We only address
feedback in the form of thermal energy, we do not consider kinetic feedback.

The TC00 model deserves some discussion: it is known to work 
very well for the formation of galaxy disks in cosmological simulations, so we
decided to test it in non-rotating or cluster-like halos.  In its original
implementation, after a star formation event the algorithm distributes SN
energy to all the SPH neighbours (typically 32 particles) and then disables
their radiative cooling for a time $\tau$=30 Myr.  This ad-hoc assumption
mimics the adiabatic phase of the propagation of a SN remnant, avoiding a
quick radiative loss of the deposited energy.  These authors found that,
compared to their previous scheme of thermal and kinetic feedback, this model
allows to better conserve angular momentum of their spiral galaxies.  A
problem with this method is that the SPH smoothing radius, and with it the
number of gas particles involved in the ``adiabatic phase'', strongly depends
on resolution.  Stinson et al. 2006 removed this resolution-dependence
problem by 
suitably computing how many gas particles have their cooling disabled. In
Governato et al. 2007, the Stinson et al. 2006 feedback recipe was
used in a cosmological 
simulation of a disk galaxy and was proven to be very efficient in reducing
the loss of angular momentum: this feedback model produces more extended disks
with smaller bulges and the right trend of star formation history with galaxy
mass.  Besides, the simulated galaxies lie close to the observed baryonic
Tully-Fisher relation.

\subsection{Numerical methods}
\label{nummeth}

We use the GADGET-2 code, a parallel Tree+SPH code (Springel 2005) with a
fully adaptive time-step algorithm. The version of the code that we use adopts
an SPH formulation with entropy conserving integration and arithmetic
symmetrisation of the hydrodynamical forces (SH03), and
includes radiative cooling computed for a primordial plasma with vanishing
metalicity.

As mentioned in the Introduction, we address three different star formation
and feedback models. The first one is the original GADGET-2 effective model
for star formation and feedback (EFF). It is based on a simple model for the
multi-phase structure of the star-forming gas on the small scales which are
not resolved in cosmological simulation.  The second one (simple star
formation, SSF) is an implementation of that described in KA96, a 
single-phase model for star formation and SNe feedback which is
described in Sec.~\ref{katz}. The third one
is an implementation of Stinson et al. 2006, starting 
from SSF, consists in turning off radiative cooling for a fixed amount of time
after a star formation episode (delayed cooling, DEL).

For simplicity, we do not consider any type of kinetic feedback, because none
of the schemes we use has a self-consistent prescription for it, so it would
be implemented as an independent prescription applied on top of the sub-grid
model. The EFF, SSF and DEL models have been
previously presented in Sec.~\ref{GDT:sf_eff}, Sec.~\ref{katz} and
Sec.~\ref{ST06}. 
In the following, we adopt
values of $\rho_{\rm tr}$ corresponding to a number density of $n_{\rm
thr}$=0.25 cm$^{-3} h^{-3}$ and $T_{\rm thr}$= $2\cdot10^{4}$ K .

\subsection{Simulations}
\label{simulsection}
We run simulations with the three star formation and feedback models for three
configurations, two isolated, non-rotating halos and a cosmological
cluster-sized halo.

For the two isolated halos, we set up initial conditions for our models as in
Viola et al. 2008\nocite{Viola}, considering objects whose DM component has a
Navarro, Frenk \& White (1996, 1997)\nocite{NA96.1}\nocite{NA97.1}
density profile. Gas pressure is computed using the
universal gas-density and temperature profiles derived by Komatsu \&
Seljak 2001\nocite{2001MNRAS.327.1353K}. They make three
assumptions: (1) the gas is in 
hydrostatic equilibrium within the gravitational potential of the DM halo
(as described in Suto et al. 1998\nocite{Suto98}; (2) the slopes of
the DM and gas density 
profiles are equal at the virial radius, thus fixing the gas thermal energy;
(3) the gas follows a constant polytrophic equation of state; we use the value
$\gamma_{p}$=1.18, for the effective polytrophic index. Initial positions of DM
and gas particles are assigned by Monte-Carlo sampling the analytical density
profiles of DM and gas.  To create an equilibrium configuration for the DM
halo, we assign initial velocities of the DM particles by assuming a local
Maxwellian approximation (Hernquist
1993\nocite{1993ApJS...86..389H}). The width of the 
distribution, which gives the velocity dispersion of the DM particles, is
obtained by solving Jeans' equation. Regarding gas particles, because the gas
is in hydrostatic equilibrium, particles are assigned zero velocities and
their internal energy is set according to the computed temperature
profile. This procedure allows to generate configurations that are in
approximate equilibrium, so, before switching on cooling, star formation and
feedback, we let the system evolve for two dynamical times, so as to start
from a truly relaxed state.

With this procedure we generate two halos, having masses of $10^{13}$ {\msun},
typical of a $z$=0 poor galaxy group, and $10^{12}$ {\msun}, typical of an
isolated galaxy.  Virial radii are computed as $r_{200}$, the radius for which
the average density is 200 times the critical density at $z$=0.
Concentrations of the two halos are chosen respectively as $c_{\rm NFW}$=6.3
and $c_{\rm NFW}$=7.25, given by the relation between mass and concentration
provided by NFW (1997).  Both halos are sampled with 6 x $10^{4}$ DM and gas
particles inside $r_{200}$; the profiles are extrapolated to several virial
radii so as to have pressure support at the virial radius. We assume the
baryon fraction to be $f_{\rm bar}$=0.19.  We set the Plummer-equivalent
softening to be $2.64\ h^{-1}$ kpc, the value suggested by Power et
al. 2003\nocite{Pow03}, for Me12.
  
For Me13, we rescale the softening to $5.7\ h^{-1}$ kpc. We assume the
minimum value for the SPH smoothing length to be 0.5 times that of the
gravitational softening; we also set the number of the SPH neighbours $N_{\rm
ngb}$ to be 32.  In all runs we set the initial angular momentum to zero.  The
main characteristics of these halos are summarised in Table~\ref{res:tab}.
Since the overdensity used to define the virial radius is constant
($\Delta\rho=200 \rho_c$), $t_{\rm dyn}$ (Eq.~\ref{tdyn}) takes a value
of 0.56 Gyr for both halos.

We evolve Me12 for 4 dynamical times and Me13 for 8 dynamical times; the
difference is due to the longer time needed by the larger object to reach a
similar evolutionary stage when compared with the smaller one.

The third configuration used in this paper is a DM halo taken from a
cosmological simulation.  We use the initial conditions of the object labelled
CL4 of Borgani et al. 2006 at the ``medium'' resolution: the mass of a
DM particle 
is $1.5 \cdot 10^9$ {\msun}, the mass of a gas particle is $2.3 \cdot 10^8$
{\msun}, and the Plummer-equivalent softening is $5$ $h^{-1}$ proper kpc from
$z$=0 to $z$=2 and is kept fixed in comoving coordinates for $z$$>$2.  The
background cosmology is a flat $\Lambda$CDM cosmological model with
$\Omega_{m}$=0.3, $h$=0.7, $\sigma_{8}$=0.8 and $\Omega_{b}$=0.04.  The
initial conditions were produced using the Zoomed Initial Condition technique
(Tormen et al. 1997\nocite{1997MNRAS.286..865T}; see Borgani et
al. 2006 for further details). 
At redshift $z$=0, our selected cluster has a mass of $\approx 1.6 \cdot
10^{14}$ {\msun} within $R_{\rm vir} \approx 1.1$ $h^{-1}$ Mpc, defined as the
radius enclosing an overdensity of $\approx 100$ times the critical density,
as predicted by the spherical collapse model in our assumed cosmology. It
therefore contains $\approx 9 \cdot 10^4$ DM and gas particles, a number which
is comparable to that adopted for our isolated runs.  In the spirit of this
work, in this cosmological run we do not model SNIa, kinetic feedback,
metalicity of the gas and metal cooling. We only keep into account the effect
of a uniform redshift-dependent UV radiation background (Haardt \& Madau
1996\nocite{1996ApJ...461...20H}).

\begin{figure}
\centering{
\includegraphics[viewport = 45 42 630 310,clip,height=6.cm]{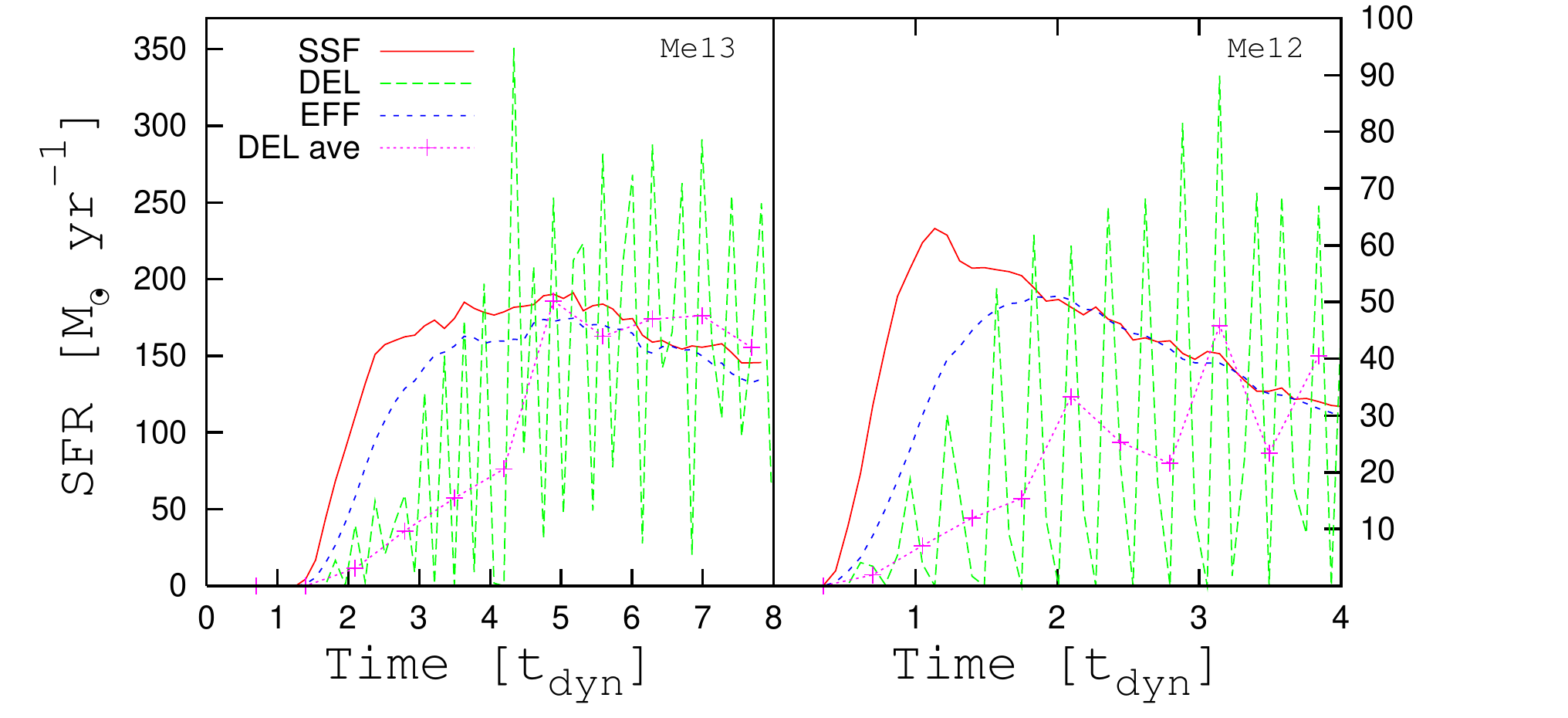}    
\captionsetup{font={normalsize,sf}, width= 0.9\textwidth}
\caption{SFRs as a function of time (in units of $t_{\rm dyn}$) for simulations
  of isolated halos with the three studied models: EFF (short-dashed line),
  SSF (solid line) and DEL (long-dashed).The left panel
  shows Me13 SFRs up to $8 t_{\rm dyn}$ and the right panel the Me12 SFRs
  up to $4 t_{\rm dyn}$(see text). We resampled the Me13 and the Me12
  DEL SFRs (dotted line) with a constant time interval equal
  respectively to 0.13 $t_{\rm dyn}$ and 0.08 $t_{\rm dyn}$.}}
\label{sfrates}
\end{figure}

\begin{table*}
\centering
\begin{tabular}{@{}c c c c c c c@{}}
\hline\hline &  \texttt{$M_{200}$} & \texttt{$r_{200}$} &
\texttt{$c_{\rm NFW}$} & \texttt{$m_{\rm DM}$} & \texttt{$m_{\rm gas}$}  \\
\hline

\texttt{Me13} & $10^{13}$ & 350 & 6.3 & $1.3\cdot10^{8}$ &  $3.1\cdot10^{7}$
\\
 \texttt{Me12}& $10^{12}$ & 162 & 7.25 &  $1.3\cdot10^{7}$ & $3.1\cdot10^{6}$
 \\
 \texttt{CL4} & $1.6 \cdot 10^{14}$ & 1100 & --- & $1.5 \cdot 10^9$ & $2.3 \cdot 10^8$ \\

\end{tabular}
\captionsetup{font={normalsize,sf}, width= 0.8\textwidth}
\caption{Main properties of the simulated halos. In all runs, we used
  6*$10^4$ DM particles and as many gas particles inside
  $r_{200}$. Column 1: halo name. Column 2: mass enclosed in within
  $r_{200}$ in {\msun}. Column 3: value of $r_{200}$ in $h^{-1}$ kpc. 
 Column 4: NFW concentration. Column 5: mass of a DM particle in
  {\msun}. Column 6: mass of a gas particle in
  {\msun}. }
\label{res:tab}
\end{table*}

\subsection{Results}

In this Section we compare SFRs, phase diagrams, density and temperature
profiles obtained simulating our Me13, Me12 and CL4 halos with the EFF, SSF ad
DEL star formation and feedback schemes described above.

\begin{table*}
\centering
\begin{tabular}{@{}c c c c@{}}
\hline\hline Run & \texttt{EFF} & \texttt{SSF} & \texttt{DEL} \\
\hline
\texttt{Me13} \scriptsize{($4t_{\rm dyn}$)} & 0.07 & 0.08 & 0.01\\

\texttt{Me12} \scriptsize{($4t_{\rm dyn}$)}&  0.33 & 0.35 & 0.08 \\

\texttt{Me13} \scriptsize{($8t_{\rm dyn}$)} & 0.22 & 0.23 & 0.08 \\

\texttt{CL4} \scriptsize{($z$=0)}& 0.27 & 0.25 &0.19 \\
\hline

\end{tabular}
\captionsetup{font={normalsize,sf}, width= 0.6\textwidth}
\caption{Gas cold fraction (mass of stars over mass of baryons) for
  Me13 and Me12 within $r_{200}$ and for CL4 within $r_{vir}$.  We
  show results for Me13 and Me12 at $4 t_{\rm dyn}$, Me13 at $8 t_{\rm
    dyn}$, CL4 at $z$=0. 
}
\label{models}
\end{table*}

\subsubsection{Isolated halos: star formation rates}
\label{sfr_text}

In Fig. 2.4 we show SFRs for EFF, SSF, and DEL simulations of the
two isolated halos Me13 and Me12, evolved for $8 t_{\rm dyn}$, and $4 t_{\rm
dyn}$ respectively.

Regarding Me13 (left panel), in both EFF and SSF schemes star formation starts
after $\approx 1.3 t_{\rm dyn}$, when the cooling flow is established and the
gas in the central part of the objects is dense and cold enough to become
star-forming. The behaviour of star formation in these two schemes is similar,
with SSF forming $\approx$10\% more stars than EFF. At $6 t_{\rm dyn}$, a
similar SFR of slightly less than $\approx$150 M$_\odot$ yr$^{-1}$ is
achieved, slowly declining with time while the gas is consumed. Feedback in
the EFF scheme is thus only slightly more efficient than in the SSF one, and
this causes a slightly slower rise of star formation.

The DEL scheme shows a different behaviour. The SFR grows much more slowly
than in EFF and SSF schemes, and at $4 t_{\rm dyn}$ it is still 30\% lower on
average, reaching the other two curves at At 8$t_{\rm dyn}$.  Clearly, the DEL
scheme is efficient in suppressing star formation in this object. But the most
noticeable property of this SFR curve is that it is intermittent and spiky.

The right panel of Fig. 2.4 shows the SFRs for Me12.  Also in this
case, the onset of star formation is quicker for the SSF scheme, and the SFR
is $\approx$25\% higher in SSF than in EFF at its peak. After $2 t_{\rm dyn}$
the EFF and SSF schemes do converge to the same SFR, which then decreases with
time.  Again, DEL shows an intermittent behaviour, and, on average, a much
slower rise in SFR; it reaches the SFR of the other two schemes, on average,
only after $3 t_{\rm dyn}$; later on, oscillations are visible even for the
average value of the star formation curve.

Note that EFF and SSF SFRs converge for Me13 and Me12 at different times.  The
dynamical time of the halos is the same, but the cooling times are shorter for
Me12 and thus its evolution is faster.

The spiky behaviour of star formation in the DEL scheme is due to the fact
that many particles simultaneously cool down and reach the star formation
threshold at the same time. These particles spawn stars simultaneously,
getting energy and stopping from cooling. All the star-forming gas is thus
prevented from further star-forming for 30 million years. After that time gas
cools and condenses, crosses again the threshold and the cycle is repeated,
involving new gas which has cooled in the meanwhile.

Cold fractions (defined as the mass in stars divided by the total baryon mass
within the virial radius) for the three schemes in the Me13 and Me12 cases
after 4 $t_{\rm dyn}$ are listed in Table~\ref{models}, where we also list the
cold fractions after 8 $t_{\rm dyn}$ for Me13.  Consistently with what
discussed above, SSF scheme is the most efficient model in converting gas into
stars, EFF scheme in only slightly less efficient.  On the other hand, due to
its intermittent behaviour, cold fractions in DEL model are very low, about
one tenth of EFF and SSF in the Me13 case and about one quarter in Me12 at the
end of the simulations.

These results confirm the well-known result that thermal feedback is not
efficient in models like EFF and SSF, while the DEL model is very effective at
suppressing star formation, but at the cost of a strongly intermittent
behaviour.

\subsubsection{Isolated halos: thermodynamics}
\label{phasediags}

To achieve a better physical understanding of the differences among our three
feedback schemes, we investigate the phase diagrams ($\rho$ versus $T$) of gas
particles We use the effective temperature for the EFF scheme, since it is the
pertinent one as far as hydrodynamics is concerned.

Fig. 2.5 and 2.6 show phase diagrams for the Me13
and Me12 halos.  We show both halos for sake of completeness, but all
conclusions that are drawn from Me13 are confirmed by Me12, so we will
concentrate on the more massive halo.  The upper left panel of
Fig. 2.5 shows the phase diagram for Me13 with the EFF model after $4
t_{\rm dyn}$.  Here we only consider gas particles lying in the inner 20
$h^{-1}$ kpc of the halo.  This diagram is populated in two main regions: hot,
low-density particles flowing toward the halo centre (the cooling flow)
populate the upper-left corner, denser and colder particles, the ones in the
multi-phase regime, are visible in a tight relation at $T\sim10^5$ K.  Only a
few particles join the two regions, tracing a cooling path at intermediate
densities and temperatures.  The effective temperature of the star-forming gas
particles is set by the weighted average of the hot phase energy (given by
Eq.~\ref{EFF_uh}) and the cold phase energy (which is fixed), so it is higher
than $T \approx 10^4$ K, which is where the cooling function drops, and is a
function of the density through the multi-phase model. This
density-temperature relation is sharply truncated at the density where the
probability of spawning a star becomes nearly unity.  Because particles heated
by feedback would populate the upper-right corner of this diagram, this
demonstrates that, in absence of kinetic winds, feedback in the EFF model only
acts in rising the temperature of the star-forming particles.

The upper right panel of Fig. 2.5 shows the same plot for the
SSF scheme. This time the star-forming particles cool down to $\approx 10^4$
K; due to the short cooling times at such high densities, SN energy given to
gas particles is almost completely uneffective at populating the upper-right
corner of the diagram.

The lower panels shows two phase diagrams for the DEL scheme.  For Me13 we
have chosen two different times: $t = 3.8 t_{\rm dyn}$ (lower left panel),
corresponding to a phase of quenched star formation, and $t = 4 t_{\rm dyn}$
(lower right panel), corresponding to a peak of star formation.  For Me12,
shown in Fig. 2.6, the chosen times are $t = 3.6 t_{\rm dyn}$
and $t = 3.9 t_{\rm dyn}$.  Consistently with the SFRs, at $3.8 t_{\rm dyn}$
all particles in the Me13 halo have temperatures higher than the threshold for
forming stars, while at $4 t_{\rm dyn}$ a large number of particles are dense
and cold enough to form stars.  At the later time a column of particles with
very high density and high temperatures, spanning the range $10^5 < T < 3
\cdot 10^6$ K, is visible. These particles have just acquired energy from a
star formation episode.  The acquired energy varies depending on whether the
same particle or a neighbour has spawned a star.  The $4 t_{\rm dyn}$ diagram
also shows the trajectory of a single gas particle. This starts at the
leftmost end, with high temperature and low density. It cools down to $T \sim
10^4$ K and spawns a stars. As a consequence it is heated up to $\sim 10^6$ K;
a phase of adiabatic expansion follows, caused by the fact that the particle
has its cooling stopped.  At the final time of the trajectory, $t=4 t_{\rm
dyn}$, cooling is switched on again.  This shows that the heated particles,
adiabatically expanding and then cooling, occupy the region at relatively high
densities, between $5 \cdot 10^7$ and $5 \cdot 10^8$ $h^2$ M$_\odot$
kpc$^{-3}$, and temperatures between $10^5$ and a few times $10^{6}$ K. During
the cooling phase of the particle its trajectory in the phase diagram shows a
``bump'' in which both temperature and density rise. This is due to a weak
shock caused by its interaction with high-density particles which are
adiabatically expanding.  These weak shocks cause the density to {\it
increase} gradually after each star formation burst.  Indeed,
Fig. 2.5 shows that DEL star-forming particles have higher
densities than EFF and SSF ones.

In the outer part of the halo, the phase diagram of Me13
(Fig. 2.5) do not show any difference among the three schemes,
confirming that in such relatively massive halos differences in star formation
and feedback do influence only the very inner, star-forming region.  This is
different in the less massive Me12 halo (Fig. 2.6), where the
DEL feedback model shows some influence on the external parts.

\begin{figure}
\centering{
\includegraphics[viewport = 60 48 600
  600,clip,height=8.cm]{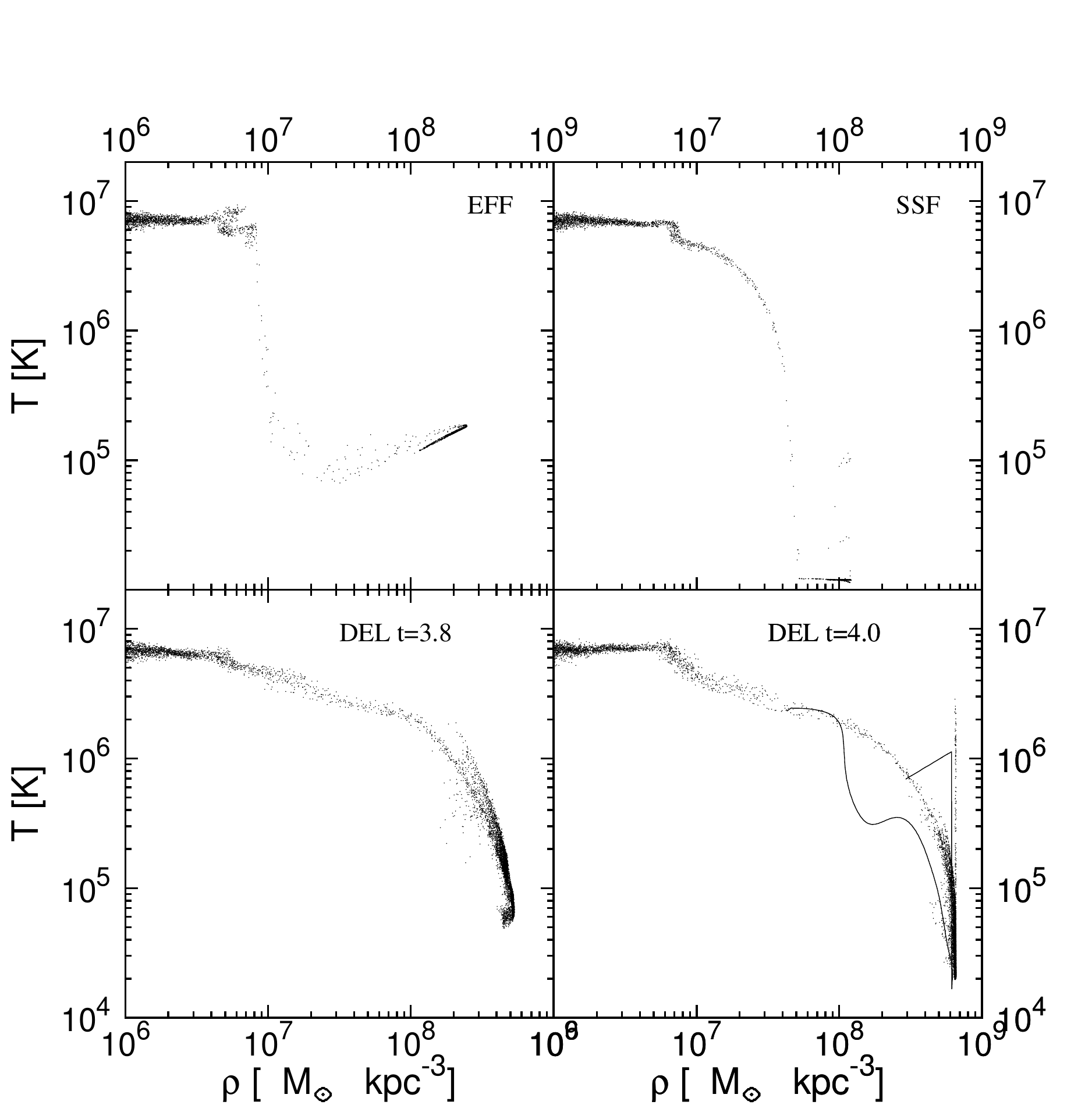}
\captionsetup{font={normalsize,sf}, width= 0.9\textwidth}
\caption{
Phase diagrams ($T$ versus $\rho$) for gas particles of Me13 simulations.
Only particles in the inner 20 $h^{-1}$ kpc are shown.
Upper panels show the EFF (left) and SSF (right) models at $4 t_{\rm dyn}$. 
Lower panels show the DEL model at $3.8 t_{\rm dyn}$ (left) and $4
t_{\rm dyn}$ (right).} }
\label{phaseMe13}
\end{figure}

\begin{figure}
\centering{
\includegraphics[viewport = 60 48 600 600,clip,height=8.cm]{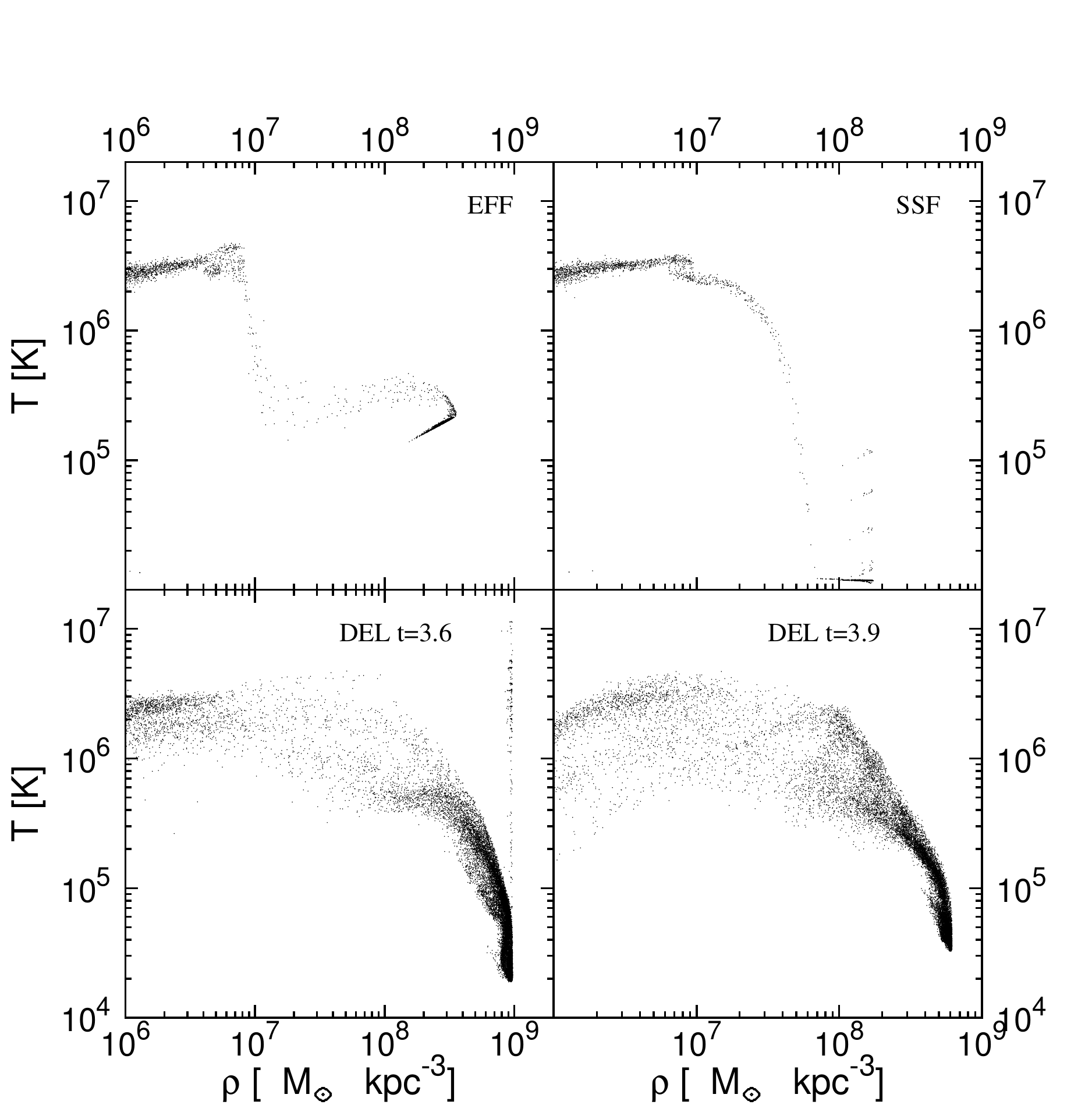}
\captionsetup{font={normalsize,sf}, width= 0.9\textwidth}
\caption{
Phase diagrams ($T$ versus $\rho$) for the gas particles for Me12 after.
Panels are as in Figure~\ref{phaseMe13}; lower panels are at $3.6
t_{\rm dyn}$ (left) and $3.9 t_{\rm dyn}$ (right).}}  
\label{phaseMe12}
\end{figure}

\begin{figure}
\centering{
\includegraphics[viewport = 60 48 550 520,clip,height=8.5cm]{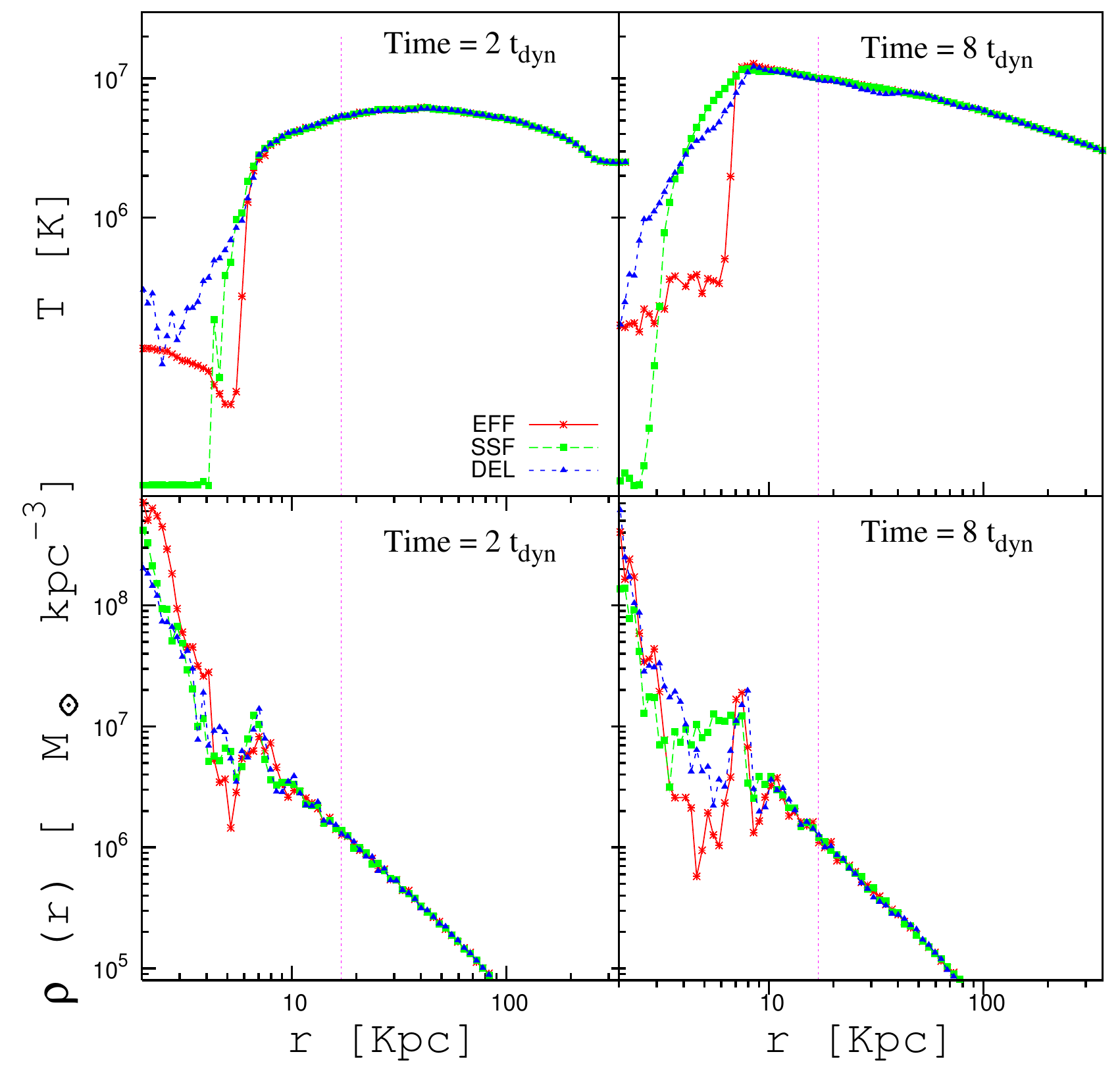}
\captionsetup{font={normalsize,sf}, width= 0.9\textwidth}
\caption{Me13 gas density (bottom panels) and temperature (top panels)
  profiles at $2 t_{\rm dyn}$ (left) and $8 t_{\rm dyn}$ (right) for
  the three models studied. The vertical line marks the scale
  corresponding to three softening lengths.}}
\label{Me13}
\end{figure}

\begin{figure}
\centering{
\includegraphics[viewport = 60 48 550 520,clip,height=8.5cm]{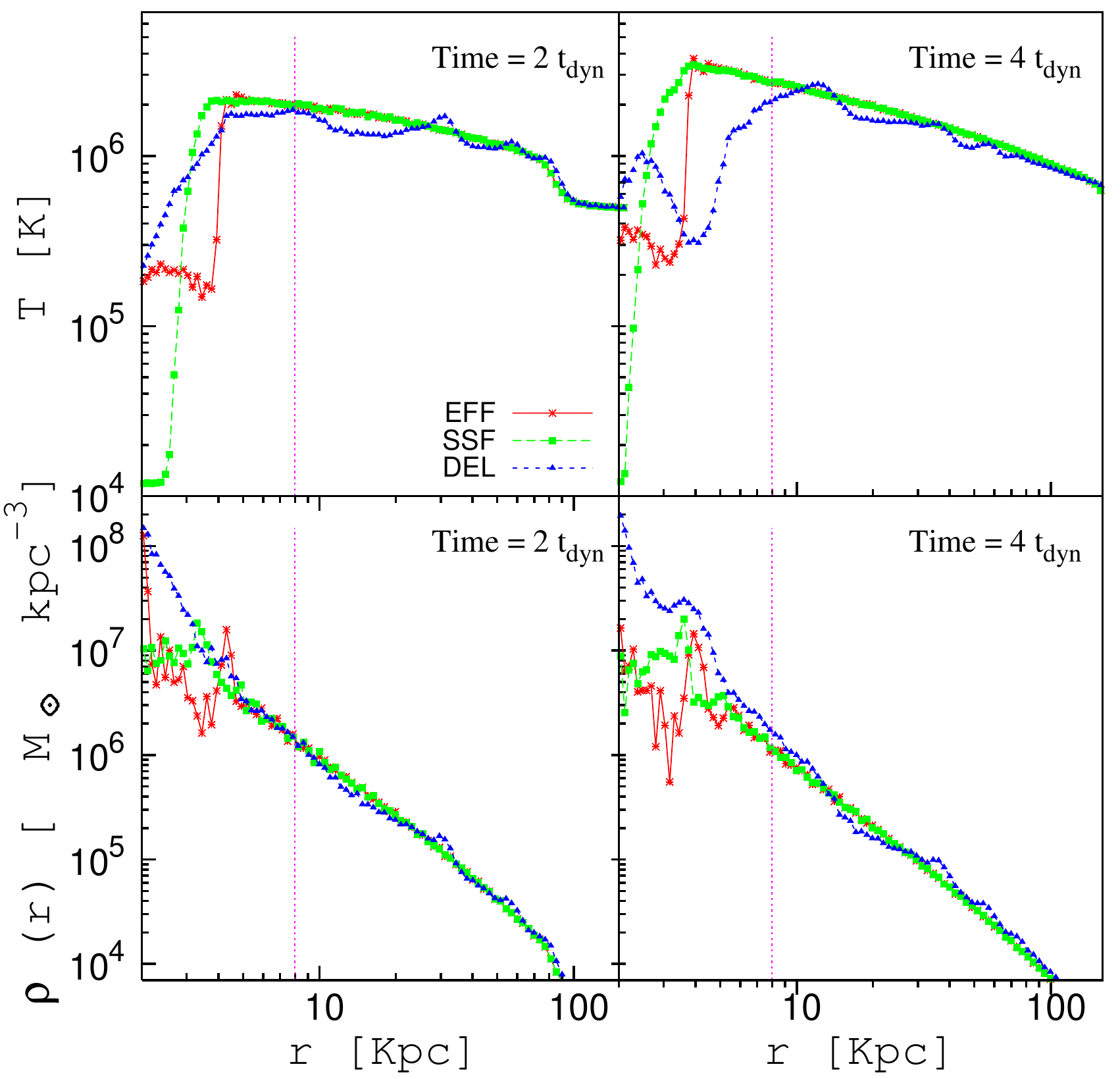}
\captionsetup{font={normalsize,sf}, width= 0.9\textwidth}
\caption{Me12 gas density (bottom panels) and temperature (top panels)
  profiles at $1 t_{\rm dyn}$ (left) and $4 t_{\rm dyn}$ (right) for
  the three models studied. The vertical line marks the scale corresponding to three softening lengths.}}
\label{Me12}
\end{figure}

In Fig. 2.7 we show gas temperature (effective temperature for the
EFF model) and density profiles for the Me13 case after $2 t_{\rm dyn}$, when
all SFRs are still growing and differences among models are more marked, and
$8 t_{\rm dyn}$, where models give convergent values.  The vertical line marks
three softening scales; smaller structures are not properly resolved.

No significant differences among the three models are visible at the scales
that are well resolved in this simulation.  At small distances temperatures
drop to $10^4$ K in the SSF and to $\sim$$10^5$ K both for EFF and DEL, in the
first case as the effect of the multi-phase model, in the second case as an
effect of the delayed cooling.  The density profiles show that at
sub-softening scales DEL density is higher than the other two models,
consistently with what inferred from the phase diagrams.

Analogously to Fig. 2.7, Fig. 2.8 shows temperature and
density profiles for the Me12 halo at {$2$} and $4 t_{\rm
dyn}$.  The general behaviour is similar to the Me13 case, with the difference
that wiggles are present in the DEL profiles.  These are compression waves
propagating through the outer halo as a consequence of the efficient and
intermittent energy injection from SNe.  These are not visible in the more
massive Me13 halo, which is also characterised by a higher virial temperature.

In conclusion, the most relevant differences among models are restricted to
the central, sub-softening regions where star formation occurs. The effects of
stellar feedback are visible on the halo gas component only for the smaller
halo and for the efficient feedback scheme.  This is in line with the
well-known finding that stellar feedback cannot influence the thermodynamics
of the intra-cluster medium (Borgani et al. 2004\nocite{Borg}, Kay et
al. 2007\nocite{Kay07}, Nagai et al. 2007\nocite{Nagai07}).

We also performed several numerical tests on our isolated halos, to assess the
stability of our results with respect to the details of the implementation and
to the resolution. In particular, we verified on the Me13 halo that using 10
times more particles does not change either the intermittent behaviour of the
star formation rate in the DEL case or its low star formation efficiency.
Therefore, also with higher resolution, the star formation is delayed when
compared with the EFF scheme.

\begin{figure}
\centering{
\includegraphics[viewport = 60 48 650 455,clip, height = 6cm]{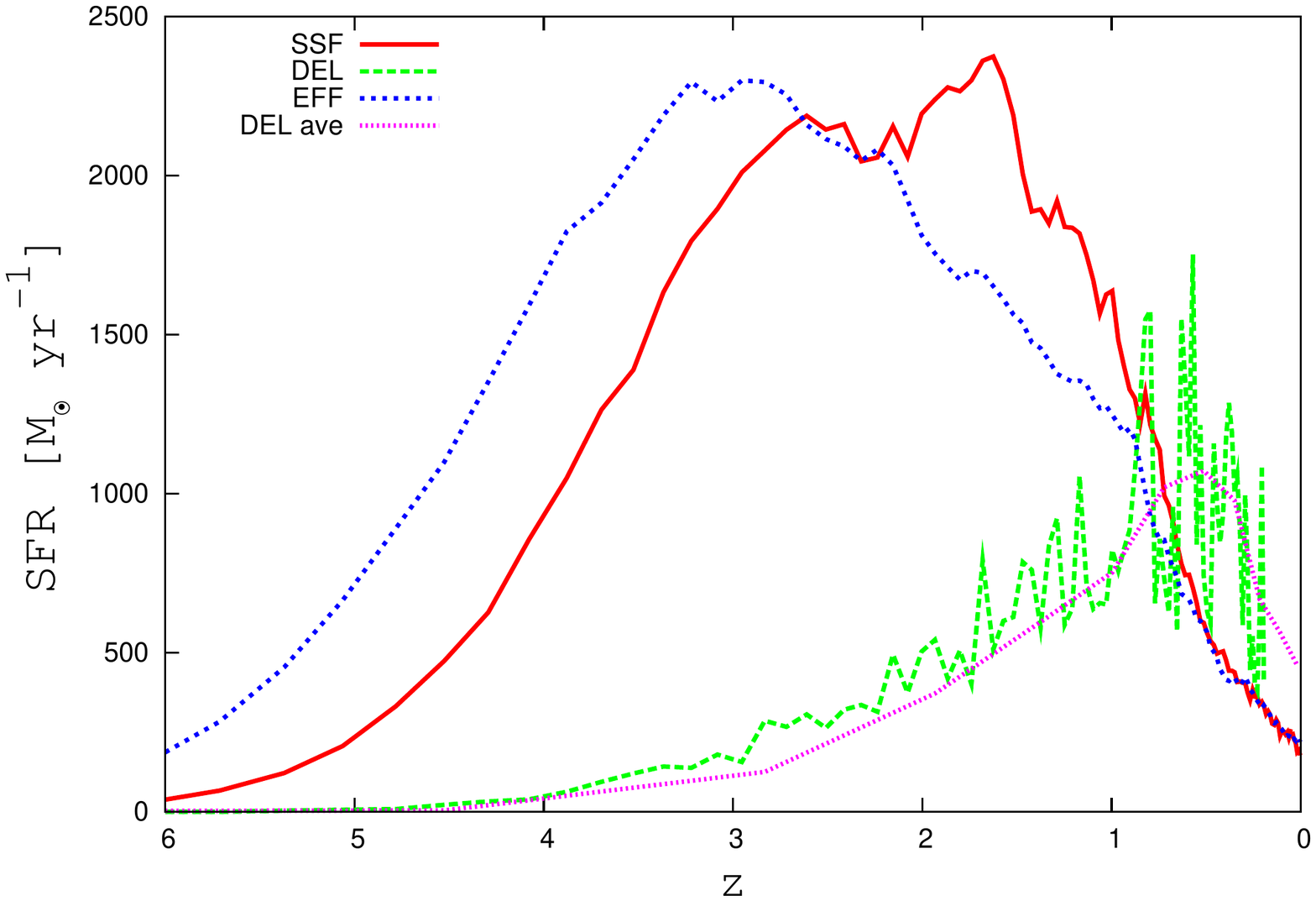}
\captionsetup{font={normalsize,sf}, width= 0.9\textwidth}
\caption{SFRs as a function of redshift in $t_{\rm dyn}$ for the
  cosmological CL4 simulations with the three models studied. Symbols
  are as in Figure~\ref{sfrates}.}}
\label{sfrates_cosmo}
\end{figure}

\subsubsection{Cosmological case}

Fig. 2.9 shows the SFRs for the CL4 cluster.  For the EFF,
SSF and DEL models, star formation peaks respectively at redshift $z$$\sim$3
$\sim$1.7 and $\sim$0.7.  Regarding this last model, the typical spiky SFR is
visible only at low redshift, and this is due to the fact that star formation
at high redshift is averaged over many progenitors; when, at low redshift, the
most massive progenitor dominates the mass distribution, the spiky character
of the star formation becomes visible.

Cold fractions at $z$=0 for the three cases are reported in
Table~\ref{models}: SSF and EFF lock many baryons (27\% and 25\%) in the
stellar component, while DEL gives a lower figure of $\sim$19\%.

Fig. 2.10 and 2.11 show density and temperature
profiles at four different redshifts, $z$=2, $z$=1, $z$=0.5 and $z$=0.  In
this last case no significant differences among the three models are visible
at scales larger than three softening lengths.  As far as the inner, poorly
resolved region is concerned, at low redshift (0 and 0.5) the gas density is
lower for EFF and SSF than the DEL cases, simply because more gas has been
transformed into stars in the two models.  More marked differences are visible
at higher redshift (1 and 2) even at resolved scales up to $\sim$40 $h^{-1}$
kpc, especially in the temperature profiles: they clearly show the signature
of the different star formation prescriptions used.  The behaviour of
temperature profiles is overall similar to that shown by the isolated cases,
with SSF having cold star-forming gas, EFF having warmer gas with temperatures
given by the effective model, and DEL showing the hottest
temperature. Moreover, DEL influences the temperature profile to several tens
of kpc, in a way that resembles Me12 (Fig. 2.8). Indeed, at such high
redshift the mass of the main progenitor is still $\sim 3 \cdot 10^{13}
h^{-1}$ {\msun}, but with a higher average density that enhances the effect of
feedback.

In most cases differences are limited to sub-softening scales, and no relevant
difference is visible in any case beyond 50 $h^{-1}$ kpc; moreover, no star
formation and feedback model produces a roughly isothermal ``cool'' core: the
temperature peaks at $\sim 20-60\ h^{-1}$ kpc and drops to low values at
smaller radii.  This is in line with the well-known result that no stellar
feedback scheme is able to significantly influence the intra-cluster medium at
large scales.

The DEL feedback scheme is able to influence the ICM at some tens of kpc, and
to limit the fraction of cold baryons from $\sim$25\%, which is incompatible
with the typical 10\% value of clusters (Balogh et
al. 2001\nocite{Balogh01}, Lin et al. 2003\nocite{Lin03}),
to a lower value of 19\%. However, this is obtained at the cost of severely
suppressing star formation at high redshift and moving the peak of SFR to a
very low redshift. This is incompatible with the estimated age of cluster
galaxies, which form the bulk of their stars at $z > 2$.  

Other groups, e.g, Governato et al. 2007, didn't find unrealistic star
formation 
histories when using SF schemes similar to our DEL one.  The main difference
with our result, is that they simulate the cosmological formation of a {\it
galaxy-sized} halo, characterised by an early formation epoch and a high spin
parameter. The gas there is shock-heated at high redshift, then it cools down
and settles in a rotation-supported cold disk with low SFR and consequently a
low level of feedback. For a cluster-sized halo, merging is active up to low
redshift and no rotation-supported disk forms.  As a consequence, the
behaviour typical of our non-rotating isolated halos is the predominant one
and this causes an unrealistically delayed star 60.

\begin{figure}
\centering{
\includegraphics[viewport = 60 48 600 600,clip,height=8.5cm]{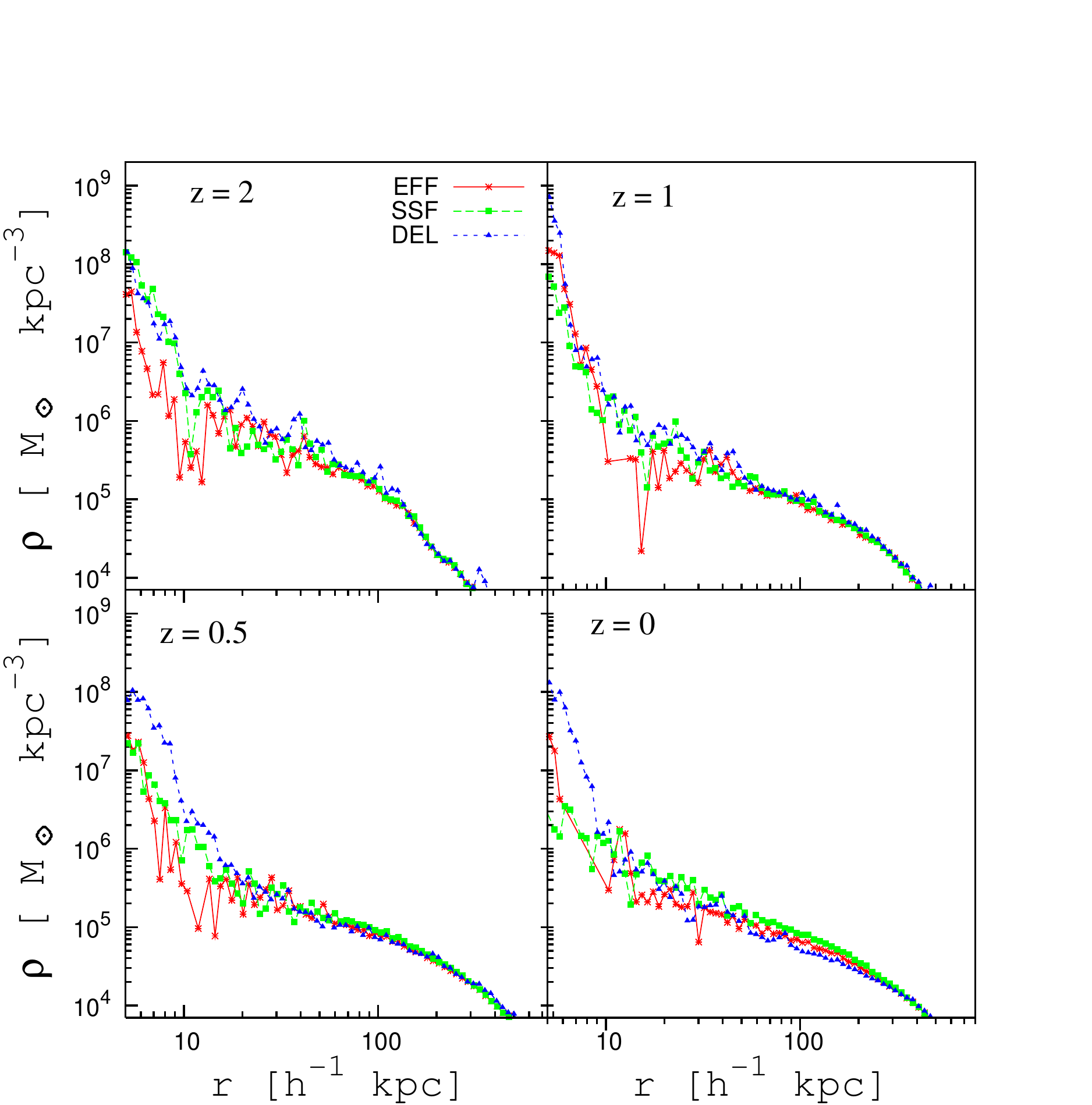}
\captionsetup{font={normalsize,sf}, width= 0.9\textwidth}
\caption{Evolution of the gas density in the cosmological run for the
  three models studied. The vertical line marks the scale
  corresponding to three softening lengths.}}
\label{Cosmo_gas}
\end{figure}

\begin{figure}
\centering{
\includegraphics[viewport = 60 48 600 600,clip,height=8.5cm]{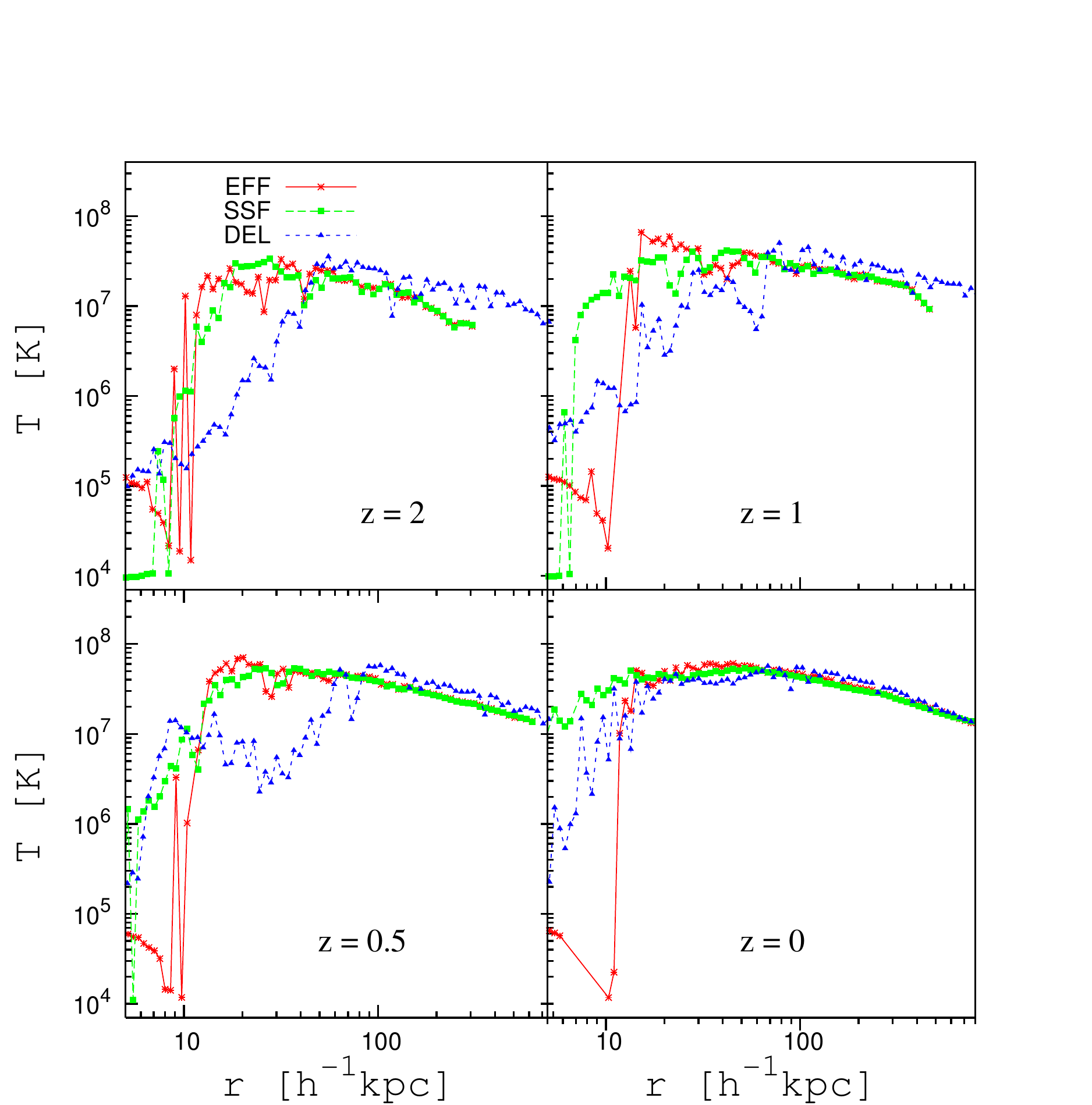}
\captionsetup{font={normalsize,sf}, width= 0.9\textwidth}
\caption{Evolution of the gas temperature in the cosmological run for the
  three models studied. The vertical line marks the scale
  corresponding to three softening lengths.}}
\label{Cosmo_temp}
\end{figure}

\subsection {Conclusions}

In this section, we presented a comparison of different star formation
and SNe feedback 
prescriptions in NFW isolated, non-rotating DM halos of mass $10^{13}$ and
$10^{12}$ {\msun} and in a $1.6\cdot 10^{14}$ {\msun} cosmological DM halo,
using the TREE+SPH code GADGET-2.  Usually, such tests of star formation
models are performed on disk galaxies, where gas is rotation-supported.  Our
settings are remarkably different: in our isolated halos a cooling flow is
established soon after the beginning of the simulation, and gas is
pressure-supported when it is not falling to the centre of the halo. This is
similar to what happens in a galaxy cluster, like in our cosmological case.
So this represents a complementary test with respect to the standard disk
formation tests.

We chose to test the GADGET effective model (EFF, SH03), which
is based on a multi-phase description of the gas contained in a particle, an
implementation of the scheme proposed by Katz et al. 1996 (SSF, KA96),
 where feedback
energy is given to star-forming particles in the form of thermal energy, and
an implementation of Thacker \& Couchman 2000 with the improvements of
Stinson et al. 2006 (DEL). In this last scheme, after a star formation event,
the gas gains energy from SNe explosions and cooling is quenched for a fixed
period of time of 30 Myr. For simplicity, we did not consider feedback in the
form of kinetic energy.

Our main conclusions are:

\begin{itemize}
\item In the EFF and SSF models thermal feedback is inefficient
  and unable both to limit star formation and to influence the gas
  dynamics beyond the gravitational softening scale. 
\item In isolated non-rotating halos, the effect of delaying
  radiative cooling in a cooling flow is to drastically reduce the
  amount of formed stars. This happens because the simultaneous
  cooling of a large amount of gas triggers an intermittency
  in star formation, with intense episodes followed by periods in
  which the whole gas is heated up and no star formation is allowed.
\item Delaying radiative cooling in a cooling flow when using thermal
  feedback has little effect on the thermodynamics of the
  intra-cluster medium at length scales larger than some softening
  lengths, as expected. Temperature and density profiles do not differ
  much between runs with delayed and normal cooling.  For the smaller
  isolated halo and for the cosmological halo at moderately high
  redshift, compression waves are visible in the temperature profiles;
  they are signatures of the intermittent star formation behaviour.
\item A cosmological simulation of formation and evolution of a $\sim
  10^{14}$ {\msun} galaxy cluster, using the DEL scheme, shows trends
  in SFRs, cold fractions and density and temperature profiles which
  resembles the one we obtained in our isolated cases.
\item In the cosmological simulation, in particular, the peak in the star
  formation history for the DEL scheme is at a redshift $z<1$, in
  disagreement with observations. The cold fraction is instead relatively
  low, 19\% at redshift $z$=0; this is due to the late star
  formation.
\item No model is able to produce a roughly isothermal core like 
  those observed in cool-core clusters.
\item EFF and SSF schemes gives very similar result also in the
  cosmological test case with too much gas locked in the stellar component.
\end{itemize}

We conclude that, while SN feedback in EFF and SSF schemes are not efficient
in countering the cooling flow and gas radiative losses at the halo centre,
the DEL scheme proves extremely effective at doing so. The cost of it is an
unrealistic delay in the star formation history.  While star formation and
feedback schemes which turn off radiative cooling have proved to be effective
at producing realistic disk galaxies in cosmological simulations, caution
should therefore being used when utilising similar schemes as general-purpose
ones.

\chapter{The origin of IntraCluster stars in cosmological simulations
  of galaxy clusters}\label{ICL}
\footnote{Murante, Giovalli, Gherard, Arnaboldi,
Borgani, Dolag, 2007, MNRAS, 377, 2}  
Observations of diffuse intracluster light and individual intracluster
stars in nearby clusters
(Arnaboldi et al. 2002\nocite{Magda02}, Arnaboldi et
al. 2003\nocite{Magda03}, Arnaboldi et al. 2004\nocite{Magda04},
Feldmeier et al. 2003\nocite{Feld03},
Mihos et al. 2005\nocite{MihosVirgo}, Gerhard et
al. 2005\nocite{Ortwin}) and at intermediate redshift (Gonzales et
al. 2000\nocite{Gonzales2000}, Feldmeier et al. 2004\nocite{Feld04},
Zibetti et al. 2005\nocite{Zibetti}) indicate 
that a substantial fraction of stars becomes unbound from galaxies as
these fall towards the densest parts of their cluster environment.

The radial distribution of the intracluster light (ICL) in galaxy
clusters is observed to be more centrally concentrated than that of
the cluster galaxies (Zibetti et al. 2005), a result which was predicted
from cosmological hydrodynamical simulations of galaxy clusters
(Murante et al. 2004\nocite{M04}, M04 hereafter). Zibetti et al. 2005
also find that the 
surface brightness of the ICL correlates both with BCG luminosity and
with cluster richness, while the fraction of the total light in the
ICL is almost independent of these quantities. Other observations
indicate an increase of the relative fraction of diffuse stars from
the mass scale of loose groups (less than 2 \%,
Castro-Rodr{\'{\i}}guez et al. 2003\nocite{Castro-Rodr},
Feldmeier et al. 2003\nocite{FeldIAU03})
to that of Virgo-like clusters ($\approx 5-10$\% Feldmeier et
al. 2003\nocite{Feld03}, Arnaboldi et al. 2003,
Mihos et al. 2005) up to the most massive clusters 
(10-20 \% or higher Gonzales et
al. 2000, Feldmeier et al. 2002\nocite{Feld2002},
Gal-Yam et al. 2003\nocite{GalYam}, Feldmeier et al. 2004, Krick et
al. 2006\nocite{Krick06}). 

The origin and evolution of this diffuse stellar component (DSC) is
currently unknown and several mechanism are being investigated. The
ICL may be produced by stripping and disruption of galaxies as they
pass through the central regions of relaxed clusters
(Byrd $\&$ Valtonen 1990\nocite{ByrdValt}, Gnedin
2003\nocite{Gned03}). Other mechanisms are the stripping of stars 
from galaxies during the initial formation of clusters
(Merritt 1984\nocite{Merritt84}); creation of stellar halos in galaxy
groups, that later fall into massive clusters, and then become unbound
(Mihos 2004\nocite{Mihos04}, Rudick et al. 2006\nocite{Rudick});
stripping of stars during high--speed galaxy 
encounters in the cluster environment (Moore et
al. 1996\nocite{Moore96}). Evidence for 
ongoing stripping from elliptical galaxies in clusters was
presented by Cypriano et al. 2006\nocite{Cypr06}.

In parallel, numerical simulations have been performed to investigate
the properties the DSC in galaxy clusters within the current
cosmological models. Napolitano et al. 2003\nocite{Napo03} used
Dark-Matter (DM) only 
simulations, and identified the stellar component using the DM
particles as tracers. For the first time, M04 used a $\Lambda$CDM
cosmological hydrodynamical simulation, including radiative cooling
and star formation, to quantify the amount and the distribution of the
DSC in a set of 117 clusters. Willman et al. 2004\nocite{Fabio} and
Sommer-Larsen et al. 2005\nocite{SommerLarsen}
found a DSC in their simulated single clusters. Willman et al. 2004
discussed the origin of the DSC: they found a correlation
between the cluster growth and the increase in the DSC mass, and that
both massive and small galaxies contribute to its formation.

Recently, Rudick et al. 2006 performed collisionless simulations where
high--resolution model galaxies were inserted in their dark matter halos
at a given redshift, and then their common evolution in a cluster was
followed from that time on. A DSC was formed, and Rudick et al. 2006 found
that the cluster DSC grows with the accretion of groups during the
cluster history.

In this work, we focus on the formation mechanism of the ICL in a
cosmological hydrodynamical simulation (Borgani et al. 2004, M04).
The formation of galaxies and their subsequent dynamical evolution in
a time dependent gravitational potential is a direct consequence of
the hierarchical assembly process of cosmic structures.  Using a large
($192^3 h^{-3}$ Mpc$^3$) volume simulation, we study a statistically
significant ensemble of galaxy clusters and follow how stars become
unbound from galaxies during the evolution of clusters as a function of
cosmic time. We also address the stability of our results against
numerical resolution by carrying out the same analysis on three
clusters from this set, which were re-simulated at a substantially
improved force and mass resolution.

The plan of the paper is as follows: in Section~\ref{simclus} we describe our
numerical simulations and in Section~\ref{skidid} we give details on the
galaxy identification and properties. In Section~\ref{dscsection} we describe
the identification of the diffuse stellar component (DSC).  In
Section~\ref{sdcorigin} we present the link between galaxy histories and the
formation of the DSC; in Section~\ref{reseffect} we discuss how resolution and
other numerical effect may affect our results; in Section~\ref{secunbound} we
discuss the dynamical mechanisms that unbind stars from galaxies in clusters
and compare with the statistical analysis of the cosmological simulation
performed in the previous Sections.  In Section~\ref{concl} we summarise our
results and give our conclusions.

\section{The simulated clusters}
\label{simclus}
The clusters analysed in this paper are extracted from the large
hydrodynamical simulation (LSCS) of a ``concordance'' $\Lambda$CDM
cosmological model ($\Omega_m=0.3$, $\Omega_\Lambda=0.7$, $\Omega_{\rm
b}=0.019\,h^{-2}$, $h=0.7$ and $\sigma_8=0.8$). This simulation is
presented in Borgani et al. 2004 and we refer to that paper for
additional details. The LSCS is carried out with the massively
parallel Tree+SPH code {\small GADGET2} (Springel et al. 2001\nocite{GADGET}, 
Springel 2005\nocite{GADGET2}), and
follows $480^3$ dark matter particles and as many gas particles in a
periodic box of size $192\, h^{-1}$ Mpc. Accordingly, the mass
resolution is $m_{\rm dm}=4.6 \times 10^9\, h^{-1} M_\odot$, $m_{\rm
gas}=6.9 \times 10^8 h^{-1} M_\odot$ and $m_{\rm star}=3.465 \times
10^8 h^{-1} M_\odot$. The Plummer--equivalent softening length for the
gravitational force is set to $\epsilon=7.5\, h^{-1}$ kpc, fixed in
physical units from $z=0$ to $z=2$, while being fixed in co-moving
units at higher redshift. The SPH softening length of the gas is
allowed to shrink to half the value of the gravitational force
softening. The simulation includes radiative cooling, the effect of a
photo--ionising uniform UV background, star formation using a
sub-resolution multi-phase model for the interstellar medium (Springel
\& Hernquist 2003\nocite{SpringHern03}), feedback from supernovae (SN)
explosions, 
including the effect of galactic outflows. The velocity of these
galactic winds is fixed to $v_w\simeq 340\vel$, which corresponds to
50$\%$ efficiency for SN to power the outflows.

Clusters are identified at $z=0$ using a standard friends-of-friends
(FOF) algorithm, with a linking length of $0.15$ times the mean dark
matter inter-particle separation. We identify 117 clusters in the
simulation with $M_{FOF}>10^{14}h^{-1}M_\odot$. Cluster centres are
placed at the position of the DM particle having the minimum value of
the gravitational potential. For each cluster, the virial mass
$M_{\rm vir}$ is defined as the mass contained within a radius
encompassing an average density equal to the virial density,
$\rho_{\rm vir}$, predicted by the top--hat spherical collapse
model. For the assumed cosmology, $\rho_{\rm vir}\simeq 100\rho_c$,
where $\rho_c$ is the critical cosmic density (e.g. Eke et
al. 1996\nocite{Eke96}). 

To test the effects of numerical resolution on the final results, we
select three clusters, having virial masses $M_{\rm vir} =1.6,2.5,2.9
\times 10^{14} h^{-1}$M$_\odot$, and re-simulate them twice with
different resolution. While the first, lower-resolution simulation is
carried out at the same resolution as the parent simulation, the
second simulation had a mass resolution $45 \times$ higher, with a
correspondingly smaller softening parameter, $\epsilon=2.1\, h^{-1}$
kpc. These re-simulations are performed using more efficient SN
feedback, with a wind velocity $v_w\simeq 480\vel$. A detailed
description of these re--simulations is provided by Borgani et
al. 2006\nocite{BorgNum}.

\section{Identifying galaxies in a cluster with SKID} 
\label{skidid}

The identification of substructures inside halos is a longstanding
problem, which is not uniquely solved.  In the present work, we need
to identify galaxies in the simulations from the distribution of star
particles which fill the volume of the cosmological simulation.

In the LSCS, ``galaxies'' are defined as self---bound, locally
over-dense structures, following the procedure in M04, which is based
on the publicly available SKID algorithm\footnote{See {\tt
http://www-hpcc.astro.washington.edu/tools/skid.html} } (Stadel
2001\nocite{SK}).  At 
a given redshift, once the star particles have been grouped by SKID,
we classify as galaxies only those groups which contain at least 32
bound star particles. There is a degree of uncertainty in the galaxy
identification by SKID, as in other similar identification algorithms,
which comes in from the assignment of those star particles which are
located in its outskirts of each self-bound object.  The main
advantage of this identification algorithm is that it provides a
dynamically--based, automated, operational way to decide whether a
star particle belongs to a gravitationally bound object or not.
Additional details of the galaxy identification algorithm and on our
tests are given in the Appendix.

We expect that, once a self-bound structure of luminous particles has
been formed at a given redshift, most of its mass will remain in bound
structures, for all subsequent redshifts. However, it may happen that
a group of particles classified as a ``galaxy'' at one output
redshift, with a number of particles just above the specified minimum
particle threshold for structure identification, may fall below this
limit at the next redshift output. This may occur, for example,
because the group is evaporated by interaction with the environment.
Following Springel et al. 2001\nocite{Springel2001MT}, we call
structures that can be 
identified only at one output redshift {\sl volatile}, and do not
consider them further.

All star particles that never belong to any galaxies identified in the
selected redshift outputs are also assigned to this volatile
class. Such star particles either do not belong to any bound structure
already at the first output redshift, at $z=3.5$, or they form in a
galaxy {\sl and} become unbound between two simulation output
redshifts. In both cases, since we cannot assign those stars to the
history of any galaxies, we cannot determine their dynamical origin.

An important issue in our study concerns the reliability of the
simulated galaxy population.  If galaxies are under-dense, they can
easily lose stars or be completely disrupted as a consequence of
numerical effects. In simulations, low-mass galaxies may have typical
sizes of the order of the adopted softening parameter, so that their
internal mass density is underestimated. Therefore at the low-mass
end, we expect that our simulated galaxies will have an internal
density which is an {\it increasing} function of galaxy mass. On the
other hand, numerical effects should be less important for the more
massive galaxies.

\begin{figure}
\centering
\includegraphics[width=0.33\linewidth]{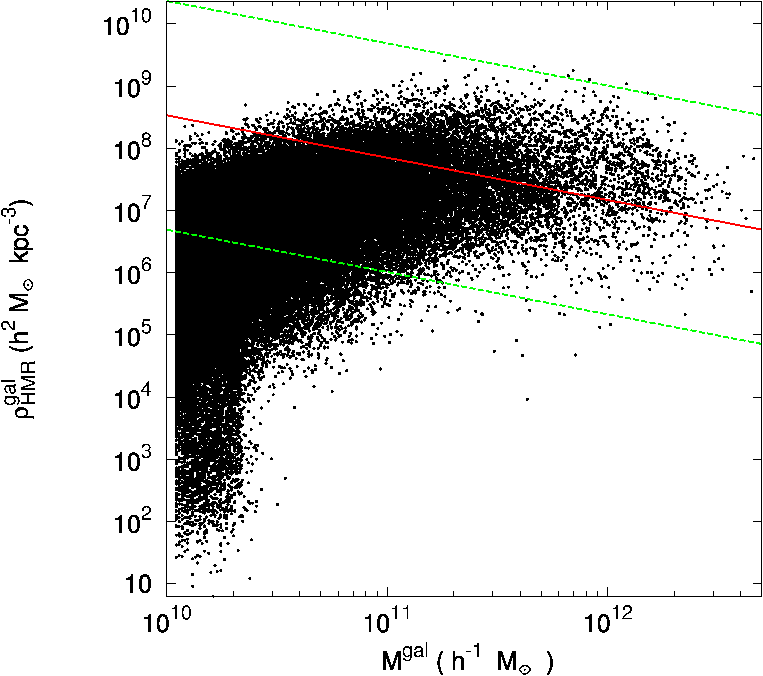}
\captionsetup{font={normalsize,sf}, width= 0.9\textwidth}
\caption{ The mean galaxy stellar density inside the half-mass-radius
  as a function of galaxy mass. All galaxies identified in the parent
  simulation at all 17 redshift outputs areshown. The solid line
  shows an estimate of the observed galaxy densities from SDSS data
  (see text). The dotted lines show the densities corresponding to the
  $3 \sigma$ scatter reported in Shen et al. 2003.  }
\label{galdens}
\end{figure}

To investigate this issue, we evaluate the stellar density of all the
simulated galaxies at the half-mass-radius, and plot these in
Figure~\ref{galdens} as a function of the galaxy mass, combining all
redshift outputs.
\nocite{Shen}
For real galaxies, the internal density of (early--type) galaxies is a
decreasing function of their mass, as shown most recently by Shen et
al. 2003, who measured the size distribution for 140,000 galaxies 
from the Sloan Digital Sky Survey. We use their measured size
distribution to estimate the observed galaxy densities within the
half-mass-radius. For this purpose, we take the expressions for a
Hernquist profile in Hernquist 1990\nocite{Hern} to relate Sersic
half-light radii to 
three-dimensional half-mass radii, and then convert the Sersic
size-stellar mass relation for early-type galaxies of Shen et al. 2003 to a
relation between stellar mass and mean density within the half-mass
radius.  The solid line in Fig.~\ref{galdens} represents the resulting
estimate of the mean galaxy density, with the dotted lines limiting
the $3 \sigma$ scatter of the size distribution as reported in Shen et
al. 2003.  The dots in Fig.~\ref{galdens} show the equivalent mean 
densities of our simulated cluster galaxies.  In what follows, we use
the lower $3 \sigma$ envelope to estimate the minimum acceptable
galaxy densities $\rho_{\rm min}(M_{\rm gal})$.  Galaxies with density
lower than this minimum density are discarded and also classified as
volatile.  From Fig.~\ref{galdens} we note that the observed trend of
decreasing density with increasing mass is recovered in our
(lower-resolution) parent simulation for galaxy masses $\magcir
10^{11} h^{-1} M_\odot$.

In order to quantify the effect of volatile galaxies on our final
results, we tested other density thresholds, namely (i) one
corresponding to $1 \sigma$ scatter in the $R_e$ distribution, (ii) a
fixed value of $\rho = 5 \cdot 10^6 h^2 M_\odot$/kpc$^3$, as well as
(iii) a galaxy mass threshold, $M = 6 \cdot 10^{10} h^{-1}
M_\odot$. Our results remain qualitatively unchanged when either of
these criteria is adopted.

\section{Identifying the Diffuse Stellar Component} 
\label{dscsection}

The star formation model implemented in our simulations is based on a
gas--density threshold criterion (Springel \& Hernquist 2003). This
ensures that stars can only form inside existing gravitational
potential wells, so that star formation does not take place outside DM
halos. Thus DSC stars must have become unbound from their parent
galaxies sometime after their formation. Therefore in our analysis, we
define as diffuse stellar component (DSC) all those star particles
which (i) do not belong to any self-bound galaxy at $z=0$, (ii) were
part of a non-volatile structure at earlier redshifts whose
density exceeded the minimum density for its mass as defined above.

In surface brightness measurements of the DSC, sometimes a
distinction is attempted between the component associated with the
halo of the central dominant (cD) galaxy and the intra--cluster light,
which fills the whole cluster region. Quoting from Uson et
al. 1991\nocite{Uson91}:
''{\it whether this diffuse light is called cD envelope or
diffuse intergalactic light is a matter of semantics: it is a diffuse
component distributed with elliptical symmetry at the canter of the
cluster potential}''. In our analysis, we will not make such a
distinction: all star particles that do not belong to any self-bound
galaxy at $z=0$, including the cD galaxy identified by SKID, are part
of the diffuse stellar component if they were once part of a
non-volatile, above minimum-density structure.

The part of the DSC contributed by galaxies which have a central
density lower than $\rho_{\rm min}$ is not considered in our analysis,
because it is most likely affected by numerical effects. These
low-density structures include a population of extremely low--density
objects found by SKID at the very low mass end, many of them
representing a mis--identification of SKID due to their small number
of particles ($<100$). However, by discarding the contribution from
low-density and volatile galaxies, we may also neglect a possibly
genuine contribution to the DSC from a population of low--mass
galaxies. Because of this, our estimate of the diffuse light fraction
in the simulated clusters may be an underestimate, although we believe
the corresponding bias to be relatively small; we shall discuss this
issue in Sec.~\ref{reseffect}.

\begin{figure}
\centering
\includegraphics[viewport=60 70 580 550,clip,
  height=8.cm]{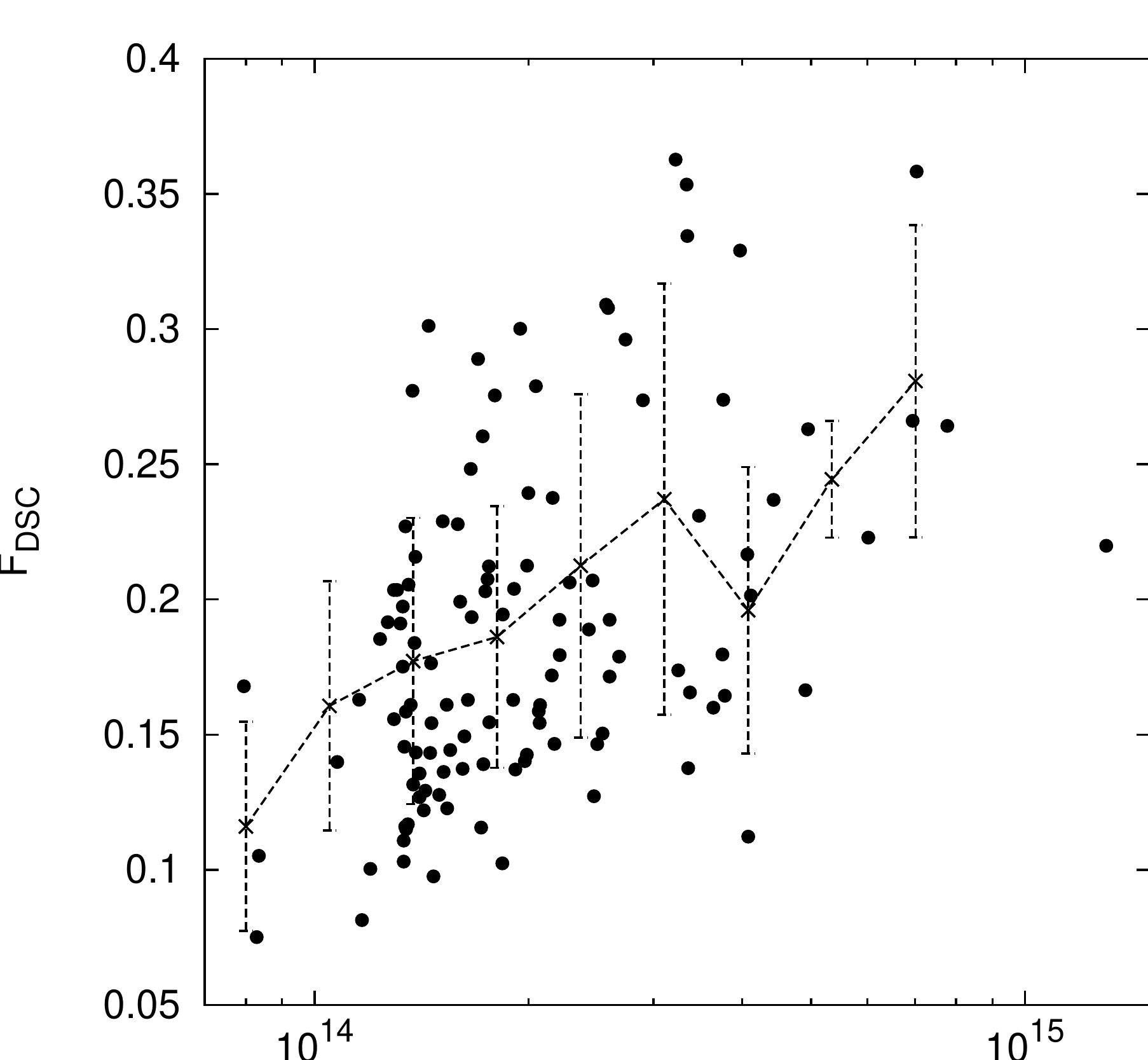}
\captionsetup{font={normalsize,sf}, width= 0.9\textwidth}
\caption{The fraction of stellar mass in the DSC relative to the total
  stellar mass as a function of the cluster virial mass. Dots are for
  the 117 clusters in our parent simulation. The crosses show the
  average values of this ratio in different mass bins, with the
  error bars indicating the r.m.s. scatter within each bin.  }
\label{newsdcfrac}
\end{figure}

Figure~\ref{newsdcfrac} shows the fraction $F_{\rm DSC} = M^*_{\rm
DSC}/M^*_{\rm tot}$, where $M^*_{\rm DSC}$ is the stellar mass
in the DSC and $M^*_{\rm tot}$ is the total stellar mass found within
$R_{\rm vir}$ for each cluster in the parent simulation, as a function
of the cluster mass. In this computation, the diffuse star particles
from volatile galaxies have been discarded. We report on the
fractions of DSC stars in all the steps of our selection procedure in
Table~\ref{clustab}.

Consistent with the results shown in M04, we find: 1) The fraction of
DSC relative to the total stellar light in clusters 
increases with cluster mass, albeit with a large scatter, and 2)
the DSC fraction in the simulated clusters is in the range
$0.1 < F_{\rm DSC}^{obs} < 0.4$. Both results are broadly consistent with
the observed trends and values (Arnaboldi et al. 2004, Aguerri et al
2005\nocite{Aguerri2005}, Aguerri et al. 2006\nocite{Aguerri2006}); 
however, a direct comparison between observed and measured
values of $F_{\rm DSC}$ is only qualitative, because simulations provide
the volume--averaged mass fraction directly, while this is not true
with the observed DSC fractions.

\section{Tracing the origin of the DSC}
\label{sdcorigin}

The large set of simulated clusters extracted from the cosmological
simulation allows us to perform a statistical study of the origin of
the DSC. From Fig.~\ref{newsdcfrac}, it is clear that clusters with
similar mass can have rather different amounts of DSC at $z=0$. For
this reason, we will first address the general trends in the origin of
the DSC that are independent of the characteristics and dynamical
history of individual clusters, such as the redshift at which the most
of the DSC becomes unbound, and from which galaxies the intracluster stars
mainly originate. We will then investigate whether significant
differences in the production of the DSC can be found between clusters
belonging to different mass classes, and discuss the robustness of our
results against numerical resolution.

We study the origin of the DSC by adopting the following strategy: we
follow back in time all the particles in the DSC component at $z=0$
within each cluster's virial radius and associate them with bound
structures present at any earlier redshifts.  For all clusters and the
17 redshift outputs (from $z=0$ to $z=3.5$), we compile the list of
all galaxies as described in Sect.~\ref{skidid}. Subsequently, for
each DSC particle at $z=0$, we check whether it belong to any of these
galaxies at earlier redshift. If no galaxy is found, the DSC particle
is discarded, because we cannot establish its origin.
If a galaxy is found, then there are three options:
\begin{itemize}  
\item This galaxy has a central density larger than the adopted
  threshold and it belongs to the ``family tree'' of a galaxy
  identified at $z=0$ (see the next subsection); the DSC particle is
  then associated with that family tree.
\item This galaxy has a central density larger than the adopted
  threshold, but it does not belong to a family tree of any galaxy at
  $z=0$; the DSC particle is then considered to come from a
  ``dissolved'' galaxy.
\item This galaxy has a central density below the adopted
  threshold and is thus considered as ``volatile'';  the DSC particle
  is then discarded.
\end{itemize}
In this way, the progenitors of all retained DSC particles can be
found.

\subsection{Building the family trees of galaxies}

We build the merger trees of all galaxies identified at $z=0$, and
refer to them as ``family trees'' to distinguish them from the standard DM
halo merger trees.  

The ``family trees'' are built as follows. For each output redshift $z_{i+1}$
of the simulations, we follow all the DM, star and gas particles within the
virial radius of the identified cluster at $z=0$.  We build catalogs of all
galaxies from the corresponding star and DM particles distributions. For a
given galaxy identified at redshift $z_{i+1}$, we tag all its star particles
and track them back to the previous output redshift $z_i$. We then make a list
of the subset of all identified galaxies at $z_i$ which contain the tagged
particles belonging to the specified galaxy at $z_{i+1}$.

We define a galaxy $G_i$, at output redshift $z_i$, to be progenitor of
a galaxy $G_{j}$ at the next output redshift $z_{i+1}$ if it
contains at least a fraction $g$ of all the stars ending up in
$G_{j}$. The definition of progenitor depends on the fraction $g$.
Our tests show that the number of galaxies identified as progenitors
is stable for $g$ values varying in the range 0.3--0.7. The value
adopted for our analysis is $g=0.5$, which is the same value adopted
in several reconstructions of the DM halo merger trees presented in
the literature (Kauffmann 2001\nocite{KauffProc}, Springel et
al. 2001, Wechsler et al. 2002\nocite{Wechsler2002}).

\begin{figure*}
\centerline{
\includegraphics[width=0.33\linewidth, height=9cm]
		{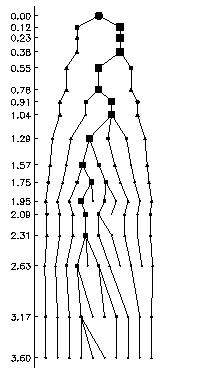}
\includegraphics[width=0.33\linewidth, height=9cm]
		{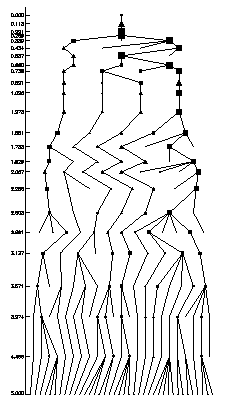}
\includegraphics[width=0.15\linewidth, height=9cm]
		{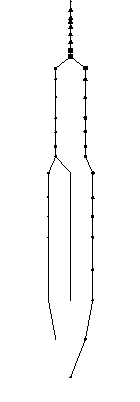}
\includegraphics[width=0.15\linewidth, height=9cm]
		{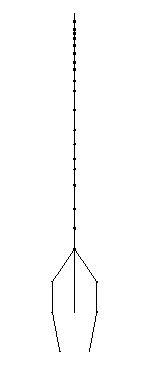}
}
\captionsetup{font={normalsize,sf}, width= 0.9\textwidth}
\caption{Left: Family tree of the cD galaxy of cluster A in the
  low-resolution simulation (see Table \ref{clustab}).  Right: Family
  trees of the cD galaxy, the third-most massive galaxy, and a
  lower-mass galaxy in the high-resolution re-simulation of the same
  cluster. -- The size of symbols is proportional to the logarithm of
  the mass of the galaxies at the corresponding redshift.  Shown on
  the vertical axis on the left are the output redshifts used to
  reconstruct the family trees; these are different in both
  simulations. A galaxy in these trees is considered a progenitor of
  another galaxy if at least 50\% of its stars are bound to its
  daughter galaxy, according to the SKID algorithm.  Many more
  galaxies can be identified in the high-resolution simulation at
  similar redshift. The cD family tree is characterised by one
  dominant branch with a number of other branches merging into it, at
  both resolutions.  Squares and triangles represent our
  classification of ``merging'' and ``stripping'' events, see
  Section~\ref{DSCmerging}. Circles correspond to redshift at which the
  galaxy is not releasing stars to the DSC. }
\label{famtreelow}
\end{figure*}

We then build the family trees for all galaxies found at $z=0$ in all
the 117 clusters of our sample. Given the adopted mass threshold of 32
star particle per galaxy, this amounts to an overall number of 1816
galaxies at redshift $z=0$, and 71648 galaxies in all redshift outputs.  

Figure~\ref{famtreelow} shows the family tree of the cD galaxy of a
cluster having virial mass $M{\rm vir}=1.6 \times 10^{14} h^{-1}
M_\odot$ (cluster A in Table~\ref{clustab}).  The cD galaxy family
tree is complex and resembles a typical DM halo merger tree, with the
cD being the result of a number of mergers between pre--existing
galaxies. Other galaxies have a much simpler formation history, with
fewer or no mergers of luminous objects.  This is
illustrated in the right part of Fig.~\ref{famtreelow} which shows
merger trees from the high-resolution re-simulation of the same
cluster, for the cD galaxy, the third-most massive galaxy in the
cluster, and a low-mass galaxy.  In more massive clusters, galaxies
whose family trees are intermediate between that of the cD and the
third-most massive galaxy can also be found. They are however
among the most massive galaxies in their cluster, and they are often
the most massive galaxy of an infalling subcluster, which has not
merged completely with the main cluster yet.

Once the family trees of all galaxies in our clusters at $z=0$ are
built, we then analyse the formation history of the DSC.

\subsection {The epoch of formation of the DSC}

\begin{figure}
\centering{
\includegraphics[viewport=130 85 630 490,clip,height=9.cm]{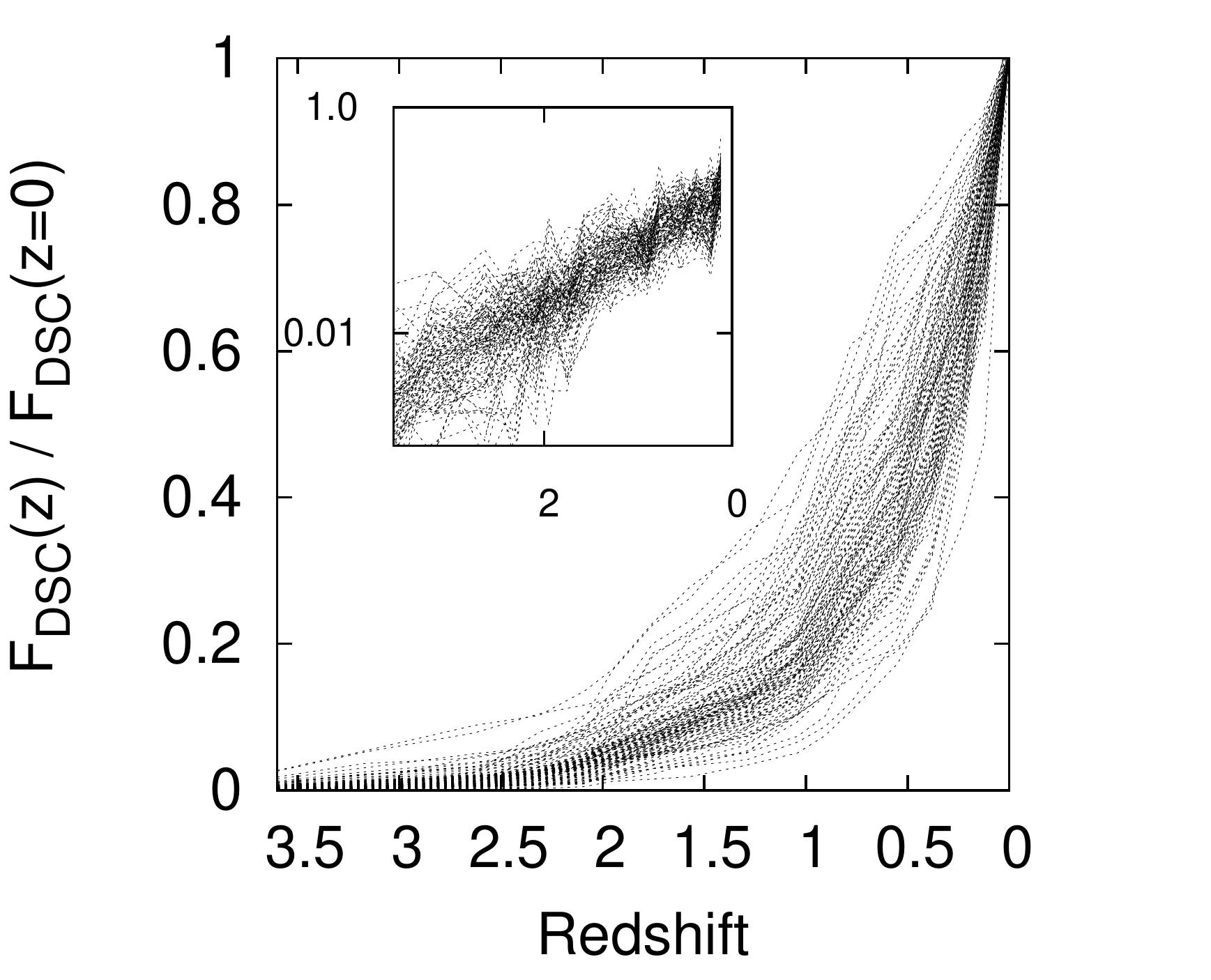}
\captionsetup{font={normalsize,sf}, width= 0.9\textwidth}
\caption{ Fraction of star particles found in the DSC at redshift
  $z=0$ which are already in the diffuse component at redshift $z$,
  for all clusters in our set. The inset show the same curves in
  logarithmic coordinates. }
}
\label{f:summstrip}
\end{figure}

As already discussed, the star formation model used in our simulations
implies that stars can only form inside existing gravitational
potential wells, so that all DSC stars must have become unbound from
their parent galaxies sometime after their formation. In
Fig. 3.4, we plot the fraction of star particles in
the DSC at $z=0$ which are already in the DSC at redshift $z$. The
bulk of the DSC is created after $z \approx 1$, when on average only
$\approx 30$ per cent of $z=0$ DSC star particles already reside 
outside their parent galaxies, with significant cluster-to-cluster
variations. However, from the inset of Fig. 3.4 we note
that the production of the DSC follows a power-law, thus implying that
it is a cumulative process which, on average, does not have a
preferred time scale. Willman et al. 2004 found a similar result based on the
analysis of their high--resolution simulation of a single cluster,
with a continuous growth of the DSC fraction and no preferred epoch of
formation.

No statistically significant correlation is found between the fraction
of DSC at $z=0$ and a number of possible tracers of the dynamical
history of the cluster, such as the concentration of the NFW profile,
the number of (DM--halo) major mergers, or the epoch of the last major
merger.  This suggests that the process of formation of the DSC is
more related to the local dynamics of the interactions between
galaxies and the group/cluster environment, rather than to the global
dynamical history of the cluster.

\subsection {DSC and the history of galaxies}

We now proceed to establish which galaxies are the main contributors to
the formation of the DSC. For each DSC star particle at $z=0$, we look
for a $G_j$ galaxy at $z_i$ to which it last belonged. When this
galaxy is found, we check whether the galaxy $G_j$ is associated with
the family tree of a galaxy $G_k$ at $z=0$. If so, then the DSC star
particle is associated with the ``family tree'' of the galaxy $G_k$.
If the $G_j$ galaxy at $z_j$ is not associated with the family tree of
any $G_k$ galaxies at $z=0$, but its family tree ends at $z_{j+m}$, then
the DSC star particle is associated with a dissolved galaxy. If no
bound structure is found, then the particle is associated with a {\sl
volatile} structure, and it is not considered in the subsequent
analysis.

As a next step, we compute what fraction of the DSC particles comes
from the family trees of galaxies at $z=0$, as a function of the
binned galaxy mass at $z=0$, $M_{\rm gal}(z=0)$. Then the DSC mass
$M^*_{\rm DSC}(M_{\rm gal})$ obtained for each $M_{\rm gal}(z=0)$ bin
is normalised by the total stellar mass of the respective cluster.
The total fraction $F_{\rm DSC}$ for each cluster is finally given by
the sum over all contributions from all galaxy masses at $z=0$.

\begin{figure*}
\centerline{
\includegraphics[width=0.31\linewidth]{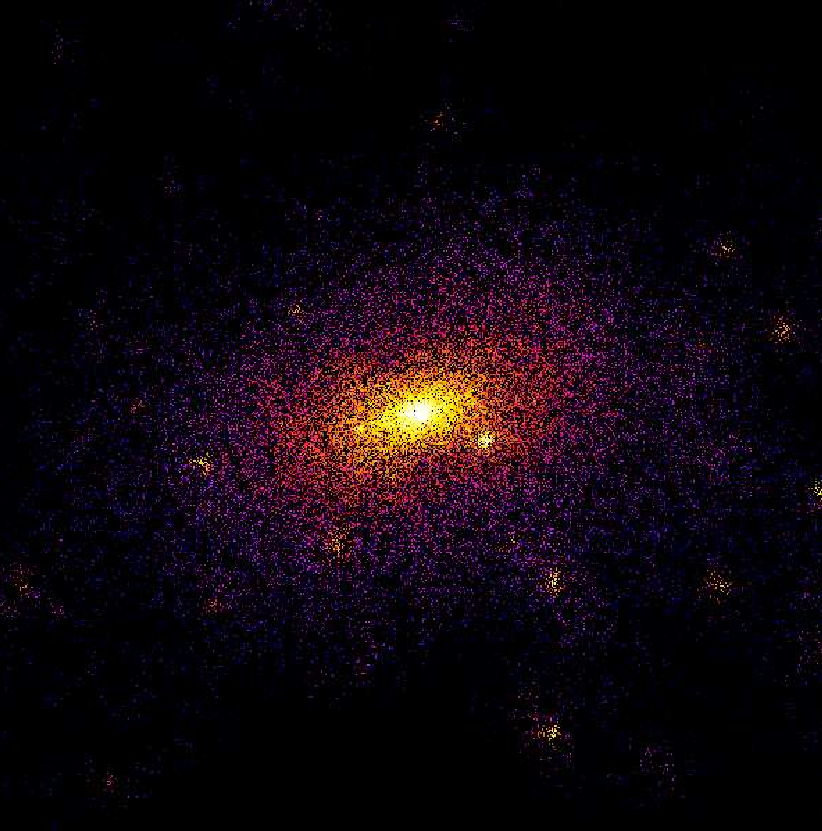}
\includegraphics[width=0.31\linewidth]{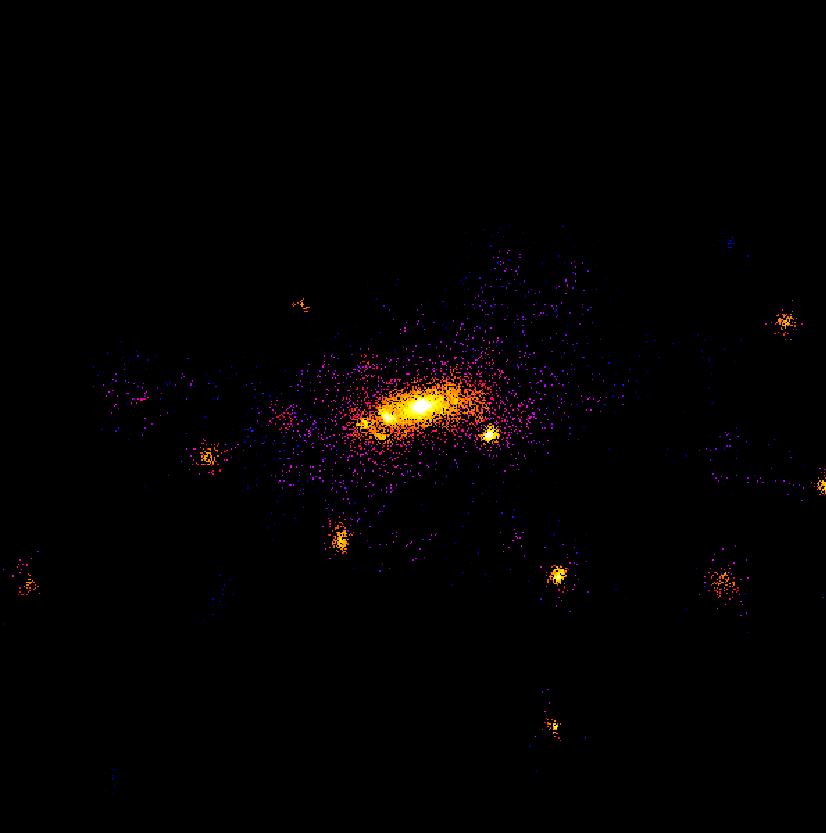}
\includegraphics[width=0.31\linewidth]{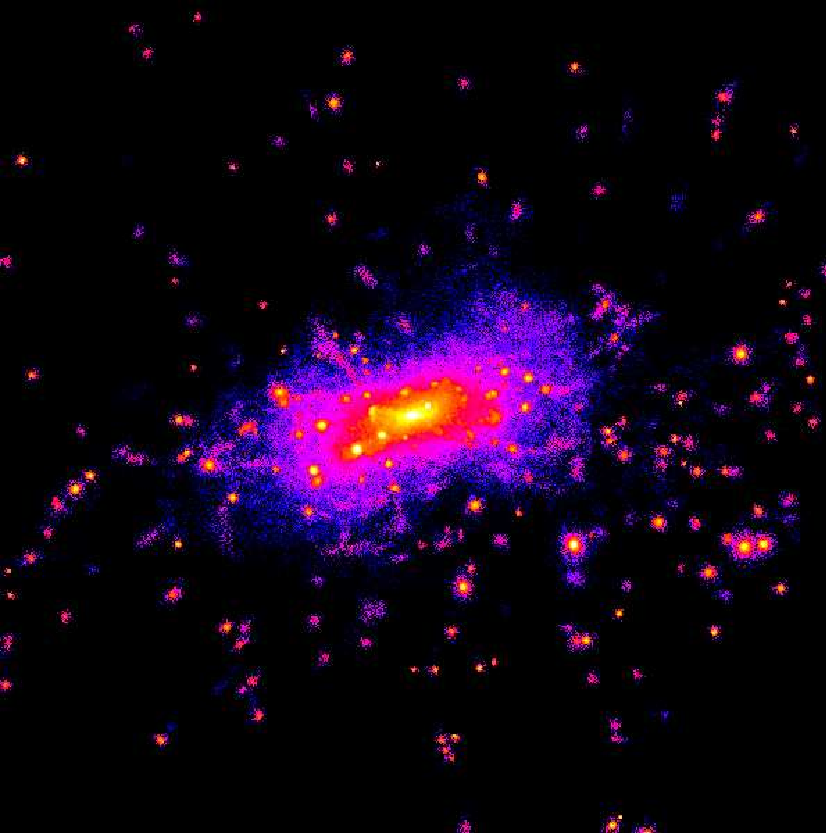}
}
\centerline{
\includegraphics[width=0.31\linewidth]{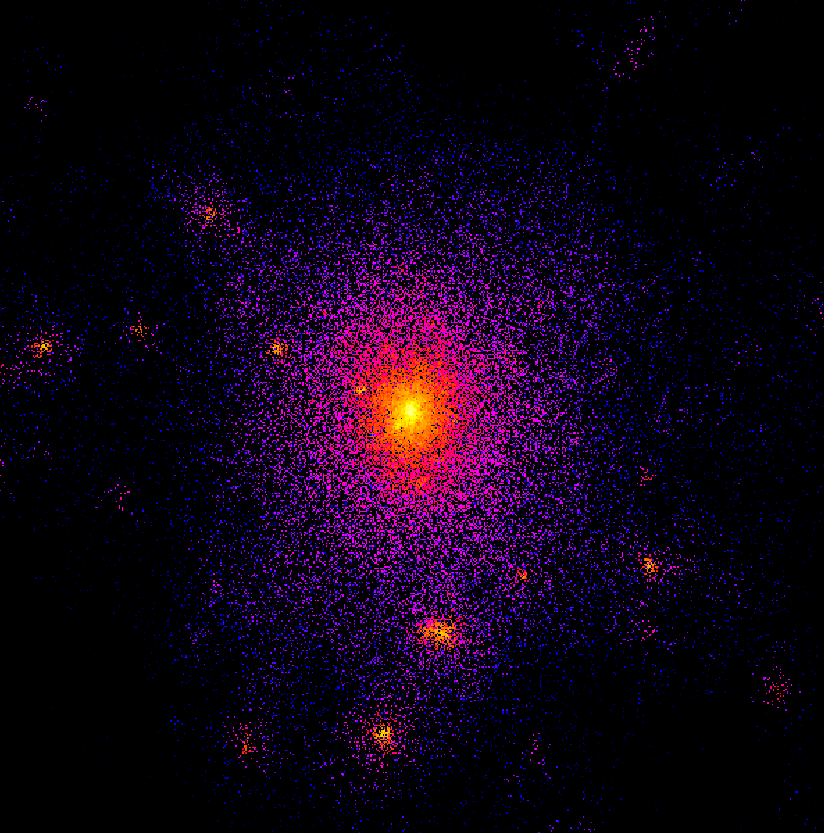}
\includegraphics[width=0.31\linewidth]{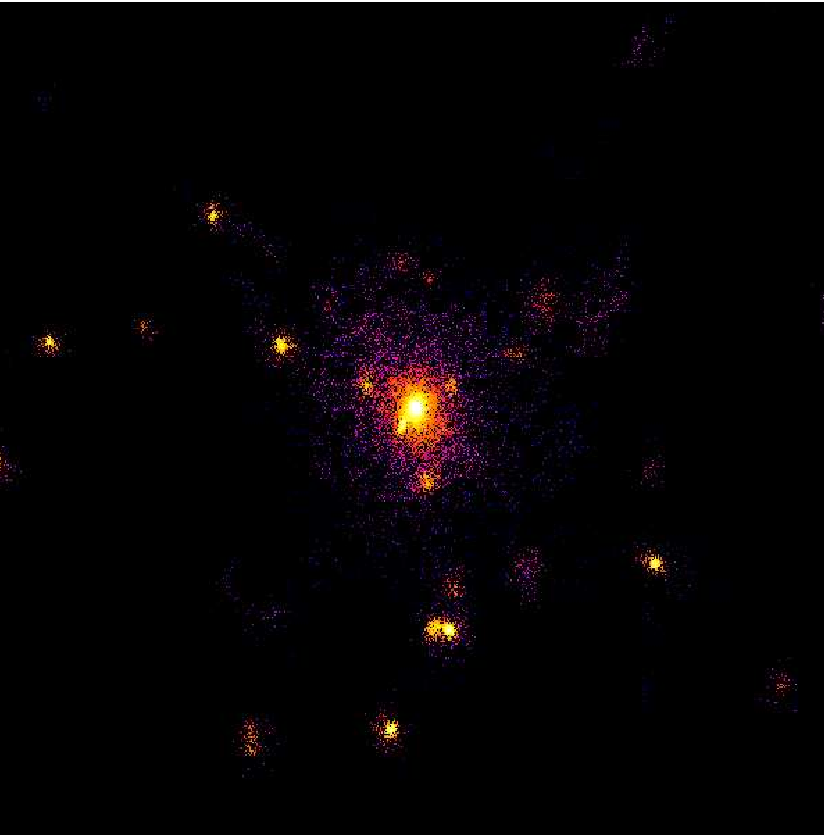}
\includegraphics[width=0.31\linewidth]{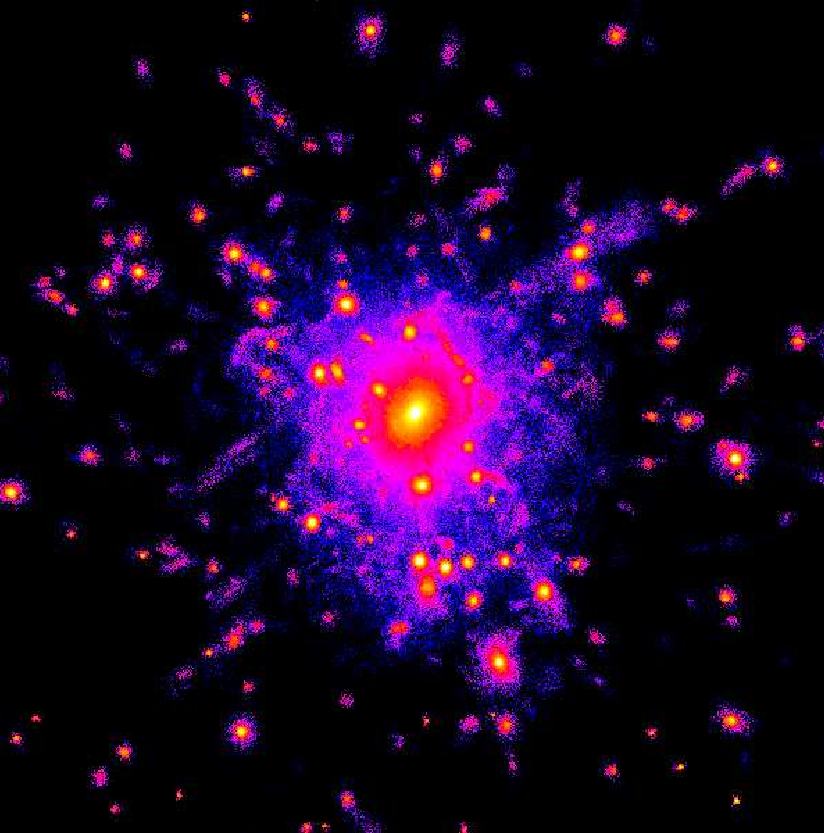}
}
\centerline{
\includegraphics[width=0.31\linewidth]{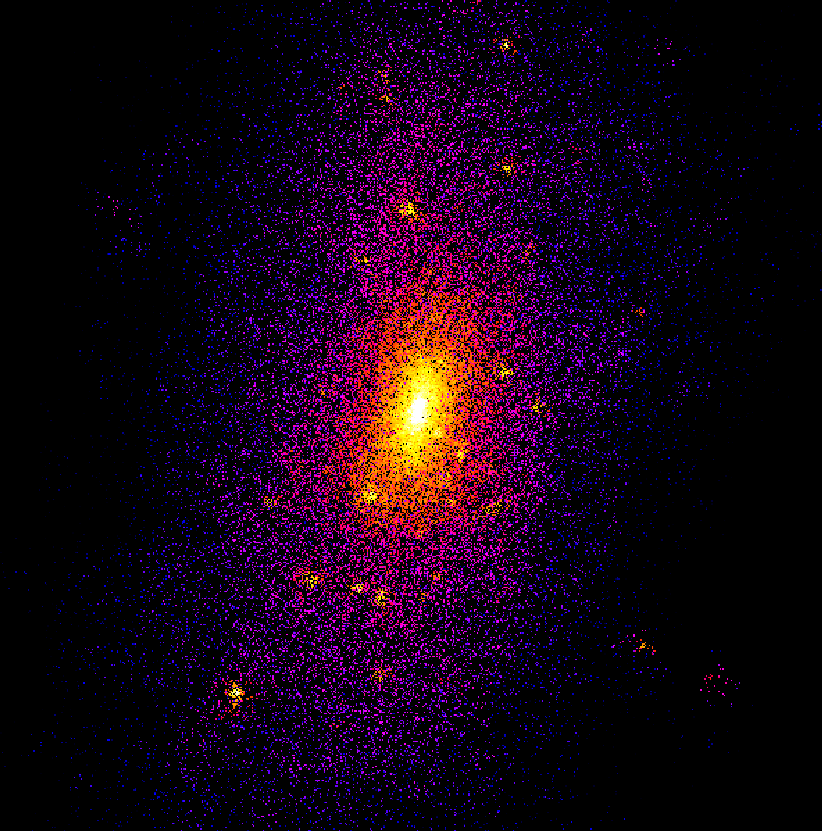}
\includegraphics[width=0.31\linewidth]{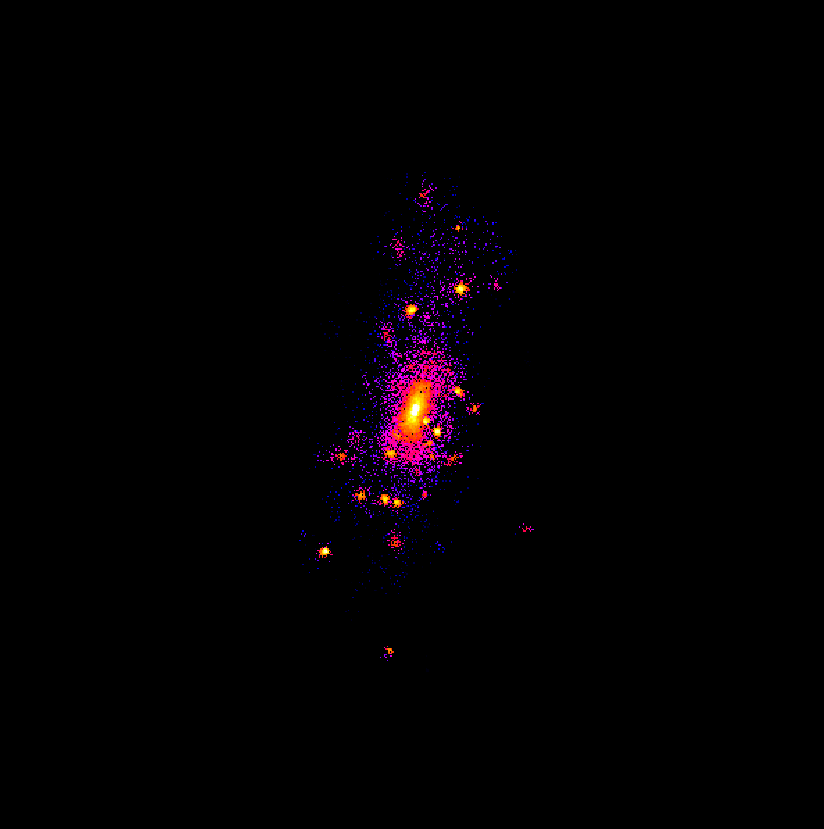}
\includegraphics[width=0.31\linewidth]{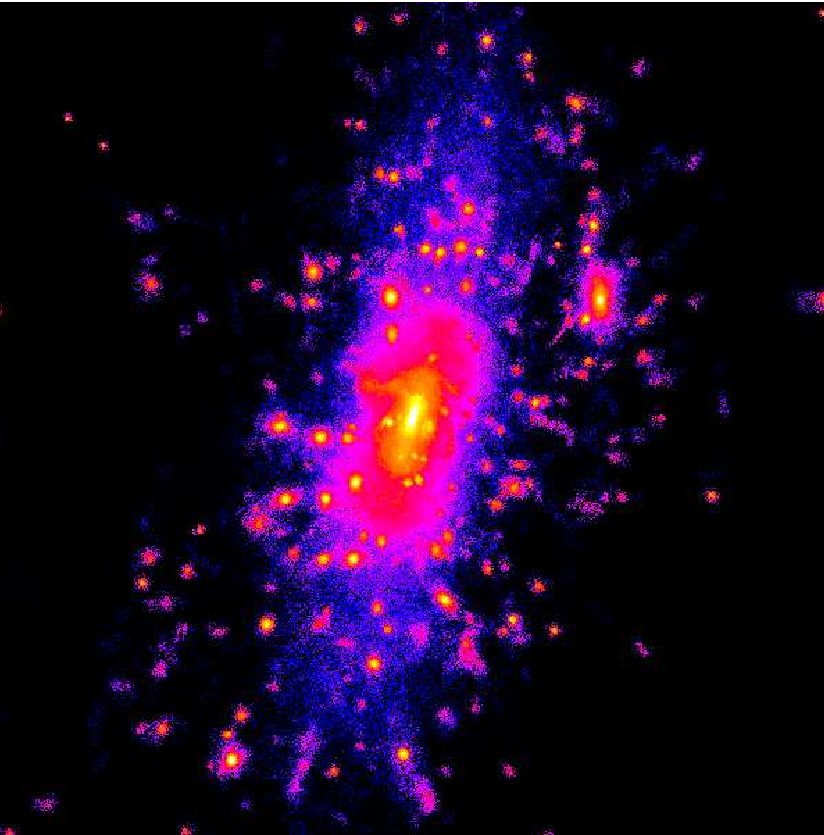}
}
\centerline{
\includegraphics[width=0.31\linewidth]{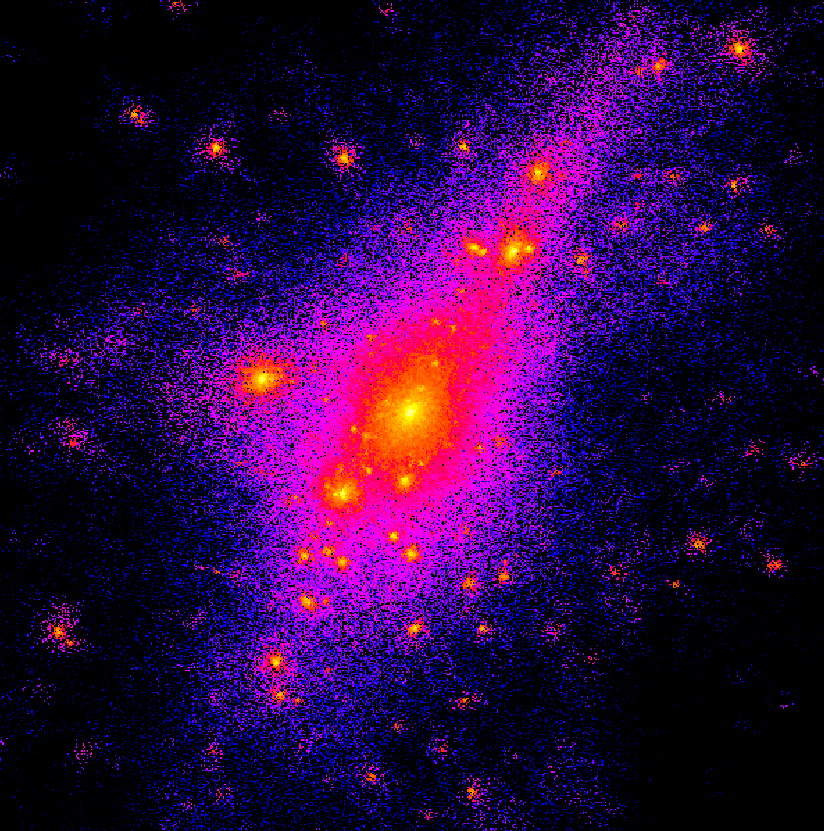}
\includegraphics[width=0.31\linewidth]{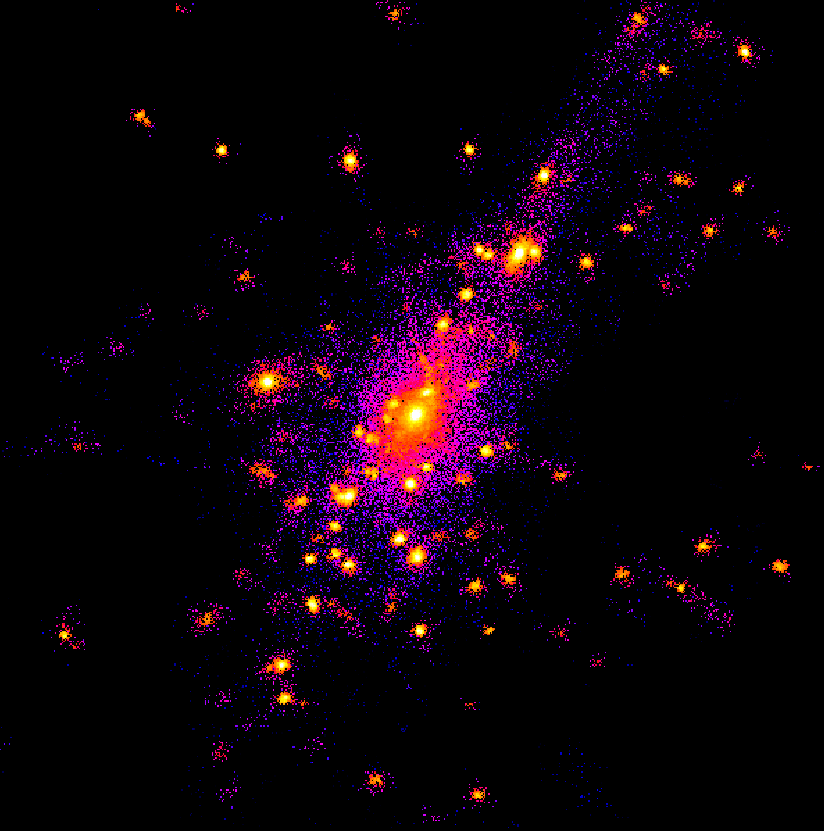}
\includegraphics[width=0.31\linewidth]{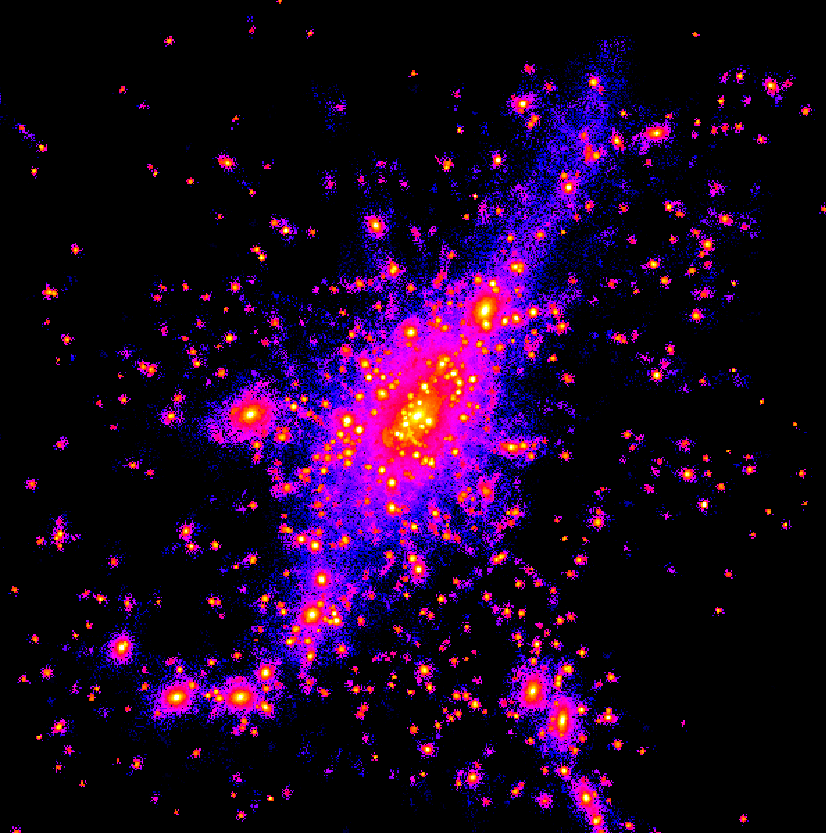}}
\caption{The distribution of the dark matter (left panels) and of the
  stars (canter panels) for clusters (A), (B), (C), (D) from the
  cosmological simulation, and the distribution of stars in the
  high-resolution re-simulations of the same four clusters (right
  panels), all at redshift $z=0$. The frames are $3 h^{-1}$ Mpc on a
  side in the first three rows and $6 h^{-1}$ Mpc in the last row,
  corresponding to $\approx 2 R_{\rm vir}$ for the four clusters (see
  Table ~\ref{clustab}).  They show density maps generated with the
 SMOOTH algorithm, applied separately to the DM and star particle
  distributions. Colour scale is logarithmic and different for DM and
  stars: from $10^{-0.5}$ to $10^5$ times critical density and from
  $10$ to $10^6$ times critical density for stars and DM
  particles, respectively.  }
\label{4clus}
\end{figure*}

\subsection {Standard resolution simulation - 4 exemplary clusters }

We discuss the results of this analysis for the four clusters shown in
Figure~\ref{4clus}. The figure shows the density distribution of DM
and star particles in the four clusters, and also the distribution of
star particles in the high-resolution re-simulation of these clusters.
The main characteristics of these clusters are given in
Table~\ref{clustab}. The galaxies identified by SKID correspond to the
densest regions plotted in yellow in this Figure.

These four clusters cover a wide range of masses (see
Table~\ref{clustab}), and the two intermediate mass clusters
B and C have very different dynamical histories: cluster B
experienced a major merger at $z \approx 1$, while cluster C is
undergoing a merger event at $z=0$  which began at $z \approx 0.2$.

\begin{figure*}
\centering{
\includegraphics[viewport= 90 70 580 530,clip, height=9cm]{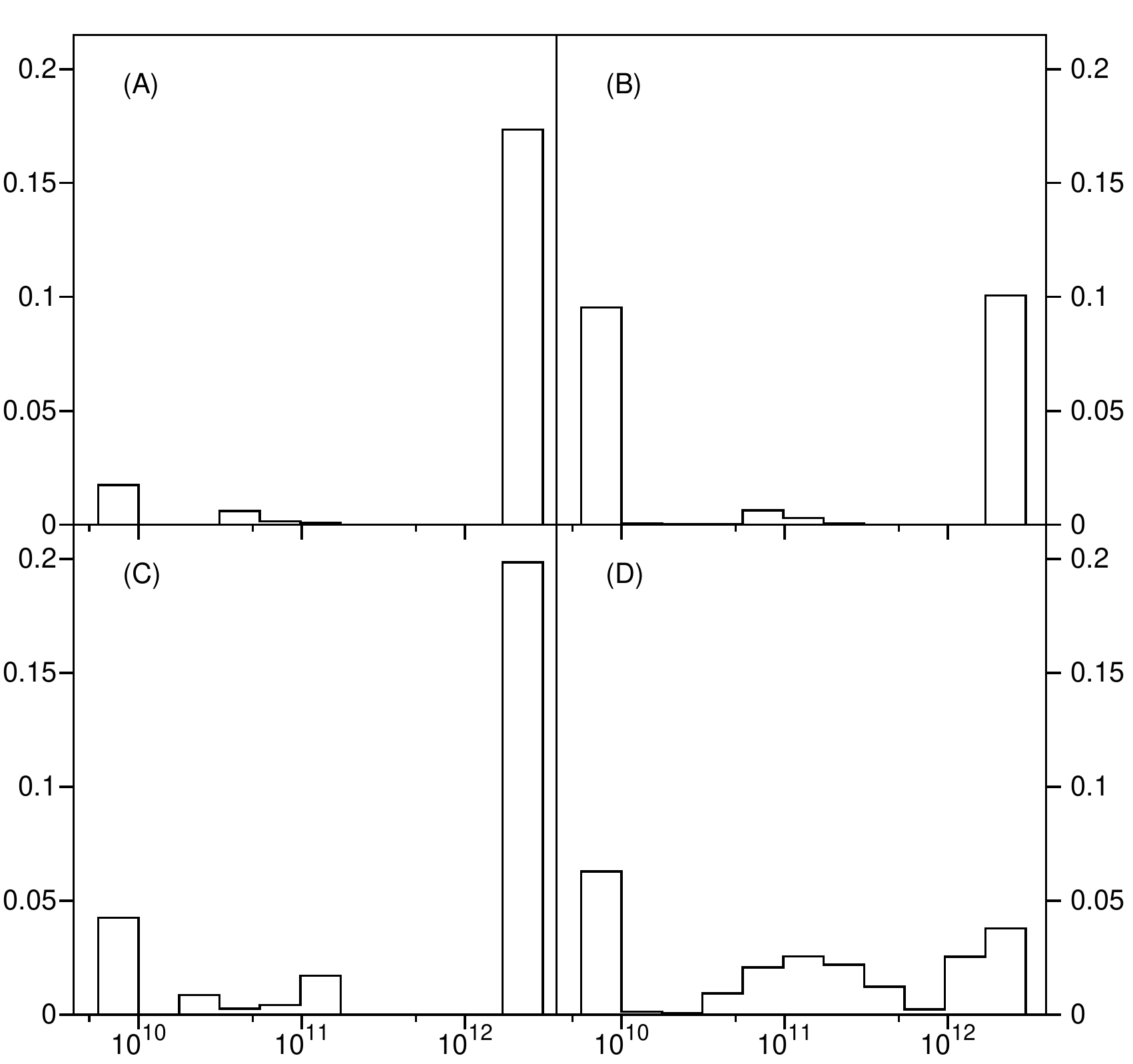}
\captionsetup{font={normalsize,sf}, width= 0.9\textwidth}
\caption{Histograms of the fraction of DSC star particles identified at $z=0$,
  associated with the family trees of $z=0$ galaxies of different masses:
  $F_{\rm DSC}(M_{\rm gal})=M^*_{\rm DSC}(M_{\rm gal})/M^*_{\rm tot}$. Results
  are reported for the four clusters shown in Figure 3.5 (see also
  Table \ref{clustab}). We use 10 galaxy mass bins, logarithmically spaced,
  from $M_{\rm min}=1.1 \times 10^{10} h^{-1} M_\odot$ to $M_{\rm
  max}=3.1 \times 10^{12} h^{-1} M_\odot$. The leftmost column in each panel
  gives the contribution from dissolved galaxies, regardless of their mass.
  It is the mass range only for sake of clarity.  For these 4 clusters, the
  only family tree contributing to the rightmost column is that associated
  with the BCG.  }}
\label{histosingleLR}
\end{figure*}

\begin{figure*}
\centering{
\includegraphics[viewport= 90 70 580 550,clip, height=9cm]{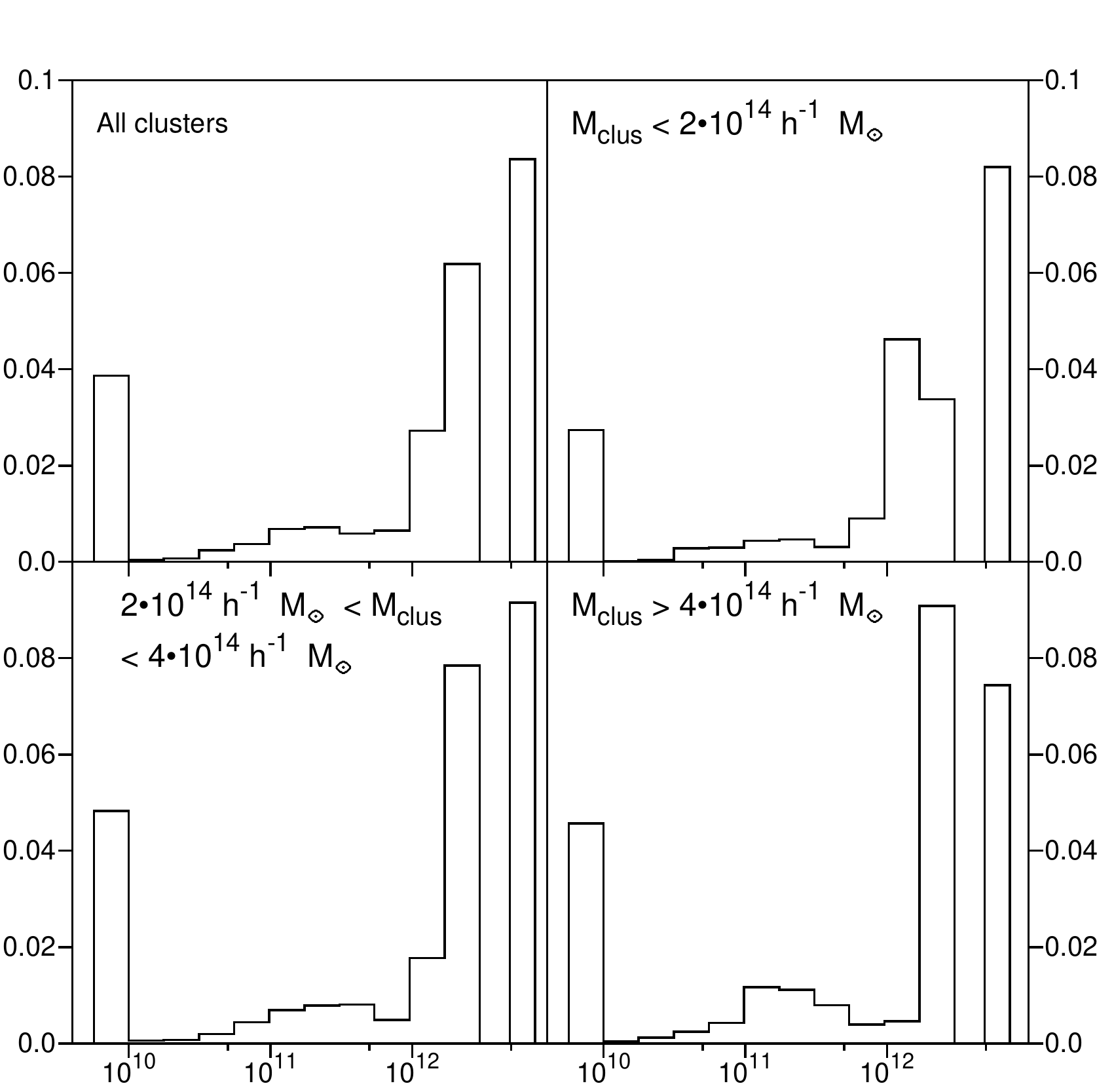} }
\captionsetup{font={normalsize,sf}, width= 0.9\textwidth}
\caption{ Histograms of the relative contribution to the DSC from the
  formation history of galaxies belonging to 10 $M_{\rm gal}(z=0)$
  mass bins, for the entire set of clusters in the simulation, and as
  a function of cluster mass.  Upper left panel: average over the
  whole 117 cluster set. Upper right panel: average over the 71 least
  massive clusters. Lower right panel: average over the 11 most
  massive clusters.  Lower left panel: average over the 35
  intermediate--mass clusters.  We use 10 $M_{\rm gal}(z=0)$ mass
  bins, logarithmically spaced from $M_{\rm min}=1.1 \times 10^{10}
  h^{-1} M_\odot$ to $M_{\rm max}=3.1 \times 10^{12} h^{-1} M_\odot$.
  The leftmost column in each panel represents the contribution from
  dissolved galaxies, regardless of their mass. The rightmost column
  in each panel shows the contribution from the history of the single
  most massive galaxy of each cluster, regardless of its actual mass.}
\label{histoallLR}
\end{figure*}

In Fig. 3.6, the histograms show the mass fractions
of the DSC associated with the family trees of $M_{\rm gal}(z=0)$
galaxy for these four clusters.  
From Fig. 3.6, we can draw the following picture for the
origin of the DSC:
\begin{itemize}
\item The bulk of the DSC comes from the formation history of the most
  massive galaxy, except in the most massive cluster D;
\item Dissolved galaxies give a significant contribution in two
 out of the four clusters (clusters B,D);
\item All other galaxy family trees provide either a small (clusters
 A--C) or modest (cluster D) contribution to the DSC.
\end{itemize}

\begin{table*}
\centering

\begin{tabular}{@{}cccccccc@{}}

\hline\hline

 Label & $M_{\rm vir}$ & $R_{\rm
 vir}$ & $F_{\rm DSC}^{\rm all}$ & $F_{\rm DSC}$ &
 $F_{\rm DSC}^{\rm vol}$ & $F_{\rm DSC}^{\rm dis}$ & $F_{\rm DSC}^{\rm
 ng}$\\ 
&  \scriptsize{[$10^{14}h^{-1} M_\odot$]} & \scriptsize{[$h^{-1}$ kpc]}& & & & & \\
\hline
A & 1.6 & 1200 & 0.33 & 0.20 & 0.11 & 0.02 & 0.02 \\ 
B & 2.5 & 1290 & 0.36 & 0.21 & 0.12 & 0.10 & 0.05 \\ 
C & 2.9 & 1350 & 0.45 & 0.27 & 0.14 & 0.04 & 0.04 \\
D & 13.0 & 2250 & 0.45 & 0.22 & 0.23 & 0.06 & 0.00 \\
Ave & -- & -- & 0.34 & 0.18 & 0.09 & 0.04 & 0.07 \\
\hline
\end{tabular}
\captionsetup{font={normalsize,sf}, width= 0.9\textwidth}
 \caption{Virial masses, viral radii, and DSC fractions for the four clusters
A--D shown in Fig. 3.5. The fraction shown in the column 4, $F_{\rm
DSC}^{\rm all}$, includes the contribution from low-density galaxies. The
$F_{\rm DSC}$ value in column 5 is obtained omitting the particles unbound
from low-density galaxies.  For completeness we report the fraction $F_{\rm
DSC}^{\rm vol}$ of discarded particles from low-density or volatile structures
in column 6; the fraction $F_{\rm DSC}^{\rm dis}$ of star particles from
dissolved galaxies in column 7, which corresponds to the leftmost columns in
the histograms of Figs. 3.6 and 3.7; the fraction
$F_{\rm DSC}^{\rm ng}$ of star particles that never belonged to any galaxy in
column 8. The last row of the table reports the average DSC fractions for the
whole sample of 117 clusters. }
\label{clustab}
\end{table*}

In the case of the most massive D cluster, a significant contribution
to the DSC comes from intermediate--mass galaxies. Willman et al. 2004 also
found in their simulation of one cluster with mass similar to our
cluster D, that galaxies of all masses contributed to the production
of the DSC. Our results suggest that when the cluster statistics is
enlarged, such cases are rare; in our set this is the case in 3
clusters out of the 11 most massive ones from the whole set of 117
clusters.

Furthermore, Figure~\ref{4clus} shows that cluster D is still
dynamically young, with a number of massive substructures both in the
DM and in the star particle distribution.  This is probably the main
reason why the DSC formation in this cluster is not dominated by the
most massive galaxy: the sub--clumps contain galaxies of various
masses which experienced several mergers in their history, producing a
significant amount of DSC. This is also confirmed by the analysis of
the family trees of the galaxies belonging to this cluster: 12 of the
85 identified galaxies had more than one merger in their history,
while usually only one or two galaxies in each cluster are found to
have a complex family tree.

In the other 3 clusters, the largest fraction of the DSC star
particles is associated with the formation history of the cluster's
most massive galaxy. Cluster B also shows a large contribution coming
from dissolved galaxies: perhaps this suggests that the tidal field
associated in this cluster was more efficient in disrupting galaxies
rather than stripping some of their stars.

Our analysis so far does not exclude that some fraction of the DSC at
$z=0$ is produced in subclusters or groups, such as suggested by
Rudick et al. 2006. 
In fact, the analysis of cluster D suggests that
this does happen. If these sub-clusters or groups migrate to the
centre of the cluster and finally merge, our procedure would associate
the DSC particles unbound from these structures with the family tree
of the cD at $z=0$.

However, if tidal stripping of the least--bound stars in all galaxies
were the main mechanism for the production of the DSC, we would expect
a more similar fraction of DSC star particles from all galaxy masses.

\subsection {Standard resolution simulation - statistics for 117 clusters }

We now turn to the statistical results for the whole set of 117
clusters.  In Figure~\ref{histoallLR} we show the contributions from
the same galaxy mass bins as before, but averaged over all clusters
and over different cluster mass ranges.  To obtain our average values,
we sum the mass of diffuse star particles in all clusters in the
appropriate galaxy mass bin and normalise it to the total stellar mass
of all clusters. This procedure creates a ``stacked-averaged''
cluster. The average value of the diffuse light fraction is $<F_{\rm
DSC}>=16 \%$ of the total stellar mass.

In the upper left panel of Fig.~\ref{histoallLR}, showing the fractional
contributions from galaxy mass bins averaged over the whole
cluster set, the rightmost column represents the
contribution from the clusters' BCGs only.  The value of the mass for
this class is arbitrary; this bin has been plotted separately since
the masses of the BCGs increase with cluster mass and, therefore, BCGs
in different clusters can belong to different mass bins. This effect is
clearly visible in the upper right panel of Fig.~\ref{histoallLR}, which
refers to the less massive clusters in our set. For these clusters,
the BCGs fall into two mass bins, with the majority of them falling
in the second most massive bin.

The other three panels of Fig.~\ref{histoallLR} show the same relative
contributions when the average is performed over (i) the 11 most
massive clusters (lower right panel, $M_{\rm vir} > 4 \times 10^{14}
h^{-1} M_\odot $ with $<F_{\rm DSC}>=19 \%$), (ii) the 35 clusters
having intermediate mass (lower left panel, $2 \times 10^{14} < M_{\rm
vir} < 10^{14} h^{-1} M_\odot$ with $<F_{\rm DSC}>=18 \%$), and (iii)
the 71 least massive clusters (upper right panel, $M_{\rm vir}< 2
\times 10^{14} h^{-1} M_\odot $ with $<F_{\rm DSC}>=13 \%$). These
average values show a weak trend with cluster mass in the production
of the DSC.  As a further test, we have divided clusters according the
amount of the DSC fraction itself. This analysis also confirms that
the dominant contribution to the DSC comes from the BCG family tree,
independent of $F_{\rm DSC}$ as expected, given the weak relation
between $F_{\rm DSC}$ and cluster mass. We also find that the
contribution from dissolved galaxies is slightly higher for clusters
with $<F_{\rm DSC}> $ greater than $25 \%$.

The results shown in Fig.~\ref{histoallLR} are consistent with the
previous analysis of the four clusters: the bulk of the DSC is
associated with the galaxies in the family tree of the most massive
galaxy of each cluster. Galaxies in the family trees of smaller $z=0$
mass bins contribute only few tenths of the fraction from the BCG
family tree.

Dissolved galaxies also contribute significantly to the DSC, but it is
possible that their estimated contributions are affected by some
numerical effects.  In fact, if the analysis is restricted to galaxies
whose density is within $1 \sigma$ of the observed galaxy density
estimate (Section~\ref{skidid}), then the contribution to the $F_{\rm
DSC}$ from the BCG rises to $\approx 76\%$, while that from dissolved
galaxies drops to $\approx 8\%$. As expected, most ``dissolved
galaxies'' are those with low densities, which indicates that their
contribution to the DSC may be affected by the limits in numerical
resolution. This needs further work to be properly understood.

\subsection {Merging and stripping in galaxy family trees}
\label{DSCmerging}

\begin{figure*}
\centering{
\includegraphics[viewport= 55 70 580
  550,clip,width=8.cm]{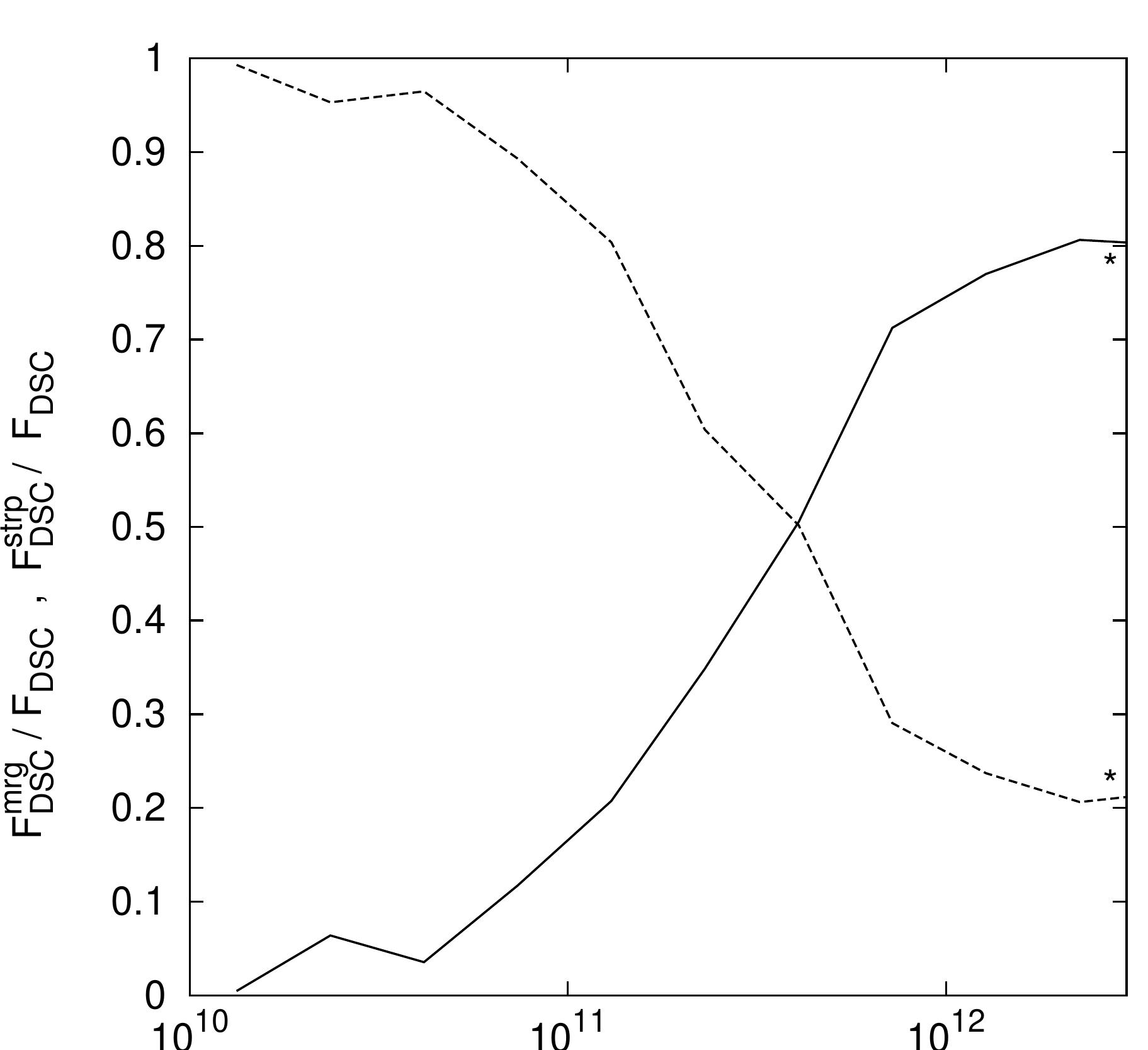}}
\captionsetup{font={normalsize,sf}, width= 0.9\textwidth}
\caption{ The fraction of DSC stars arising in the ``merger'' part,
  $F_{\rm DSC}^{\rm mrg}/F_{\rm DSC}$, and in the ``stripping''
  part, $F_{\rm DSC}^{\rm strp}/F_{\rm DSC}$, of the galaxy family
 trees, as a function of their $z=0$ galaxy mass class (solid and
  dashed lines, respectively).  The asterisks mark the values for the
  contribution from the BCG family tree only. See text for details.}
\label{f.mrgstr}
\end{figure*}

As shown in Fig.~\ref{famtreelow}, the BCG is the galaxy that
experiences the largest number of luminous mergers during its
assembly, and it often has the most complex family tree in a cluster.
Our result that a large fraction of DSC is associated with the BCG
supports a scenario where mergers release stars from their parent
galaxies to the intracluster space. 

To investigate this further, we estimate the fraction of DSC particles
associated with the {\sl merger} part of the family trees, and with the {\sl
stripping} part of the family trees, as follows.  We define a DSC star
particle to arise from a {\sl merger} at redshift $z_j$ if the galaxy it was
last bound to has more than one progenitor at
$z_{j-1}$\footnote{It may happen that the final phase of the merger is not
detected by our galaxy identification procedure, because the two merging
structures are very close to each other and have small relative velocities. In
this case, the SKID algorithm generally merges the two objects into a single
one, even if the merger is not yet completed. For this reason, we assign a DSC
star particle to the merger part of its parent galaxy family tree even if it
becomes unbound two family tree levels after the merger, at $z_{j+1}$
(i.e. from the ``offspring of the offspring'' of a merger).}.  The DSC
particles coming from the progenitors at $z_{j-1}$ are also defined as arising
from a merger. We take a DSC star particle to be unbound through {\sl
stripping} if the galaxy it was last bound to at redshift $z_j$ has only
one progenitor at $z_{j-1}$.  The different parts of the family trees are
indicated in Fig.~\ref{famtreelow}, where the squares represent the {\sl
merger} part of the tree, triangles the {\sl stripping} part, while circles
mark the part of the tree where no stars are released to the DSC.

In Fig. 3.8, we show the fraction of the DSC star
particles which arise from the {\sl merger} part of the family tree,
as a function of the final galaxy mass.  For high--mass galaxies,
most of the DSC originates from the {\sl merger} part of their family
trees. Low mass galaxies, on the other hand, lose stars only
via {\sl stripping}. After each {\sl merger} between massive galaxy
progenitors, up to $30\%$ of the stellar mass in the galaxies involved
has become unbound. This large fraction perhaps indicates that many of
these mergers take place either in strong tidal fields generated by
the mass distribution on larger scales, or just before the merger
remnants fall into their respective cluster.

Combined with the result that most of the DSC star particles are
associated with the family trees of the most massive galaxies, the
fact that most of the DSC is released during {\sl merger} events
implies that the bulk of the DSC originates in the merger assembly of
the most massive galaxies in a cluster.  The more standard picture for
the formation of the DSC, in which all galaxies lose their outer stars
while orbiting in a nearly constant cluster gravitational potential,
is not confirmed by current cosmological hydrodynamic simulations. It
appears that strong gravitational processes, linked to the formation
of the most massive galaxies in the cluster and to mergers between
luminous objects, are the main cause for the creation of the DSC.

A further mechanism possibly at work is the complete disruption of
galaxies, which also takes place preferentially in the cluster central
regions. In our cosmological simulation this formation mechanism for
the DSC is likely to be enhanced by numerical effects, which tend to
produce under-dense galaxies. We address this issue below when we
discuss our high-resolution simulations of the clusters in
Fig.~\ref{4clus}.

\subsection {cD Halo {\it vs} Intracluster Light}
\label{cdvsicl}

So far, we have made no attempt to distinguish between a component of DSC
associated with the unbound halos of the central cD galaxies, and a more
cluster-wide DSC.  Our definition of the DSC includes stars in the cluster
central regions and a part of these may well be in the form of cD halos (see
the Appendix).  Independent of how well a distinction between these two
components can be made, one might expect that the fraction of DSC stars that
comes from the merging tree of the BCG would be most concentrated towards the
cluster centre.  To shed some light on this question, we show in
Fig. 3.9 the same analysis for the average over all galaxy clusters
in the simulation as in Fig. 3.7, but now excluding all DSC
particles residing in the central $250 h^{-1}$ kpc around their cluster
centres.  The remaining total DSC fraction drops to about $6 \%$, less than
half of the total, reflecting the steep radial profile of the DSC
(Murante et al. 2004, Zibetti et al. 2005.  For $R>250 h^{-1}$kpc, the family trees
of the most massive 
galaxies still provide the largest contribution to the DSC (per mass bin), but
the cumulative contributions from family trees of less massive galaxies now
dominates the BCG component by a factor $\sim 2$. At the same time, the
relative contribution from dissolved galaxies increases. The lower panel of
Fig. 3.9 shows the same analysis but now excluding DSC particles
within $0.5 R_{\rm vir}$. In this case, the $<F_{\rm DSC}>$ drops to $\sim
1\%$, with the fraction from the BCG family tree now similar to that from
other galaxies.

Since the BCG halo is likely to be less extended than $250 h^{-1}$ kpc, our
interpretation of the results in Fig. 3.9 is that the {\sl merger}
part of the most massive galaxy family tree in each cluster contributes
substantially to the DSC also outside the cD halo. However, at radii $ 250
h^{-1} \mbox{kpc} < R < 0.5 R_{\rm vir}$, the cumulative contribution from the
family trees of other massive galaxies dominates the DSC.  Presumably, these
are the most massive galaxies within subgroups, which fell into the cluster
and brought in their own DSC, but which have not yet had time to merge with
the BCG. This interpretation is consistent with the simulation results of
Willman et al. 2004, Rudick et al. 2006. Only in the outskirts of clusters, at $R> 0.5
R_{\rm vir}$, we find that the DSC particles come preferentially from the {\sl
stripping} part of family trees from all galaxy mass bins.

The relevance of {\sl merger} events for the formation of the DSC may
explain why diffuse light is more centrally concentrated than
galaxies, in both observations (Zibetti et al. 2005, Arnaboldi et
al. 2002) and in 
simulations (M04, Willman et al. 2004, Sommer-Larsen et al. 2005).  Stars from
accreted satellite galaxies form extended luminous halos around
massive galaxies (Abadi et al. 2006\nocite{Abadi06}), and if these
massive galaxies end up 
concentrated to the cluster centre, their diffuse outer envelopes
would preferentially contribute to the DSC in the cluster centre.

Our results on the origin of the DSC are also consistent with the
predictions by D'Onghia et al. 2005\nocite{Elena} that simulated
fossil galaxy groups 
have a larger amount of intra--group stars than normal groups. Indeed,
if fossil groups are the dynamically most evolved groups, then their
galaxies had more time to interact and build up the central elliptical
galaxy ( see e.g. D'Onghia et al. 2005). The number of galaxy--galaxy mergers
in groups appears to be closely related to the amount of DSC liberated
(Sommer-Larsen 2006\nocite{SommerGroups}).  A direct comparison of our
results with Willman et al. 2004 is difficult because of their
re-normalisation of the 
simulated galaxy luminosities in order to fit the observed luminosity
function, but these authors also concluded that luminous galaxies
provide a substantial contribution to the DSC.

\begin{figure*}
\centering{
\includegraphics[viewport= 50 70 350
  550,clip,height=9.cm]{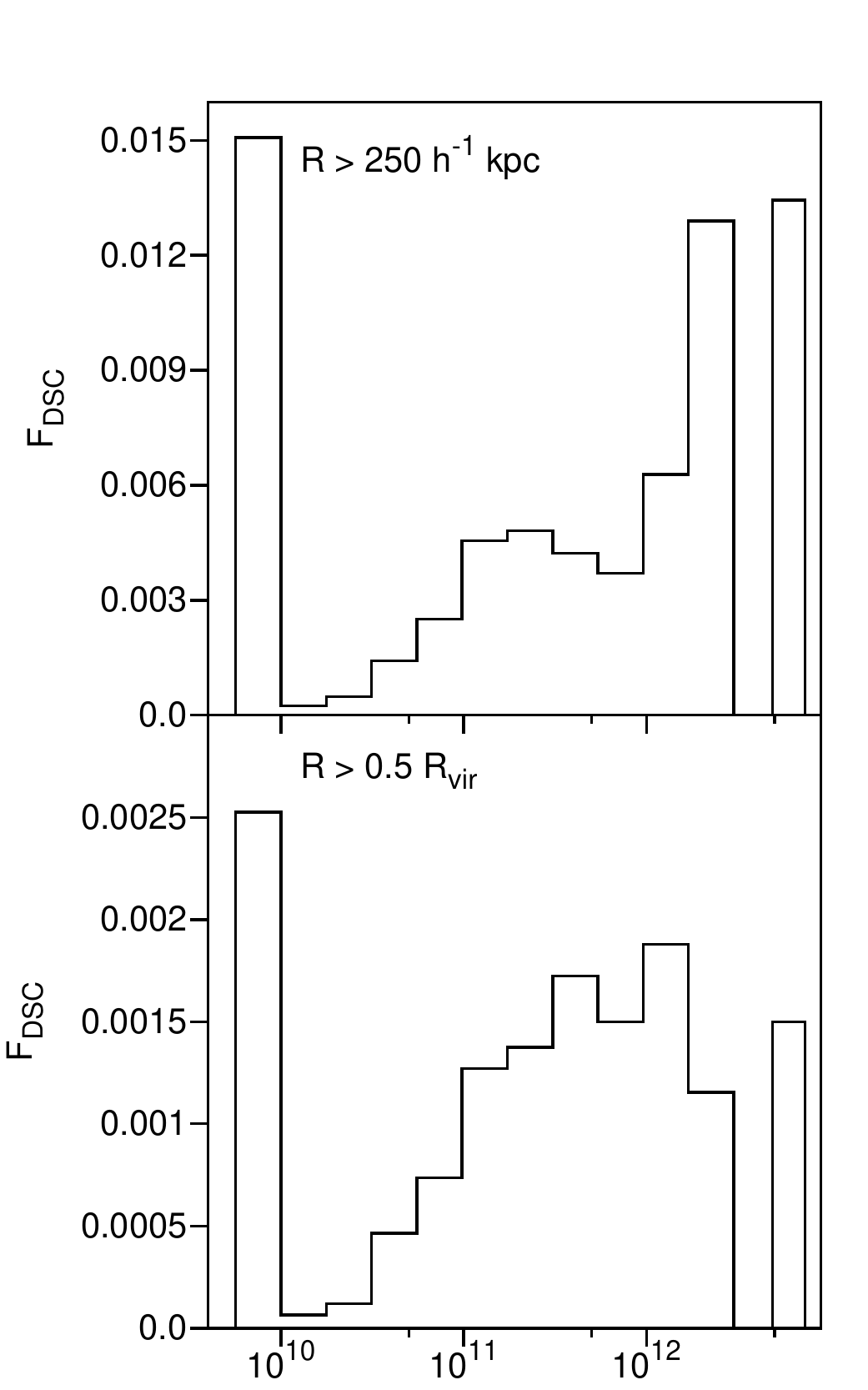}
\captionsetup{font={normalsize,sf}, width= 0.9\textwidth}
\caption{ The same as in Fig. 3.6, averaging over
all clusters, but excluding the DSC star particles inside $R=250
h^{-1}$ kpc (upper panel) and inside $0.5 R_{\rm vir}$ of each cluster
(lower panel).  }}
\label{nohalo}
\end{figure*}

\section {High resolution simulations and the effects of numerical
resolution on the formation of the DSC}
\label{reseffect}

To address the stability of our results against mass and
force resolution, we have re-simulated three clusters extracted from
the cosmological box (A,B,C from Table \ref{clustab}) with 45 times
better mass resolution and $\simeq 3.6$ times smaller gravitational
softening.\footnote{Cluster D has been re-simulated at only 10 times
better mass resolution and its analysis is not discussed here.}  A
detailed presentation of these re-simulations is given in Borgani
et al. (2006). Galaxies formed in these high-resolution simulated
clusters have densities similar to those shown in Fig.~\ref{galdens}
for masses larger than $\approx 10^{11} h^{-1} M_\odot$, and the
low--density tail seen in Fig.~\ref{galdens} is shifted towards
lower masses, in accordance with the better resolution.

\begin{figure*}
\centering{
\includegraphics[viewport= 50 55 600 530,clip,height=9cm]{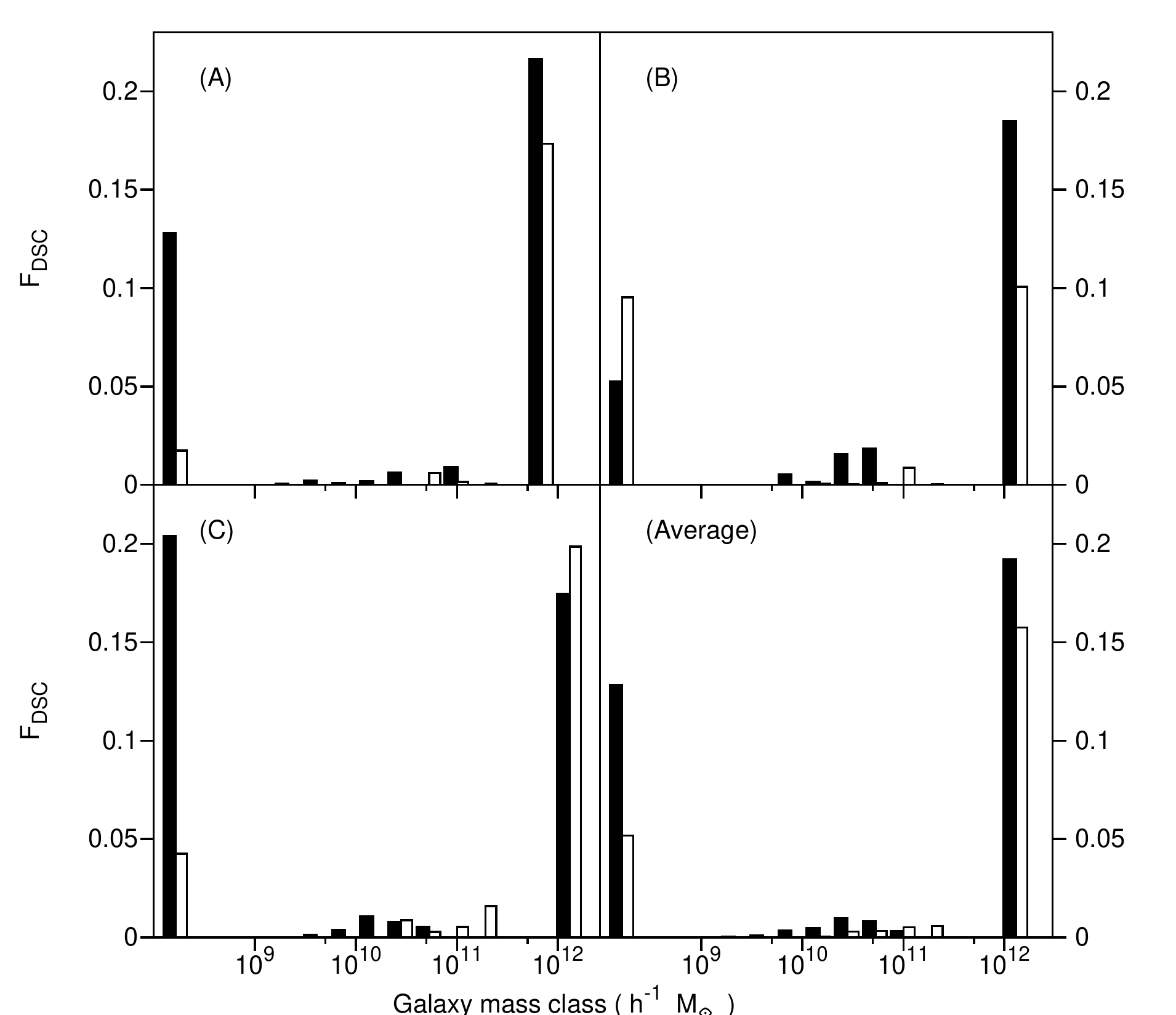}}
\captionsetup{font={normalsize,sf}, width= 0.9\textwidth}
\caption{Histograms of the fraction of the $z=0$ DSC particles
associated with galaxy family trees in 15
 $M_{\rm gal},(z=0)$ mass bins, for the three clusters re--simulated at
high--resolution (filled columns), and in the standard--resolution
(empty columns, cf.~Fig.~\ref{histosingleLR}).  The mass bins are
logarithmically spaced, from $M_{\rm min}=2 \times 10^{8} h^{-1}
M_\odot$ to $M_{\rm max}=3.1 \times 10^{12} h^{-1} M_\odot$.  The
leftmost columns show the contribution from dissolved galaxies,
regardless of their mass.  Upper-left, upper-right and lower-left
panels show the comparison for clusters (A), (B), (C), respectively;
the lower-right panel show the average for these three clusters.  }
\label{histrescomp}
\end{figure*}

We carry out our study of the DSC in these three clusters following
the procedure described previously, discarding DSC star particles
contributed from under-dense and volatile structures\footnote{For the
high--resolution clusters we use 24 different redshifts, starting from
$z=5$.}.  In Fig.~\ref{histrescomp}, we show the fractions of DSC star
particles identified at $z=0$ and originating from the history of
galaxies belonging to different mass bins. The full columns refer to
the analysis of the high--resolution simulations of the three
clusters, while the open columns show the results from
Fig. 3.7 for the low-resolution simulations.

In the clusters simulated with high resolution, the results on the origin
of the DSC are consistent with what we found for the standard
resolution.  The DSC builds up in parallel with the formation of the
most massive galaxies in the cluster. The amount of DSC star particles
produced during the history of all other galaxies is still negligible
when compared with the contribution from the most massive cluster
members. With the increase in resolution, more DSC stars now come from
dissolved galaxies in clusters A and C, and an increasing number comes
from the family tree of the cD in clusters A and B.

Another question is whether the results on the DSC are affected by the
efficiency of the kinetic feedback from SNe. To address this point, we
re-simulate cluster A at the resolution of the cosmological box, with
(i) the same feedback efficiency and (ii) the speed of the galactic
ejecta set to zero.  We find very similar results in the strong
and weak feedback cases: no significant contribution from intermediate
mass galaxies, 7.5\% and 8.7\% of the DSC coming from dissolved
galaxies, and 89.7\% and 87.1\% of the DSC originating from the
history of the BCG, respectively.

However, the overall fraction of intracluster stars in these three
clusters changes between our low-resolution and high-resolution
simulations.  Once the star particles from volatile and under-dense
galaxies are discarded, the fraction of DSC within the virial radius
of clusters A, B and C is $F_{\rm DSC}=0.37, 0.28, 0.41$ in the high
resolution simulations, compared with $F_{\rm DSC} = 0.20, 0.21, 0.27$
in the standard resolution case. The overall increase of the DSC
fraction at high-resolution is mostly, but not only, related to
an increase in the fraction of DSC from dissolved lower-mass galaxies
(see Fig.~\ref{histrescomp}).

This result is not in contradiction with Sommer-Larsen 2006 finding
that the DSC fraction in his simulations remains constant when the
resolution is increased. This is because in the Sommer-Larsen 2006
simulations the numerical resolution is increased without adding the
corresponding higher frequencies in the initial power spectrum. Then
the number of low-mass galaxies and (small--object) mergers does not
increase significantly, and one expects an approximately constant DSC
fraction. This suggests that the increase in the DSC fraction in our
high-resolution simulations is related to the addition of small objects
through the added high-frequency part of the power spectrum.

We expect that the effect of numerical overmerging is reduced at
higher resolution (e.g. Borgani et al. 2006 and references therein for a full
discussion of this issue), so we must look for other effects
that could dominate the disruption of the smaller galaxies.  The
results from Sommer-Larsen 2006 also rule out a significant effect
from stronger tidal shocks during high-speed collisions with
low-impact parameters, when galaxies become denser at higher
numerical resolution. One possibility is that, when the resolution is
increased, the number of numerically resolved mergers increases. On
the basis of the results reported above, this could turn into an
increased efficiency in the production of the DSC. If so, a solution
to the problem could lie in a more realistic feedback mechanism.

Another issue related to the numerical resolution concerns the number
of small ($\lg 10^{11} h^{-1} M_\odot$) galaxies identified in these
simulations. Recent determinations of the K--band luminosity function
for galaxies in clusters give a faint--end slope between $\alpha =
-0.84$ and $\alpha=-1.1$ Lin \& Mohr 2004\nocite{LinMohr04}. The faint
end slope of 
the stellar mass function in our cosmological simulation is flatter
than the observed luminosity function: we obtain $\alpha \approx
-0.7$, thus implying that we miss a number of small galaxies. A
shallower slope of the faint end of the luminosity function is a
general problem of numerical simulations like those presented here
( e.g. Willman et al. 2004).

\begin{figure}
\centering{
\includegraphics[viewport= 55 33 630
  550,clip,height=9cm]{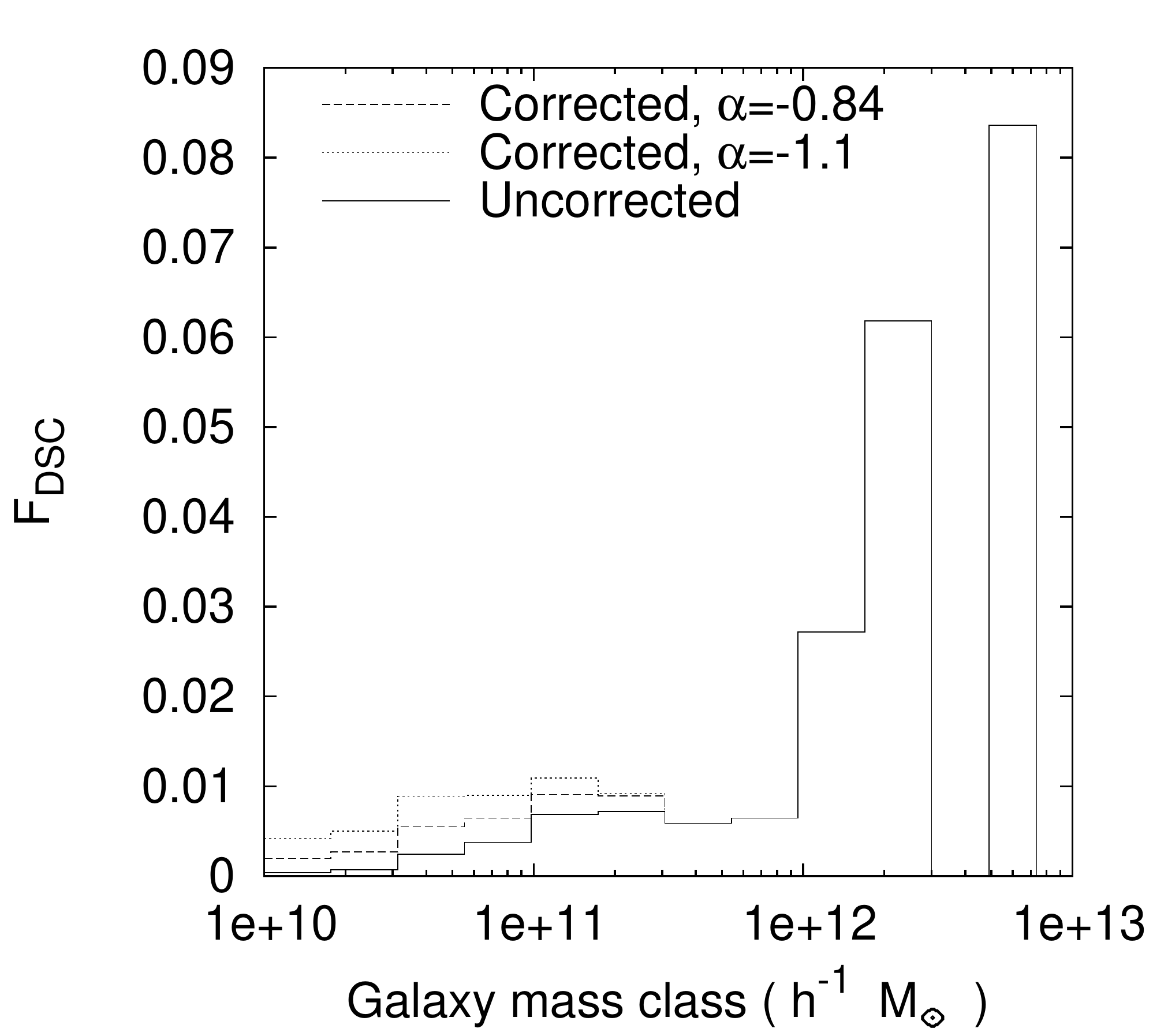}}
\captionsetup{font={normalsize,sf}, width= 0.9\textwidth}
\caption{Corrections to $F_{\rm DSC}$ to account for the
low--mass end of the galaxy mass function.   The solid line is the
uncorrected $F_{\rm DSC}$; the dashed and the dotted lines give the
corrected value when the low--mass end of the galaxy cluster stellar
mass function has a slope $\alpha=-0.84$ and $\alpha=-1.1$,
respectively.}
\label{normfdsc}
\end{figure}

If the number of low--mass galaxies is underestimated in the simulations, then
their contribution to the DSC would be affected in the same way. To estimate
this effect, we computed how many galaxies we would expect in each mass bin of
Fig.~\ref{histoallLR}, if the faint end of the stellar mass function was given
by $n(M) = K \cdot (M/M_*)^\alpha$, with $\alpha=-0.84$,$-1.1$ and
$M_*=5\cdot10^{11} h^{-1} M_\odot$. The constant is fixed by requiring that
the number of galaxies for a mass $M=2\cdot 10^{11} h^{-1} M_\odot$ is the
same as in the simulation.  This method is similar to the re-normalisation of
the luminosity function in  Willman et al. 2004.  We then assume that the missed
galaxies contribute the same relative fraction of their mass to the DSC as the
present-day galaxies of similar mass in the simulation, and multiply the
fraction $F_{\rm DSC}(M)$ by the ratio of $N(M)/N_s(M)$, where $N_s(M)$ is the
number of simulated galaxies found in the bin and $N(M)$ is the integral of
$n(M)$ in the same bin. This correction is applied to each mass bin up to
$2\cdot 10^{11} h^{-1} M_\odot$.  Fig.~\ref{normfdsc} shows the result of this
correction when it is applied to the average distribution of $F_{\rm DSC}$ for
the whole set of three clusters (lower right panel of Fig.~\ref{histrescomp}).
The DSC production is still dominated by the contribution coming from the BCGs
in the clusters.  The effect of such correction is to bring the contribution
of the mass bins corresponding to masses $M<10^{11} h^{-1} M_\odot$ to the
same levels of the others. Correction is stronger for the smaller mass
bins. Nevertheless, the contribution of these mass classes to the global
$F_{\rm DSC}$ remain small. Note that the increases in $F_{\rm DSC}(M)$ values
in the mass bin $\sim 2\cdot 10^{11} h^{-1} M_\odot$ is due to a few galaxies
with mass $1.5<M<2 \cdot 10^{11} h^{-1} M_\odot$, whose contribution to the
DSC has been corrected.  The contribution $F_{\rm DSC}(M)$ from galaxy having
masses smaller than $\approx 1.1 \cdot 10^{10} h^{-1} M_\odot$ in the three
re-simulated clusters, where they are resolved, is very small.

A further issue to be considered is that all galaxies are spheroidal at the
numerical resolution of these cosmological simulations; indeed, the
self--consistent formation of disk galaxies is still a challenge in
hydrodynamic $\Lambda$CDM simulations.  Are our conclusions on the origin of
the DSC likely to be affected by the absence of disk galaxies in our
cosmological simulation? Generally, disk galaxies are more vulnerable to tidal
forces, but the amount of matter lost in tidal tails is small, unless the
tidal field is very strong, whereas elliptical galaxies lose their outer stars
more easily. We do not expect that it would make a lot of difference for the
amount of DSC released in the merging processes leading to the formation of
the cluster BCG galaxies and most of the DSC, if a fraction of the
participating galaxies were disk galaxies. However, this needs to be checked
once simulations can reproduce disk galaxies.  Independent arguments based on
tidal stripping from disk galaxies in a semi--analytical model of galaxy
formation (Monaco et al. 2006\nocite{PG1}, Monaco et
al. 2006\nocite{MORGANA}) suggest that at most $\sim 10$ 
per cent of the total 
stellar mass of each cluster is contributed to the DSC by the `` quiet tidal
stripping'' mechanism, even for the most massive clusters where observations
points toward a larger amount of diffuse stars.

As already discussed in Section~\ref{skidid}, our force resolution is
not enough to resolve the inner structure of the simulated
galaxies. As a consequence, their internal density is likely to be
underestimated, so that these galaxies are more vulnerable to tidal
stripping and disruption than real galaxies. This numerical artifact
is not completely removed even in our high resolution re-simulations,
where the Plummer--equivalent softening is $\simeq 2 h^{-1}$ kpc. Even
when simulated galaxies with central densities lower than a chosen
threshold are removed, we still find that less massive galaxies are
less dense than massive galaxies, at variance with observational
results. This probably accounts for most of the contribution to the
DSC from dissolved galaxies.  If we had more realistic, denser small
galaxies in the simulation, this might decrease their contribution to
the DSC even more.

We can summarise the discussion of systematics effects in our analysis
related to numerical resolution as follows:
\begin{enumerate}
\item 
Our conclusion, that the formation of the DSC
is intimately connected with the build--up of the cluster's BCG, is
confirmed in higher numerical resolution simulations;
\item
this conclusion appears insensitive to the limitations of current
simulations in reproducing low-mass galaxies, disk galaxies, and
the faint-end luminosity function;
\item resolving galaxies with smaller masses in the high-resolution
simulations does not have a strong effect on the formation and
evolution of the DSC;
\item  the global value of $F_{\rm DSC}$ depends on resolution, and it
increases in simulated clusters with higher resolution. The value of
this fraction has not converged yet, in the range of numerical
resolution we examined.
\end{enumerate}

\section{How do stars become unbound? On the origin of the diffuse
stellar component}
\label{secunbound}

\begin{figure}
\centerline{
\includegraphics[viewport= 55 40 600
  550,clip,height=6cm]{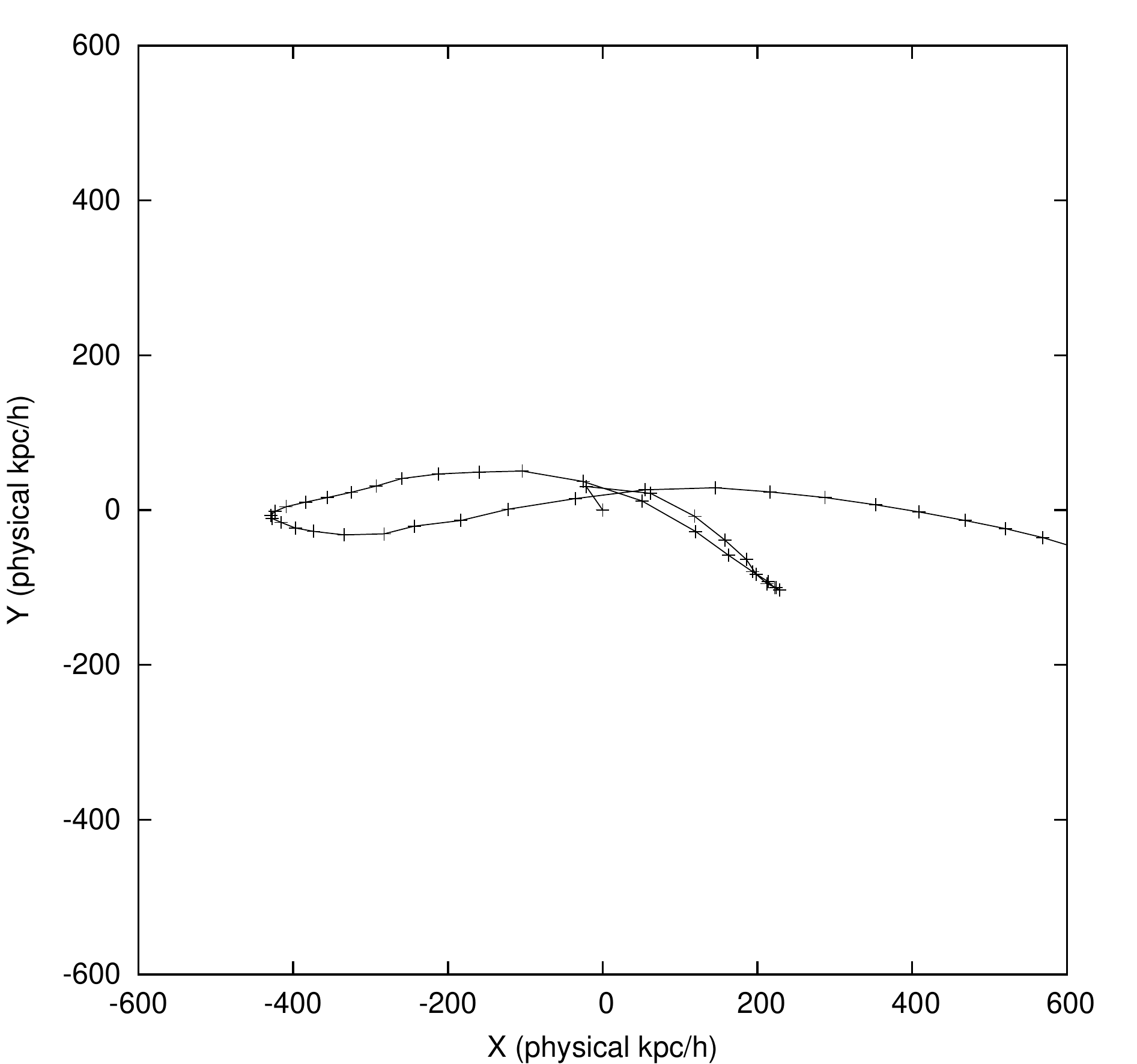}}
\captionsetup{font={normalsize,sf}, width= 0.9\textwidth}
\caption{Typical projected orbit of a galaxy ending in a major merger
with the cD galaxy; here in cluster A at
redshift 0.269. }
\label{figorbit}
\end{figure}

In cosmological hydrodynamic simulations, stars form in galaxies. The
DSC is built up from stars that are dissolved from their parent
galaxies. This is an ongoing process linked to the accretion of
substructure (see above and also Willman et al. 2004); in fact, most
of the DSC originates at relatively recent redshifts.  Stars may be
added to the DSC through a number of dynamical processes, listed here
in the approximate time sequence expected during the infall of a
significant subcluster:

\begin{enumerate}
\item
Tidal stripping of the preexisting diffuse stellar population from the
in-falling subcluster or group: The DSC in the substructure, generated
by dynamical processes in the substructure, is added to the DSC of the
main cluster when both structures merge.
\item
Stripping from extended galaxy halos created in substructures: In
subclusters or galaxy groups, galaxy interactions occur with lower
relative velocities and are thus generally more damaging than in
interactions with the high relative velocities typical for galaxy
clusters.  Interactions or mergers within substructures may create
loosely bound stellar halos which are stripped from their parent
galaxies and the substructure when entering the cluster tidal field
(Mihos et al. 2004, Rudick et al. 2006). This stripping may be delayed when
the merging in the subcluster happens before its accretion, or
immediate when the galaxy interactions occur already deep in the tidal
field of the main cluster.
\item 
Tidal shocking and stripping during merger with the cD galaxy: The
massive galaxies in the substructure generally interact with the
cluster centre and the cD galaxy on near-radial orbits (see
Figure~\ref{figorbit}). In a high-speed encounter of a massive galaxy
with the cD, stars from both galaxies may gain sufficient energy to be
(almost) tidally unbound from their parent galaxies. The tidally
shocked stars from the intruder are then subsequently unbound by the
ambient tidal field or further tidal shocks, remaining at similar
orbital energies as their mother galaxy had at the time. Those from
the cD galaxy become part of the cD envelope. The process may happen
several times as the intruder galaxy orbit decays by dynamical
friction.  This mechanism is related to the cannibalism scenario for
the growth of the cD galaxy, described in Ostriker \& Hausman
1977\nocite{Ostr77}, Merritt 1995\nocite{Merritt95}
 and others; however, here the dynamical
friction appears to be more effective, presumably due to the large
dark matter mass associated with the infalling substructure.
\item 
Tidal dissolution of low-density galaxies: These galaxies may enter
high-density regions of the cluster along their orbits, such as the
dark matter cluster centre, and dissolve completely if of sufficiently
low density. 
\item 
Tidal stripping in galaxy interactions: Stars may be torn out from
galaxies during tidal interactions along their orbits in the cluster,
and be dissolved from their parent galaxies by the cluster tidal
field. The participating galaxies survive as such. Galaxies of all
masses are affected. The last two processes together are often
described as harassment (Moore et al. 1996).
\end{enumerate}

The statistical results of the previous sections allow us to put
some constraints on the relative importance of these various 
processes, and to identify further work needed to clarify the
origin of the diffuse stellar population in galaxy clusters. 
These results can be summarised as follows. 
\begin{itemize}
\item
Most of the DSC is liberated from galaxies in the merger tree of the
most massive, central galaxy in the cluster, i.e., simultaneously with
the build-up of this galaxy.
\item
If only the fraction of the DSC outside $250{\rm kpc}$ from the
cluster centre, i.e., outside the cluster dominant galaxy's halo, is
considered, the contribution from the BCG family tree is comparable to
that from other massive galaxies.  Only outside $\sim 0.5 R_{\rm vir}$
do galaxies of all masses contribute to the DSC.
\item
There is a further, sizable contribution to the DSC in the
simulations, from dissolved galaxies. However, the fraction of DSC
stars from dissolved galaxies depends directly on the simulations'
ability to faithfully represent the lower mass galaxies, and is seen
to vary strongly with the resolution of the simulation.  This
contribution to the DSC is thus currently uncertain; the prediction
from the simulations is likely to overestimate the contribution of
dissolved galaxies to the ICL in observed clusters.
\item
About 80\% of the DSC that comes from the merger tree of the cD
galaxy, is liberated shortly before, during, and shortly after major
mergers of massive galaxies. About 20\% is lost from these galaxies
during quieter periods between mergers.
\item In each significant merger, up to $30\%$ of the stellar
mass in the galaxies involved becomes unbound.
\item Most of the DSC is liberated at redshifts $z=0-1$.
\end{itemize}

These results imply that the main contribution to the DSC in our
simulations comes from either tidal shocking and stripping during
mergers with the cD galaxy in the final cluster (mechanism iii),
and/or from merging in earlier subunits whose merger remnants later
merge with the cD (i, ii).  The traditional tidal stripping process
(v) appears to contribute only a minor part of the DSC but may be the
dominant process for the small fraction of the DSC that ends up at
large cluster radii.

Further work is needed to see which of the channels (iii) or (i,ii) is
the dominant one, and whether in the latter the contribution of the
preexisting DSC in the accreted subclusters (i) dominates over that
from extended galaxy halos (ii).  The work of Willman et al. 2004 and
Rudick et al. 2006 
shows that the contribution from infalling
groups is important but the division between the channels (i) and (ii)
is not clear. In this case, our results imply (a) that merging must
have been important in these groups, and (b) that the massive galaxies
in the infalling groups that carry most of the final DSC will mostly
have merged with the BCG by $z=0$, so that the tidal shocking and
stripping process (iii) will play some role as well. The description
of D'Onghia et al. 2005 and Sommer-Larsen 2006 of fossil groups
as groups that are older than other groups, in a more advanced
evolutionary stage with a dominant elliptical galaxy, and with a
larger fraction of DSC, suggests dense, evolved groups as promising
candidates for contributing significantly to the cluster DSC.  One
recent observational result that also fits well into this picture of
accreting groups that have already had or are having their own merger
events, is the observation of Aguerri et al. 2006 that the DSC
fraction in Hickson groups correlates with the elliptical galaxy
fraction in these groups.

Certainly it is clear from our results that the formation of the cD
galaxy, its envelope, and the DSC in galaxy clusters are closely
linked. Further analysis is required to determine whether these
components are dynamically distinct, and what kinematic signatures can
be used to distinguish between them in observations of cD clusters.

\section{Conclusions}
\label{concl}
In this paper, we have studied the origin of the diffuse stellar
component (DSC) in galaxy clusters extracted from a cosmological
hydrodynamical simulation.  We identified galaxies in 117 clusters
with the SKID algorithm, tracing each of them back in time at 17
different redshifts from $z=3.5$ to $z=0$. This allowed us to build
the family tree of all galaxies identified at $z=0$ in all clusters. We
find that all BCGs are characterised by complex family trees, which
resemble the merging trees of DM halos. At the resolution of our
simulation, only a small number of massive galaxies other than the
BCGs undergoes several mergers during their past history. The majority
of galaxies never have mergers, or only one at very early redshift.

Because of the star formation criteria employed in the simulation, all
stars found in the DSC at $z=0$ were born in galaxies and later
dissolved from them. We track each DSC star particle back to the last
redshift when it still belonged to a galaxy, and thus link it to the
dynamical history of this galaxy. We exclude all DSC star particles from
the analysis which arise from volatile and under-dense galaxies; the
latter being defined relative to the observational mass--radius
relation of early type galaxies by Shen et al. 2003.
The main results of our analysis can be summarised as follows.
\begin{itemize}
\item The formation of the DSC has no preferred redshift and is a
  cumulative power--law process up to redshift $z=0$. We find that
  $\simeq 70$ per cent of the DSC is formed after redshift $z=1$.
\item We find a weak increase of the final amount of DSC stars with
  the mass of the cluster, but no significant correlation with the
  global dynamical history of the clusters.
\item For all but the 3 most massive clusters, DSC star particles come
  mainly from the family tree of the most massive (BCG) galaxy. I.e.,
  the formation of the DSC goes largely in parallel with the build-up
  of the BCG galaxy.
\item Most DSC star particles become unbound during merging phases along
  the formation history of the BGCs, independent of cluster mass.
\item Masking the inner 250 $h^{-1}$ kpc of each cluster, in order to
  exclude the cD halo from the analysis, does not qualitatively change
  the emerging picture.
\end{itemize}

>From these results we conclude that the bulk of the DSC star particles
are unbound from the galaxy in which they formed by the tidal forces
acting before, during, and shortly after merging events during the
formation history of the BCGs and other massive cluster galaxies.
Only in the outskirts of clusters, $R > 0.5 R_{\rm vir}$, we find that
galaxies of many different masses provide comparable contributions to
the DSC , which is similar to a ``quiet stripping'' scenario, but the
actual mass in DSC stars in these regions is small.

The formation of the BCG in these simulations is related to many
mergers which begin early in the history of these galaxies and
continue all through $z=0$.  As discussed in the previous section, it
is reasonable to infer that the massive elliptical galaxies, which
merge with the BCGs, are contributed by infalling groups, which have
already generated their own DSC and/or loosely bound halos, as found
by Willman et al. 2004 and Rudick et al. 2006. Part of the cluster
DSC will also be generated by the tidal shocking and stripping during
the merger of these massive galaxies with the BCG itself; the relative
importance of these processes is yet to be established.

Since the fraction of diffuse light stars contributed by each
accreting group depends on the details of the dynamical history of the
group itself, such a mechanism for the generation of the DSC can hide
a direct link with the formation history of the clusters. This may be
the reason why we do not detect a clear correlation between the $z=0$
DSC fraction and the cluster formation history.

At the resolution of our (and other similar) simulations, it is not
yet possible to resolve the inner structure of low-mass galaxies. We
have taken this into account in our analysis by discarding all DSC
particles from galaxies with densities below a threshold set by
observations. In addition, there are well-known problems in
cosmological simulations with forming disk galaxies, and with
reproducing the galaxy luminosity function. These issues clearly
introduce some uncertainty in the discussion of the origin of the DSC
in hydrodynamic $\Lambda$CDM simulations.  We find that the global
amount of DSC in our simulations {\sl increases} with numerical
resolution and has not yet converged in the best simulations. Thus a
straightforward comparison of observed DSC fractions with numerical
simulations is not possible yet. On the other hand, massive galaxies
are well-resolved in our simulations, and we believe that our main new
result, that a major fraction of the DSC in galaxy clusters is
dissolved from massive galaxies in merging events, is a robust one.

\newpage

\section*{APPENDIX A: Identifying cluster galaxies with SKID} 
\label{AppA}

For the purposes of the present work, we need a dynamical and
automated way to identify galaxies in the simulations. Our galaxies
must be self---bound structures, locally over-dense, and we need an
operational procedure to unambiguously
decide whether a star particle at a given redshift belongs to an
object or not. For this reason, we follow the procedure adopted in M04
and use the publicly available SKID algorithm (Stadel 2001) and
apply it to the distribution of star and dark matter particles. 

\begin{figure*}
\centerline{
\includegraphics[viewport= 55 40 600 550,clip, height
  =9cm]{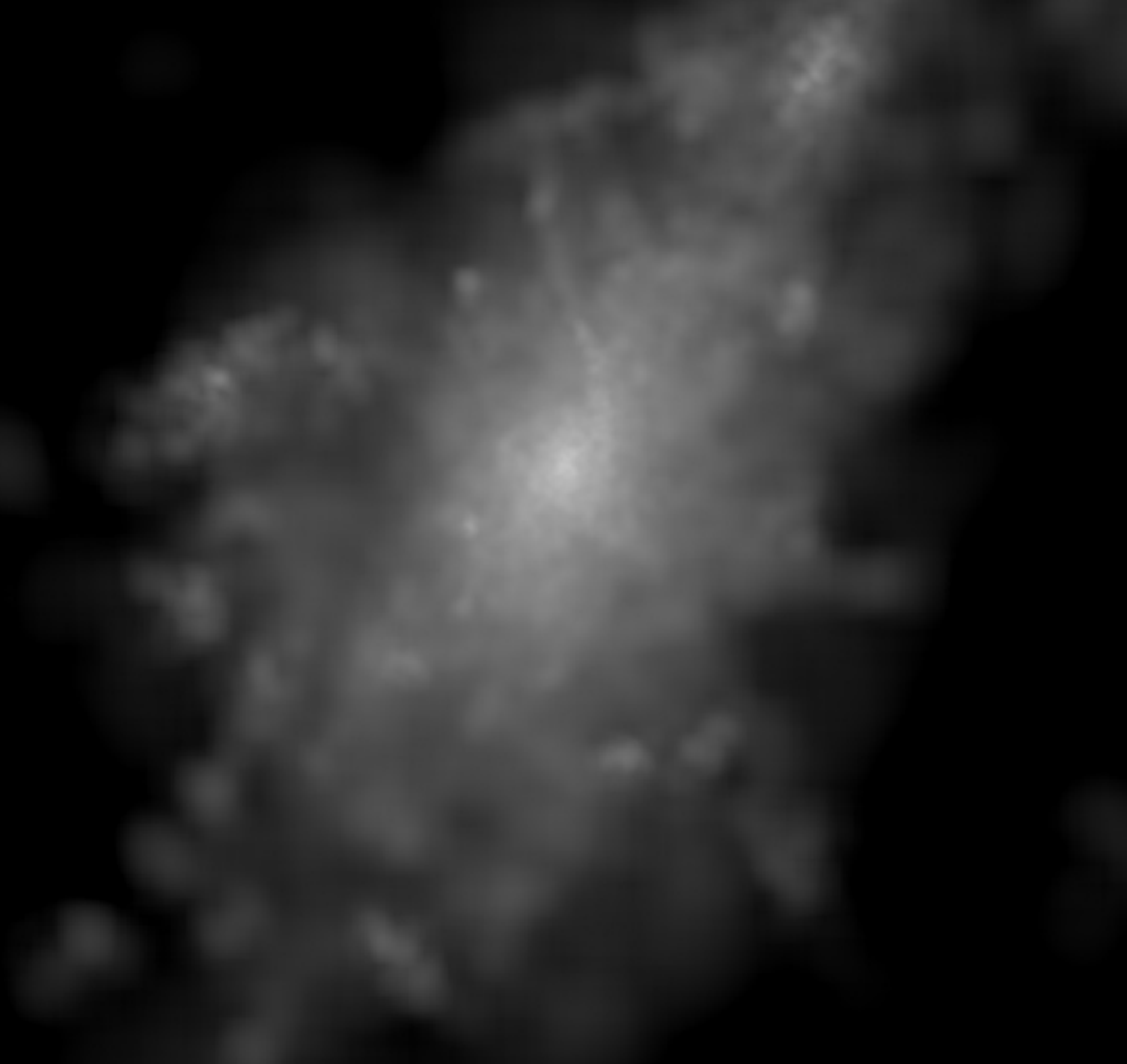}
\includegraphics[viewport= 55 40 600 550,clip, height =9
  cm]{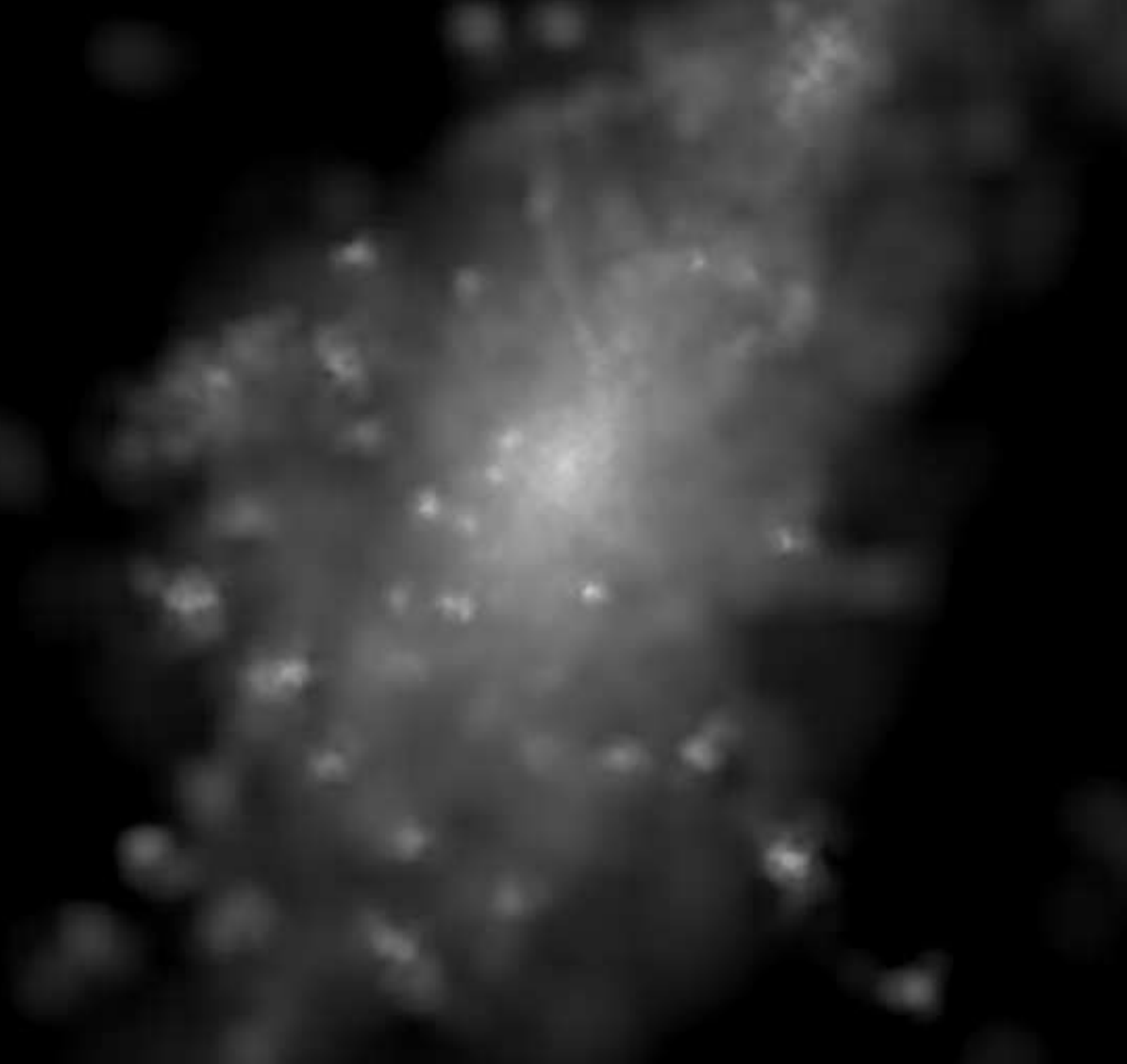}}

\captionsetup{font={normalsize,sf}, width= 0.9\textwidth}
\caption{ Surface density map of the DSC found in cluster D, when
three values of $N_{\rm sm}$ are used (left panel) and when only one
is used (right panel).  }\label{appmaps}
\end{figure*}

\begin{figure}
\centerline{
\includegraphics[viewport= 65 50 310 580,clip, height =13cm]{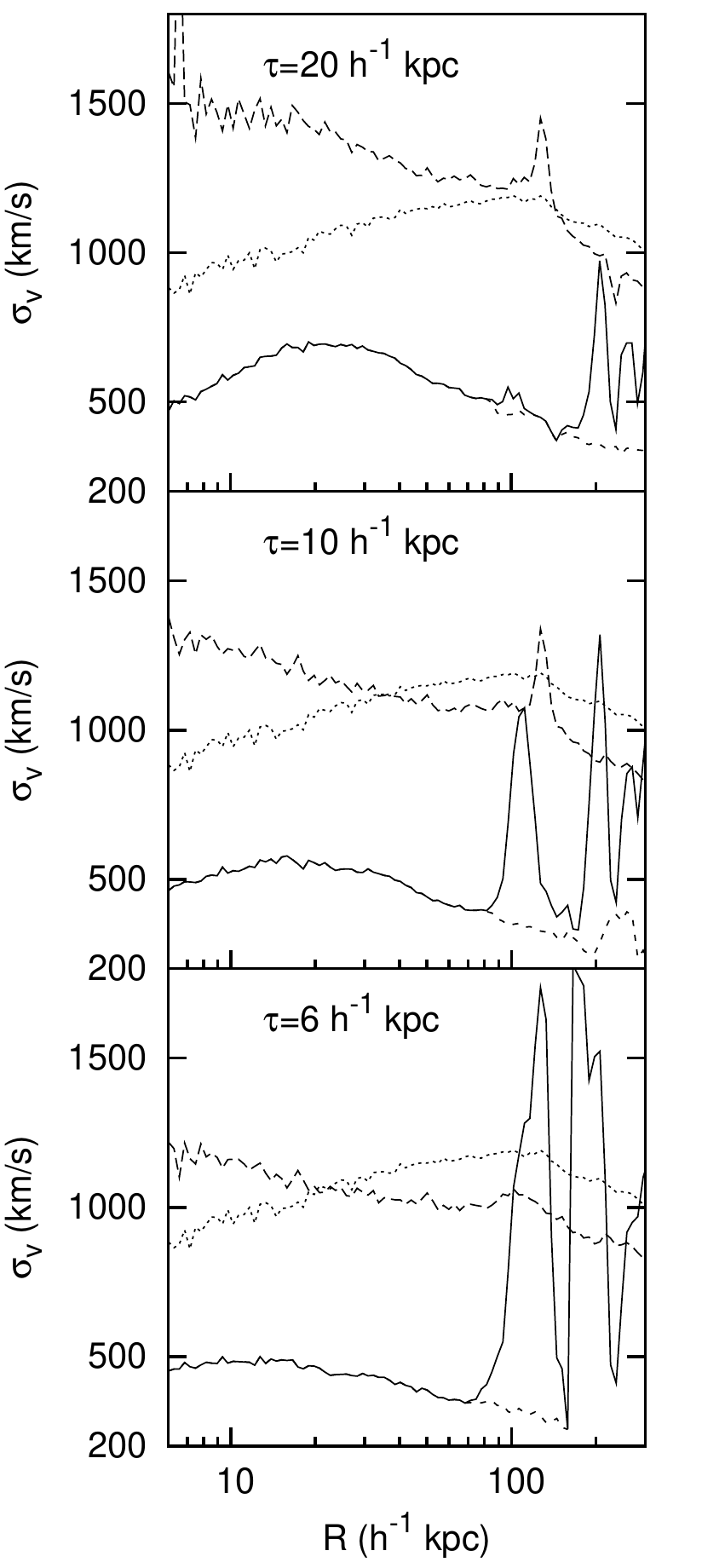}
}
\captionsetup{font={normalsize,sf}, width= 0.9\textwidth}
\caption{ 
Velocity dispersions in the central part of cluster A, simulated at our
higher resolution. Solid lines: stars belonging to SKID galaxies;
long-dashed lines: DSC stars; dotted lines: DM particles; short-dashed
lines: cD stars. Upper panel shows the results for the SKID analysis
performed with a value of $\tau=20 h^{-1}$ kpc, centre panel for
$\tau=10 h^{-1}$ kpc, lower panel for $\tau=6 h^{-1}$ kpc.
}
\label{sigmavel}
\end{figure}

The SKID algorithm works as follows:
\begin{itemize}
\item An overall density field is computed from the distribution of
  all available particle species, generally DM, gas and star
  particles.  The density is estimated with a SPH spline--kernel,
  using a given number $N_{\rm sm}$ of neighbour particles. In the
  following we only include DM and star particles.
\item The star particles are moved along the gradient of the density
  field in steps of $\tau/2$. When a particle begins to oscillate
  inside a sphere of radius $\tau/2$, it is stopped. In this way,
  $\tau$ can be interpreted as the typical size of the smallest
  resolved structure in the distribution of the star particles.
\item When all star particles have been moved, they are grouped using a
  friends-of-friends (FOF) algorithm applied to the moved particle
  positions. The linking length is again $\tau/2$.
\item The gravitational potential and binding energy of each group
  identified in this way is computed by accounting for all the
  particles inside a sphere centred on the centre of mass of the
  group and having radius $2\tau$ (for the moved star particles, their
  initial positions are used in the computation of the potential). The
  binding energies of individual particles are then used to remove
  from the group all the star particles which are recognised as
  unbound, in an iterative way: the centre of mass of the group and
  its potential are recomputed after a particle has been discarded.
\item Finally, we retain such a SKID--group of stars as a galaxy if it
  contains at least 32 particles after the removal of unbound
  stars. The exact value of this number threshold is unimportant, but
  the smaller the threshold is, the higher is the probability of
  identifying as ``galaxy'' a random set of neighbouring star
  particles. Using 32 particles correspond to a mass threshold of
  $M=1.1 \times 10^{10} h^{-1} M_\odot$ for the cosmological simulation. 
\end{itemize}

The resulting set of objects identified by SKID depends on the choice
of two parameters, namely $\tau$ and $N_{\rm sm}$. After many
experiments and resorting to visual inspection in a number of cases,
we find that a complete detection of bound stellar objects requires
the use of a set of different values of $N_{\rm sm}$. Using only one
value for $N_{\rm sm}$ results in ``missing'' some galaxies. We use
$N_{\rm sm}=16,32,64$, and define a {\sl galaxy} to be the set of star
particles which belong to a SKID group for any one of the above
$N_{\rm sm}$ values. If a star particle belongs to a SKID group for
one value of $N_{\rm sm}$ and to another group for a different $N_{\rm
sm}$, then the groups are joined and are considered as a single
galaxy. All star particles not linked to any galaxies are considered
to be part of the diffuse stellar component in the cluster. The left
panel of Fig. \ref{appmaps} shows the surface density map of the DSC,
as identified for our cluster D when all the three values of $N_{\rm
sm}$ are used. In the right panel, we show the same map obtained using
only $N_{\rm sm}=32$. In the latter, the bright spots correspond to
``missed'' galaxies.

$\tau$ roughly corresponds to the size of the
smallest resolved structure, and we adopt
$$
\tau \approx 3 \epsilon \eqno(A.1)
$$ 
which is the scale where the softened force becomes equal to the
Newtonian force. We have tested this choice by
visual inspection in a number of clusters, and by performing an
analysis of the velocity dispersions for the stars belonging to SKID
galaxies and to the DSC. Fig. \ref{sigmavel} shows the velocity
dispersions for various components, namely the stars in galaxies, the
stars in the DSC, the stars in the cD galaxy and the DM particles, for
our cluster A (re-simulated at our higher resolution) when the value of
$\tau$ is varied. The velocity dispersions are computed in spherical
shells centred on the cluster centre, defined as the position of the
DM particle having the minimum gravitational potential. For this
high-resolution simulation, our fiducial choice is $\tau=6 h^{-1}$
kpc. In the bottom panel of Fig. \ref{sigmavel}, the spikes in the
velocity distribution of the stars in galaxies (the solid curve) at
$R>100 h^{-1}$ kpc correspond to SKID objects; no prominent spikes or
bumps appear in the velocity dispersion profile for the DSC
(long-dashed line), meaning that no structures in velocity that might
correspond to ``missed'' galaxies are present in this component. Also,
the value of the velocity dispersions for DSC and DM particles (dotted
line) stay within $\approx 20$ \% from each other, as expected when
both component sample the same gravitational cluster potential.

When a larger value is used ($\tau=10 h^{-1}$ kpc), spikes begins to
appear in the DSC velocity dispersion curves, indicating that some
objects, or part of them, are missed by the algorithm. This is
especially clear for the spike at $R \approx 100 h^{-1}$ kpc, which is
present both in the velocity dispersion of stars in galaxies and in
the velocity dispersion of the DSC, and clearly indicates that a
fraction of stars in some galaxies have been mis-assigned to the
diffuse component. Also, the discrepancy between DSC and DM velocity
dispersions begins to grow, and the velocity dispersion of cD stars
gets unrealistically large, $\sigma_v> 500$ km/s. This happens because
some low--speed DSC stars begin to be assigned to the cD, whose
typical star particle velocities are even lower, thus increasing the
velocity dispersion of the cD star population. The situation gets
worse it the value of $\tau$ is increased to $20 h^{-1}$ kpc.

We have performed a similar analysis on some clusters taken from our
cosmological set, where the fiducial value from eq. (A.1) is $\tau=20
h^{-1}$ kpc. Again, increasing $\tau$ to larger values results in
having structures in the velocity space of the DSC, presumably due to
missed objects. We conclude that the scaling (A.1) gives good results
for the SKID galaxy identification, while keeping $\tau$ fixed to a
given value (e.g. $20 h^{-1}$ kpc) when the force resolution is varied
is not a good choice.

When analysing particle distributions at redshift $z>0$, we keep fixed
the value of $\tau$ in co-moving coordinates, thus allowing the
minimum physical size of our object to decrease with increasing
redshift. While this does not obey equation~(A1), $\tau$ never becomes
less than $\epsilon$. The effect is probably to slightly increase the
amount of ``volatile'' galaxies at higher redshifts.  Again, this
choice was tested by visual inspection and by analysing the velocity
dispersion distributions.

Also, we note that at high redshift ($z>1$) the distribution of gas
particles inside star--forming proto--clusters often contains adjacent
clumps of star particles. Applying SKID to such distribution results
in ``galaxies'' composed of two or more of such clumps, which instead
should be considered as separate galaxies. For this reason, we choose
to use only DM and star particles for the galaxy identification.

\chapter{MUPPI - Multiphase particle integrator}
\label{MUPPI_chap}
In chapter 2, we presented various algorithms to include star
formation and SN energy feedback in cosmological simulation
(Sec.~\ref{sfsimu}) and in semi-analytical models (Sec.~\ref{MO04}).
In this chapter, we present a novel algorithm, MUPPI
(\textbf{MU}lti\textbf{P}hase \textbf{P}article \textbf{I}ntegrator), a
modified version of the Monaco 2004 (MO04, Sec.~\ref{MO04}) analytic ISM, Star
Formation and SNe feedback model, whose implementation in the 
GADGET-2 code (as a routine substituting the original star
formation function, see Sec.~\ref{GDT:sf_eff}) has
been the principal work of the present PhD Thesis. The version of the
GADGET code we use adopts an SPH formulation with entropy conserving
integration and arithmetic symmetrisation of the hydrodynamical forces
(see Sec.~\ref{GDTentro}) and includes radiative cooling computed for
a primordial plasma with vanishing metallicity.\\ 
One of the major problems of current simulations of galaxy formation
is how to address the complex hydrodynamical
processes in the multiphase interstellar medium (Ceverino \& Klypin,
2008?). The insertion of a physically motivated model for star
formation and SN feedback in place of a set of phenomenological
recipes is the main novelty brought by MUPPI to the current SF models for
simulations of galaxy formation. In MUPPI, in fact, the ISM dynamics depends
only on \textit{local} thermodynamical conditions, i.e. we don't use
equilibrium solutions to describe the ISM evolution. The
``equilibrium'' approach is instead taken for example in the GADGET-2 effective
star formation model (Sec.~\ref{GDT:sf_eff}). This means that MUPPI
not only follows the non 
equilibrium phase at the onset of multi-phase regime, but also the
response to changes in the local thermodynamics, e.g. pressure and/or
temperature changes due to compressions/rarefactions and shocks.
 A detailed description is given in the
following sections.\\
The organisation of the present chapter is as follows. In
Sec.~\ref{MUP:ini}, we report how we initialise the model; in
Sec.~\ref{MUP:core} we describe the model core and thus all the
processes involved in regulating the ISM during and after star
formation; finally, we account for the SNe energy redistribution in
Sec.~\ref{MUP:fin}. In Appendix ~\ref{param}, we list the employed parameters;
in Appendix ~\ref{flowchart} we depict the flow chart of the MUPPI code and of the
GADGET-2 code, highlighting the routines that have been changed by the MUPPI
 implementation.  

\section{Model equations and initialisation}
\label{MUP:ini}
The ISM gas is found in at least two forms: cold clouds of neutral
atomic or molecular hydrogen and hot ionised hydrogen near hot young
stars. These two inter-penetrating components have similar pressure but
very different temperature and density ranges
(McKee \& Ostriker 1977, Efstathiou 2000).\\
\begin{figure}
\centering{
\includegraphics[width=0.4\linewidth]{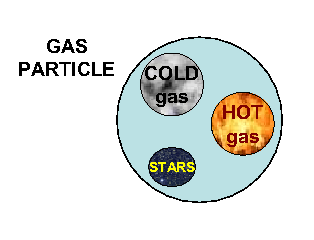}
\captionsetup{font={normalsize,sf}, width= 0.9\textwidth}
\caption{Schematic illustration of a MUPPI gas particle.}}
\label{MUPPI:part}
\end{figure}

 We follow the assumption that each gas
 particle is composed by a star component and by two gas phases, a hot and a
 cold one (denoted respectively by subscripts $h$ and $c$), which are in
 thermal pressure equilibrium, i.e. 
\be
n_h\cdot T_h = n_c \cdot T_c
\ee
Here and in the following $n$ denotes the number density, e.g. $n_h =
M_h/(\mu_h\cdot mp)$, where $m_p$ is the proton mass, $\mu_h = 4 / (5
\cdot f_{HI} + 3) =  0.6$ and
$\mu_c = 4./ (3. \cdot f_{HI} + 1.) = 1.2$ are the molecular
weights, with $f_{HI}$ the fraction of neutral hydrogen. 
Note that we kept the 
temperature of the cold phase fixed to $T_c = 10^3$K.    
At a given timestep, a gas particle enters the {\it multiphase regime} if it
full fills both a density, $\rho>\rho_{\rm thr}$, and a 
temperature threshold, $T<T_{\rm thr}$. In the present work, we adopt
values of $\rho_{\rm thr}$ corresponding to a number density of $n_{\rm
thr}$ = 0.25 cm$^{-3}$ and $T_{\rm thr}$ = $5\cdot10^{4}$
K. Multiphase particle quantities are then initialised using the
current values for the gas particle mass $M_{p}$, specific internal energy
$E_{int}$ and volume 
\be
V_{p} = \frac{M_{p}}{\rho}
\label{MUPPI:vol}
\ee 
The hot and cold
mass fraction ($F_h$= 0.999 and $F_c$ = 0.001) set initial values for
the hot, $M_h$, and the cold mass, $M_c$. The particle should be
 completely in the hot phase at the onset of multi-phase regime; we set
 the value of $F_c$ to 0.001 only to avoid numerical instabilities in
 Ordinary Differential Equations (ODE)
 integration. Then, the initial 
temperature and, consequently, the energy of the hot phase are
evaluated as:  
\be
T_h = \frac{E_{int}\cdot M_p}{\frac{3}{2}k\frac{M_h}{\mu_h m_p}} -
\frac{M_c}{M_h}\frac{\mu_h}{\mu_c}T_c 
\ee
\be
E_h =  \frac{3}{2}k \cdot \frac{M_h}{\mu_h m_p} T_h
\ee
 At the first time step, MUPPI only performs initialisation, then the particle
exits the model and goes back to GADGET. The next timestep, if it
 still fulfils the 
multiphase thresholds, it is allowed to enter the MUPPI
core routine if the first
\textit{exit} condition is not met:
\be
\rho_{out}<\frac{\rho_{\rm thr}}{1.5} \qquad \qquad \textrm{}
\ee  
This is to avoid gas particles which are almost outside the star
forming region (because part of a galactic fountain for example) to
evolve the model and thus update their stellar mass component,
besides all the other multiphase quantities. We
verified, in fact, that without this exit condition, gas particles
may spawn stars when they are already far away from the star forming
region. \\
If no other exit conditions are met, a particle stays in the
multi-phase stage for a time $t_{fin}$ = 2$t_{dyn}$, where $t_{dyn}$
is defined below (Eq.~\ref{MUPPI:tdyn}). After $t_{fin}$, the giant
molecular cloud to which the particle belongs is considered to be
destroyed. We use as an estimate of $t_{dyn}$ its value when the cold
fraction reaches 99$\%$; after that Eq.~\ref{MUPPI:tdyn}describes the
dynamical time of the rising molecular gas phase, rather than that of
the parent GMC. 

Similarly to MO04, we model the interplay among the three phases with a
system of ODEs describing the various
mass and energy flows which involve each phase. From the same work, we
adopt the following equations: 
\be
\dot{M}_{\star} = \dot{M}_{SF} - \dot{M}_{re} \qquad \qquad \qquad 
\textrm{\small{MASS IN STARS}}
\label{sf}
\ee
\be
\dot{M}_c = \dot{M}_{cool} - \dot{M}_{sf} - \dot{M}_{ev} \qquad \qquad
\textrm{\small{COLD MASS}}
\label{mc}
\ee
\be
\dot{M}_h = -\dot{M}_{cool} + \dot{M}_{re} + \dot{M}_{ev} \qquad \qquad
\textrm{\small{HOT MASS}} 
\label{mh}
\ee
A schematic view of the mass flows among the different phases is given
in Fig. 4.2.\\
The energy of the cold phase is kept constant at a temperature $T =
100K$. The energy of the hot phase instead evolves according to:
\be
\dot{E}_h = \dot{E}_{SN} - \dot{E}_{cool} + \dot{E}_{hyd} \qquad \qquad
\label{eh}
\ee
where the first term on the right hand side describes the
non-gravitational energy injected by exploding supernovae, the second
term the radiative energy losses by cooling ($\dot{E}_{cool} =
E_{h}/t_{cool}$) and the third term the 
energy change due to the hydrodynamics, introduced by the
thermodynamical conditions of the medium after the gas particles
evolved of our model (see Sec.~\ref{MUPPI:interface}).\\
We describe each term of the above equations (~\ref{sf} - ~\ref{eh}) 
in the next sections.

\section{Model core}
\label{MUP:core}
After initialisation, the multiphase gas particle
evolution is followed by integrating the ODEs for mass and energy flows
using an adaptive step 
size fourth-order Runge-Kutta (RK4) algorithm. The RK4 routine
propagates a solution 
over an interval ($x_2 - x_1$) by combining the informations from
several Euler-style steps and then using this information to match a
Taylor series expansion up to some higher order (see e.g ``Numerical
recipes in C. The art of scientific
computing''\nocite{1992nrca.book.....P} for a technical account). The
integration is taken from $x_1 = 0$ to the 
current GADGET timestep $x_2 = \Delta t$ with initial stepsize $h = \Delta
t/100$.  In particular situation, it could happen that the RK4 stepsize
becomes similar to the cooling time, $t_{cool}$, thus compromising the
code ability in closely follow the hydrodynamical changes. Therefore
before entering the RK4 routine, we check if $h > t_{cool}/5$, then
the stepsize is set to $h = t_{cool}/5$.\\ 
Once inside the RK4 derivative function, the code computes over the
timestep $h$ the solutions $M_c$, $M_c$, $M_{*}$ and $E_h$ to the MUPPI
differential equations describing star formation and feedback related
physics. This is
achieved at first by directly computing the current values for the
following quantities: the hot and cold filling factors,
\be
 f_h = \frac{1}{ 1.0 + \frac{F_c}{F_h}\cdot \frac{\mu_h}{\mu_c} \cdot \frac{T_c}{T_h}} \qquad \qquad
f_c = 1 - f_h
\ee
the volumes occupied by the hot and the cold phases,
\be
V_h =  V_p \cdot f_h   \qquad \qquad V_c = V_p \cdot f_c
\ee
and finally the densities of the hot and cold component
\be
\rho_h =  M_h / V_h  \qquad \qquad \rho_c = M_c / V_c
\ee
 We use the GNU-Scientific Libraries (GSL) to perform the integration
over the whole code timestep $\Delta T$.
When convergence is reached, the model exits the RK4 routine and goes
back to the main code. During such integration, two more
\textit{exit} conditions are posed: 
\be
M_h <  M_c/10^9 \qquad \qquad
\ee
which check if the gas particle is ``freezing'', thus becoming
depleted of hot phase and not able to sustain a 
multi-phase gas anymore, and 
\be
f_c <  10^{-18} \qquad \qquad 
\ee
which controls if the gas particle is depleted of the
  cold phase. If these cases are true, the gas particle exists the RK4
  routine and goes back to the main code. Soon after, all its
  multiphase quantities are set at zero and the particle exits MUPPI. \\

In the next paragraphs, we describe the physical processes and flows 
between the gas phases and the star mass. 
\subsection{The formation of molecular clouds}
The original MO04 paper tries to model the formation of
giant molecular clouds (GMC). In numerical simulations, often one gas
particle has a mass lower than that of a typical GMC, and thus it
makes no sense to follow the formation of GMCs inside it. Thus, we use
instead a phenomenological prescription for describing the amount of
molecular gas and the consequent star formation.\\

Blitz \& Rosolowsky (2006) showed that the ratio of atomic to
molecular gas in galaxies is primarily determined by the interstellar gas
pressure $P_{ext}$. Following their work, the fraction $f_{\rm mol}$ of
atomic hydrogen (which is directly connected with the star formation
rate) present in the cold phase is calculated as:
\be
f_{\rm mol} = \frac{1}{1 + 4(\frac{P_{0}}{P_{ext}})}
\label{fcoll}
\ee 
where $P_{0} = 10^{4}$ (normalised to observed values found in the Milky Way) and
$P_{ext} = P_{th} + 
P_{kin}$ is the total pressure exerted by the gas on the ISM.
The thermal pressure $P_{th}$ is mainly powered by the thermal
energy injected by SN explosions (Mc Low 2002)
 while the kinetic term $P_{kin}$ is mainly driven by cold cloud
 turbulence. We modelled the kinetic pressure in MUPPI with a simple
 prescription, thus considering separate contributions from thermal and kinetic
pressure, and verified that results are substantially unchanged with
respect to the formulation (~\ref{fcoll}). We dropped such kinetic pressure
model not to add one more free parameter in MUPPI. \\

\subsection{Star formation}
\label{MUP:sf}
Once the fraction $f_{\rm mol}$ has been computed, star formation
begins. The rate at which
cold gas transforms into stars, i.e. the star formation rate (SFR), is
evaluated as: 
\be
\dot{M}_{SF} = f_{\star} \cdot \frac{f_{\rm mol}\cdot M_{c}}{t_{dyn}}
\label{SFR}
\ee
where $f_{\star}$ gives the fraction of star forming gas effectively
converted into stars.
The fraction $f_{\rm mol}$ multiplied by the cold mass $M_{c}$ gives the
amount of gas available for star formation, i.e. the molecular
  gas, $M_{mol}$. 
 We assume that star formation proceeds on the order of the dynamical
 time given by:  
\be
t_{dyn} = \sqrt{\frac{3\pi}{32G\rho_{\rm c}}} \backsimeq 5.15\cdot 10^{7}
\sqrt{\mu_{c}\cdot n_{c}} yr
\label{MUPPI:tdyn}
\ee
A fraction $f_{re}$ (0.2 in our runs) of cold gas
involved in star formation is immediately restored to the hot phase:
\be 
\dot{M}_{re} = f_{re} \cdot \dot{M}_{SF}  
\ee
 Note that here we just build up
stellar mass, we don't spawn star particles. This is eventually done
after exiting the MUPPI routine, when the computation goes back to
GADGET and the probabilistic method described by Eq.~\ref{prob_EFF} is
invoked to form new star particles. The star formation rate calculated
by MUPPI is adopted in the probabilistic method. \\
\begin{figure}
\centering{
\includegraphics[width=1\linewidth]{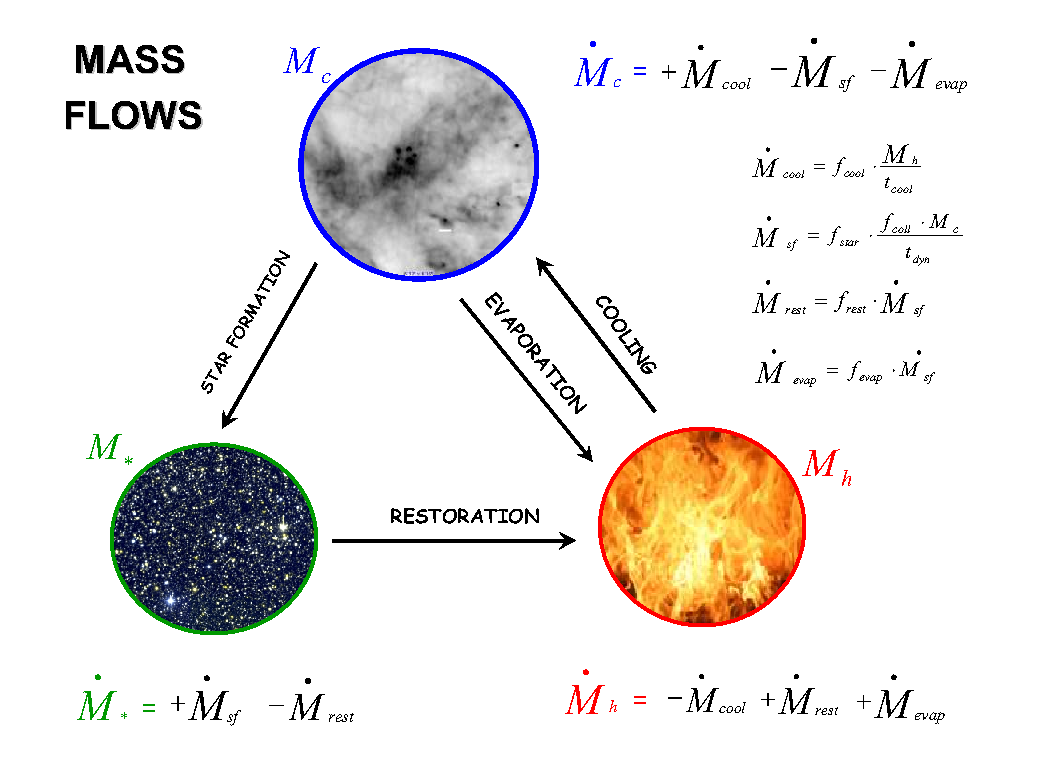}
\captionsetup{font={normalsize,sf}, width= 0.9\textwidth}
\caption{Schematic illustration of the mass flows between the
  different phases.}}
\label{MUPPI:mflows}
\end{figure}
\subsection{The cold and the hot masses}

The replenishment of the cold phase is counteracted by star formation
(Eq.~\ref{SFR}), which depletes the cold gas reservoir, 
 and by evaporation of the cold collapsing gas back to the hot
phase due to SN explosions.\\
 The rate at which gas cools from the hot phase is assumed to be: 
\be
\dot{M}_{cool} =  \frac{M_h}{t_{cool}}
\ee
where $t_{cool}$ is the radiative cooling time derived from the GADGET-2
cooling function which uses tabulated cooling rates given by 
Sutherland \& Dopita (1993)\nocite{1993ApJS...88..253S}. 
If the multiphase gas particle under exam has a temperature lower than
$T = 10^{4} K$, we fit the cooling time with the following expression: 
\be
t_{cool}^{fit} = t_{cool} \cdot (\frac{T}{T_{cf,0}})^{-3}
\ee
where $T_{cf,0} = 1.78 \cdot 10^{4} K$. We made this approximation
because the original GADGET cooling function dies at $T = 10^{4} K$.\\

A future development of MUPPI will consist in consistently including
molecular and atomic cooling, down to $T$ = 100 K. Moreover, when
metallicity of gas will be available thanks to the integration of
MUPPI with the chemical evolution model implemented by Tornatore et al.
2004\nocite{2004MNRAS.349L..19T} and already present in the code, we
will also include the dependence of gas cooling upon its metallicity.
The fit implemented here to follow cooling at temperatures lower than
1000 K is only needed to avoid instabilities at the onset of
multi-phase regime, when the particle may have low temperatures and the
energy feedback from SNe is not yet active. We verified that using different
kind of fits doesn't substantially change our results.

As a first source of early SN feedback, we suppose that collapsing
clouds are evaporated back to the hot phase by exploding
supernovae with a rate given by:
\be
\dot{M}_{ev} = f_{ev} \cdot \dot{M}_{SF}   
\ee
where $f_{ev}$ is 0.1 in our runs.
This term is inserted in the equation regulating the hot gas mass
evolution in order to take into account the effects of thermal
conduction at the interface between the cold and the hot mass.
 Again, the integration of MUPPI and the Tornatore et al. 2004 scheme will
allow us to also consider SNIa energy contribution to the hot phase.
\subsection{Supernova feedback energy}

We suppose that SNe exploding within a star forming gas particle give rise to a
super-bubble which propagates in the more pervasive hot
phase. As previously depicted in Sec.~\ref{fateSN}, we consider two
possible feedback regimes following naturally from 
this setting: the \textit{pressure confinement} regime, modulated by
the fraction $f_{\rm fb,i}$, which
gives the amount of thermal energy stopped by the external pressure
within the star forming gas particle and the \textit{blow out} regime,
modulated by the parameter $f_{\rm fb,o}$, which describes
the energy blowing out of the gas particle which is redistributed
among neighbours.\\ 
 The
 redistribution of the fraction $f_{\rm fb,o}$ of energy which blows \textit{outside} the
 particle is treated by a separate routine, called just after MUPPI
 (see Sec.~\ref{MUP:fin}).\\ 
 In the ODE integrator core, we thus take into account only the fraction of
 SN energy which remains trapped \textit{inside} the star forming particle.
The heating rate due to SNe is:
\be
\dot{E}_{SN} = E_{51} \cdot f_{\rm fb,i} \cdot \frac{\dot{M}_{sf}}{M_{\star,sn}}
\ee
where $E_{51} = 10^{51}$ erg is the canonical value for the energy
released by one supernova and $M_{\star,sn} = 120 M_{\odot}$ is the
mass of formed stars per SN.\\

\subsection{The SPH/model interface and the term $\dot{E}_{hydro}$}
\label{MUPPI:interface}
\begin{figure}
\centering{
\includegraphics[width=0.8\linewidth]{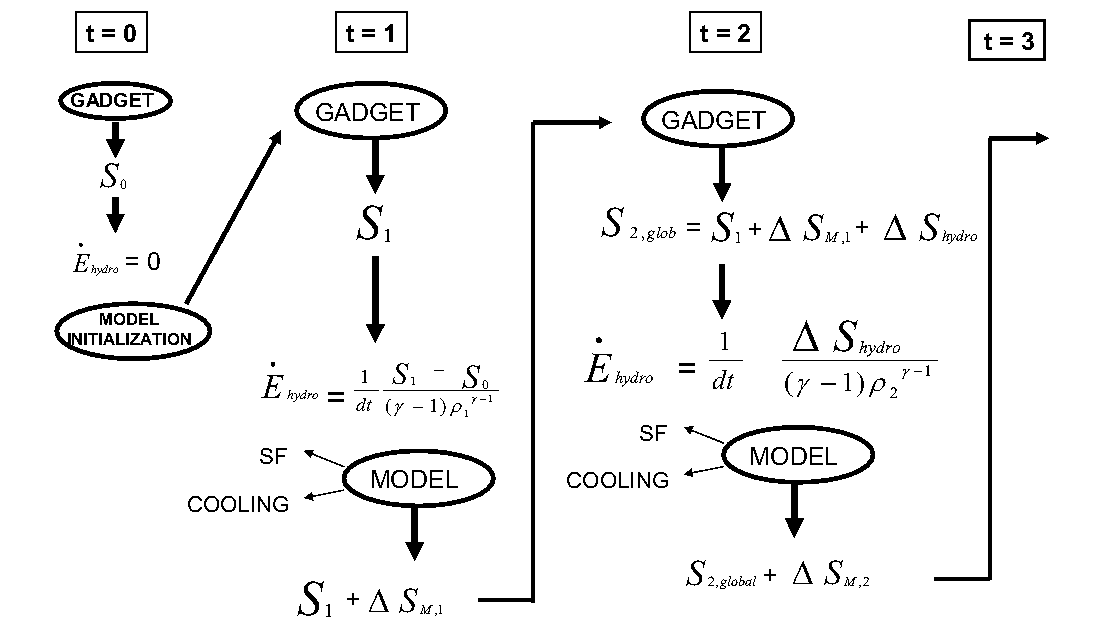}
\captionsetup{font={normalsize,sf}, width= 0.9\textwidth}
\caption{Schematic illustration of the interaction of MUPPI
   with the GADGET SPH; see text for details.}}
\label{MUPPI:ehydro}
\end{figure}

 Star formation and cooling change the hydrodynamical conditions
 of the multiphase particle after the ODE integration. We thus need
 to introduce a term in which we can store these changes to later
 inform the SPH.\\ 
 When a gas particle with entropy $S_n$ fulfils the multiphase
 thresholds and enters MUPPI, its entropy is updated to
 $S_n + \Delta S_{muppi}$. The term $\Delta S_{muppi}$ is
 evaluated from the mass weighted average energy $E_{ave}$ of the hot
 and the cold phases, given by: 
\be
 E_{ave} = k \cdot \frac{T_{ave}}{(\gamma -1 )\cdot \mu_{med} \cdot m_p}
\ee
where
\be
T_{ave} = \frac{\mu_{med}}{M_c + M_h} \cdot \frac{M_h \cdot T_h}{\mu_h} 
	  \frac{M_c \cdot T_c}{\mu_c}
\ee
and $\mu_{med}= (\mu_c \cdot M_c + \mu_h \cdot M_h)/(M_c + M_h)$. \\
 
At the following timestep ($n+1$) the SPH evaluates the new thermodynamical
 conditions of the medium $S_{n+1}$, increasing thus the entropy by:
\be
\Delta S_{hyd}=  S_{n+1} - S_n - \Delta S_{muppi} \label{shydr}
\ee
The rate of energy change due to the hydrodynamics is thus:
\be
\dot{E}_{hyd}= \frac{1}{dt}\frac{\Delta S_{hyd}}{(\gamma -
    1)\rho^{\gamma -1}}
\ee
where $\gamma$ is the adiabatic index. The term $1/(\gamma -
    1)\rho^{\gamma -1}$ is the SPH factor to convert entropies into
    energies.
The term $\Delta S_{muppi}$ is subtracted in the evaluation of Eq.~\ref{shydr} 
because it would otherwise be computed twice, at timestep $n-1$ directly
by MUPPI, and at timestep $n$ since included in $\Delta S_{hyd}$. We note
that gas entropy changes in the entropy conserving version of
GADGET can happen due to two processes: viscous energy dumping, caused
by the artificial viscosity, and hydrodynamic variations in pressure
and temperature, due to interaction with surrounding particles. The
latter process is physical, e.g. it is extreme during shocks. This
term is the one which enables MUPPI to respond to the local
hydrodynamical properties of the gas and to their evolution.
For a schematic representation, see Fig. 4.3.

\section{Final steps: storing and redistributing the ``blow-out
  regime'' energy}
\label{MUP:fin}
After having found the solutions to the system of ODEs, the
computation proceeds in the main code by computing the amount of
SNe energy that blows outside the particle (``blow-out
regime'') and thus has to be redistributed among
particle neighbours.\\  
The total out-flowing thermal energy per particle that needs to be
redistributed at a given timestep is calculated as:
\be
\Delta E_{SN,o} = E_{51} \cdot f_{\rm fb,o}\cdot \frac{\Delta
  M_{sf}}{M_{\star,sn}} 
\ee
where $E_{51} = 10^{51}$ erg is the energy released by one supernova,
 $\Delta M_{sf}$  is the mass in stars formed in the timestep
and $M_{\star,sn} = 120 M_{\odot}$ is the mass of formed stars per SN.\\
Before exiting the main code, we update the SPH entropy
by considering the change in internal energy introduced by the
evolution of the model core. Afterwards, the code goes back to GADGET
and later to the function to redistribute the blow out energy is
called.
\subsection{Thermal energy redistribution} 
Following the SPH formalism, we define neighbouring particles as those
lying within the sphere of radius given by the star forming particle
smoothing radius $h_{\star}$. For each SN explosion event, the maximum
number of particles that can receive the SN energy is fixed by the
number of SPH neighbours ($N_{ngb} = 32$). Among the SPH
neighbours, we select those which lie within 
the cone with vertex at the star forming gas particle, axis parallel
to the local density gradient $\vec{\nabla}\rho$ and aperture
$2\theta= 140\pi$ (see Fig. 4.4).

\begin{figure}
\centering{
\includegraphics[width=0.5\linewidth]{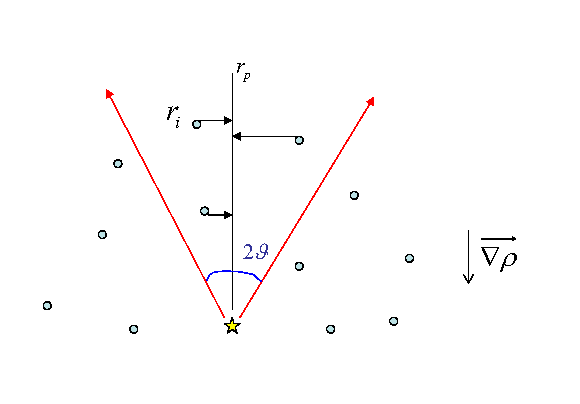}
\captionsetup{font={normalsize,sf}, width= 0.9\textwidth}
\caption{Schematic illustration of the energy redistribution mechanism.}}
\label{MUPPI:ered}
\end{figure}

We then distribute the energy 
$\Delta E_{SN,o}$ on the basis of the distance from the neighbour $i$
to the straight line $r_{\vec \rho}$, traced by the star 
forming particle in the opposite direction of the local density
gradient. We choose this scheme of energy redistribution to mimic the hot gas
flow along the direction of smallest resistance. For simplicity, we
only computed the energy exchange and drop the mass flow.

With this scheme, each particle keeps part of its SN energy and gives
another part to its neighbours. Thus, the pressure confinement or blow
out regime are accounted for by the spatial distribution of
particles. In the centre of a disk, for example, particles are heated
up by neighbours, their temperature increases, and as a consequence
they would tend to evaporate by buoyancy. But they are confined by the
pressure of other particle of the disk and the energy stays
localised. At the boundary of the disk, a particle taking energy will
not be confined, thus it will evaporate, with a velocity which 
depend by the energy it gains. We will show examples of this process
in the next Chapter.
We also implemented other energy redistribution schemes, e.g. we tried
to assign energy isotropically, or to all neighbours which resides in
the semi-sphere opposed to the density gradient. The scheme described
here has shown to be the most effective in reproducing the
confinement/blow out behaviours of the heated gas. Also, we tried
different energy assignment kernel; for example, using the distance
from the star forming particle centre instead of $r_p$ in
Eq.~\ref{ered_eq}. Within one 
energy redistribution scheme, changing the relative energy given to
each neighbours does not influence our results. 

The SN energy fraction received by a neighbour $i$ is computed as:
\be
\Delta E_{SN,i} = \frac{m_i \cdot W(|r_i - r_{\vec \rho}|,h_{\star})\Delta
  E_{SN,o}}{\rho_i}  \label{ered_eq}
\ee 
Note that we use the same SPH kernel used in the
hydrodynamical calculations, and therefore farther particles
get an energy fraction which is significantly
lower than the ones lying closer the density gradient vector $ r_{\vec
  \rho}$. 
\section{Summary}
In this Chapter we have presented MUPPI, MUltiPhase Particle
Integrator, our original sub-grid model of the Inter 
Stellar Medium (ISM) based on the analytical model by Monaco 2004 (see
Sec.~\ref{MO04}), which has 
been implemented on top of the GADGET-2 code (see Chapter 2,
Sec.~\ref{GDT}). \\ 
An illustration of changes in the GADGET code is given in
Fig. 4.5; we completely re-wrote the star formation
and cooling function, while changes in the hydrodynamics and time stepping
functions are less important.\\

The main feature of this model is that it does not use equilibrium
solutions for the ISM evolution (as instead does the original GADGET
star formation routine) but, rather, it assumes the ISM dynamics depends
only on \textit{local} properties. Each gas particle, in fact, evolves the model
and all its equations \textit{by its own}: the differential
equations for mass and energy flows are thus solved taking in direct
account the \textit{local} thermodynamic conditions.\\ 
Of course, since MUPPI changes the thermodynamical properties of gas
particles, the SNe energy injection also influences the SPH part of the code.
Gas cooling is treated by MUPPI, and only the hot phase
gas can cool, thus SNe energy is {\b not} immediately radiated away in our
model. In fact, the hot phase density is typically very low, since the
majority of the gas mass is in the cold phase (see next Chapter);
cooling is not very efficient in this condition. Moreover, the hot
phase is {\it conserved} from a time step to the next one and its
physical evolution is computed; e.g., SH03
effective model instead re-evaluate the equilibrium, thus allowing a
lot of gas to change its phase as a consequence
of SNe energy injection. This makes MUPPI feedback process self
consistent and quite effective without the need of introducing an
ad-hoc kinetic energy injection.\\

The implemented chain of processes are as as follows:
\begin{itemize}

\item [I] Each gas particle is assumed to be made of hot gas, cold gas and
  stars where gas phases are in pressure equilibrium;

\item [II] The amount of gas available for star formation is given by
  a phenomenological prescription proportional to the thermal pressure
  (Blitz \& Rosolowsky, 2006)
\item [III] The star formation rate gives the number of SNII, whose
  energy is partly given to the hot phase.
 \item [IV] The evolution of the
  system is obtained by numerically integrating the system of equation for the
  mass and energy flows.
\item [V] SN energy is released both inside and outside each gas
  particles. The redistribution of out-flowing energy to neighbouring particles 
 takes place along the least resistance path. 
\end{itemize}
\newpage
\begin{subappendices}
\setlength\topmargin{-0.5in}
\section{APPENDIX I: List of parameters}
\label{param}
\begin{center}
            \vspace{2 cm}
\end{center}
\begin{center}
\begin{tabular}{@{}ll}
$n_h$ / $n_c$ &  hot/cold number density \\
$T_h$ / $T_c$  & hot/cold temperature \\
$m_p$ & Proton mass \\
$f_{HI}$ & fraction of neutral hydrogen \\ 
$\rho_{thr}$ & density threshold for multiphase state \\
$T_{thr}$ & temperature threshold for multiphase state \\
$n_{thr}$ &number density threshold for multiphase state \\
$V_p$ &  total volume of the gas particle\\ 
$V_h$ / $V_c$ &  hot/cold volume \\ 
$M_h$ / $M_c$ & hot/cold phase mass \\
$F_h$ / $F_c$   & hot/cold mass fraction \\
$\mu_h$ / $\mu_c$  & hot/cold phase molecular weight \\
$E_{int}$  & internal energy coming from GADGET \\
$k$ & Boltzmann constant \\
$\rho$ &  total density of the gas particle \\
$t_{cool}$ & cooling time \\
$f_h$ / $f_c$ & hot/cold filling factors \\
$f_{\rm mol}$  & fraction of atomic hydrogen \\
$P_0$ & pressure normalised to the Milky Way \\
$P_{ext}$  & total pressure (here just thermal) \\

\end{tabular}

\begin{tabular}{@{}ll}
$M_{*}$ & mass in stars \\
$\dot{M_{*}}$ &  rate of change of the mass in stars \\
$\dot{M_{SF}}$ & star formation rate \\
$f_*$ & efficiency of star formation \\ 
$t_{dyn}$ & dynamical time \\
$\dot{M_{re}}$ & restoration rate \\
$f_{re}$ & fraction of restored mass \\
$\dot{M_{cool}}$ & cooling rate \\ 
$t_{cool}^{fit}$ & fit to the cooling time data\\
$\dot{M_{ev}}$ & evaporation rate \\
$f_{ev}$ & evaporated fraction of the star forming cloud\\
$\dot{E_{SN}}$ & SNe heating rate \\
$f_{\rm fb,i}$ & fraction of SNe energy trapped inside the gas particle\\
$f_{\rm fb,o}$ &fraction of SNe energy blowing outside the gas
  particle\\
$\dot{M_{\star, sn}}$ & mass of formed star per SN \\
$\dot{M_{c}}$ & rate of change of cold mass \\
$\dot{M_{h}}$ & rate of change of hot mass\\
$\dot{E_h}$ & net of energy flux of the hot phase \\
$\dot{E_{cool}}$ & rate of energy loss by cooling \\
$\dot{E_{hyd}}$ & rate of energy change due to hydrodynamics \\
$\Delta E_{SN,o}$ & SNe energy to be redistributed in the timestep \\
$\Delta M_{sf}$ &  mass in stars formed in the timestep\\
$\Delta E_{SN,i}$ & SNe energy received by neighbour $i$ in the timestep \\
\end{tabular}
\end{center}

\newpage
\section{APPENDIX II: Flow charts}
\label{flowchart}
\begin{center}
\begin{figure}[!b]
\centering{
\includegraphics[width=0.8\linewidth]{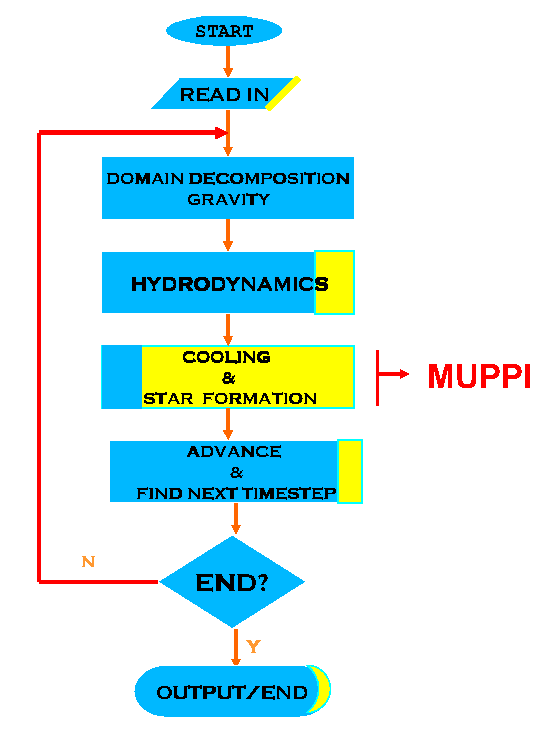}
\captionsetup{font={normalsize,sf}, width= 0.9\textwidth}
\caption{Flow chart of the GADGET-2 code. In yellow we highlight how
  much of the selected routine has been changed due to the insertion
  of the MUPPI model.}}
\label{MUPPI:flowGDT}
\end{figure}
\newpage

\begin{figure}
\centering{
\includegraphics[width=0.9\linewidth]{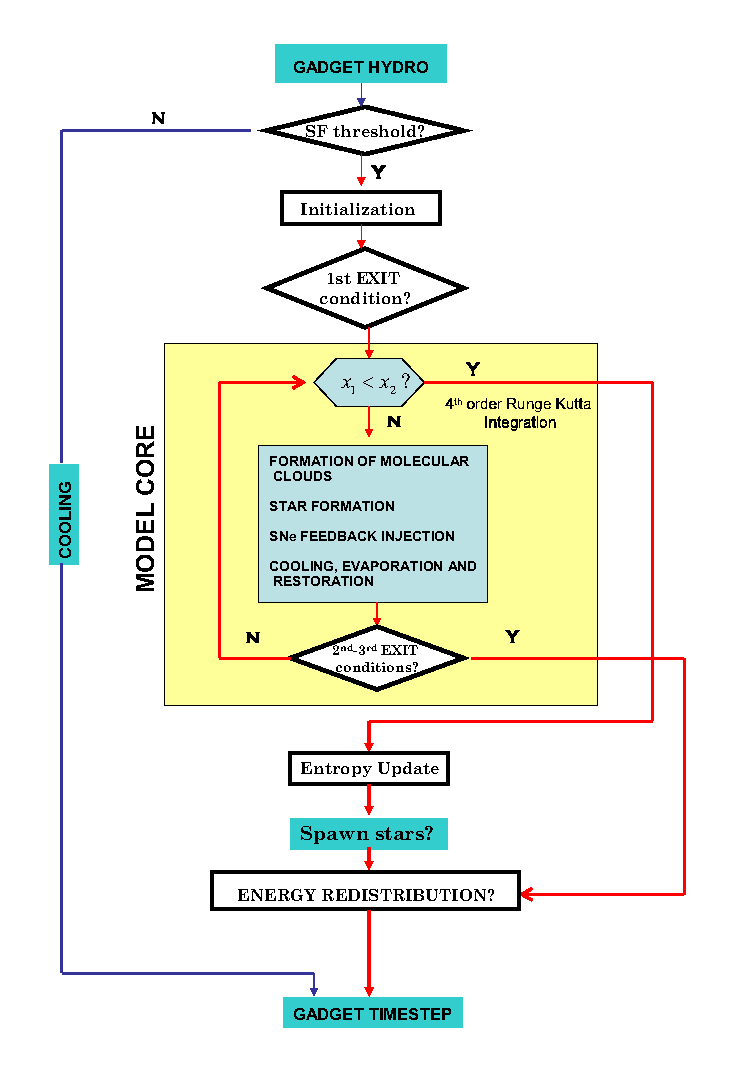}
\captionsetup{font={normalsize,sf}, width= 0.9\textwidth}
\caption{Flow chart of the MUPPI code. The illustrated processes are
  fully described along the present chapter. In particular, processes
  depicted above the ``model core'' section are outlined in
Sec.~\ref{MUP:ini}; the model core processes are described in
  Sec.~\ref{MUP:sf}; the last part is accounted in Sec.~\ref{MUP:fin}.}
\label{MUPPI:part}}
\end{figure}

\end{center}

\end{subappendices}

\chapter{Results}
In this section, we present and discuss the results obtained from a set
of simulations (see Table ~\ref{RES:run}) performed with the code
described in Chapter ~\ref{MUPPI_chap}. The discussion proceeds as
follows. Since one of the fundamental properties that a
star formation and feedback 
model must be able to reproduce is that in Milky Way (MW) like
simulations, properties of simulated interstellar medium (ISM) must
reproduce observations; in Sec.~\ref{RES:MW} we describe the simulation runs
conducted with a MW model and we describe the resulting
behaviour.
In order to test the MUPPI code behaviour on different initial physical
conditions, in Sec.~\ref{RES:other} we describe the simulations runs
done using a dwarf galaxy like model (DW) and two isolated halos, one
typical of the MW (CFMW) and one of a dwarf galaxy (CFDW). These last
two runs have been carried on to verify if our code works well
in very different physical conditions such those found at the centre
of cooling flows. We also use a more massive halo, with mass $M$ =
$10^{13}M_{\odot}$, which we already described in Chapter 2. In
Sec.~\ref{RES:numtest} we discuss the simulations 
conducted with the purpose of assessing the stability of our results
with respect to the details of the implementation and to the
resolution. In particular we simulated the CFDW with ten times more
(HR) and ten times less (LR) gas and dark matter particles (see
Tab.~\ref{RES:run}), and the MW in LR. Finally we discuss and
summarise our conclusion in Sec.~\ref{RES:concl}.

\begin{table*}
\centering{ 
\begin{tabular}{@{}rcccccccccccccccccc@{}}\toprule &
\multicolumn{3}{c}{$f_{fb_0}$} & \phantom{abc} &
\multicolumn{3}{c}{$n_{thr}$ [cm$^{-3}]$} &\phantom{abc} &
$f_{\rm fb,i}$ & $f_{\star}$ && \multicolumn{2}{c}{\rm{res}} &&
\multicolumn{2}{c}{\rm{GDT}} \\
\cmidrule{2-4}  \cmidrule{6-8} \cmidrule{10-10} \cmidrule{11-11}
\cmidrule{13-14}  \cmidrule{16-17}& 0.0
& 0.3 & 0.7 &&
0.05 & 0.1 & 0.25 &&  0.05 & 0.1 && HR & LR && EFF & EFF+wind\\	
\textbf{\sf MW} & $\bigstar$ & $\bigstar$ & $\bigstar$ && $\bigstar$ &
$\bigstar$ & $\bigstar$ && $\bigstar$  & $\bigstar$ && & $\bigstar$
&& $\bigstar$ & $\bigstar$\\ 
\textbf{\sf CFMW} &  $\bigstar$ & $\bigstar$ & $\bigstar$ && &
&$\bigstar$ && &  && & &\\
\textbf{\sf DW} & $\bigstar$ & $\bigstar$ & $\bigstar$ && $\bigstar$ &
$\bigstar$ & $\bigstar$ && &  && & & & &$\bigstar$ \\
\textbf{\sf CFDW} &  $\bigstar$ & $\bigstar$ & $\bigstar$ && & &
$\bigstar$ && & && $\bigstar$ & $\bigstar$ && $\bigstar$ & $\bigstar$\\
\bottomrule
\end{tabular}}
\captionsetup{font={normalsize,sf}, width= 0.9\textwidth}
\caption{Simulation summary. We show the
 labels of the initial condition models (see Sec.\ref{RES:ic}) {\it
 vs} 
the characteristics of the simulation runs: the model
 parameters that have been varied ($f_{fb_0}$, $n_{thr}$, $f_{\rm fb,i}$
 and $f_{\star}$), the
 numerical tests done with 
 varying resolution (HR: 10 times better mass resolutionl LR: ten
 times worse mass resolution) and 
 finally runs done with the original GADGET SF code, with (EFF+wind) and
 without (EFF) winds (see Sec.~$\ref{GDT:sf_eff}$).}
\label{RES:run}
\end{table*}

\section{Initial Conditions (ICs)}
\label{RES:ic}
\subsection{Isolated galaxy models (MW, DW)} 
\label{RES:ic_glx}
Testing the effect of sub-grid models in galaxy formation
simulations is best done in numerical realizations of isolated
galaxies. These models are constructed on purpose to resemble observed
galaxies, with a disk of gas and stars, and optionally a stellar bulge,
all embedded in an extended dark matter halo with structural
properties (e.g. mass, spin, density profile) consistent with
cosmological simulations of the hierarchical growth of CDM halos. \\
The galaxy models used in this work have been created by Simone
Callegari, following the prescription depicted in Springel, Di Matteo
\& Hernquist 2005\nocite{2005MNRAS.361..776S}. In what follows we
review their basic features, for a full description see the referenced paper.\\
Both $\texttt{MW}$ and $\texttt{DW}$ are generated with near
equilibrium distributions of collisionless particles consisting in
rotationally supported disc of gas and stars (with spin parameter $\lambda$), a star bulge (not in
$\texttt{DW}$) and a dark matter halo. These different structural
components are described by independent parameters (see Tab.). 
The halo dark matter mass distribution of both $\texttt{MW}$ and
$\texttt{DW}$ galaxy models has been modelled with an Hernquist (1990) 
profile. 
The disc components of gas and stars have been modellel with an
exponential surface density profile having scalelenght $r_{s,d}$. The
total mass of the disc, $M_{disk}$, has been computed as a
fraction of the total mass of the galaxy.
The $\texttt{MW}$ model has also of a spherical star bulge, modelled with 
an Herquist profile. The bulge scalelenght $r_{s,b}$ is treated as
a free parameter parametrized in units of $r_{s,d}$, while the bulge
mass $M_{bul}$ is specified as a fraction of the total galaxy mass. 
In both galaxy models, a fraction $f_{gas}$ of the disc is assumed to
be in gaseseos form, the rest in stellar form.

We evolved all galaxy model with non-radiative physics only,
  i.e. no cooling and star formation, for 10 dynamical times, and
  verified that our models are numerically stable.
  We use the non-radiatively evolved models after 4 dynamical times as
  initial conditions for MUPPI, so as to be
  sure that MUPPI evolution is not polluted by numerical instabilities.
\begin{table*}
\centering
\begin{tabular}{@{}l l l l l l l l@{}}
\hline\hline &  \texttt{$f_{bar}$} & \texttt{$r_{s,d}$} &
\texttt{$M_{disk}$} &  \texttt{$r_{s,b}$} & \texttt{$M_{bul}$} &
\texttt{$f_{gas}$} & \texttt{$\lambda$}\\ 
\hline\\
\texttt{MW}& 0.06& 2.9 & $2.6 \cdot 10^{10}$ &  0.58 & $6.6 \cdot
10^{9}$ & 0.1 & 0.04\\
\texttt{DW} & 0.05 & 3.5 & $5.6 \cdot 10^{9}$& 0 & 0 & 0.2 & 0.04\\
\end{tabular}
\captionsetup{font={normalsize,sf}, width= 0.8\textwidth}
\caption{Main properties of the galaxy models. Column 1: galaxy
  name. Column 2: radial disc scalelenght in Kpc. Column 3: total mass in the
  disc (gas + stars) in $M_{\odot}$. Column 4: bulge scalelenght in
  Kpc. Column 5: mass in the bulge in $M_{\odot}$. Column 6: gas fraction
  in the disc. Column 7: spin parameter.}
\label{RES:glx}
\end{table*}

\subsection{Isolated halos (CFMW, CFDW)}
The procedure used to generate the initial conditions for the isolated,
non-rotating halos is the same reported at the beginning of
Sec.~\ref{simulsection}.
We use a NFW density profile for the DM distribution, instead of
  an Hernquist one, since the former is suggested by cosmological
  simulation. 
We refer the reader to Chap.2  for a full description on the halo ICs
generation.\\
\begin{table*}\centering
\begin{tabular}{@{}l l l l l l@{}}
\hline\hline &  \texttt{$M_{200}$} & \texttt{$r_{200}$} &
\texttt{$c_{\rm NFW}$} & \texttt{$m_{\rm DM}$} & \texttt{$m_{\rm gas}$} \\
\hline
\\
\texttt{CFMW}& $6.6 \cdot 10^{11}$ & 197 & 12 & $3.1\cdot10^{6}$ &
$4.2\cdot10^{5}$ \\
 \\
\texttt{CFDW} & $1.1 \cdot 10^{11}$ & 80 & 13 & $5.7 \cdot 10^5$ &
$1.4 \cdot 10^5$  \\
\end{tabular}
\captionsetup{font={normalsize,sf}, width= 0.8\textwidth}
\caption{Main properties of the simulated halos. See
  Sec.~\ref{simulsection} and text for more
  significant details. Column 1: halo name. Column 2: mass enclosed in within
  $r_{200}$ in $M_{\odot}$. Column 3: value of $r_{200}$ in kpc. 
 Column 4: NFW concentration. Column 5: mass of a DM particle in
  $M_{\odot}$. Column 6: mass of a gas particle in
  $M_{\odot}$. }
\label{RES:CF}
\end{table*}
With the procedure depicted in Chapter 2, we generated two halos,
having the same DM and barionic mass of the Milky Way
($\texttt{CFMW}$) and of a dwarf galaxy
($\texttt{CFDW}$); see Tab.~\ref{RES:CF} for a summary on halos main
properties.
$\texttt{CFMW}$ halo is sampled with 2.5 x $10^{5}$ DM and 1 x $10^{5}$ gas
particles inside $r_{200}$, while $\texttt{CFDW}$ with 1.9 x $10^{5}$
DM and 5 x $10^{4}$ gas particles inside $r_{200}$. The
baryon fraction is $f_{\rm bar}$=0.05 in $\texttt{CFMW}$ and
$f_{\rm bar}$=0.06 in $\texttt{CFDW}$. We set the Plummer-equivalent softening, following Power et
al. 2003\nocite{Pow03}, to be $1$ kpc in $\texttt{CFMW}$ and
$0.6$ kpc in $\texttt{CFDW}$ for the DM and half those values for the
gas. We further assume the 
minimum value for the SPH smoothing length to be 0.5 times that of the
gravitational softening. The number of the SPH neighbours has been set
to $N_{\rm ngb}$ to be 32.  In all runs we set the initial angular
momentum to zero. 

We evolved the $\texttt{CFDW}$ runs for 5.5 Gyr, while the
$\texttt{CFMW}$ runs have been evolved just for 1 Gyr due to the higher
computational cost.

\begin{table*}[b]
\centering
\begin{tabular}{@{}l l l l l l l l l l@{}}
\hline
  \texttt{$n_{thr}$}  & \texttt{$T_{thr}$} &
\texttt{$f_{\star}$} & \texttt{$f_{ev}$} & \texttt{$f_{\rm re}$} &
\texttt{$T_{c}$} &  \texttt{$t_{\rm fin}$} &  \texttt{$P_{0}$} &
\texttt{$f_{b_{\rm in}}$} &\texttt{$f_{b_{\rm out}}$} \\
\hline\hline
0.25 & 5 $\cdot 10^4$ & 0.01 & 0.1 & 0.2 & 1000 & 2\texttt{$t_{dyn}$}
& 1000 & 0.02 & 0.3 \\
\end{tabular}
\captionsetup{font={normalsize,sf}, width= 0.9\textwidth}
\caption{Reference set of parameters. Column 1: number density
  threshold in \texttt{cm$^{-3}$}. Column 2: temperature threshold
  in \texttt{K}. Column 3: star formation efficiency. 
 Column 4: evaporated fraction of cold cloud. Column 5: restored mass
  fraction.  
Column 6: temperature of the cold phase. Column
 7: exit condition from the multi-phase state. Column 8: external
  pressure normalization. Column 9: fraction of SN energy feedback
 trapped inside the particle. Column 10: fraction of SN energy feedback
 blowing outside the particle.} 
\label{RES:param}
\end{table*}

\section{Milky Way}
\label{RES:MW}
In this section we investigate te effects of MUPPI on simulations of the
$\texttt{MW}$ model runned with our reference set of MUPPI parameters (see
Tab.~\ref{RES:param}). Here we want to describe the behaviour of the
gas when MUPPI is used. We will discuss only the effect of varying
$f_{\rm fb,o}$ 
in this section: a parameter study will be presented in
Sec.~\ref{RES:paramtest}. \\

\subsection{The Inter Stellar Medium }
\label{MUPPI:ISM_SINGLE}
\begin{figure*}
\centerline{
\includegraphics[width=0.4\linewidth]{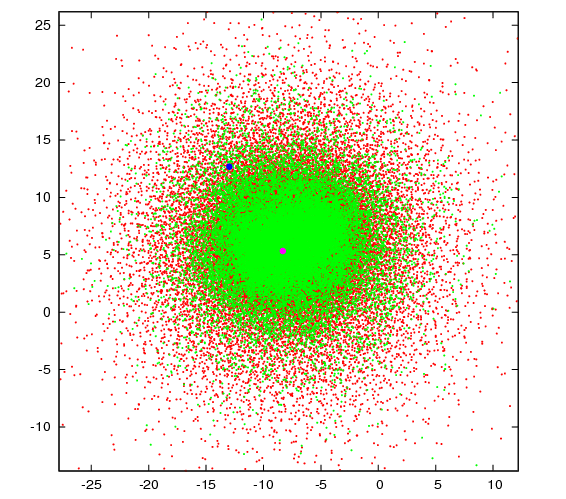}
}
\captionsetup{font={normalsize,sf}, width= 0.9\textwidth}
\caption{The distribution of stars (green) and gas (red) in the xy
    plane in the initial conditions for simulation of the \texttt{MW} galaxy where we
    highlight the position of the \textit{core} (purple) and
    \textit{disc} (blue) multi-phase particles studied in
    Sec.$~\ref{MUPPI:ISM_SINGLE}$} 
\label{RES:pos_singlepart}
\end{figure*}

We start by showing the evolution of the thermodynamical properties
of single gas particles lying $\texttt{(a)}$ in the centre of the galaxy and
$\texttt{(b)}$ on the disc, at the edge of the star forming region
(see Fig.~\ref{RES:pos_singlepart}).

\begin{figure*}[!]
\centerline{
\includegraphics[width=0.4\linewidth]{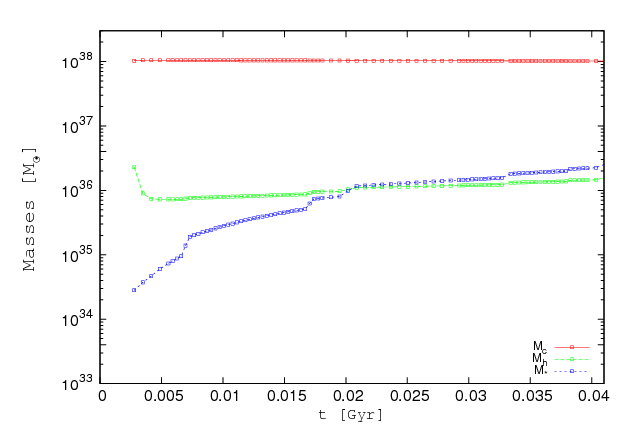}
}
\centerline{
\includegraphics[width=0.4\linewidth]{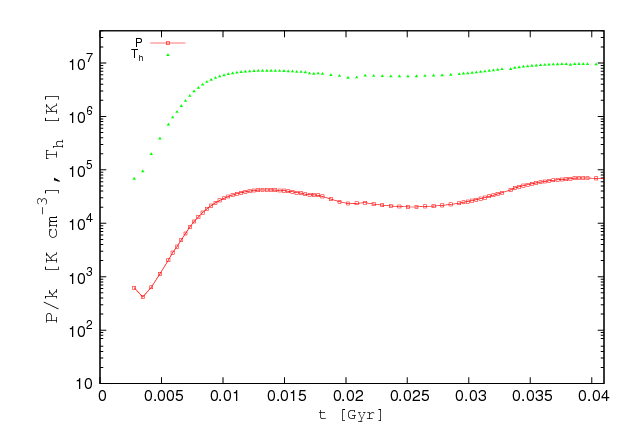}
}
\centerline{
\includegraphics[width=0.4\linewidth]{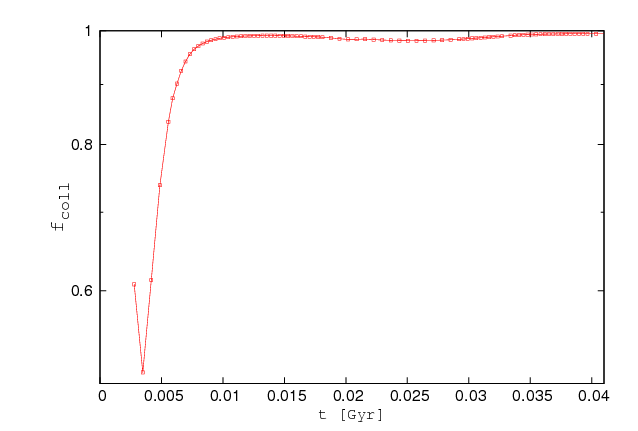}
}
\centerline{
\includegraphics[viewport= 0 8 655 420,clip, height =4.3cm]{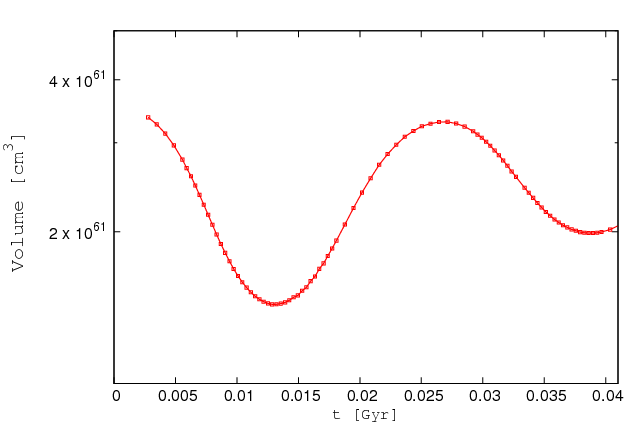}
}
\centerline{
\includegraphics[width=0.4\linewidth]{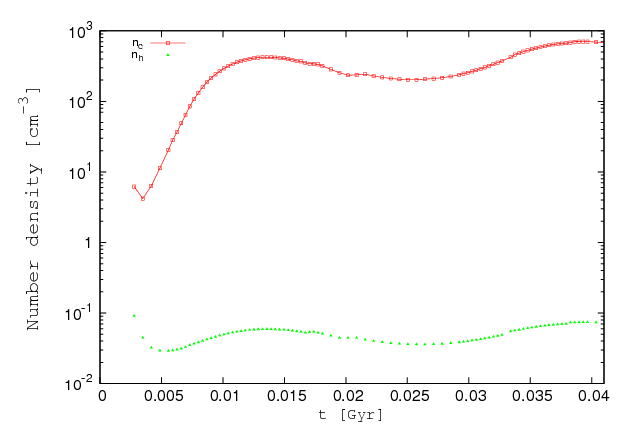}
}
\captionsetup{font={normalsize,sf}, width= 1\textwidth}
\caption{The ISM evolution of a gas particle lying in the galaxy
  core. From the top to bottom, the panels shows the evolution of the
  cold, hot and star masses in $M_{\odot}$ (first 
  panel); the hot phase temperature in $K$ and the
  pressure in $K cm^{-3}$ (second panel); the fraction of molecular gas
  (middle panel); the gas particle volume in $cm^{3}$ (second-last panel); the
  cold and the hot number densities in $cm^{-3}$ (bottom panel). See
  Sec.~\ref{MUPPI:ISM_SINGLE} for details.}
\label{RES:ism_core}
\end{figure*}

\begin{figure*}[!]
\centerline{
\includegraphics[width=0.4\linewidth]{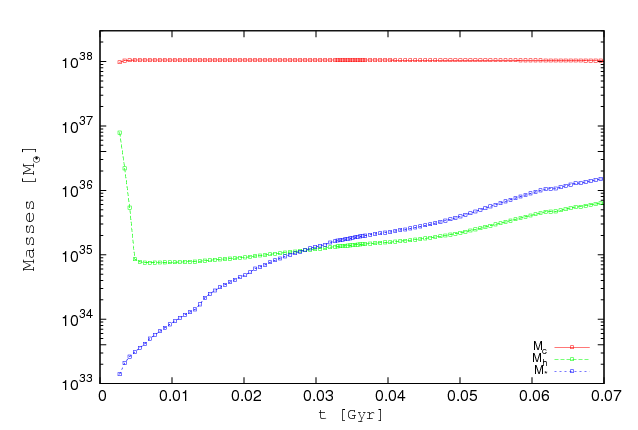}
}
\centerline{
\includegraphics[width=0.4\linewidth]{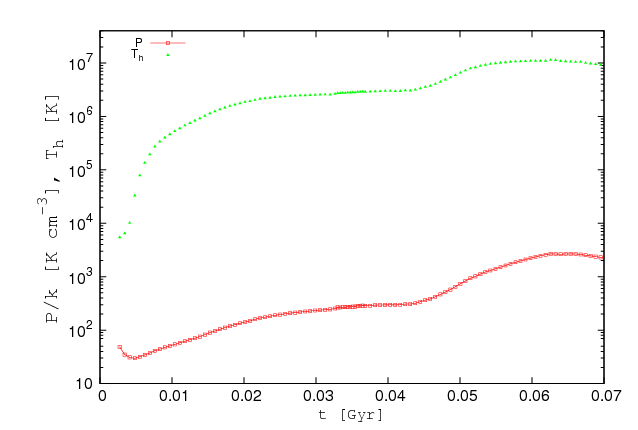}
}
\centerline{
\includegraphics[width=0.4\linewidth]{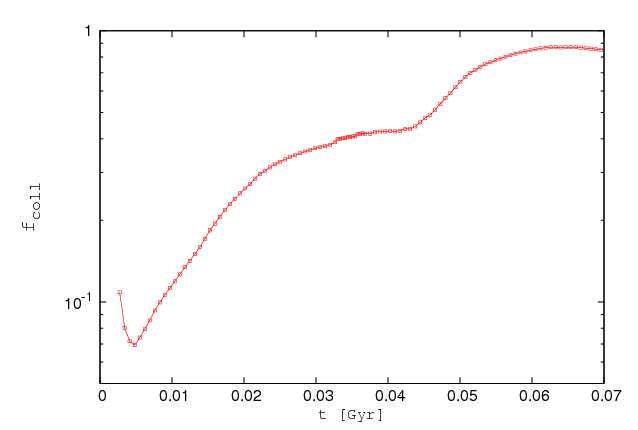}
}
\centerline{
\includegraphics[viewport= 0 8 655 420,clip, height =4.3cm]{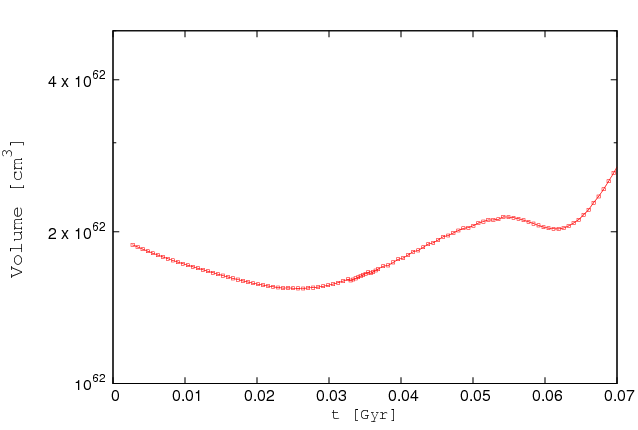}
}
\centerline{
\includegraphics[width=0.4\linewidth]{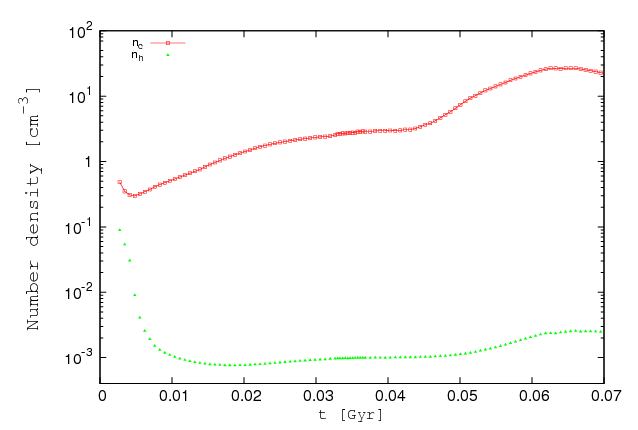}
}
\thispagestyle{empty}
\captionsetup{font={normalsize,sf}, width= 0.9\textwidth}
\caption{Same as in Fig.~\ref{RES:ism_core} but for a gas particle
  lying in the disc.}
\label{RES:ism_disc}
\end{figure*}

 In Fig.~\ref{RES:ism_core}, we show properties of the \textit{central}
 gas particle and in Fig.~\ref{RES:ism_disc} those of the 
\textit{disc} gas particle.\\
At the onset of the simulation, the gas particles undergo a
rapid collapse as the cooling begins to act.
As soon as the \textit{core} gas particle fulfills the star formation
thresholds and enters MUPPI, its initially hot gas rapidly cools
and feeds the cold phase. This is clearly shown by the top panel of
Fig.~\ref{RES:ism_core} where we plot the evolution of the
cold, hot and stellar mass. Corresponding to the rapid filling of the cold
phase (which soon reaches an equilibrium value) and to the increase of
 the molecular gas fraction, shown in the central panel, the stellar mass
undergoes a fast growth and thus a substantial amount of SNe energy
feedback is injected in the medium. This causes a quick jump in
the particle hot phase temperature and pressure (shown in the second
 panel from the top), which respectively reaches $T_h \sim 10^7$ K
 and $P/K = 7 \cdot 10^4$ K cm$^{-3}$. 

In the
third panel from the 
top, we show the evolution of the 
fraction of molecular gas ($f_{coll}$, Eq.~\ref{fcoll}): after a
 sudden decay corresponding to the initial lack of pressure support, this
fraction rapidly increases with the pressure and stabilises to a value
around unity, with small fluctuations due to the pressure changes. 
This means that, for this particle, all the cold gas is in the
 molecular, star-forming form. This is due to the higher value of
 $P/K$, which is between 10$^4$ and 10$^5$ K cm$^{-3}$. The plot of
 Fig.~\ref{RES:ism_core} clearly show how the physics of the ISM
 inside the particle is driven by the pressure. When $P/K$ increases,
 $f_{coll}$ and $n_c$ increase, giving an higher star formation rate
 and consequently a stronger energy feedback which causes the
 temperature of the hot phase to also increase. \\
On the other hand, the pressure responds to volume changes as it can
 be seen by comparing trends in the second and in the fourth
 panel. When volume decreases (compression) the pressure increases and
 vice-versa. But the volume, as defined by Eq.~\ref{MUPPI:vol}, is
 determined by the SPH evolution; here we have the mechanism which causes
 the response of the ISM to the local hydrodynamics. 
 In the bottom panel, we show the
evolution of the cold and hot number density: the cold one rises as
the cold phase is feeded by cooling of hot gas; on the other
side the hot number density diminishes. The whole multi-phase stage
 lasts $\sim$ 40 Myr, then the particle exits the multi-phase regime
as imposed by our exit condition (see Sec.~\ref{MUP:ini}).\\

\begin{figure*}[!]
\centerline{
\includegraphics[width=0.35\linewidth]{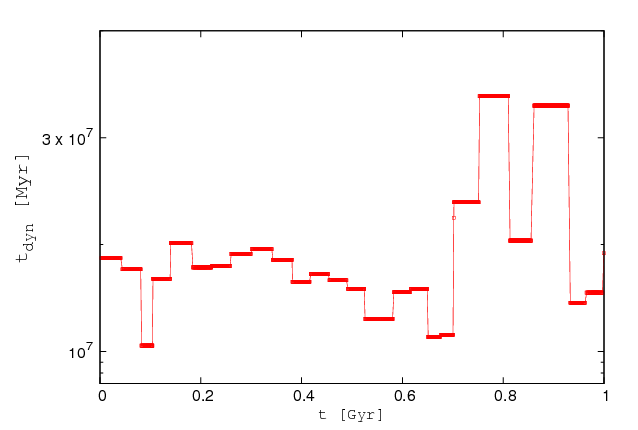}
\includegraphics[width=0.35\linewidth]{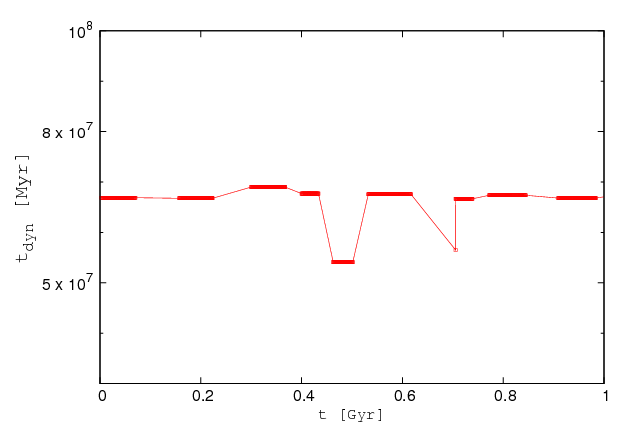}
}
\captionsetup{font={normalsize,sf}, width= 0.9\textwidth}
\caption{Dynamical timescale in Myr for a gas
  particle residing in the galaxy core (left panel) and for a gas
  particle in the disc (right panel), in different multi-phase stages
  during a period of $1$ Gyr.}
\label{RES:tdyn_core}
\end{figure*}

As depicted in Fig.~\ref{RES:ism_disc}, the typical evolution of a
gas particle lying in the disc is characterised by a different
behaviour with respect to a gas particle lying in the 
dense central galaxy regions. First of all, being the particle in a
 colder and less dense zone, its cold phase forms in more
timesteps (even if this effect is small), while the general interchange
between the various mass phases is instead analogous. The particle
pressure (second panel) results much less smaller than that in
 ``core'' case. Values of $P/K$, $n_h$, $n_c$ for this case resamble
 those observed in the solar neighbourhood. The gas particle we chose
 is at 6.2 Kpx from the galaxy center. Note that the model generated
 such an ISM behaviour self-consistently; we don't impose particular
 values for $P/K$, $n_h$ and $n_c$. After 30 Myr since the gas
 particle entered in multi-phase stage, the volume decreases causing a
 small increase in pressure which then stays almost constant. Thus
 star formation increases giving some SNe feedback. This could be due
 e.g to a spiral arm passing through the position of the
 particle. The pressure wave causes a rapid 
increases in the fraction of molecular gas (middle panel) and in the
cold number density (bottom panel).\\

After about 50 Myr, the pressure shows a more significant increase
due to the slow piling up of thermal energy given by the star
 formation. The hot mass increases and this pressurize the
 particle. The local thermodynamics is not responsable for this
 effect: from the fourth panel, we can see taht the particle is in an
 expansion phase. At this point, the star formation boosts and further
 pressurize the particle, until, at the very end of the multi-phase
 stage, an expansion driven by the SPH stops the runaway process
Such an expansion is caused by MUPPI itself. In fact, this gas
 particle has an average distance of 0.35 Kpc from the disc plane,
 and (almost) does not move form this position. Towards the end of the
 mulit-phase stage, it has an high temperature and enough mass in the
 hot phase to cause it to flow away from the disc. At the end of the
 multi-phase stage its distance from the disc plane is already of
 0.49 Kpc and the particle has ``gone into wind''.\\
Finally we show in Fig.~\ref{RES:tdyn_core}, the behaviour of the
dynamical time over one Gyr in both the above cases. Both particles
 enter the multi-phase stage several times during this period. The central gas
particle dynamical time takes values between 10 and 30 Myr and it is
 subjected to changes from one multi-phase stage to the following. On
 the contrary, the disc gas particle dynamical time is between 50 and
 70 Myr, keeping almost a constant value in the various multi-phase
 stages, due to the calm
environment. The only visible oscillation is probably reconducible to the
crossing of a spiral arm, which augments the local density and
diminishes the dynamical time. The difference in the dynamical time is
 mainly due to the different initial pressure in the two environments;
 particles in the hot, dense core shows a faster
 ``metabolism''. However, since the time-step in GADGET is individual
 and adaptive, and is obviously shorter for particles lying in the
 center of the galaxy, the number of time-steps spent in MUPPI is of
 order 100 for all particles, and does not depend much on the environment.    
\begin{figure*}[!]
\centering{
\includegraphics[width=0.5\linewidth]{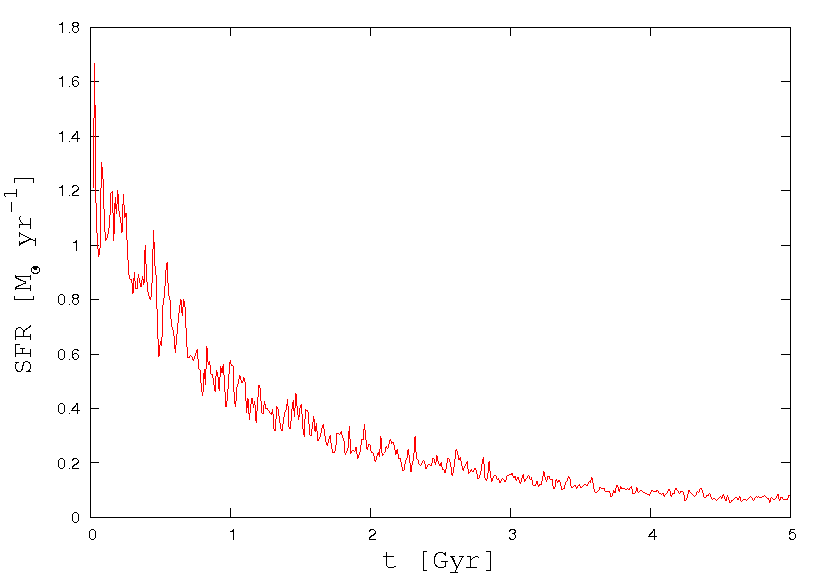}
\captionsetup{font={normalsize,sf}, width= 0.6\textwidth}
\caption{Star formation rate as a function of time for the \texttt{MW}
  model.  }}
\label{FIG:sfr_mw_std}
\end{figure*}

\subsection{Global properties of the gas particles}

Immediately after the entrance in MUPPI, i.e. when gas particles
fulfills both the thresholds in density $n_{thr}$ and in temperature $T_{thr}$
(see Tab.~\ref{RES:param}), thermally unstable gas in the disk and in
the bulge collapses, feeding the cold phase and thus causing a large
burst of star formation. Such a burst is due to the sudden
  turning on of cooling and star formation physics.
After about 2 Gyr, the $\texttt{MW}$ galaxy
settles down in a quiescent, self-regulated state with a star formation rate of
nearly 0.2 $M_{\odot}$ yr$^{-1}$, as can be seen in
Fig. 5.5. The SFR then slowly decreases as the cold gas is
consumed by star formation. From the SFR plot, we thus deduce that
MUPPI (with the parameters listed in Tab.~\ref{RES:param}) is able to
lead to a self-regulated cycle of SF, where the growth of 
  the cold phase and thus the star formation production are
  counterbalanced by the SNe feedback effects accounted in MUPPI
  (see Sec.~\ref{MUPPI:core}). \\

\begin{figure}[!]
\centering
\includegraphics[width=0.8\linewidth]{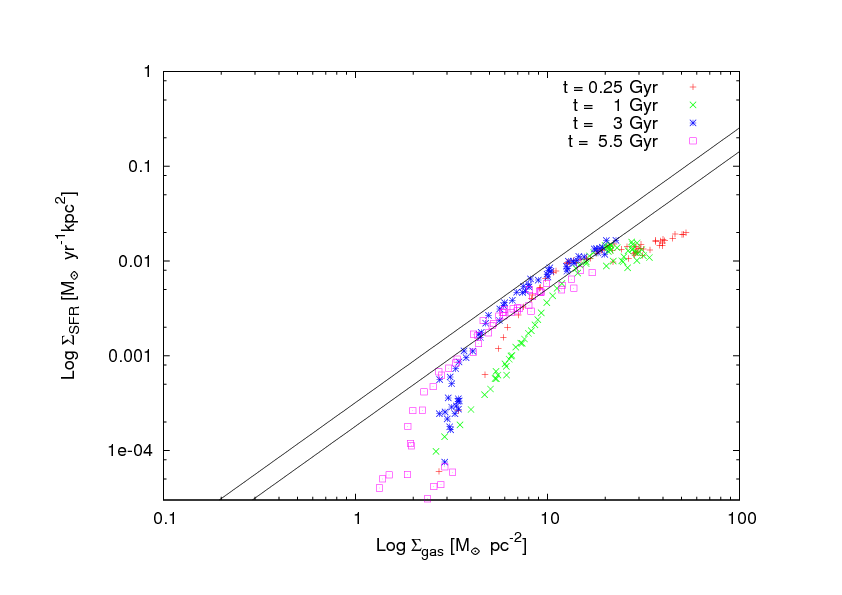}
\captionsetup{font={normalsize,sf}, width= 0.9\textwidth}
\caption{Star formation rate density as a function of cold gas surface
  density for the \texttt{MW} model. The solid lines mark the
  Schmidt-Kennicut law (1998).  }
\label{FIG:kenni_mw}
\end{figure}

\begin{figure}[!]
\centering
\includegraphics[width=0.8\linewidth]{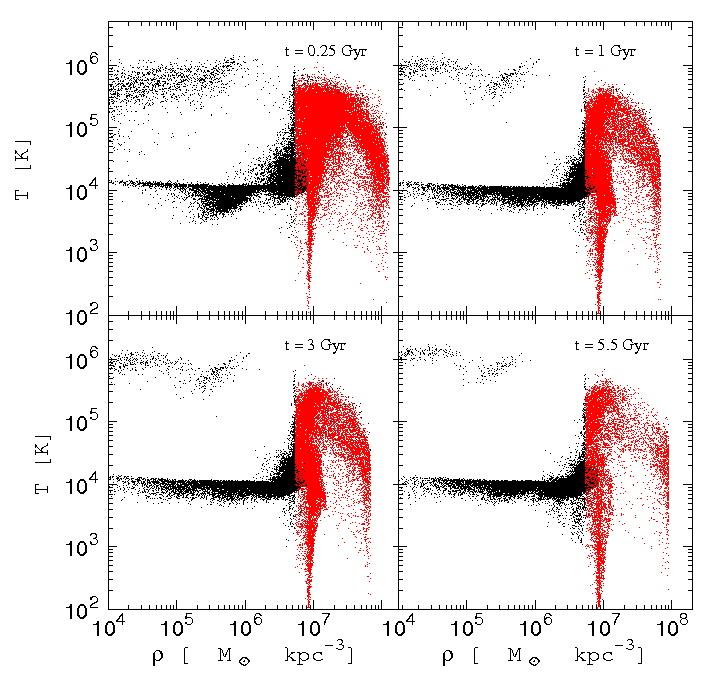}
\captionsetup{font={normalsize,sf}, width= 0.9\textwidth}
\caption{\texttt{MW} density-temperature phase diagram for SPH gas
  particle (black) and MUPPI gas particles (red). See text for details.}
\label{FIG:phase_mw}
\end{figure}
Observationally, the surface density of the star formation rate in
different galaxies scales non-linearly with the surface density of the
total gas. This relation is known as the Kennicutt-Schmidt star
formation law: reproducing this relation in galaxy formation
simulation is one of the main challenges for galaxy formation models.
In Fig.~\ref{FIG:kenni_mw} we plot the projected star formation rate
density as a function of the cold gas surface density, at different times
during the evolution of the $\texttt{MW}$ galaxy
model. 

At the beginning of the simulation ($\approx$1 Gyr), the simulated Kennicut
relation is slightly below the observed one, due to the high density
reached in the galaxy centre after the prominent initial
collapse. Then, the gas is consumed by star formation; in the
meanwhile, runaway star formation is prevented by SNe energy feedback,
which prevent too much hot gas to radiatively cool down to the cold
phase.  At later times, our simulated \texttt{MW} is in good agreement
with the observed Schimdt-Kennicut (1998) relation, indicated on this
plot by two solid lines: a further proof that MUPPI is able to
generate a self-regulated ISM. The relation is reproduced at all
times, up to the end of the simulation at 5.5 Gyr.  Note that we are
able to reproduce the star formation cut at low surface densities,
while the GADGET effective model with kinetic feedback (i.e., winds)
does not (see Fig.~\ref{RES:kenni_eff}), as we shall see in the
following section.\\

In Fig.~\ref{FIG:phase_mw} we show the SPH gas density vs average
temperature phase diagram
for the \texttt{MW} model at four times for the SPH gas particles
(black) and the MUPPI gas particles (red). The diagram is populated in
three main regions: gas lying on the disc is visible in a tight
relation at $T$ = $10^4$ K; in the upper-left corner we see gas
particles, once multi-phase, that have been driven outside the galaxy
by the SNe feedback and that are now falling back to the galaxy
while radiatively cooling. This is the typical behaviour of a {\it
  galactic fountain}. MUPPI gas particles, instead, occupy the dense region
of the plot and take different temperatures, from $\sim$10$^2$ K to
$\sim$5$\cdot10^{5}$K. In the upper-left diagram (0.25 Gyr), at high
densities and low temperatures, we see 
multi-phase particles which are cooling and probably will soon reach
the conditions to spawn a star; note in fact how the number of
multi-phase gas particles diminishes through time, due to the
  ongoing star formation which consumes the gas. Moreover, in the
``multi-phase'' region are evident gas particle which have
been heated by SNe energy feedback but that are
still in the multi-phase regime. In the four diagrams the
density-temperature relation is sharply truncated at the density where the
probability of spawning a star becomes nearly unity, i.e. beyond
$\sim$9$\cdot$10$^{7}$$M_{\odot}$kpc$^{-3}$.\\

\begin{figure}[!]
\centering
\includegraphics[width=0.8\linewidth]{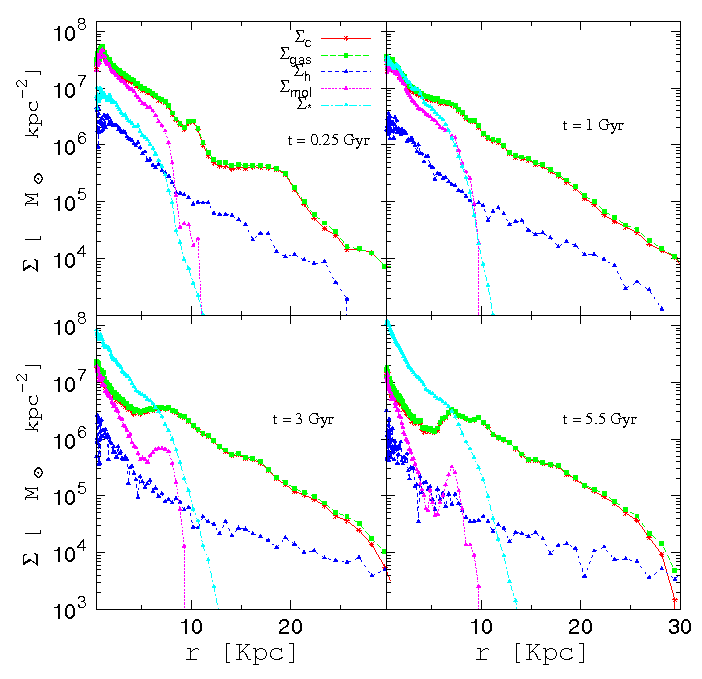}
\captionsetup{font={normalsize,sf}, width= 0.9\textwidth}
\caption{\texttt{MW} surface density profiles for stars ($\rho_{\star}$) and
  our various gas phases, i.e. cold ($\rho_c$), molecular ($\rho_{mol}$), hot
  ($\rho_h$)and total gas density ($\rho_{gas}$)}
\label{FIG:dens_mw}
\end{figure}
In Fig.~\ref{FIG:dens_mw}, we show the radial surface density
profiles for various components of the MUPPI gas particles at four
different times. Here and in the following, we evaluated density
profiles using 100 radial bins, equally spaced in $log(r)$, and ranging from 0.5
to 50 kpc. When galaxy models are concerned, we projected the particle
positions on the disk plane to obtain surface density profiles. For
the isolated halos cases, we instead used 3D radial density profiles.
We weighted hot phase gas temperature and numerical density profiles
by the hot phase gas mass; cold phase numerical temperature by the
cold pase gas mass, and pressure by the total gas mass.

In general, the global behaviour of
the gas particles reflect that of the single particles we described in
the previous section. At $t$ = 0.25 Gyr (top left panel),
the initial strong burst of star formation has just begun, thus
the star density $\rho_{\star}$ is low and corrispondently the density
of the cold $\rho_{c}$ and molecular phase $\rho_{mol}$ is high,
reaching $\sim 3 \cdot 10^7$ $M_{\odot}$ Kpc$^{-3}$. Going further in
time, the cold gas phase (together with the molecular phase) slowly
undergoes depletion due to star formation while the SN energy injected by the
newly born stars rises the temperature of the hot phase thus making
radiative cooling 
\begin{figure}[!]
\centering{
\includegraphics[width=0.7\linewidth]{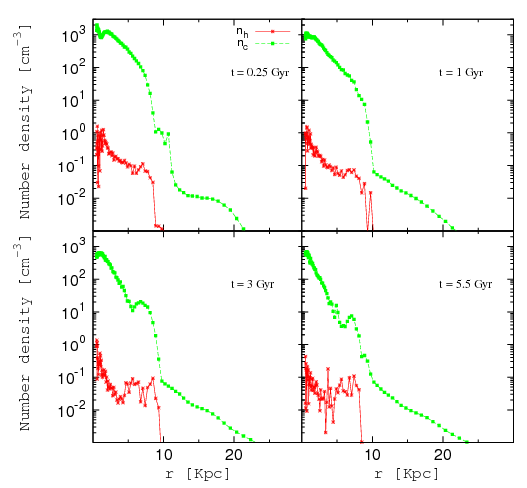}
\captionsetup{font={normalsize,sf}, width= 0.9\textwidth}
\caption{\texttt{MW} number density profiles for the cold phase
  (green) and the hot phase (red).}}
\label{FIG:numdens_mw}
\end{figure}
less efficient and limiting the cold phase replenishment. As observed
in Milky Way-like galaxies, the surface gas density become exponential
along the disc plane. The
behaviour of the molecular phase (regulated by $f_{coll}$) is as
expected: in the galaxy core, where the pressure is high, the gas at
disposal for star formation is high, while this quantity diminishes
while going further along the disc. The hot phase surface density
follows the cold phase trend, but 2 orders of magnitudes below: in the
core, where the star formation is high and thus is the 
SNe energy feedback, the gas is efficiently warmed up; towards the
galaxy outskirts, the pressure declines leading to a lower star
formation rate and thus a less efficient SN feedback injection. 
The density profile of the stellar mass, while increasing 
with time, slightly move outwards and, at 5.5 Gyr, extends till 15
Kpc. As shown in Fig.~\ref{MUPPI:face_mw_s}, MUPPI give rise to a thin
stellar disk and to a stellar bulge.\\
\begin{figure}[!]
\centering
\includegraphics[width=0.6\linewidth]{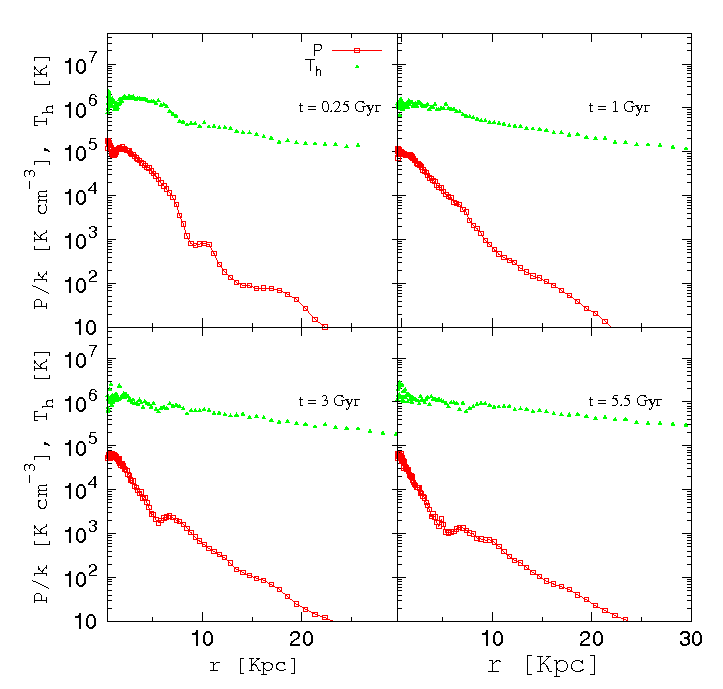}
\captionsetup{font={normalsize,sf}, width= 0.9\textwidth}
\caption{Hot temperature profiles (green) and gas pressure profiles for the
  \texttt{MW} simulation. See text for details.}
\label{FIG:thermo_mw}
\end{figure}
In Fig. 5.9 we show the number density profiles for
the hot and the cold phase at the the usual times. As expected from
the surface density profiles described above, the cold number density
is much higher than the hot one: it reaches its maximum value in the galaxy
centre and rapidly declines till $\sim$ 12 Kpc 
(nearly the extension of the star forming region), where $n_c$
  changes slopes and slowly declines 
along the disc ($n_c$ may be defined also for non multi-phase,
cold gas). The presence of the bulge reflects in $n_c$ values well above
  $10^2$ at scales $r < 5$ kpc.
Note that at 8 Kpc from the \texttt{MW} centre
(approximately the distance of the Sun from the Milky-Way centre), the
cold number density value is between 10--100 cm$^{-3}$, in agreement
with the observed values in the Solar System neighbourhood. For what
concerns the hot number density, $n_h$ peaks in the centre at $\sim$ 1
cm$^{-3}$ and slowly declines till $\sim$ 10 Kpc beyond which $n_h$ sharply
dimisishes.\\

\begin{figure*}[!]
\centerline{
\includegraphics[width=0.31\linewidth]{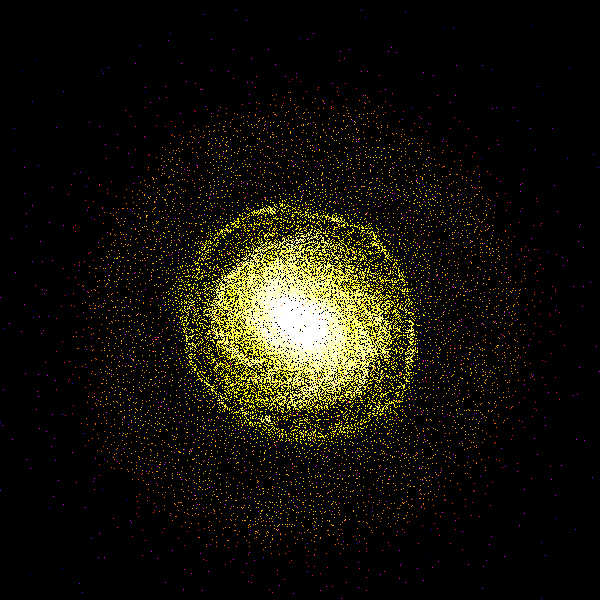}
\includegraphics[width=0.31\linewidth]{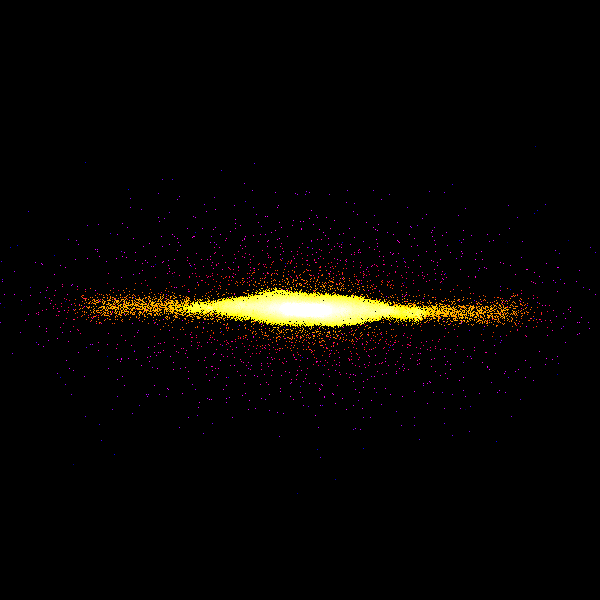}
}
\centerline{
\includegraphics[width=0.31\linewidth]{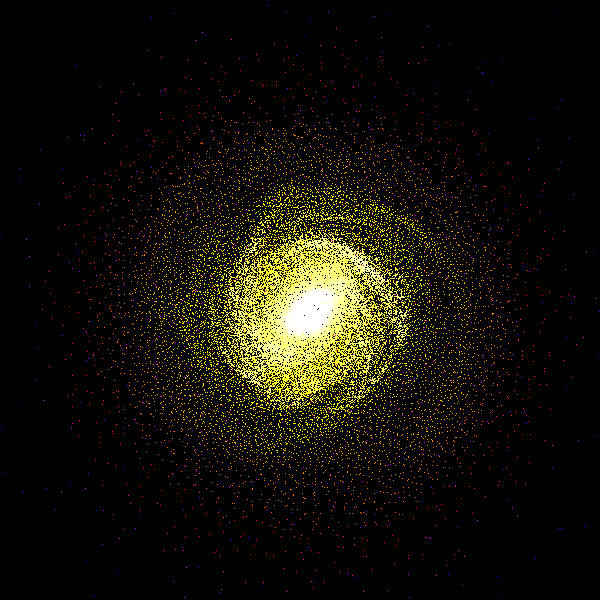}
\includegraphics[width=0.31\linewidth]{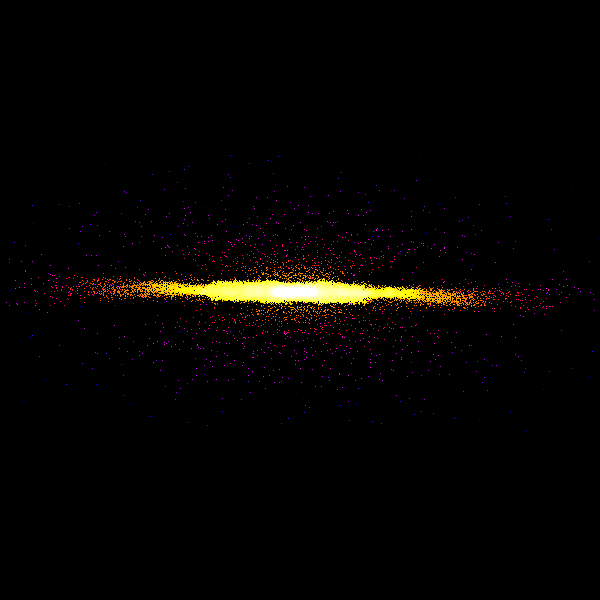}
}
\centerline{
\includegraphics[width=0.31\linewidth]{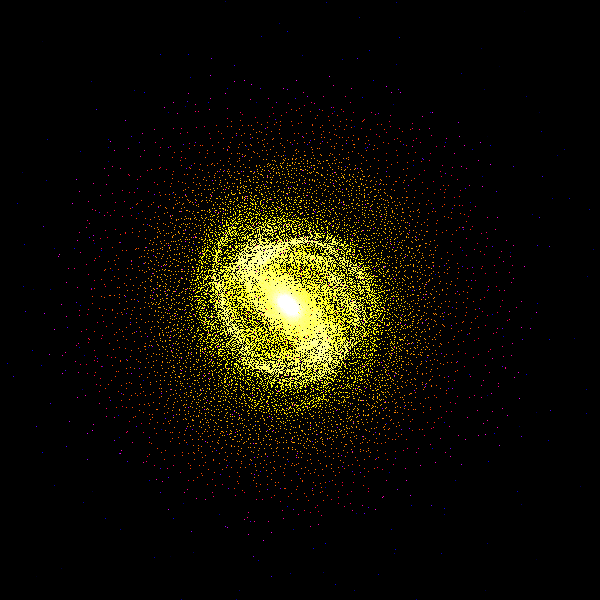}
\includegraphics[width=0.31\linewidth]{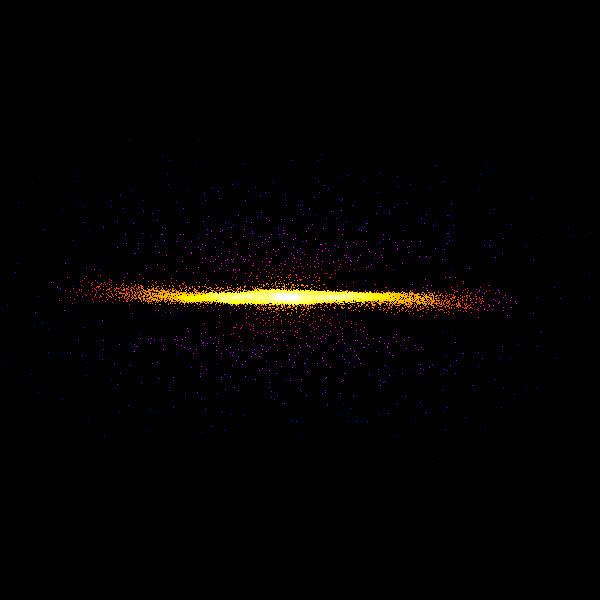}
}
\centerline{
\includegraphics[width=0.31\linewidth]{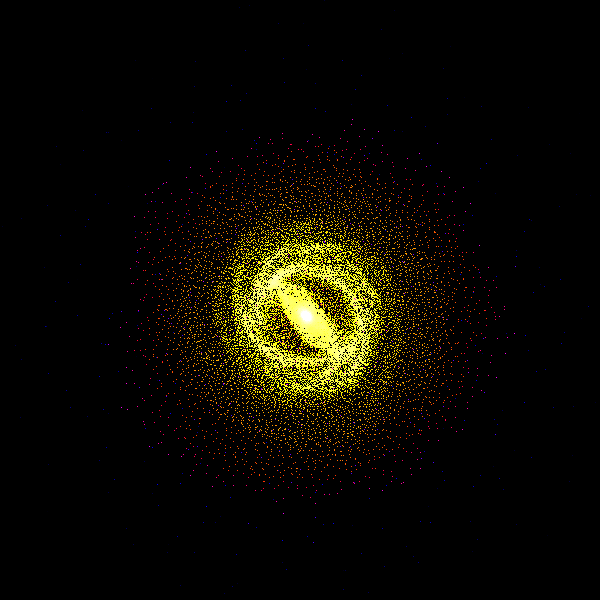}
\includegraphics[width=0.31\linewidth]{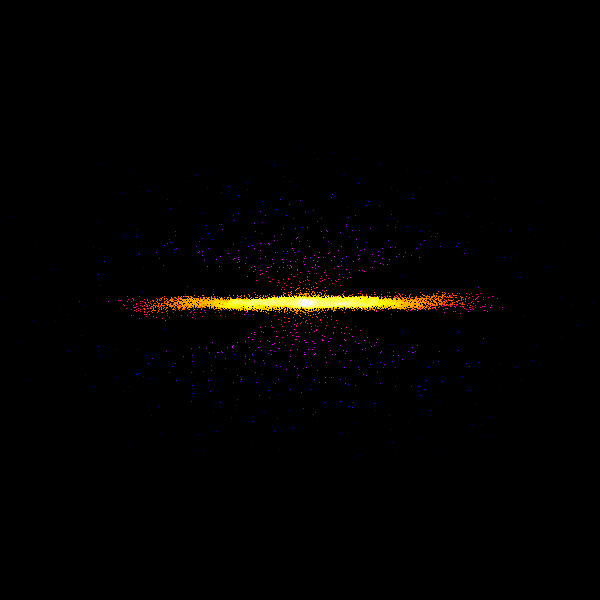}
}

\caption{The distribution of gas in the xy plane (left panels) and
  in the xz plane (right panels) from the simulation of the
  \texttt{MW} galaxy, at 0.25, 1, 3 and 5.5 Gyr (from top to bottom).
 The frames are $60$ Kpc on a side. Colour scale is
  logarithmic and scales from $10^{-0.5}$ to $10^5$ times critical
  density.}
\label{MUPPI:face_mw_g}
\end{figure*}

\begin{figure*}[!]
\centerline{
\includegraphics[width=0.31\linewidth]{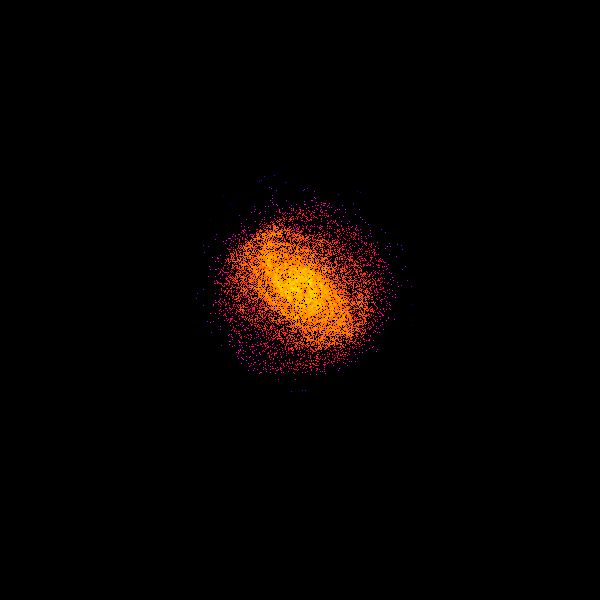}
\includegraphics[width=0.31\linewidth]{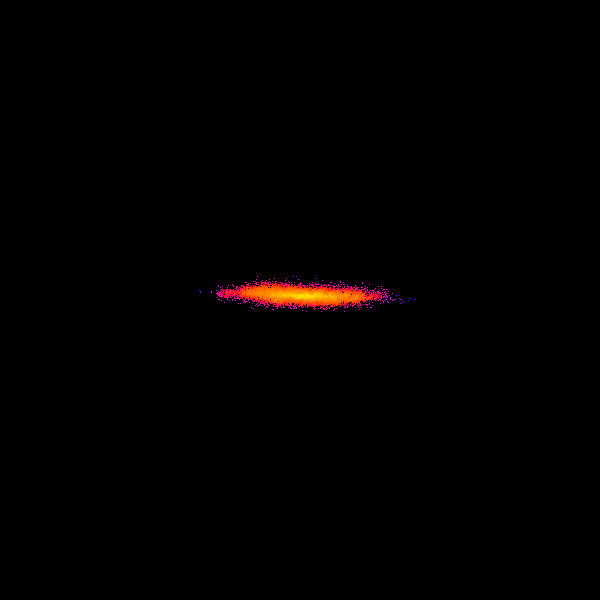}
}
\centerline{
\includegraphics[width=0.31\linewidth]{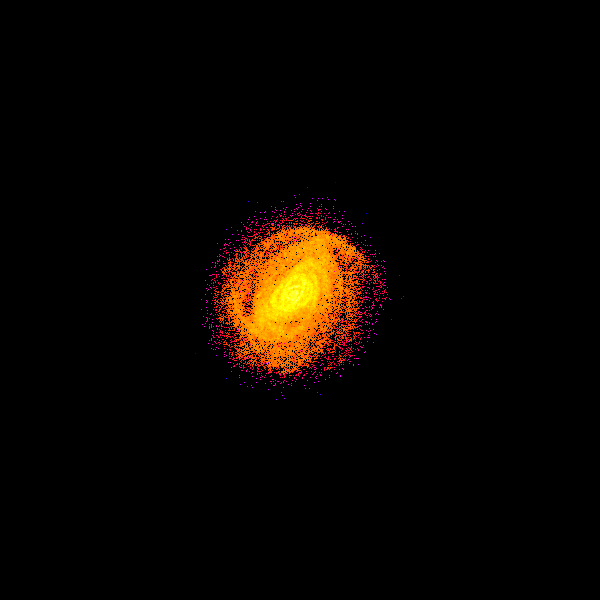}
\includegraphics[width=0.31\linewidth]{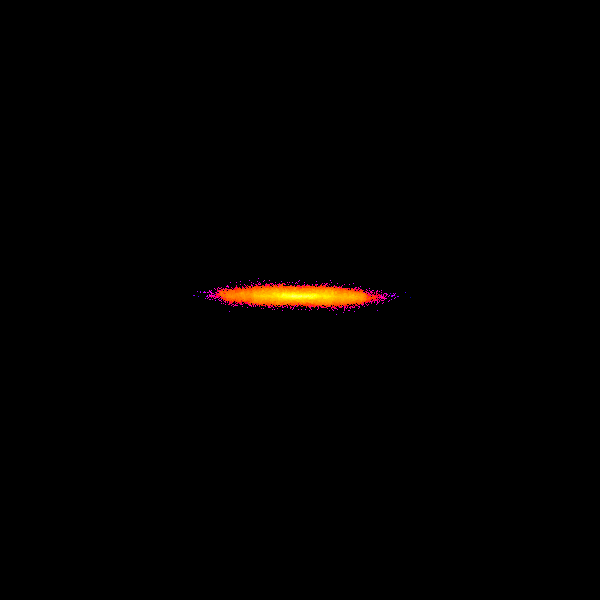}
}
\centerline{
\includegraphics[width=0.31\linewidth]{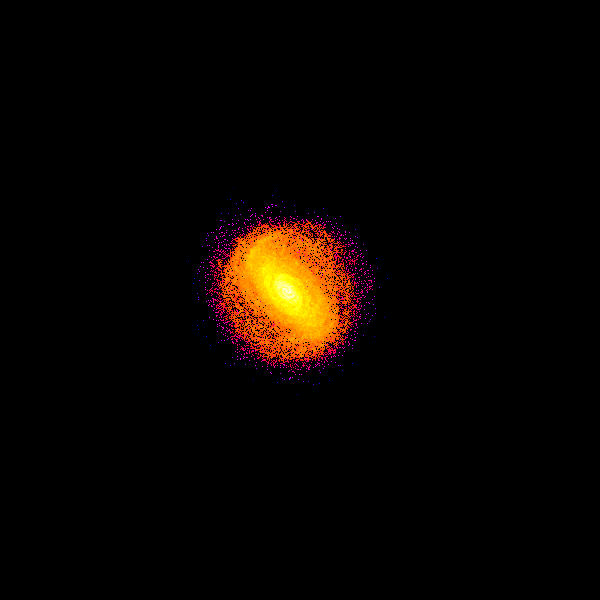}
\includegraphics[width=0.31\linewidth]{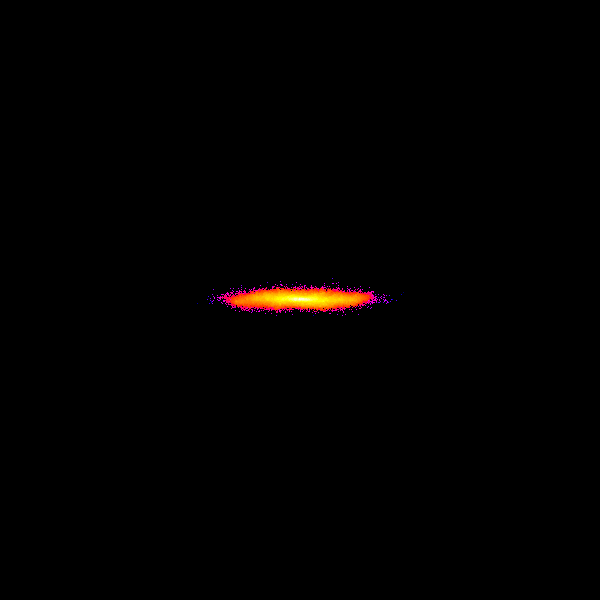}
}
\centerline{
\includegraphics[width=0.31\linewidth]{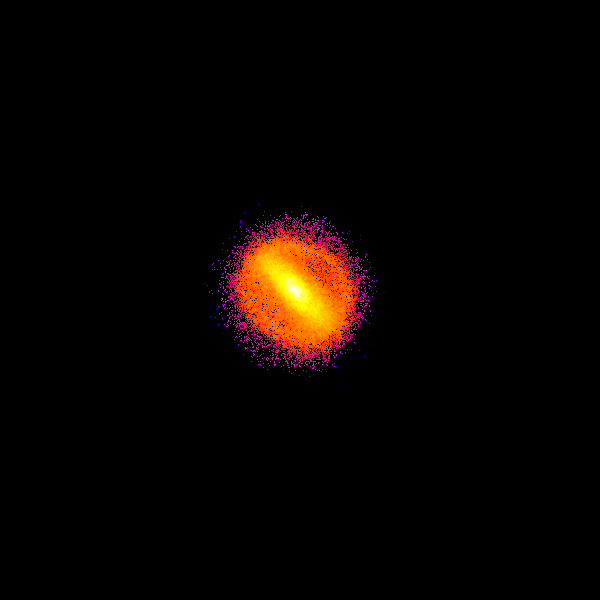}
\includegraphics[width=0.31\linewidth]{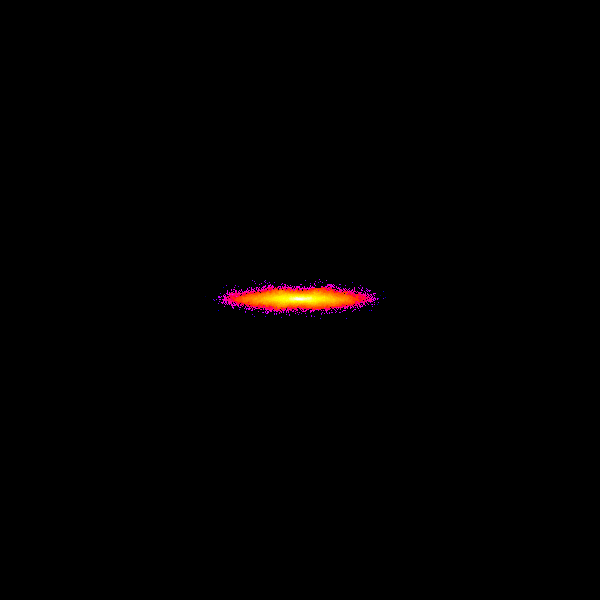}
}
\caption{The distribution of stars in the xy plane (left panels) and
  in the xz plane (right panels) from the simulation of the
  \texttt{MW} galaxy, at 0.25, 1, 3 and 5.5 Gyr (from top to bottom).
 The frames are $60$ Kpc on a side. They show density maps
  generated with the SMOOTH algorithm, applied separately to the star particle 
  distributions. Colour scale is logarithmic and scales from
  $10^{0.5}$ to $10^7$ times critical density. }
\label{MUPPI:face_mw_s}
\end{figure*}
The hot gas in our \texttt{MW} galaxy is almost isothermal with
 an average 
 temperature of 10$^5$--10$^6$K, as shown in Fig.~\ref{FIG:thermo_mw}.
In this figure we also plotted the total pressure of the SPH gas
particles which, at variance with temperature, does vary
depending on the position from the centre: while in the bulge the
average pressure ranges between 10$^{3}$--10$^5$ K cm$^{-3}$, in the disc
$P/K$ value is below $\sim$ 5$\cdot$10$^2$. Anyway, in the single particle
plots (in Fig.~\ref{RES:ism_core}--~\ref{RES:ism_disc}) we already
noticed the gas isothermal behaviour (in fact, both core and disc
particles has a $T_h \sim$10$^7$) and the differences in pressure
depending on the position from the galaxy centre.
\clearpage
\subsection{Comparison with the GADGET effective model}
\begin{figure}
\centering
\includegraphics[width=0.7\linewidth]{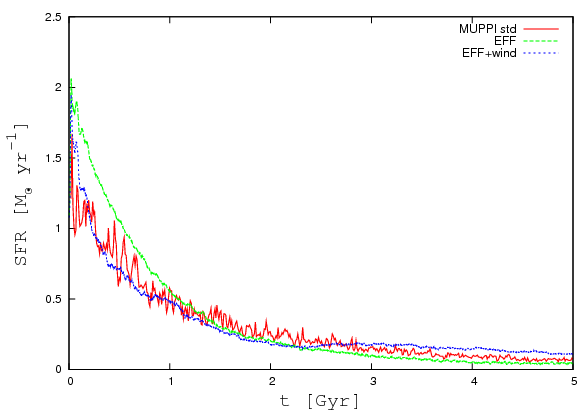}
\captionsetup{font={normalsize,sf}, width= 0.8\textwidth}
\caption{Star formation rate as a function of time for the \texttt{MW}
  model with MUPPI standard set of parameters (red solid line, see
  Tab.~\ref{RES:param}), and for the GADGET effective model with
  (blue short dashed line) and
  without (green dashed line) winds.}
\label{FIG:sfr_com_mw_eff}
\end{figure}
\begin{figure*}
\centerline{
\includegraphics[width=0.6\linewidth]{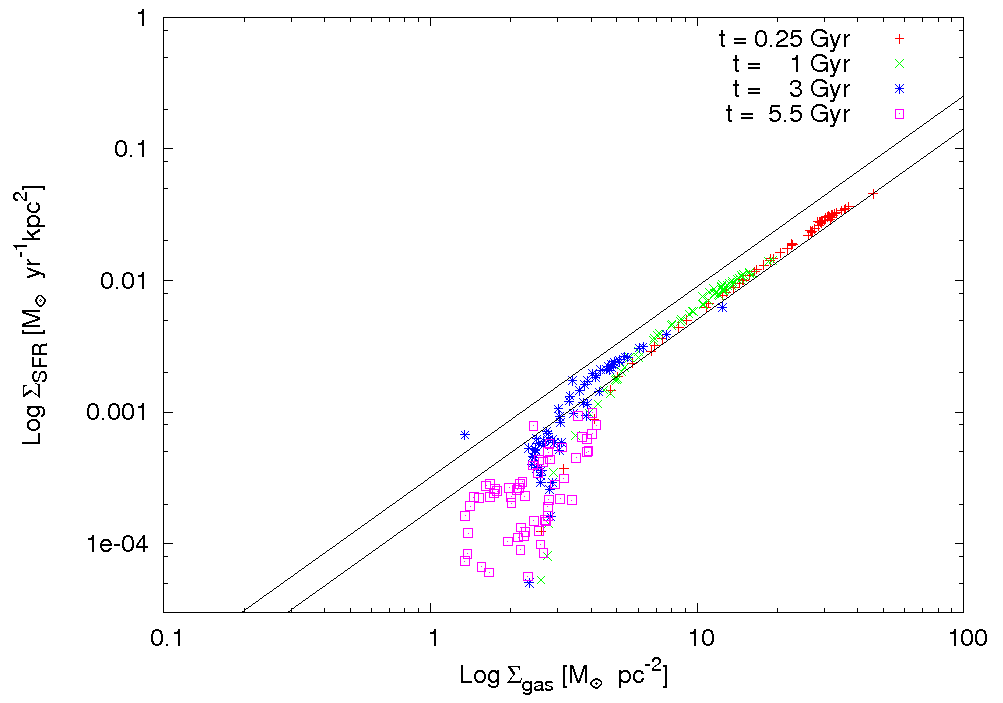}
\includegraphics[width=0.6\linewidth]{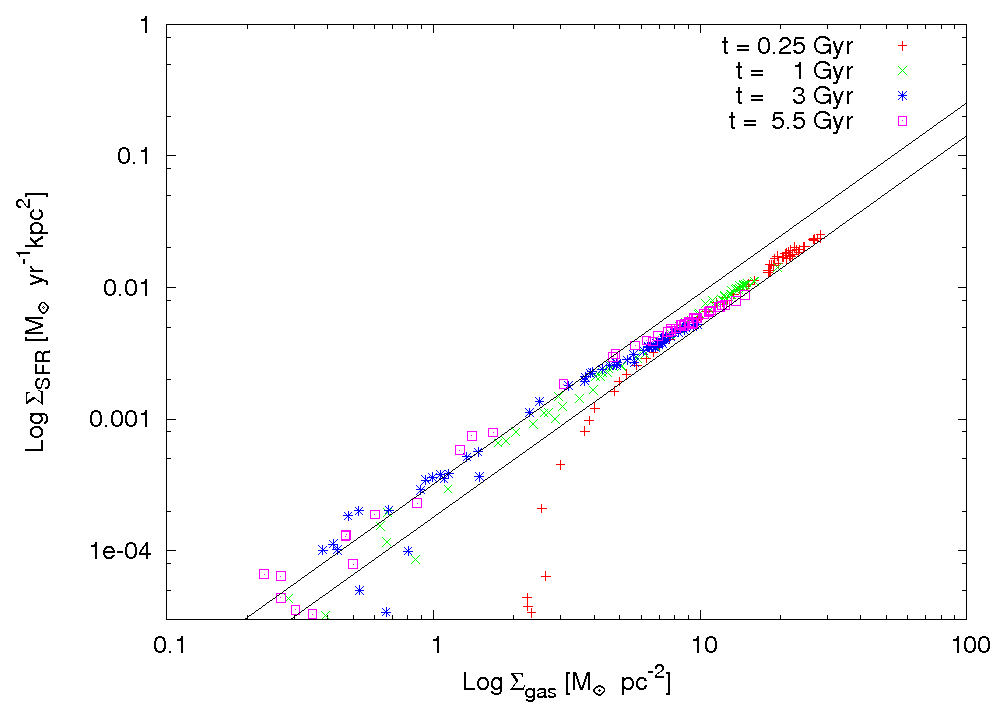}
}
\captionsetup{font={normalsize,sf}, width= 1\textwidth}
\caption{Star formation rate density as a function of gas surface
  density for the \texttt{MW} model with the effective model with
  winds (right panel) and without winds(left panel). The solid lines mark the
  Schmidt-Kennicut law (1998). }
\label{RES:kenni_eff}
\end{figure*}

 
We performed a simulation of the $\texttt{MW}$ model using the GADGET
effective model (see Sec.~\ref{GDT:sf_eff}) in order to compare MUPPI
results (described in the previous section) with those obtained with the
original GADGET star formation function.\\
In Fig.~\ref{FIG:sfr_com_mw_eff} we compare the star formation rates
obtained simulating the $\texttt{MW}$ model with MUPPI, the effective
model(EFF) and the effective model with winds (EFF+W) having velocities
equal to 340 km s$^{-1}$ 
(see Eq.~\ref{GDT:eq:wind}). In all three
schemes, star formation starts  
 instantaneously because thermally unstable gas in the disk and in the
 bulge collapses becoming immediately star-forming. The behaviour of
 star formation in MUPPI and in EFF+W models is similar, with MUPPI
 star formation being more spiky then EFF+W. At the onset of the
 simulation, MUPPI and EFF+W have a SFR of $\approx$ 1.5 M$_\odot$
 yr$^{-1}$ while EFF star formation rate is slightly more prominent 
being $\approx$ 2 M$_\odot$ yr$^{-1}$. After $\approx$ 1 Gyr, the
three models converge to the same SFR, which then decreases with
time. Note that at final times, the SFR trends are reversed with
respect to the beginning: now the EFF+W SFR is slightly higher than EFF
while MUPPI stays in the middle: this is due to the fact the higher
the SFR the higher the velocity of gas consumption, and thus the
faster the star formation is quenched due to the lack of gas supply.\\
As expected, the EFF model perfectly reproduce the Schmidt-Kennicut
star formation relation, since here this relation is \textit{imposed}
by means of the star formation timescale (see
Eq.~\ref{GDT:sf_timescale}), as shown in Fig.~\ref{RES:kenni_eff}. 
Introducing winds in the EFF model changes the results: now the EFF
model is able to generate galactic fountains but in the meanwhile is
no more able to reproduce the Schmidt-Kennicut law at low 
surface densities. As we have already discussed, MUPPI can generate
galactic fountains \textit{and} the Schmidt law even at low densities
self-consistently (see Fig.~\ref{FIG:kenni_mw}).
As a last comparison with the EFF model, we present in
Fig. 5.15 the phase diagrams ($T$ vs $\rho$) of the
$\texttt{MW}$ with standard EFF star formation (i.e. no winds). We
precedently described the same diagram obtained with MUPPI in the
previous section. The EFF model phase diagram results drastically
 different from that of MUPPI (see Fig.~\ref{FIG:phase_mw}): first of
 all, in the dense region of the plot, the gas particles follows a
 tight relation between $T$ and $\rho$, differently with MUPPI gas
 particles which instead show a larger variety of temperatures
 corresponding to the density.
\begin{figure}
\centering{
\includegraphics[width=0.8\linewidth]{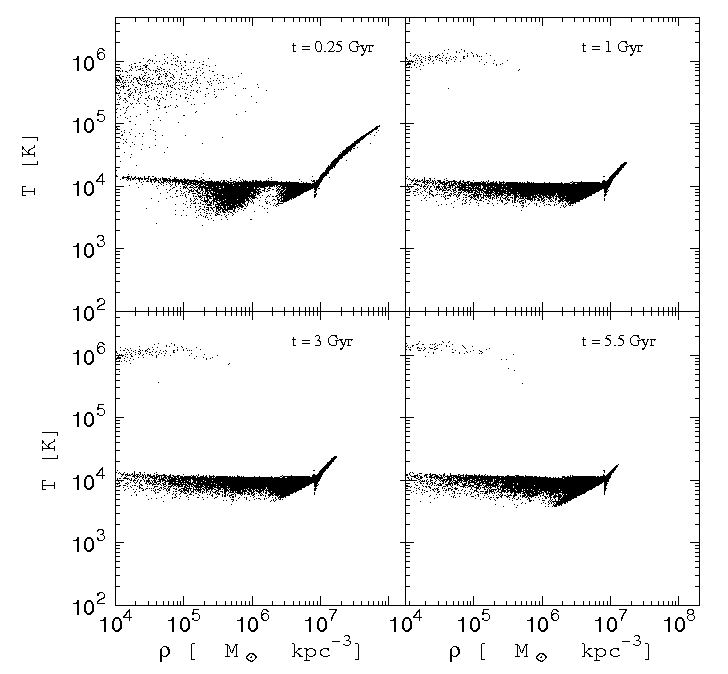}
\captionsetup{font={normalsize,sf}, width= 0.8\textwidth}
\caption{Density-temperature phase diagram for gas
  particle for the \texttt{MW} with the effective model without winds. See text
  for details.} }
\label{FIG:phase_mw_eff}
\end{figure}

\subsection{Varying the blow-out efficiency}
\label{MW_var_fb}
\begin{figure}
\centering{
\includegraphics[width=0.7\linewidth]{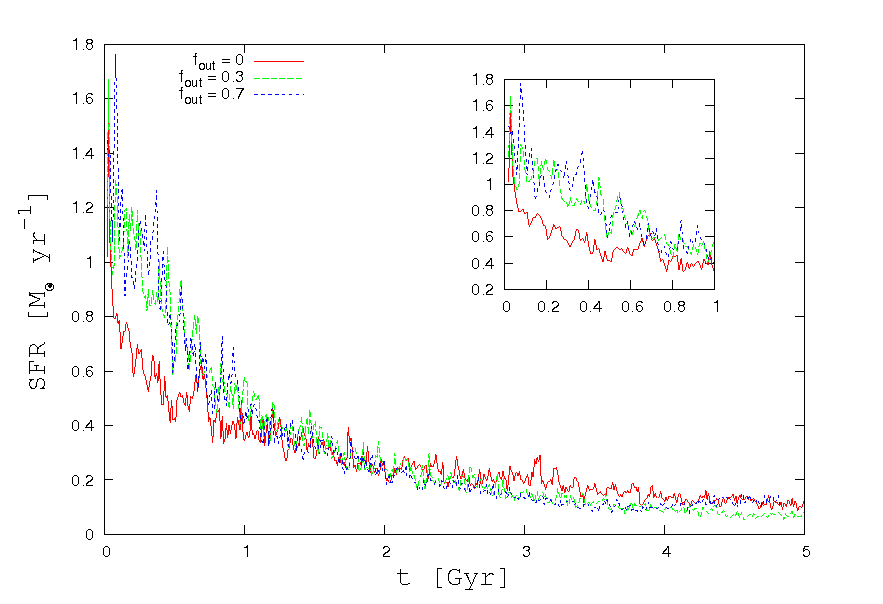}
\captionsetup{font={normalsize,sf}, width= 0.9\textwidth}
\caption{Star formation rate as a function of time for the \texttt{MW}
  model with varying fraction of SNe energy assigned to neighbouring
  particles. In the insert we focus the plot on the SFR till 1 Gyr since
  the onset of the situation. See text for details.}} 
\label{FIG:sfr_mw_var_fbout}
\end{figure}
The effect of changing the fraction $f_{fb_0}$ of the SNe energy that
blows outside the gas particle in the SFRs is demonstrated in
Fig. 5.16. While the SFR follows the same
general trend when increasing $f_{fb_0}$ from 0.3 (FB03) to 0.7
(FB07), in the case 
of zero blow-out efficiency (FB00) the SFR is slightly
different. In fact, MUPPI star formation is driven by pressure. Within
each MUPPI particle, SNe energy pressurises the hot phase. Part of such
energy is provided by SNe exploding inside the particle, but another
part comes from neighbouring multi-phase particles. When we set
$f_{fb_0}=0$, this source of pressurisation is not present and 
as a consequence, the star formation
proceeds much slower.
\begin{figure}
\centering
\includegraphics[width=0.6\linewidth]{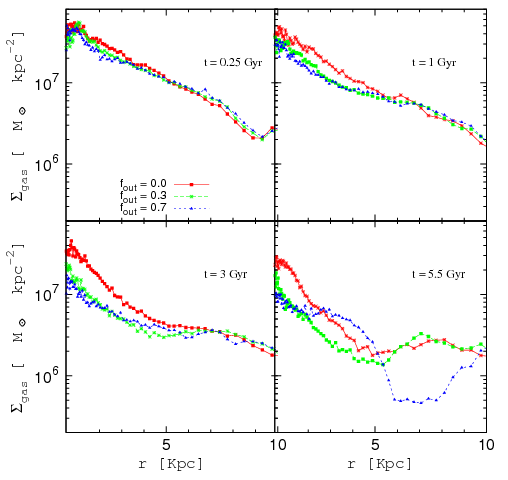}
\captionsetup{font={normalsize,sf}, width= 0.9\textwidth}
\caption{Surface density profiles for the \texttt{MW} model with 
  $f_{\rm fb,o}=0.$, $0.3$, $0.7$ for stars ($\rho_{\star}$) and
  different gas phases, i.e. cold ($\rho_c$), molecular ($\rho_{mol}$), hot
  ($\rho_h$) and total gas density ($\rho_{gas}$)}
\label{FIG:FB_rho_mw}
\end{figure}
As we can see from
Fig. 5.16, runs with non-zero blow-out fraction
 have an higher star formation efficiency and thus
consume much faster the gas supply. At final times the, the trends are
reversed with the FB00 run having a larger reservoir of gas to convert in
stars than the other two cases. This difference in gas supply between
the three runs is confirmed in Fig.~\ref{FIG:FB_rho_mw}
where we show the surface density profiles at the usual times. If at
0.25 Gyr the profiles are very similar, their behaviour changes with
time, with the ``blow-out'' runs being less dense in the first
$\approx$ 8 Kpc from the centre than the FB00 simulation. Note that
at 5.5 Gyr the FB07 run shows a density drop between 5 and 10
Kpc, due to the onset of an instability leading to the
formation of a very strong bar. Apart the strong bar formation,
there is not such a difference at this level in choosing $f_{fb_0}$ to
be 0.3 or 0.7. Anyway, if one plots the density-star formation rate
relationship for the FB07 case as shown in
\begin{figure*}
\centerline{
\includegraphics[width=0.6\linewidth]{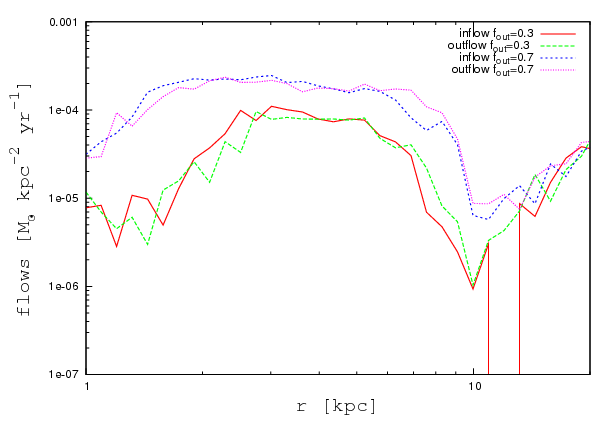}
}
\captionsetup{font={normalsize,sf}, width= 0.8\textwidth}
\caption{Inflow and outflow star rates for the
  \texttt{MW} model with $f_{\rm fb,o}$ = 0.3 and
  $f_{\rm fb,o}$ = 0.7.  }
\label{RES:flows_MW}
\end{figure*}
We also evaluated gas outflows and inflows generated by the hot gas
particles heated by SNe energy and floating away from the disk plane.
To do this, we calculated how many gas
particles lie in a slice with z coordinates $-2<z-1$ kpc and $1<z<2$
kpc. We assign to ``outflows'' those gas mass particles having negative
velocities when z<0 and positive velocities when z>0, and to ``inflows''
the gas particles having opposite behaviour.
In Fig.~\ref{RES:flows_MW} we show the resulting mass outflow and inflow as a
function of the distance from the centre of the disk for both cases
FB03 and FB07. Outflows and inflows have very similar values, being
higher for FB07 as expected. This is a tipi-Cal signature of galactic
fountains, with heated gas floating away but not escaping, cooling
and falling again on the disk.

\begin{figure*}
\centerline{
\includegraphics[width=0.6\linewidth]{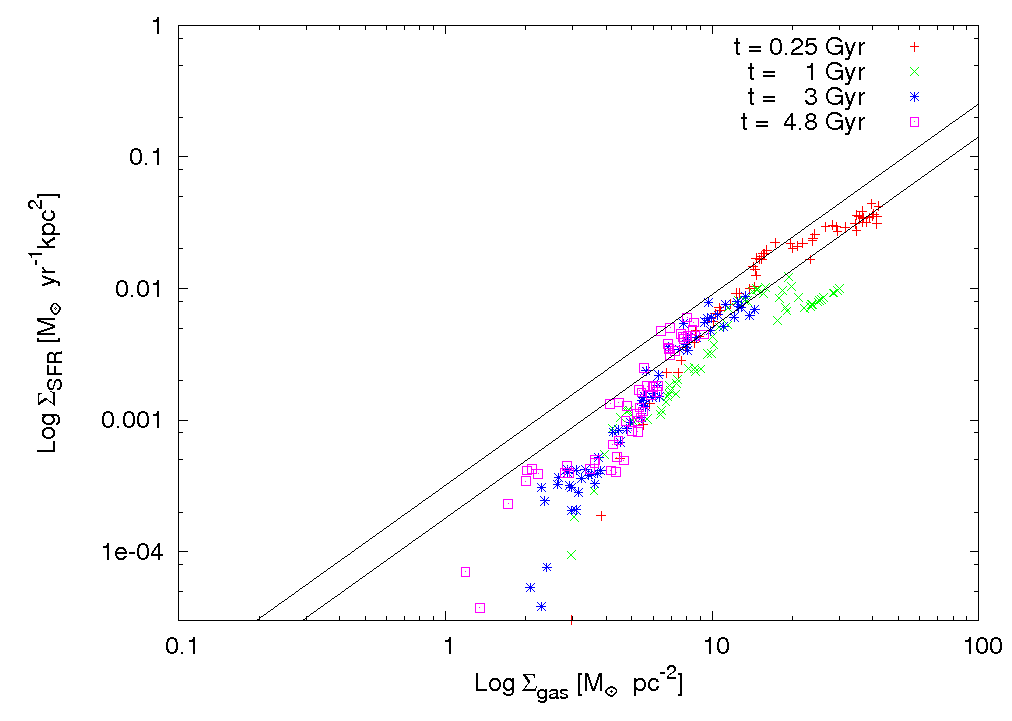}
}
\captionsetup{font={normalsize,sf}, width= 0.8\textwidth}
\caption{Star formation rate density as a function of gas surface
  density for the \texttt{MW} with $f_{\rm fb,o}$ = 0.7. The solid lines mark the
  Schmidt-Kennicut law (1998). }
\label{RES:kenni_mw_fb07}
\end{figure*}

Fig.~\ref{RES:kenni_mw_fb07}, an important difference with the FB03
run rises: in fact, in FB07 we reproduce the observed Schmidt-Kennicut law
worser than in FB03, being the cut at low surface densities too high
with respect to observations. This is the main motivation why we have
chosen  $f_{fb_0}$ = 0.3 to be our reference value for modulating the
blow-out energy.

\clearpage

\section{Other cases}
\label{RES:other}
\subsection{Dwarf galaxy}
\label{RES:dwarf_g}
\begin{figure}
\centering{
\includegraphics[width=0.8\linewidth]{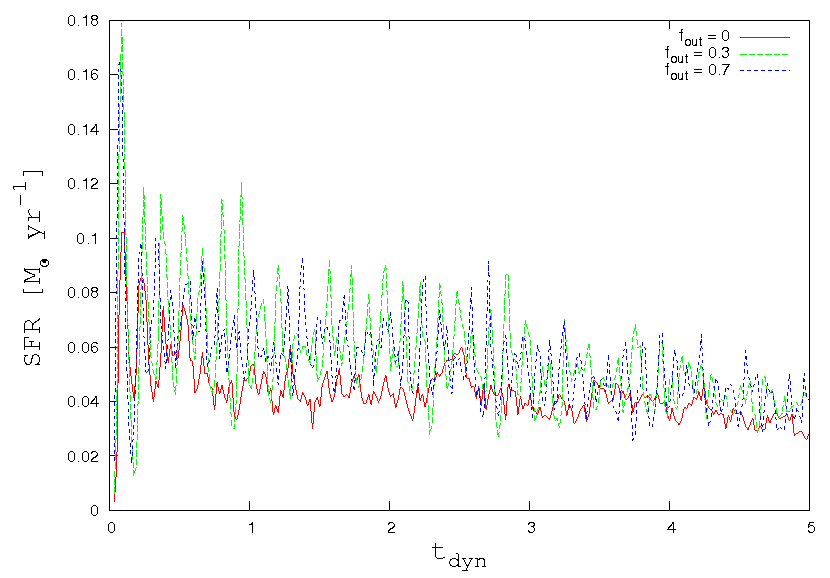}
\captionsetup{font={normalsize,sf}, width= 0.9\textwidth}
\caption{Star formation rate as a function of time for the \texttt{DW}
  model with  $f_{\rm fb,o} = 0$, $0.3$ and $0.7$.
  See text for details.}} 
\label{FIG:sfr_dw_fb}
\end{figure}
 We run a set of simulations with the purpose of assessing how MUPPI
 behaves on different 
initial physical conditions. All the numerical parameters of the
simulations are the same as in the  $\texttt{MW}$ case, except mass 
resolution and softenings (see Tab.~\ref{RES:glx}).

\begin{figure}
\centering
\includegraphics[width=0.7\linewidth]{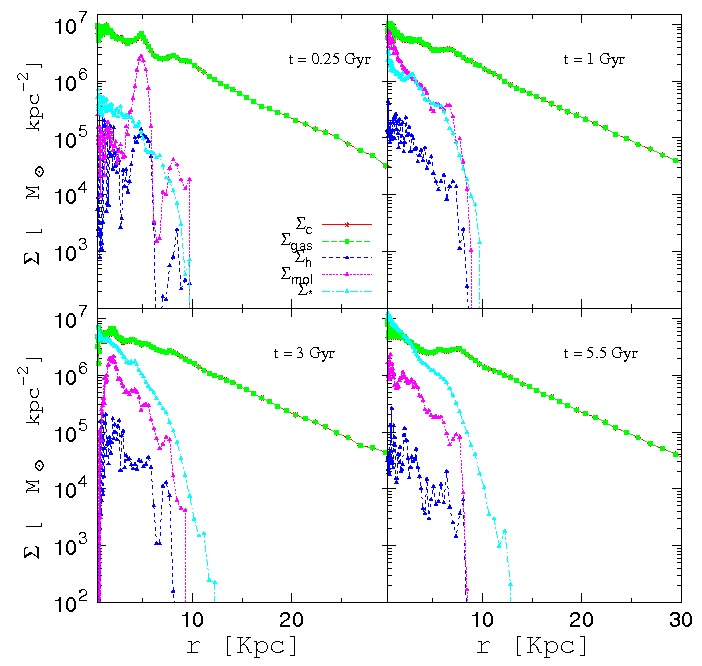}
\captionsetup{font={normalsize,sf}, width= 0.9\textwidth}
\caption{\texttt{DW} surface density profiles for stars ($\rho_{\star}$) and
  different gas phases, i.e. cold ($\rho_c$), molecular ($\rho_{mol}$), hot
  ($\rho_h$)and total gas density ($\rho_{gas}$)}
\label{FIG:dens_dw_03}
\end{figure}

Here we show our results for the  $\texttt{DW}$ run.
As we did for the $\texttt{MW}$ one, we investigate the SFR history and
compare results obtained varying the blowing-out SN
energy, as shown in Fig. 5.20. As expected, the SFRs are lower
than in the $\texttt{MW}$ run, being the $\texttt{DW}$ galaxy less
massive and dense, and thus less pressurised. 
Again, the FB00 run has the lowest SFR while the FB03 and FB07 show a similar
behaviour, with a higher SFR. As in the $\texttt{MW}$ then, turning
off the blow-out regime 
lead to a low pressure ISM and a corresponding low star formation
efficiency: the fact the ISM is less pressurised also causes the
fraction $f_{coll}$ of molecular gas to decrease. This is similar to
what happens at the edges of the MW disk (see Fig.~\ref{FIG:dens_mw}),
and is shown in Fig.~\ref{FIG:dens_dw_03}.

Here  we show for the FB03 run gas surface density profiles
  for cold and hot 
phase, stars, molecular phase and the whole gas;
the molecular density profile $\Sigma_{mol}$ evolution is shown as purple
lines with triangles. In this case, the amount of molecular gas is
always a relatively small fraction of the cold gas, while in the
$\texttt{MW}$ case the molecular gas fraction was almost unity in the
centre and declined towards the edge of the star forming zone (see
Fig.~\ref{FIG:dens_mw}).   

From Fig.~\ref{FIG:dens_dw_03}, we
see that the $\texttt{DW}$ interplay among
different phases and the resulting self-regulated regime is very
similar to that found in the $\texttt{MW}$ case, even if all surface
densities reach lower values, due to the shallower gravitational
potential well produced by the halo in this case.
\begin{figure}
\centering
\includegraphics[width=0.7\linewidth]{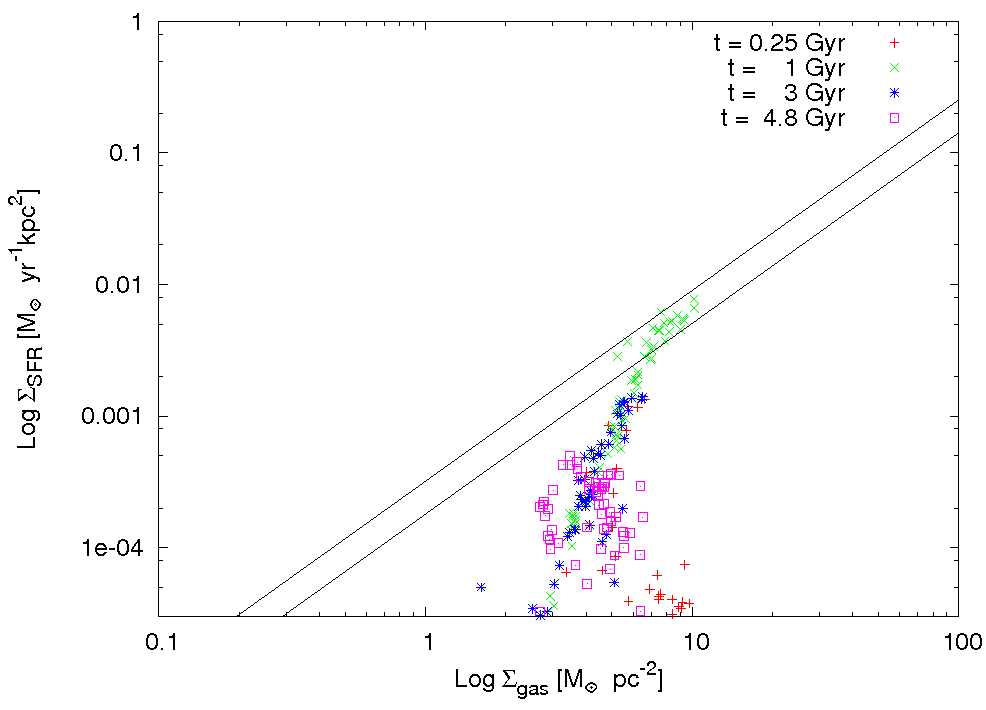}
\captionsetup{font={normalsize,sf}, width= 0.9\textwidth}
\caption{Star formation rate density as a function of gas surface
  density for the \texttt{DW} model. The solid lines mark the
  Schmidt-Kennicut law (1998).  }
\label{FIG:kenni_dw_03}
\end{figure}
The density-SFR relation
for the FB03 $\texttt{DW}$ galaxy does reproduce only the declining
part of the Schmidt-Kennicut
relation (see Fig.~\ref{FIG:kenni_dw_03}); in this case, cold gas
surface density high enough to be significantly compared with local
observation are only occasionally reached.
\begin{figure}
\centering{
\includegraphics[width=0.7\linewidth]{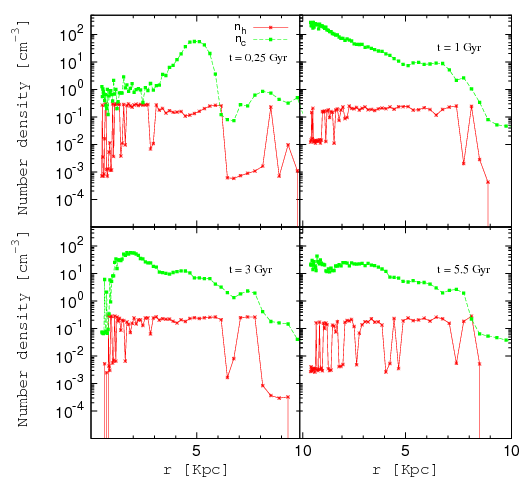}
\captionsetup{font={normalsize,sf}, width= 0.9\textwidth}
\caption{\texttt{DW} number density profiles for the cold phase
  (green) and the hot phase (red).}}
\label{FIG:numdens_dw}
\end{figure}
In order to complete the comparison with the $\texttt{MW}$ galaxy we
show in Fig. 5.23 and in Fig. 5.24 the
number density profiles and the pressure-temperature evolution. 
The number density profiles follow very closely the behaviour of the
surface density profiles, discussed above and showed in
Fig.~\ref{FIG:dens_mw}. At 0.25 Gyr, a large fluctuation in $n_c$
is visible around 5 Kpc from the centre: this bump likely originates due to the
initial large burst of star formation which generates a pressure wave
propagating through the galaxy disk. This pressure wave is clearly
visible from the pressure plot in Fig. 5.24. In
the top-left panel of Fig.~\ref{DW_faces_g} the pressure wave
instability generates a  ring-like
structure around the galaxy centre which disappears later on. This
behaviour is caused by the sudden turn-on of cooling and star
formation physics and should be regarded as a numerical effect.

As in the $\texttt{MW}$, the hot gas in our $\texttt{DW}$ galaxy is
almost isothermal with an average 
 temperature oscillating between some 10$^5$ to 10$^6$ K, as shown in
 Fig. 5.24. The reason why the temperature profile is
 sharply interrupted is that beyond 10 Kpc there are not multi-phase
 gas particles. This does not happen to the pressure trend because
 pressure can be estimated on all the SPH particle.\\
All the ISM properties we showed in this section closely resemble our
result for the MW {\it disk}, far from its bulge.
Overall, the simulations of the
$\texttt{DW}$ galaxy, 
correctly reproduce the general properties expected for a
quiet lighter galaxy, with a less active ISM.

\begin{figure*}
\centering{
\includegraphics[width=0.6\linewidth]{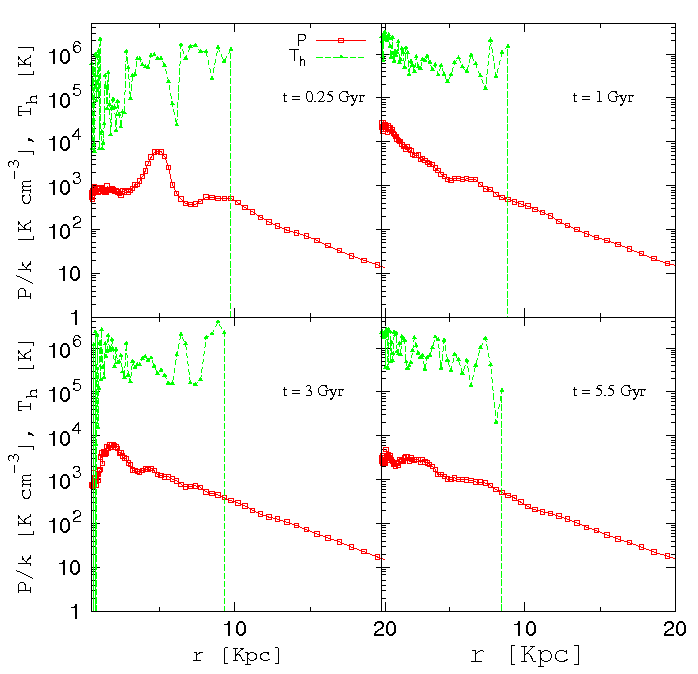}
\captionsetup{font={normalsize,sf}, width= 0.9\textwidth}
\caption{Gas pressure profile (red) and temperature profile
  (green) of the hot phase for the \texttt{DW} galaxy model.}}
\label{FIG:thermo_dw}
\end{figure*}

Comparing the distribution of gas at various times for the MW
and DW disk (figs...), a difference appears. In the MW case, the
gas disk develops long lasting, sharp spiral arms (xy view) and part
of the gas is expelled from the disk, and then falls back in fountains
(xz view).  In the DW case, such an outflow is almost unappreciable, and
the xy view shows a more irregular, disturbed structure.

This is due to the difference in the feedback strength. The high
pressure, and thus high SFR level and energy feedback, present in the
MW, heats the hot phase enough to drive gas particles outside the
disk, forming a ``thermal wind''. Such particles bring energy away
from the plane, then cool down,
after exiting the multi-phase stage, and can thus originate fountains.

The pressure in the DW case is not high enough for this to happen; on
the other hand, hot gas in the disk still exerts hydrodynamical pressure on
neighbours, destabilising the disk structure and carving cavities
(resembling SNe super-bubbles) in it.

\begin{figure*}
\centerline{
\includegraphics[width=0.31\linewidth]{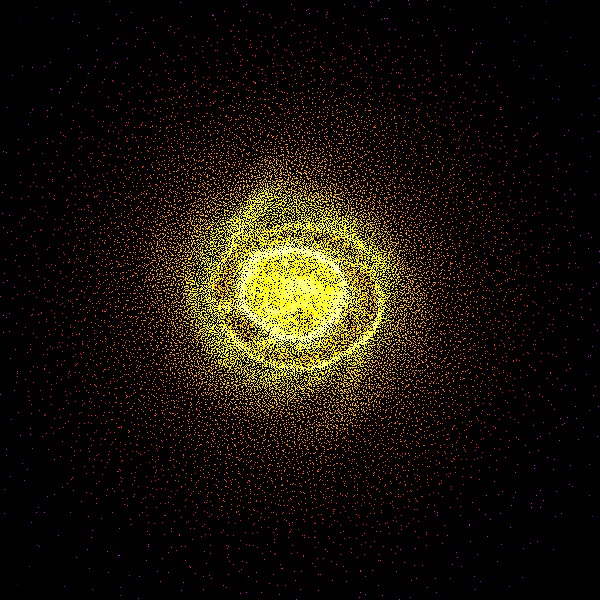}
\includegraphics[width=0.31\linewidth]{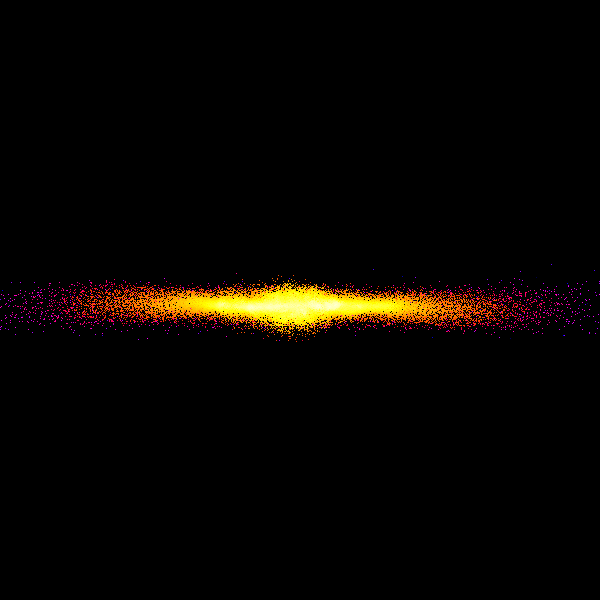}
}
\centerline{
\includegraphics[width=0.31\linewidth]{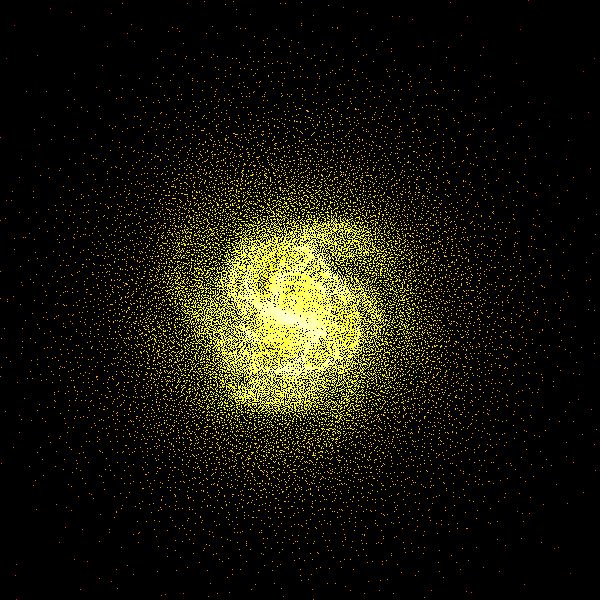}
\includegraphics[width=0.31\linewidth]{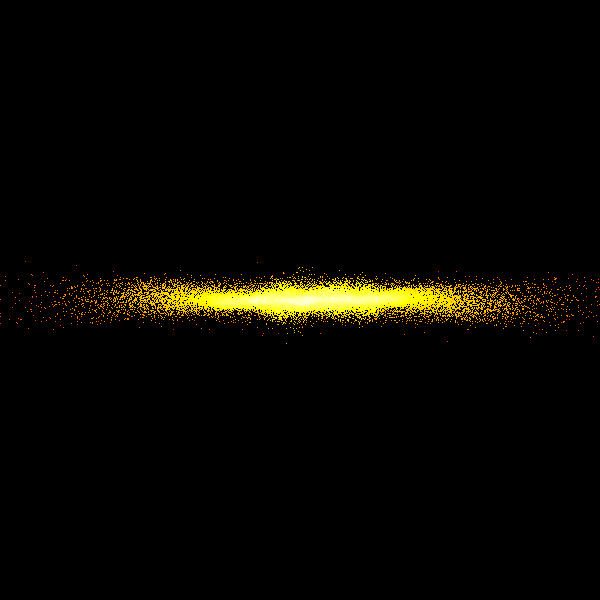}
}
\centerline{
\includegraphics[width=0.31\linewidth]{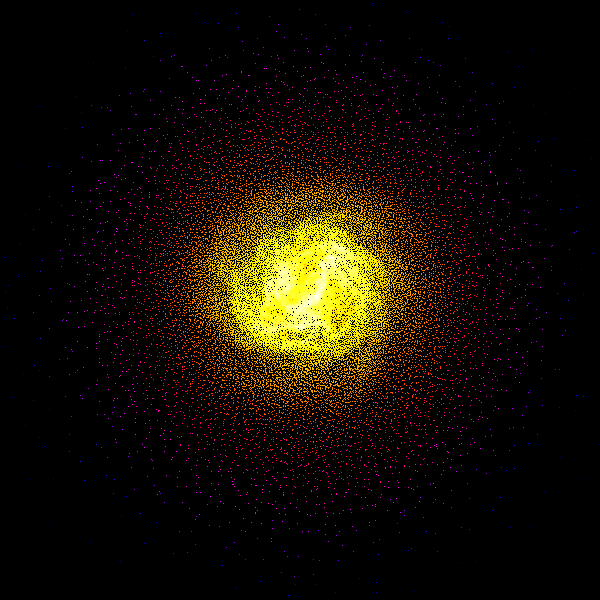}
\includegraphics[width=0.31\linewidth]{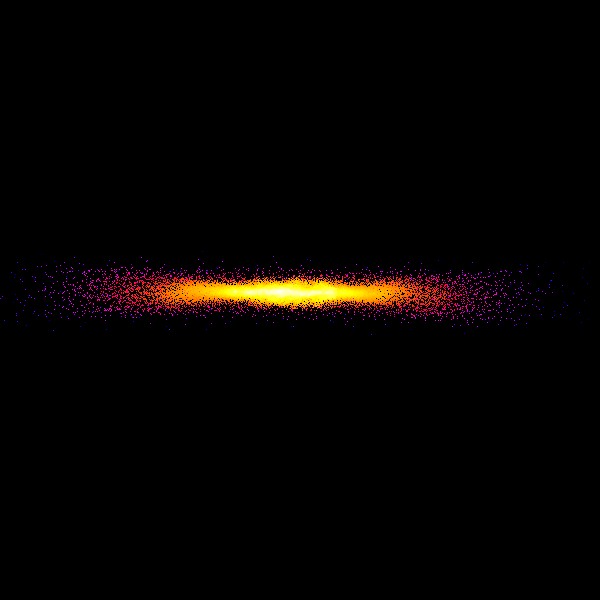}
}
\centerline{
\includegraphics[width=0.31\linewidth]{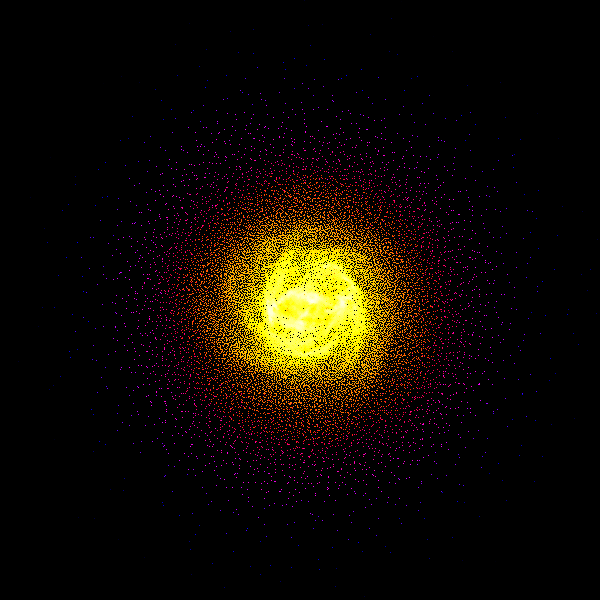}
\includegraphics[width=0.31\linewidth]{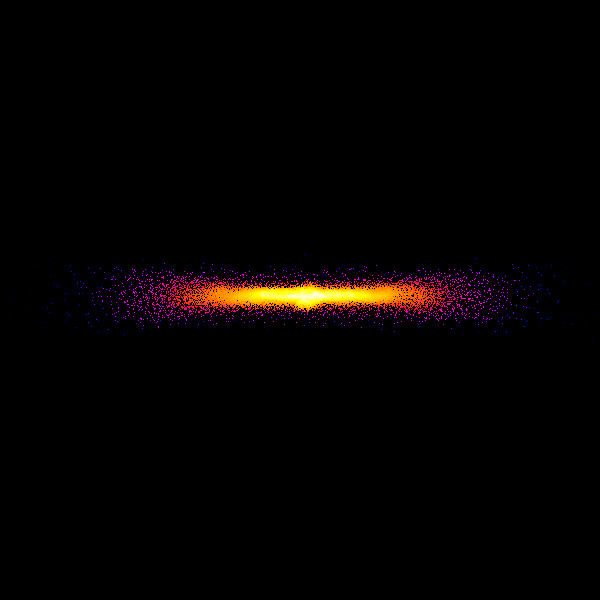}
}
\caption{The distribution of gas in the xy plane (left panels) and
 in the xz plane (right panels) from the simulation of the
 \texttt{DW} galaxy, at 0.25, 1, 3 and 5.5 Gyr (from top to bottom).
 The frames are $80$ Kpc on a side. Colour scale is
  logarithmic and scales from $10^{-0.5}$ to $10^5$ times critical
  density.}
\label{DW_faces_g}
\end{figure*}

\begin{figure*}
\centerline{
\includegraphics[width=0.31\linewidth]{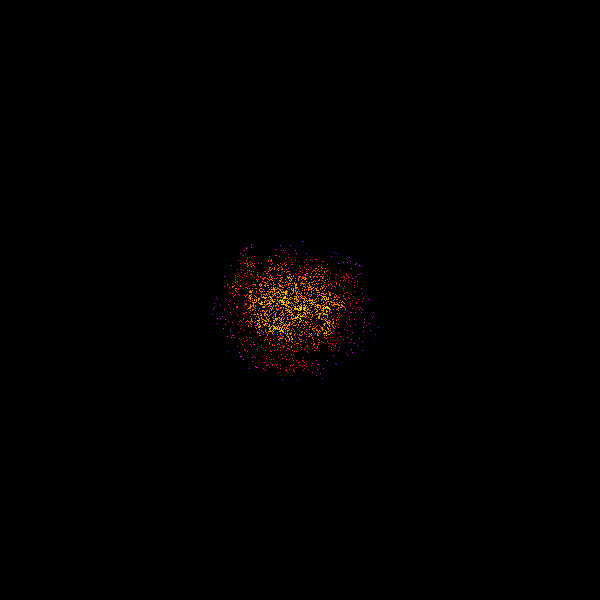}
\includegraphics[width=0.31\linewidth]{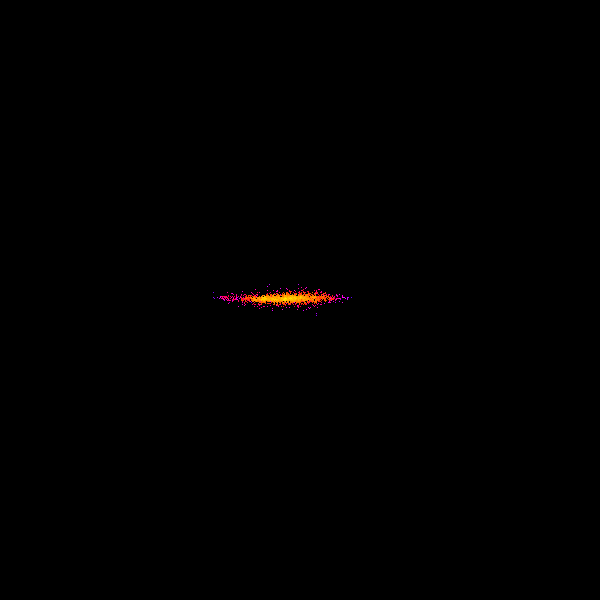}
}
\centerline{
\includegraphics[width=0.31\linewidth]{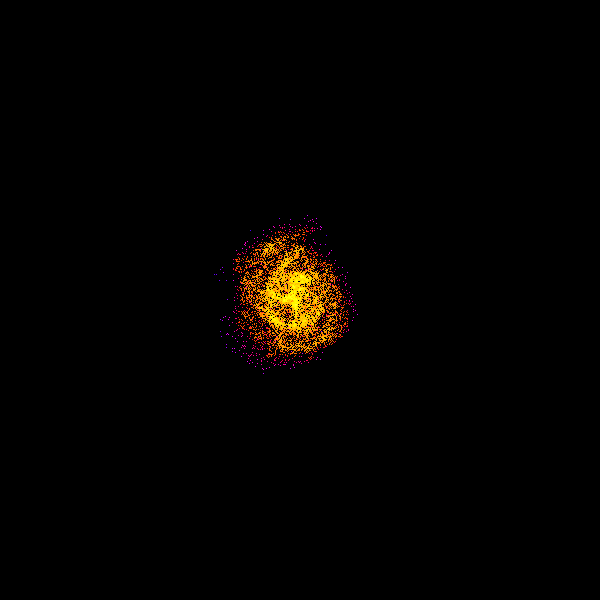}
\includegraphics[width=0.31\linewidth]{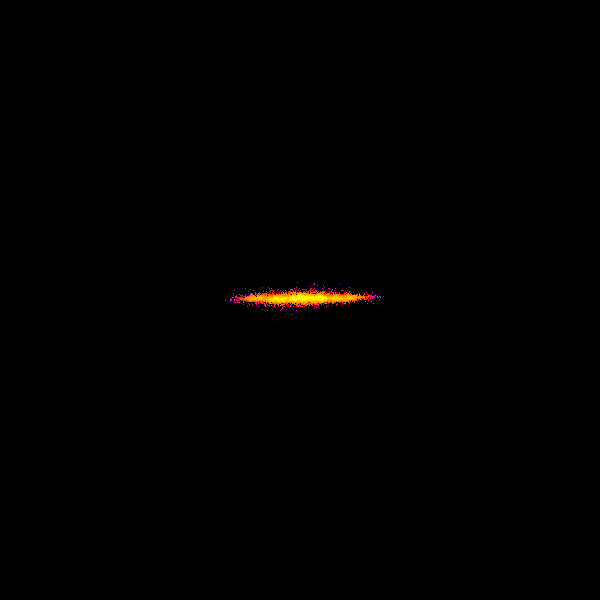}
}
\centerline{
\includegraphics[width=0.31\linewidth]{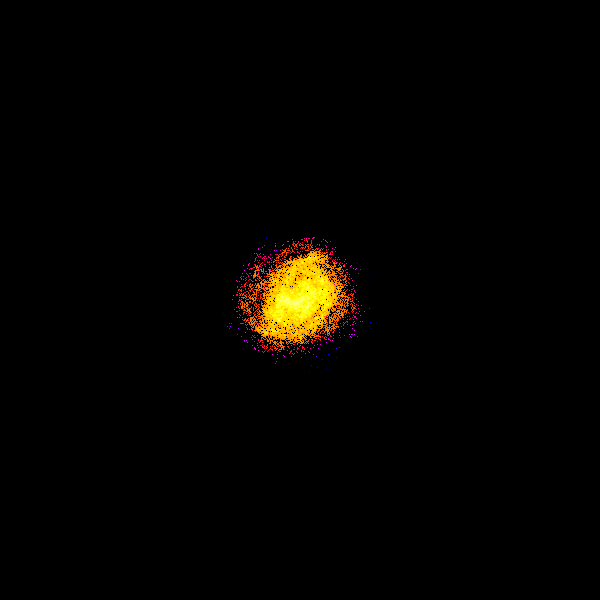}
\includegraphics[width=0.31\linewidth]{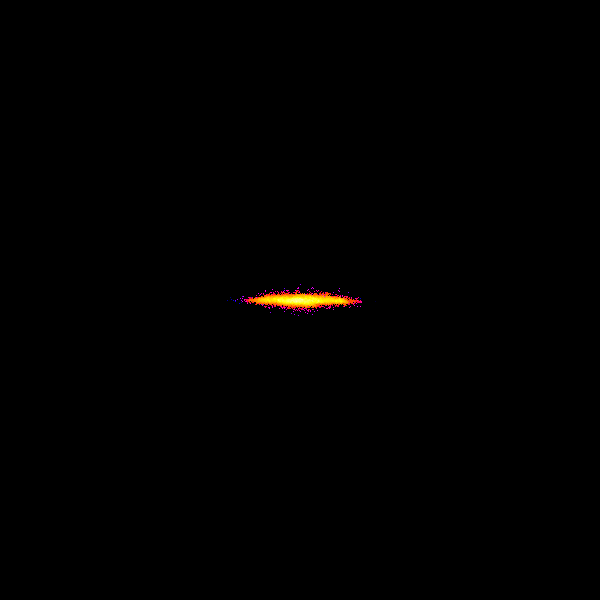}
}
\centerline{
\includegraphics[width=0.31\linewidth]{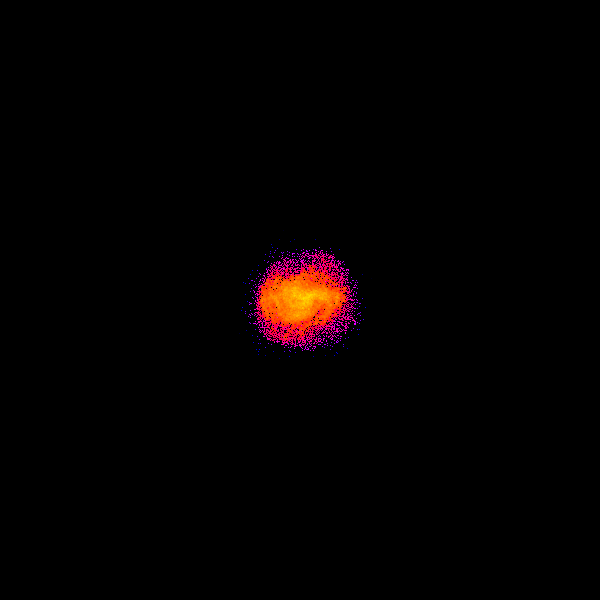}
\includegraphics[width=0.31\linewidth]{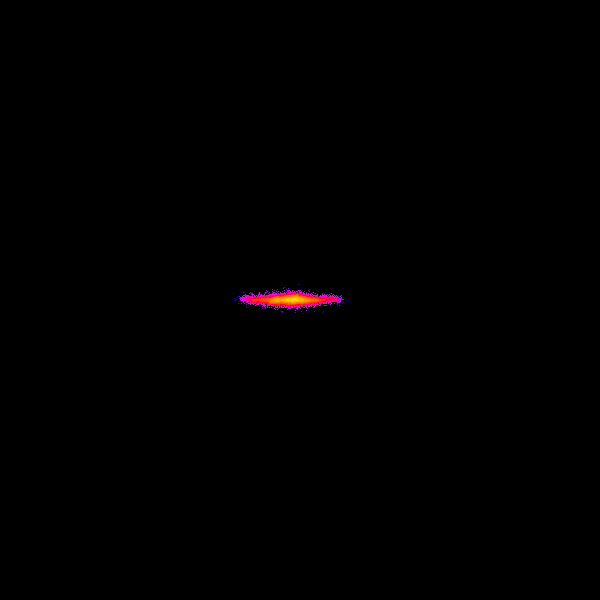}
}
\caption{The distribution of stars in the xy plane (left panels) and
  in the xz plane (right panels) from the simulation of the
  \texttt{DW} galaxy, at 0.25, 1, 3 and 5.5 Gyr (from top to bottom).
 The frames are $80$ Kpc on a side. They show density maps
  generated with the SMOOTH algorithm, applied separately to the star particle 
  distributions. Colour scale is logarithmic and scales from
 $10^{0.5}$ to $10^7$ times critical density. }
\label{DW_faces_s}
\end{figure*}

\clearpage
\newpage
\subsection{Isolated non-rotating haloes}
In this section, we describe simulations of isolated
non-rotating haloes described in Sec.~\ref{RES:ic_glx}, evolved using MUPPI.
\subsubsection{CFDW}
In Fig. 5.27 we show the SFRs obtained by
for the $\texttt{CFDW}$ halo varying the fraction of SNe energy blowing
out of the gas particle ($f_{\rm fb,o}$ = 0.0, 0.3,0.7) in the MUPPI code. For
comparison we also show the SFRs for the GADGET effective model
(EFF) and  the effective model with winds (EFF+wind). In all runs,
star formation starts few Myrs after the onset of the simulation,
when the cooling flow is established and core gas is dense and cold
enough to fulfil star formation thresholds. \\
\begin{figure}[!]
\centering{
\includegraphics[width=0.7\linewidth]{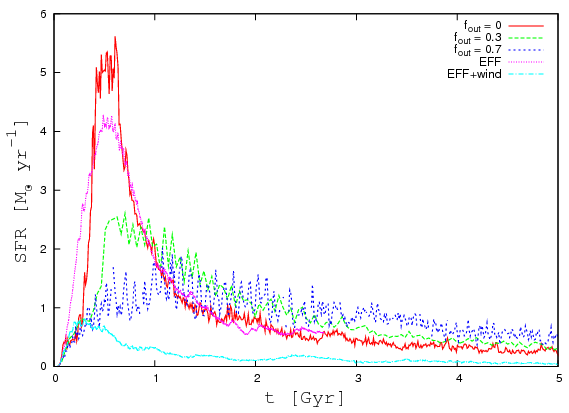}
\captionsetup{font={normalsize,sf}, width= 0.9\textwidth}
\caption{Star formation rate as a function of time for the \texttt{CFDW}
  model simulated with MUPPI,
  with $f_{\rm fb,o}$ = 0.0; 0.3;0.7 and, for comparison,
  with the GADGET effective model without winds (EFF) and with 
  winds (EFF+wind). See text for details.}} 
\label{FIG:sfr_cfdw_fb_eff}
\end{figure}
The behaviour of star formation when using MUPPI is reversed, with
  respect to
\begin{figure}[!]
\centering{
\includegraphics[width=0.7\linewidth]{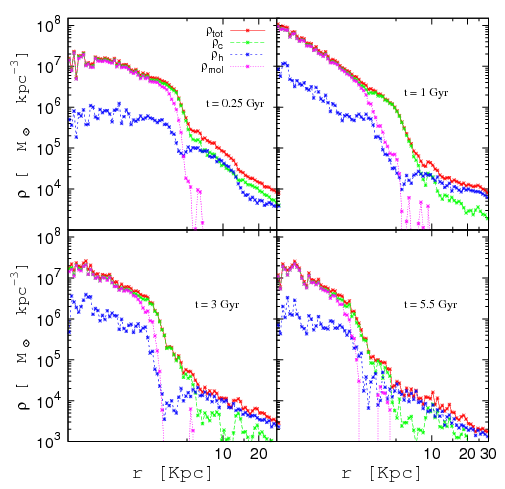}
\captionsetup{font={normalsize,sf}, width= 0.9\textwidth}
\caption{\texttt{CFDW} density profiles for cold ($\rho_c$), molecular
  ($\rho_{mol}$), hot ($\rho_h$)and total gas density ($\rho_{gas}$).}}
\label{FIG:dens_cfdw_fb}
\end{figure}
the results already obtained for the galaxy cases: the more the
energy blowing out the multi-phase gas particles, the more the cooling
  flow is quenched as thus 
the star formation suppressed. The FB00 run has a large burst of
star formation at $\sim$ 0.7 Gyr, leading to a SFR of $\sim$ 5
M$_{\odot}$ yr$^{-1}$ which, due to gas consumption, rapidly
decreases with time. Note that the SFR history of our FB00 model
  is very similar to that obtained using SH03 effective model without
  kinetic feedback. 
In the FB03 and FB07 runs the bursts of star
formation are less intense than in the FB00 case. The SFR peak is
reduced of a factor $\approx$ 2.5 using FB03 and a factor $\approx 5$
using FB07. Thus, the feedback energy injected in the medium is thus
very efficient in countering the cooling flow. As already noted in the
``galaxy'' tests, these  trends are reversed at final times: at 5 Gyr, the FB07 SFR is
\begin{figure}[!]
\centering{
\includegraphics[width=0.7\linewidth]{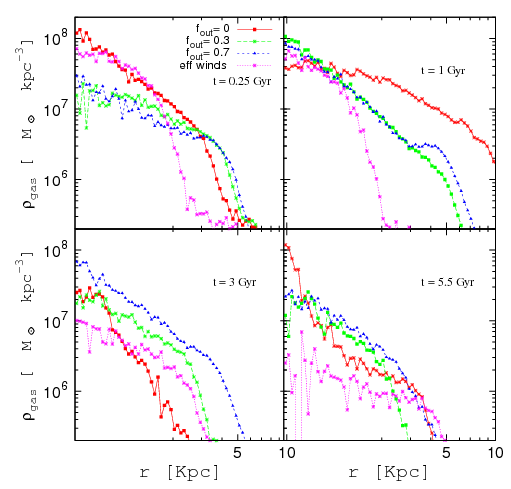}
\captionsetup{font={normalsize,sf}, width= 0.9\textwidth}
\caption{Density profiles for the \texttt{CFDW} model with varying
  fraction of blowing SNe energy for cold ($\rho_c$), molecular
  ($\rho_{mol}$), hot ($\rho_h$)and total gas density ($\rho_{gas}$).}}
\label{FIG:dens_cfdw_fb_var}
\end{figure}
slightly higher than FB00 and FB03, being larger the amount of gas
which has not yet been converted in stars. \\
Note  adding kinetic winds to the EFF model drastically reduces the
efficiency of star formation at all times. The reason for the low
EFF+wind SFR is that here the winds are very effective in ejecting gas
particles outside shallow potential well of this small halo. Applying
the wind scheme to such a low mass halo simply clears it of its gas content,
while ``thermal'' winds generated self-consistently by MUPPI do allow
sustained star formation, if at a low rate, till the end of the simulation.\\
In Fig. 5.28 we show the multi-phase gas particles
density profiles at the same times already shown in previous cases,
for the FB03 model. 
\subsubsection{CFMW}

\begin{figure}[!]
\centering{
\includegraphics[width=0.5\linewidth]{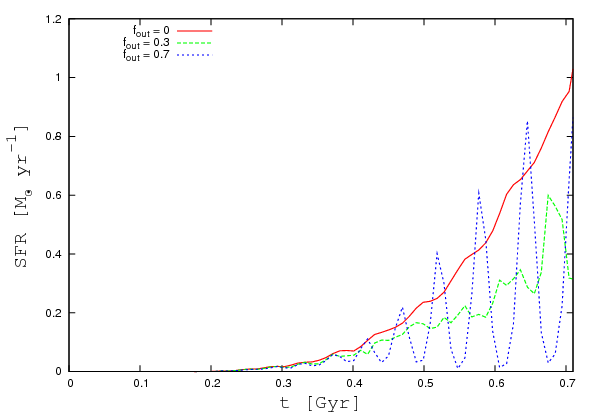}
\captionsetup{font={normalsize,sf}, width= 0.9\textwidth}
\caption{Star formation rate as a function of time for the \texttt{CFMW}
  model simulated with MUPPI, varying 
  $f_{\rm fb,o}$ = 0.0; 0.3;0.7 and, for comparison,
  with the GADGET effective model without winds (EFF) and with 
  winds (EFF+wind). See text for details.}} 
\label{FIG:sfr_cfmw_sfr}
\end{figure}  The general behaviour for the different gas components (i.e. cold,
molecular, hot) is that already  found in the ``galaxy'' models, with
cold and molecular gas growing in 
the halo core where a high pressure is reached and hot phase
increasing due to the SN feedback injection. The fraction of
  molecular gas, in particular, resembles what we found for the MW
  bulge and it is always of order unity up to the edge of the
  star-forming, multi-phase region. Note how
all gas densities grow between 0.25 Gyr and 1 Gyr: in
this interval of time, a strong cooling flow is established which
feeds the cold and molecular phase, thus igniting a burst of star
formation which pressurises the whole region. At 3 Gyr, most of the
gas previously collapsed in the  galaxy core has been already converted into stars.
Since then on, the density profiles of the various gas phases
  stay approximately constant, in a self-regulated fashion. 
Also, comparing density profiles of hot and cold gas phases with those
obtained in the galaxy runs, we can see that a larger fraction of gas
is in the hot phase, similarly to what happens in the bulge of our MW
run, while such fraction is much lower in MW disk and in the DW run.
This is due to the fact that, in this configuration, external pressure
of infalling gas forbids a larger fraction of hot, multi-phase
particle to escape the star 
forming region via buoyancy. More SNe energy thus remains in the zone,
 and heats a larger amount of gas away from the cold phase. At the same
time, each multi-phase gas particle has a higher amount of hot gas and
less gas in the cold phase; this is the reason why the trend of SFR
with $f_{\rm fb,o}$ is reversed.  
\begin{figure}[!]
\centering{
\includegraphics[width=0.4\linewidth]{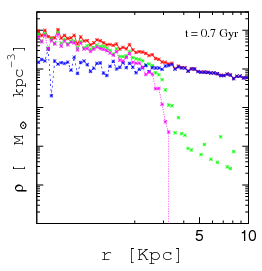}
\captionsetup{font={normalsize,sf}, width= 0.9\textwidth}
\caption{\texttt{CFMW} density profiles for ($\rho_c$), molecular
  ($\rho_{mol}$), hot ($\rho_h$) and total gas density ($\rho_{gas}$).}}
\label{FIG:dens_cfmw_dens}
\end{figure}

 In Fig. 5.29 we can further appreciate the
  efficiency of thermal feedback in MUPPI for the CFDW case and its
  dependency on the amount of energy transferred outside the 
  multi-phase gas particles to their neighbours. Here we show the
  total gas density profiles for FB00, FB03, FB07 and EFF+wind  runs at
  four different times. FB00 quickly consumes a large amount of gas,
  its density profile decreases and gets steeper. The larger the
  energy given to neighbouring particles, the smaller is the gas
  consumption, and also the slope of the profile is shallower.

Due to the high cost in computational time, the simulations runs
of the $\texttt{CFMW}$ halo has been evolved just till $0.7$
Gyr.\\
In Fig. 5.30 we present the SFRs obtained by varying
the SNe energy blowing out efficiency: the general behaviour is
similar to that already found for the CFDW case. The FB07
SFR is intermittent and spiky, because many gas particles
reach the conditions for entering the multi-phase star formation
regime at the same time and in a high pressure environment, and thus a
huge fraction of SN energy 
feedback is injected in the medium simultaneously, depressing the
SFR. When this heated gas cools and condenses, fulfils again the
thresholds and the cycle is repeated. We already found a similar behaviour
in the simulations ran with an implementation of the Stinson et
al. 2006 star formation and feedback scheme, which uses the Thacker
$\&$ Couchman (2000) feedback (see Sec.~\ref{sfr_text}); the 
difference is that MUPPI is able to ``stop'' radiative cooling
self-consistently, \textit{without} any ad-hoc artificial assumptions
as instead the Stinson 2006 recipe does.
However, this spiky SFR behaviour is almost not present when lower
fraction of energy are ejected to neighbouring particle.

\begin{figure}[!]
\centering{
\includegraphics[width=0.7\linewidth]{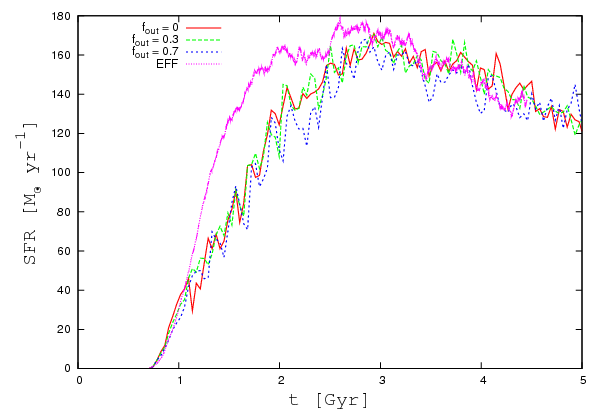}
\captionsetup{font={normalsize,sf}, width= 0.9\textwidth}
\caption{Star formation rate as a function of time for the \texttt{Me13}
  model simulated with MUPPI, varying $f_{\rm fb,o}$ = 0.0; 0.3;0.7, and, for comparison,
  with the GADGET effective model without winds (EFF).}} 
\label{FIG:sfr_cfme13}
\end{figure}

As in Fig. 5.28, in Fig. 5.31 we
show the density profiles for the different gas components considered
in MUPPI. We show here the latest time of out CFWM run only. Here, the
star forming zone, where the cold gas phase is dominant and the
molecular fraction is high, is confined to the very inner regions of
the halo, near to the softening length. 
But the pressure exerted by infalling, cooling gas is large
enough to trap almost all the multi-phase particles heated by the SNe
energy feedback. As a consequence, in the centre of the halo a large
fraction of the gas is in the hot phase. Such hot gas exerts enough
pressure to counter that of the infalling gas: it is again an example
of the efficiency of thermal feedback we obtain with MUPPI as far as
this mass scales are concerned.

We also run our Me13 cooling-flow halo (See Cap. 2) using MUPPI with
our standard parameter set and varying $f_{\rm fb,o}$. 
In Fig. 5.32 we show the star formation history for
cases FB00, FB03, FB07 and for the EFF model already show in Cap. 2.
In this case, the effect of increasing the SNe energy fraction transferred
to neighbouring particle is not strong. We can appreciate a decrease
in the star formation rate at its onset, between 1 and 3 Gyr, with
respect to the EFF model case. However, gas density profiles (not
shown) are similar for all of such four runs. At this mass scales, SNe
energy feedback is not very effective in countering the cooling flow
even when MUPPI is used. The general ISM properties we obtain in the Me13 
case are similar to those shown for the CFMW one.

\clearpage

\section{Parameter tests}
\label{RES:paramtest}
\begin{figure}[!]
\centering
\includegraphics[width=0.5\linewidth]{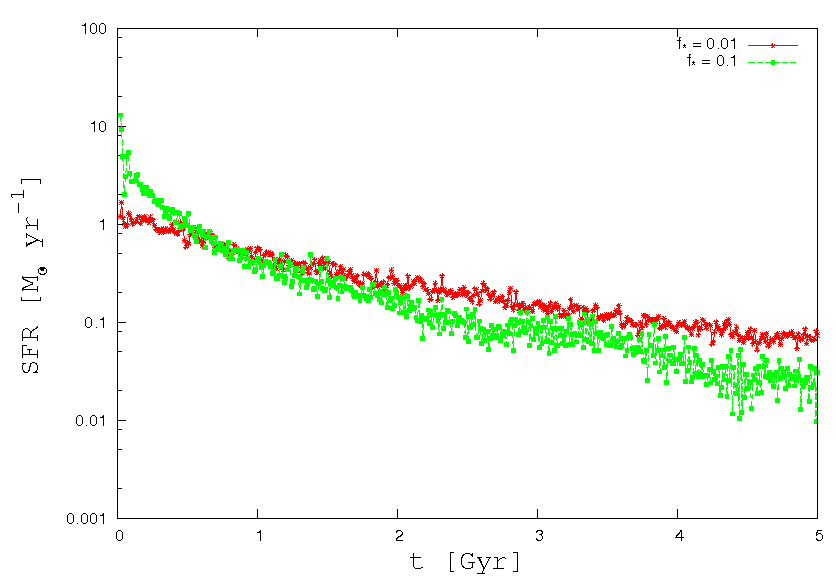}
\captionsetup{font={normalsize,sf}, width= 0.9\textwidth}
\caption{Star formation rate as a function of time for the \texttt{MW}
  model with our reference star formation efficiency $f_{\star}$ = 0.01
  and with $f_{\star}$ = 0.1. See text for details.}
\label{FIG:sfr_mw_fs}
\end{figure}

In order to assess the response of the $\texttt{MW}$ and $\texttt{DW}$
galaxies to the adopted MUPPI parameters we ran a set of
simulations varying three of the ten free parameters itemised in
Tab.~\ref{RES:param}, i.e. $f_{\star}$, $f_{\rm fb,i}$, $n_{thr}$.
We already studied the effect of having different values for 
 $f_{\rm fb,i}$. We set the remaining parameters on the basis of the work
of M04, on which our model is based. 
\\

\begin{figure*}[!]
\centering{
\includegraphics[width=0.6\linewidth]{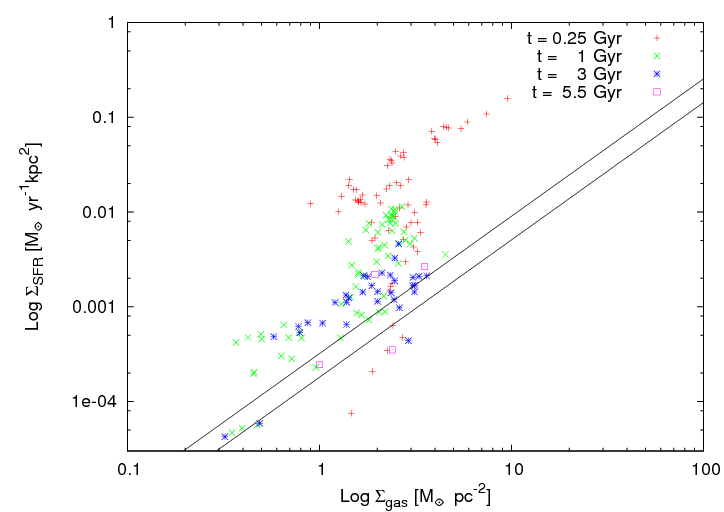}
\captionsetup{font={normalsize,sf}, width= 0.9\textwidth}
\caption{Star formation rate density as a function of gas surface
  density for the \texttt{MW} model with $f_{\star}$ = 0.1. 
The solid lines mark the 
  Schmidt-Kennicut law (1998). }}
\label{FIG:kenni_mw_fs}
\end{figure*}

When we vary one of such parameters, we keep constant all the
  remaining ones to their {\it standard} values.
We first discuss the additional tests done on the $\texttt{MW}$
galaxy. The effect of increasing the star formation efficiency
$f_{\star}$ by one order of magnitude (from 0.01 to 0.1) is as
expected. In Fig.~\ref{FIG:sfr_mw_fs} we compare the SFRs obtained by
varying $f_{\star}$: at the onset of the simulation, the
$f_{\star}$0.1 run is more efficient in forming stars, but at final
times, the SF efficiency decreases due to greater consumption of gas
if compared to the reference $f_{\star}$ run (i.e. 0.01). In
Fig.~\ref{FIG:kenni_mw_fs} we moreover show the density-SFR relation
in the $f_{\star}$0.1 case: the star formation is much more efficient
than is observed after 0.25 Gyr since the onset of the simulation;
thus the gas supply is consumed rapidly and the density-SFR relation
is no more able to satisfy the Schmidt-Kennicut law, forming too many
stars at any given surface gas density. Thus, for MUPPI low values of
this parameters have to be preferred.\\
The last test we performed on the $\texttt{MW}$ galaxy deals with the
SN feedback energy. We tested the effect of increasing the fraction
$f_{\rm fb,i}$ of SN energy which remains trapped inside the star forming
gas particle rather than being ejected outside the gas particle. {
Note that the the two parameters  $f_{\rm fb,i}$,  $f_{\rm fb,o}$ are not
related, the only requirement being that their sum must be lesser than
unity. We consider the SNe energy not accounted for by the sum of  $f_{\rm fb,i}$ and
$f_{\rm fb,o}$ to be radiated away and lost as far as the ISM is
concerned. 
In Fig.~\ref{FIG:kenni_mw_fi} we show the density-SFR relation obtained
from a simulation where we set $f_{\rm fb,i}= 0.05$ (reference value is
0.02). The plot shows that we don't reproduce the Schmidt-Kennicut law
as well as in the reference simulation (see
Fig.~\ref{FIG:kenni_mw}). Our SFR at any given cold gas surface density is
now lower than observed.
Since the SF history has no appreciable differences from the
case  $f_{\rm fb,i}=0.02$, we deduce that coupling more SNe energy to
the ISM inside the particle pressurises it more, giving a higher
hot gas phase fraction and a smaller SFR.
We thus prefer our reference value  $f_{\rm fb,i}= 0.05$.

Notice that increasing the star formation efficiency parameter and increasing
the amount of energy coupled to the particle ISM modify our Kennicut
relation in opposite directions. However, this does not obviously
translate in the possibility to vary such two parameters in opposite
direction while keeping the star formation history and the resulting
Kennicut relation unchanged. E.g., a test run with $f_{\rm fb,i}= 0.05$
and  $f_{\star}=0.05$ could {\it not} give a Kennicut relation in
agreement with observations. This is due to the non-linear behaviour
of our model, which couples the effect of varying the parameters in a
non-straightforward way.

\begin{figure*}[!]
\centering{
\includegraphics[width=0.6\linewidth]{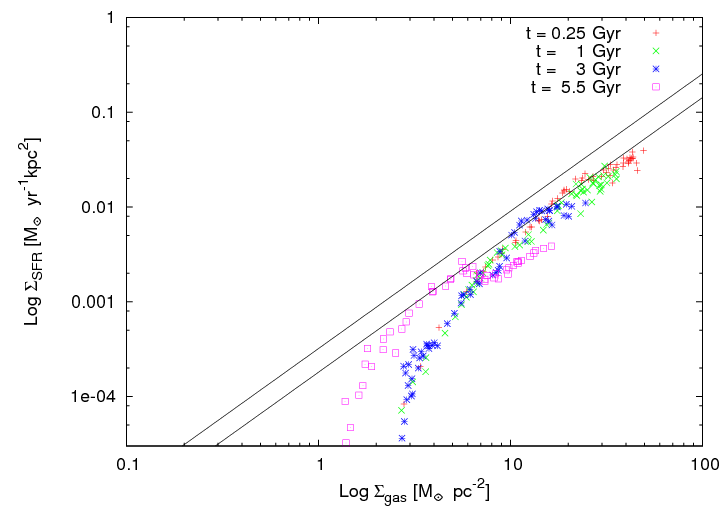}
}
\captionsetup{font={normalsize,sf}, width= 0.9\textwidth}
\caption{Star formation rate density as a function of gas surface
  density for the \texttt{MW} model with $f_{\rm fb,i}$ = 0.05. 
The solid lines mark the 
  Schmidt-Kennicut law (1998). }
\label{FIG:kenni_mw_fi}
\end{figure*}
We now discuss an additional test done on the $\texttt{DW}$
galaxy. An important factor in star formation models is the
density threshold at which star formation starts. In MUPPI, such a threshold 
regulate the initial particle pressure, and the degree of activity of
the ISM and the duration of the multi-phase stage. In fact, the lower
the initial pressure is, the more ``quiet'' the behaviour of the ISM
and the longer the initial cold phase dynamical time are.
We thus
ran two additional simulations decreasing the number density threshold
$n_{thr}$ from 0.25 cm$^{-3}$ to 0.1 and 0.05 cm$^{-3}$. 

\begin{figure}[!]
\centering
\includegraphics[width=0.5\linewidth]{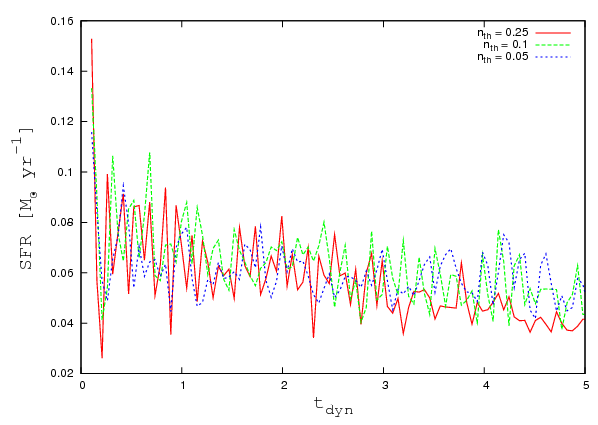}
\captionsetup{font={normalsize,sf}, width= 0.9\textwidth}
\caption{Star formation rate as a function of time for the \texttt{DW}
  model with varying threshold number density for entering the
  multi-phase state.  We resample the SFRs with a constant time interval
  equal to $\sim$ 0.03 t$_dyn$. See text for details.}
\label{FIG:sfr_dw_nth}
\end{figure}
In Fig.~\ref{FIG:sfr_dw_nth} we show the resulting SFR histories: the reference
SFR (with $n_{thr}$ from 0.25 cm$^{-3}$) is initially the
highest. This is simply explained by considering that increasing the
density threshold, the central star forming region is smaller and thus
need more time to accrete gas from the surroundings. After such a
  transient period, the SFR is very similar for the three different
  thresholds; in this regard, MUPPI shows to be quite insensitive to
  it.
Fig.~\ref{FIG:dens_dw_nth} shows the total gas surface density for our
three different thresholds at four different times. Even the surface density
profiles does not substantially vary when the density threshold for
the onset of multi-phase regime is changed. 

As we already show in Sec.~\ref{RES:dwarf_g}, the cold gas density in the
$\texttt{DW}$ case is never high enough to sample the power-law part
of the Schmidt-Kennicut law, and also with our standard parameter set,
we only reproduce its low-density decline. However, we show in 
Fig.~\ref{FIG:kenni_dw_nthre} surface density-SFR relation we obtain
when varying the density 
threshold. A slight trend to get a lower SFR at given density for
lower threshold values can be seen. This is related with the slower
``metabolism'' of multi-phase particles when their initial pressure
is lower. The trend is not clear enough to give clear indication on
the best value for this parameter; we selected as our default the one
most used in the literature, and also selected by the SH03 star
formation scheme.

\begin{figure*}[!]
\centering{
\includegraphics[width=0.42\linewidth]{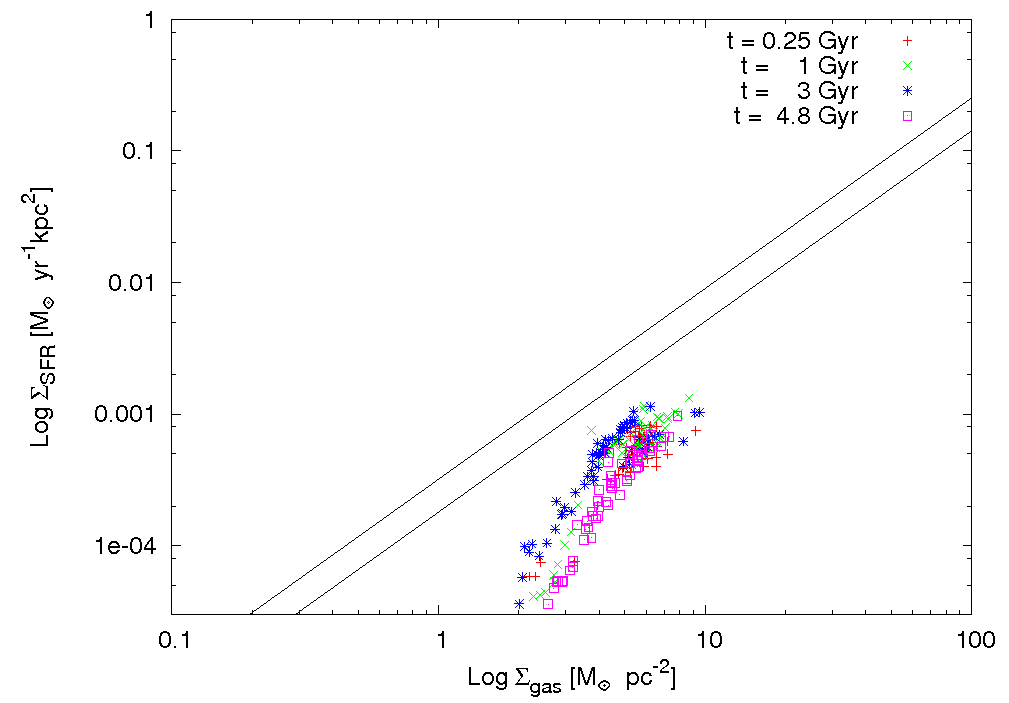}
\includegraphics[width=0.42\linewidth]{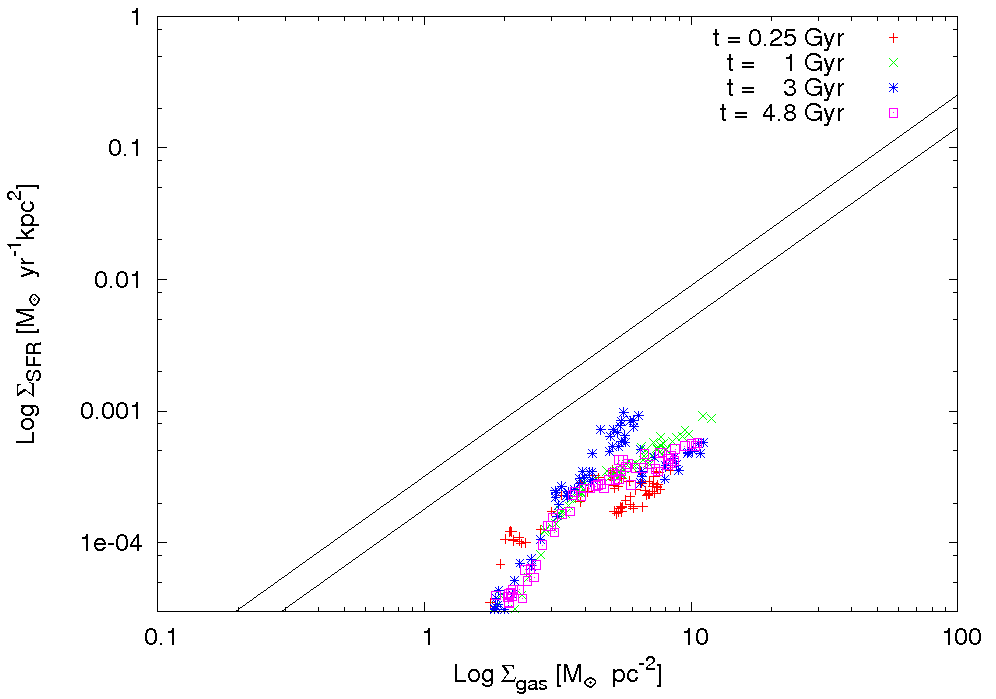}
}
\captionsetup{font={normalsize,sf}, width= 0.9\textwidth}
\caption{Star formation rate density as a function of gas surface
  density for the \texttt{DW} model with $n_{thr}$ = 0.1
  cm$^{-3}$ (left panel) and $n_{thr}$ = 0.05
  cm$^{-3}$ (right panel). The solid lines mark the 
  Schmidt-Kennicut law (1998). }
\label{FIG:kenni_dw_nthre}
\end{figure*}

\begin{figure}[!]
\centering
\includegraphics[width=0.7\linewidth]{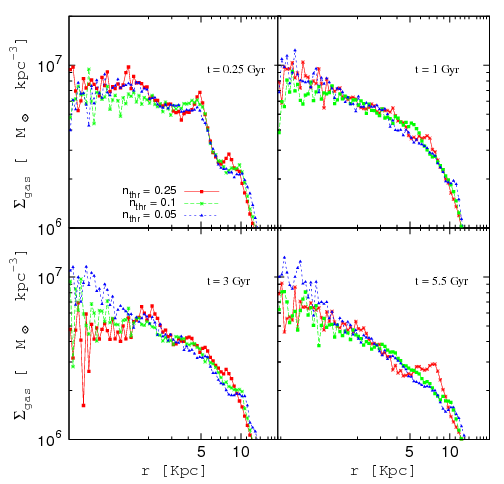}
\captionsetup{font={normalsize,sf}, width= 0.9\textwidth}
\caption{Surface density profiles for the \texttt{MW} model with different
  number density threshold $n_{thr}$, for stars ($\rho_{\star}$) and
  different gas phases, i.e. cold ($\rho_c$), molecular ($\rho_{mol}$), hot
  ($\rho_h$)and total gas density ($\rho_{gas}$)}
\label{FIG:dens_dw_nth}
\end{figure}

\clearpage

\section{Numerical tests}
\label{RES:numtest}
\begin{figure}[!]
\centering{
\includegraphics[width=0.7\linewidth]{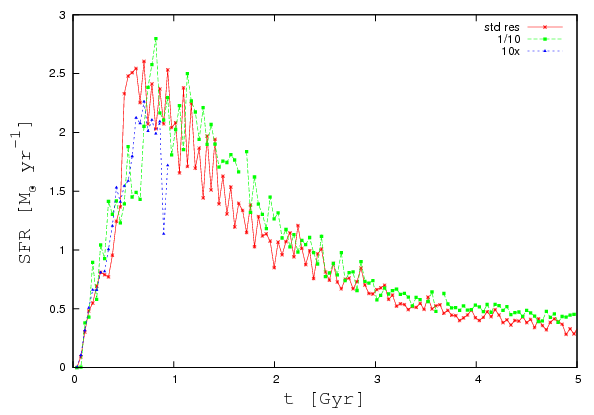}
\captionsetup{font={normalsize,sf}, width= 0.9\textwidth}
\caption{Star formation rate as a function of time for the
  \texttt{CFDW} model at standard resolution (std), low resolution
  (1/10) and high resolution (10x), using the reference set of MUPPI
  parameters. We resampled the SFRs with a constant time interval
  equal to $\sim$ 0.01 t$_dyn$.}}  
\label{FIG:sfr_cfdw_resdens}
\end{figure}
Finally, it is important to test how much the results presented in this Chapter
depend on the numerical resolution of the simulations.
To this aim, we carried out one simulation with the $\texttt{MW}$
galaxy decreasing the number of gas and dark matter particles by a
factor of ten (LR). Moreover, we
performed two simulations of the $\texttt{CFDW}$ halo in HR and LR. 
We run all the resolution tests with our reference set of MUPPI
parameters (see Tab.~\ref{RES:param}).\\
\begin{figure}[!]
\centering{
\includegraphics[width=0.7\linewidth]{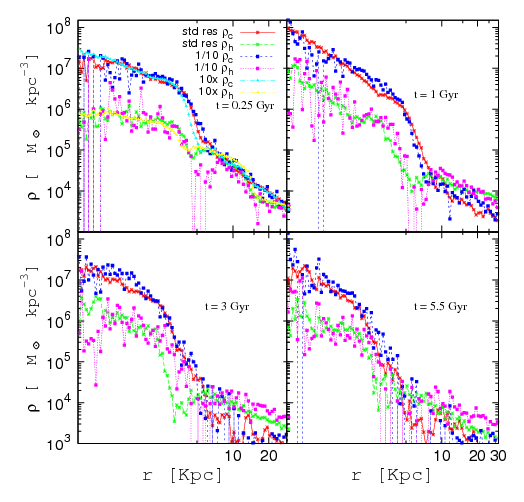}
\captionsetup{font={normalsize,sf}, width= 0.9\textwidth}
\caption{\texttt{CFDW} density profiles for the cold and the hot
  phases at standard  resolution (std res), low resolution (1/10) and
  high resolution (10x).}}  
\label{FIG:dens_cfdw_resdens}
\end{figure}
In Fig. 5.39 we show the evolution of the SFR
obtained with 
standard, low and high resolution for the  $\texttt{CFDW}$ halo. We evolved the HR run
just till 1 Gyr due to the high cost in computational time. From this
figure we can see that resulting SFR history does not show significant
differences when the resolution is varied of a factor 100.

\begin{figure}[!]
\centering{
\includegraphics[width=0.7\linewidth]{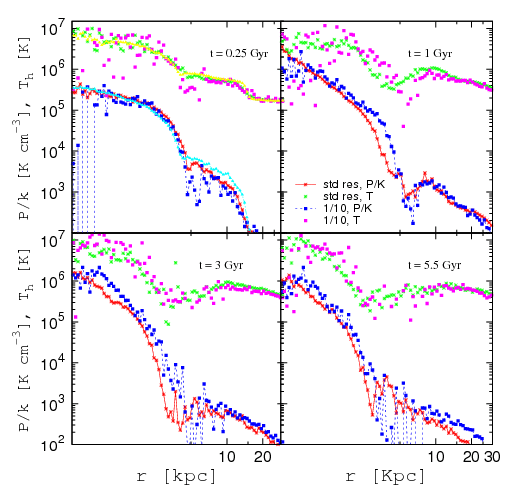}
\captionsetup{font={normalsize,sf}, width= 0.9\textwidth}
\caption{\texttt{CFDW} pressure and hot temperature profiles 
  at standard  resolution (std res), low resolution (1/10) and
  high resolution (10x).} }  
\label{FIG:thermo_cfdw_res}
\end{figure}
To further probe the stability of our ISM 
model with varying resolution, we plot in
Fig. 5.40 the radial density profiles of both
the hot and  cold phase. The three different resolution simulations
once again behave in a very similar way. The only appreciable
difference is that scatter in the profiles reduces with
increasing resolution: the HR simulations are in fact described with a
greater number of particles and thus have a much larger covering
factor. This trend is confirmed in Fig.~\ref{FIG:thermo_cfdw_res},
where we plot pressure and temperature radial profiles: results at
different resolutions are well convergent, thus confirming the
numerical stability of our code when we change resolution by two order
of magnitude for this halo. \\

\begin{figure}[!]
\centering
\includegraphics[width=0.5\linewidth]{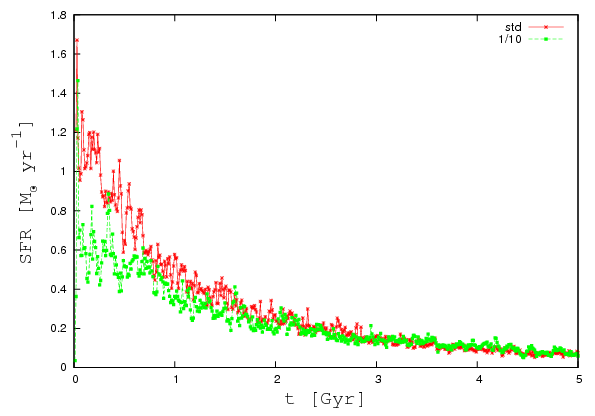}
\captionsetup{font={normalsize,sf}, width= 0.9\textwidth}
\caption{Star formation rate as a function of time for the \texttt{MW}
  model at standard resolution (res)
  and at low resolution (1/10), using the reference set of MUPPI
  parameters. See text for details.}
\label{FIG:sfr_mw_res_sfr}
\end{figure}

For what concerns the $\texttt{MW}$ galaxy, we present in
Fig.~\ref{FIG:sfr_mw_res_sfr} a comparison of the SFRs obtained at
standard ($\rm std$) and low resolution ($\rm 1/10$). This numerical
resolution test is quite different from the CFDW one. In the latter,
gas is initially in hydrostatic equilibrium with the DM gravitational
potential; in the core, temperature, density and pressure initial
profiles are quite flat. Thus, varying the resolution influences how the
physical quantities are sampled, but does not change their average
value per particle. In the MW case, instead, we have an
exponential surface density profile. Thus, under-sampling the gas
particle distribution also means assigning {\it lower} density to low
resolution gas particles at the same distance from the disk centre of
the corresponding particles in high resolution case. 
\begin{figure}[!]
\centering
\includegraphics[width=0.7\linewidth]{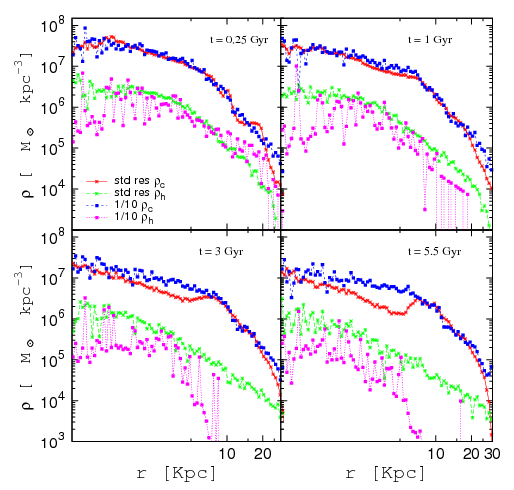}
\captionsetup{font={normalsize,sf}, width= 0.9\textwidth}
\caption{\texttt{MW} density profiles for the cold and the hot
  phases at standard  resolution (std res) and low resolution (1/10).}
\label{FIG:dens_mw_res}
\end{figure}

At the onset of the simulation, the star formation activity is
stronger in the $\rm std$ run, since a larger amount of gas cools here
than in the lower resolution run. This implies that more gas particles
fulfils the multi-phase regime thresholds and as a consequence the
star formation efficiency is increased.  This effect is given by the
fact that gas particles in our standard resolution run reach higher
densities, and cooling can begin before; it is not related to the star
formation model.  After less than 1 Gyr, the two star formation
histories do converge to similar values and becomes identical after
$\approx 2$ Gyr.

In Fig,~\ref{FIG:dens_mw_res} we show the cold and hot gas surface density profiles
for our standard and LR runs. While the cold phase is extremely stable
against varying the numerical resolution, the hot phase surface density
show a clear decrease when resolution is lowered. A similar effect is
clear in Fig.~\ref{FIG:thermo_mw_res}, where the temperature of the
hot phase proves stable against resolution but $P/K$ decreases when
resolution is lowered.
\begin{figure}[!]
\centering
\includegraphics[width=0.7\linewidth]{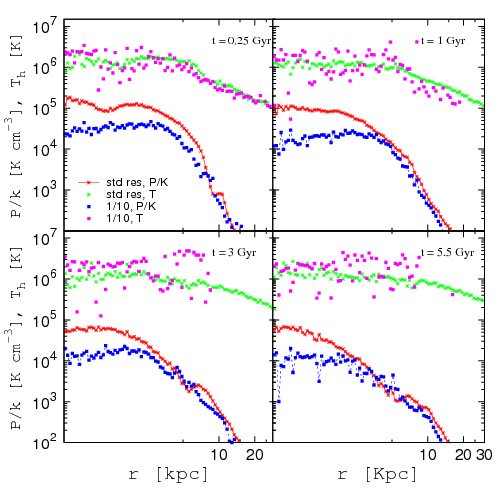}
\captionsetup{font={normalsize,sf}, width= 0.9\textwidth}
\caption{\texttt{MW} pressure and hot temperature profiles 
  at standard  resolution (std res) and at low resolution (1/10).}
\label{FIG:thermo_mw_res}
\end{figure}
This is due to the fact that particle sampling of the gas distribution
in the LR run is poor; in presence of an exponentially declining
profile, less gas particles match our density threshold for entering
the multi-phase regime. Quantities related to multi-phase regime, as the
hot gas density and pressure (which is weighted on the {\it total} gas
mass, but generated mainly by our hot gas phase}, therefore show a
decrease. The cold gas density and the hot temperature (which is
mass-weighted over the {\it hot} mass) are instead stable.
Therefore, MUPPI is still producing ISM properties which are stable
against the numerical resolution. Care must however be taken,
obviously, to resolution effect that doesn't came out of the model but
directly out of the SPH treatment of the hydrodynamics.
We however verified that, after $5.5$ Gyr of evolution, the total gas
surface density profile in our LR and standard
similar and the surface density profile of the formed stellar
component are remarkably similar.

Overall, the model we present proves to be very stable against
variation in mass and force resolution.

\clearpage
\section{Conclusions}
\label{RES:concl}
In this section we presented and discussed the results obtained by
evolving MUPPI (described in Chapter 4), our new sub-grid model
of the ISM, using our reference set of parameters, in various
physical conditions, i.e. a model of the Milky-Way, a model of
a typical dwarf galaxy and two isolated halos, one chacteristic of the
Milky-Way and one of a dwarf-like galaxy.

The main feature of our model is that it solves, for each
multi-phase gas particle, a system of differential equations
aimed to modelling the main physics of the ISM. It 
does not use equilibrium
solutions for calculating ISM properties, i.e. they depend on the
interplay between hot and cold gas phases, star formation and SNe
energy feedback, and local thermodynamical properties, as given by the
SPH code. 
This means that, besides following the
non-equilibrium phase at the onset of the multi-phase regime, MUPPI is
able to respond to changes in the local thermodynamics such as
pressure/temperature changes due to compressions/rarefactions and
shocks, as we showed along this Chapter. 

The simulation of the Milky-Way galaxy shows good agreement with the
observed Schmidt-Kennicut law (1998) and is thus able to lead to a
self-regulated cycle of star formation, where mass flows in the cold
phase and star formation are efficiently counterbalanced by the SNe
feedback effects. We showed the evolution of single gas particles lying in
different positions in the galaxy and verified how MUPPI efficiently
follows the diverse evolution of their physical properties. The
characteristics of the ISM we found for a particle lying at about solar
distance are in reasonable agreement with observations.
The gas disk develops long lasting, sharp spiral arms 
and part of the gas is expelled from the disk in the vertical direction, generating galactic
fountains without using any ad-hoc kinetic feedback prescriptions. 
These results arise as a natural consequence of 
 the ISM physics implemented in MUPPI. 

Simulation of the dwarf galaxy reproduce the main physical properties
expected from a quiet and lighter galaxy, with a less active ISM.
 Compared to the Milky-Way in fact,
the star formation rate is lower, being the dwarf galaxy less massive
and dense and thus less pressurised. In this case, we showed that we
don't have enough pressure to drive a ``thermal wind'' outside the
galaxy plane. 

We moreover verified that MUPPI works well in the very different
physical conditions found at the centre of cooling flows. 
In both Milky-Way and dwarf like halos, MUPPI is able to counter the
cooling flow by efficiently distributing the fraction of SNe energy in
the blow out regime to neighbours: the higher the fraction of blowing
out energy, the more the star formation is suppressed. 

In order to assess how MUPPI respond to the choice of its parameters, we
ran a set of simulations varying them in the galaxy models. 
Finally we tested our code against numerical resolution by simulating the
Milky-way galaxy with ten times less gas and dark matter particles, as
well as the dwarf-like halo with ten times more and ten times less gas and dark
matter particles. We found our results are numerically stable since
the general properties of the ISM does not significantly
change when we vary resolution.  

 The model we presented here will be particularly useful in
cosmological simulations of formation and evolution of isolated
galaxies and galaxy clusters. In fact, it is able to capture the
main ISM properties and produce an effective SNe thermal feedback,
without resorting on the extreme numerical resolution needed to
directly simulate the ISM. 

The first application of the present Ph.D. work will therefore be to
apply MUPPI to cosmological simulations, with the aim of determine how
an improved treatment of star formation and feedback astrophysical
processes impacts on many open issues, from the properties of simulated
disk galaxies to the properties of cold baryons (galaxies and diffuse
stellar component) in galaxy clusters, to the properties of the
Intra-Cluster Medium in presence of an effective SNe thermal
feedback.

\addcontentsline{toc}{chapter}{CONCLUSIONS}
\chapter*{Conclusions}
\vspace*{1cm}
\textsf{
If we aim in comparing galaxy formation simulations with observations, the
collisionless dynamics of the dark matter particles must be coupled to gas
dynamics and small-scale astrophysical processes.  White \& Rees in 1978
reported ``\textit{An important issue in theories of galaxy formation is the
relative importance of purely gravitational processes (as N-Body effects,
clustering, etc..) and of gas-dynamical effects involving dissipation and
radiative cooling}``: the significant point they revealed 30 years ago in
hierachical simulations is still an issue.}

\textsf{In Chapter II, we presented a comparison of
various star formation and SNe feedback 
prescriptions in two NFW isolated, non-rotating Dark Matter halos having
mass typical of a galaxy and of a poor cluster of galaxies, 
 using the TREE+SPH code GADGET-2. The aim of this work thus was to
 study the behaviour of different star formation and feedback models,
 previously tested on disk galaxies, in very different physical
 conditions such those found at the centre of cooling flows.\\
We tested the GADGET effective star formation model (EFF), which
is based on a multi-phase description of the gas contained in a particle, an
implementation of the simple scheme proposed by Katz et al. 1996 (SSF),
 where feedback energy is given to star-forming particles in the form
 of thermal energy, and an implementation of Thacker \& Couchman 2000
 with the improvements of Stinson et al. 2006 (DEL), which consists in
 turning off radiative cooling for a fixed period of time of 30 Myr when the SN
energy is deposited, thus mimiking the effect of SNe super-bubbles on
the ISM.\\
Our main conclusion is that, while SN feedback in the EFF and SSF models are
 not efficient in countering the cooling flow and gas radiative losses
 at the halo centre, the DEL scheme proves quite effective at
 doing so. The cost of it is an unrealistic delay in the star
 formation history. While star formation and 
feedback schemes which turn off radiative cooling have proved effective
in producing realistic disk galaxies in cosmological simulations, caution
should therefore being used when utilising similar schemes as general-purpose
ones, e.g. in galaxy cluster simulations.}

\textsf{In Chapter III, we have studied the origin of the diffuse
stellar component in 
galaxy clusters taken from a cosmological hydrodynamical simulation. We
found that the formation of the diffuse stellar component has no preferred
redshift and is a cumulative power--law process up to redshift $z=0$. Moreover
we found no correlation between the final amount of stars in the diffuse
component and the global dynamical hystory of the clusters. For all but the
three most massive clusters, the formation of the diffuse component has been
found to go largely in parallel with the build-up of the brighest cluster
galaxy. The most important result we found in this analysis is that most of
the diffuse star particles become unbound during merging phases (and
\text{not} by tidal stripping) along the formation history of the brighest
cluster galaxy, independent of cluster mass. Such work has been performed
using the standard Springel \& Hernquist (2003) prescription for star
formation and SNe energy feedback. An open point remains, if a more realistic
treatment of the ISM physics (and thus of star formation and SNe energy
feedback), leading for instance to the build-up of {\it disk} galaxies inside
simulated clusters at resolutions that simulation of this kind can routinely
reach, would change our conclusions.}

\textsf{This is only an example of how important in many astrophysical
problems a 
proper treatment of the ISM physics in cosmological simulations may be.
The main aim of this Ph.D. Thesis work has been introducing a sofisticated
model for following the complex astrophysical processes acting in the
interstellar medium in numerical simulations involving galaxy formations.}

\textsf{We fullly describe our model, called MUPPI, $\boldsymbol{MU}$lti$\boldsymbol{P}$hase
$\boldsymbol{P}$article $\boldsymbol{I}$ntegrator, in
Chapter IV. MUPPI follows the ISM physics
using a system of ordinary differential equations, describing mass and energy
flows among the different gas phases in the ISM inside each gas particle. The
model also includes the treatment of SNe energy transfer from star-forming
particles to their neighbours.}

\textsf{Along Chapter V, we presented and discussed the results
obtained evolving 
various physical cases with MUPPI. We simulated an isolated  model of the Milky-Way, a
model of a typical dwarf galaxy and two isolated cooling-flows halos without
galaxies, one having the same chacteristic of the Milky-Way halo and another
of a dwarf-like galaxy.  The overall result we found in these simulations is
that MUPPI is successfull in properly following the inter stellar medium
evolution in very different physical situations. In particular, the simulation
of the Milky-Way galaxy shows good agreement with the observed
Schmidt-Kennicut law (1998) and is thus able to lead to a self-regulated cycle
of star formation, where mass flows in the cold phase and star formation are
efficiently counterbalanced by the supernova feedback effects. The
characteristics of the interstellar medium we found for a particle lying at
about solar distance are in reasonable agreement with observations. Moreover,
heated gas is expelled from the disk in the vertical direction and then falls
back, generates galactic fountains.  These results arise as a natural
consequence of the interstellar medium physics as implemented in MUPPI.
Thermal feedback is thus very effective in our implementation, a result that
other sub-grid star formation recipes don't get, or obtain only with ad-hoc
parametrization (as e.g. in Stinson et al 2006 work) or resorting to
ad-hoc kinetic feedback (e.g. Springel \& Hernquist 2003).}

\textsf{In order to test the MUPPI code behaviour on different initial physical
conditions, we runned a simulation of the dwarf galaxy and we found that
results reproduce the main physical properties expected from a quiet and less
massive galaxy, with a less active intestellar medium.\\
Moreover MUPPI works well even in the very different physical conditions found
at the centre of isolated halos, where our code has been found to be able to
efficiently counter the cooling flow.\\
In order to assess how MUPPI respond to the choice of its parameters, we
ran a set of simulations varying them in the galaxy models. Finally we
tested our code against numerical resolution by simulating the Milky-way
galaxy with a ten times worse mass resolution, as well as the dwarf-like halo
with ten times more and ten times less gas and dark matter particles. We found
our results are numerically stable: the general properties of the simulated
interstellar medium does not significantly change when we vary resolution.
This is a particularly important point, as our model is intended for use in
simulations where an extreme mass and force resolution can't be reached given
the present-day available computing power.}

\textsf{We believe the model we presented here will be particularly useful in
cosmological simulations of formation and evolution of isolated galaxies and
galaxy clusters. For this reason, the first application of the present
Ph.D. work will therefore be to apply MUPPI to cosmological simulations, with
the aim of determine how an improved treatment of star formation and feedback
astrophysical processes impacts on many open issues, from the properties of
simulated disk galaxies to the properties of cold baryons (galaxies and
diffuse stellar component) in galaxy clusters, to the properties of the
Intra-Cluster Medium in presence of an effective supernovae thermal feedback.
}

\bibliographystyle{acm}
\bibliography{master_phd}

\begin{thebibliography}{100}

\bibitem{Abadi06}
{\sc {Abadi}, M.~G., {Navarro}, J.~F., and {Steinmetz}, M.}
\newblock {Stars beyond galaxies: the origin of extended luminous haloes around
  galaxies}.
\newblock {\em \mnras 365\/} (Jan. 2006), 747--758.

\bibitem{1958ApJS....3..211A}
{\sc {Abell}, G.~O.}
\newblock {The Distribution of Rich Clusters of Galaxies.}
\newblock {\em \apjs 3\/} (May 1958), 211--+.

\bibitem{1965ARA&A...3....1A}
{\sc {Abell}, G.~O.}
\newblock {Clustering of Galaxies}.
\newblock {\em \araa 3\/} (1965), 1--+.

\bibitem{1989ApJS...70....1A}
{\sc {Abell}, G.~O., {Corwin}, H.~G., and {Olowin}, R.~P.}
\newblock {A catalog of rich clusters of galaxies}.
\newblock {\em \apjs 70\/} (May 1989), 1--138.

\bibitem{Aguerri2006}
{\sc {Aguerri}, J.~A.~L., {Castro-Rodr{\'{\i}}guez}, N., {Napolitano}, N.,
  {Arnaboldi}, M., and {Gerhard}, O.}
\newblock {Diffuse light in Hickson compact groups: the dynamically young
  system HCG 44}.
\newblock {\em \aap 457\/} (Oct. 2006), 771--778.

\bibitem{Aguerri2005}
{\sc {Aguerri}, J.~A.~L., {Gerhard}, O.~E., {Arnaboldi}, M., {Napolitano},
  N.~R., {Castro-Rodriguez}, N., and {Freeman}, K.~C.}
\newblock {Intracluster Stars in the Virgo Cluster Core}.
\newblock {\em \aj 129\/} (June 2005), 2585--2596.

\bibitem{2001MNRAS.328L..37A}
{\sc {Allen}, S.~W., {Schmidt}, R.~W., and {Fabian}, A.~C.}
\newblock {The X-ray virial relations for relaxed lensing clusters observed
  with Chandra}.
\newblock {\em \mnras 328\/} (Dec. 2001), L37--L41.

\bibitem{Magda02}
{\sc {Arnaboldi}, M., {Aguerri}, J.~A.~L., {Napolitano}, N.~R., {Gerhard}, O.,
  {Freeman}, K.~C., {Feldmeier}, J., {Capaccioli}, M., {Kudritzki}, R.~P., and
  {M{\'e}ndez}, R.~H.}
\newblock {Intracluster Planetary Nebulae in Virgo: Photometric Selection,
  Spectroscopic Validation, and Cluster Depth}.
\newblock {\em \aj 123\/} (Feb. 2002), 760--771.

\bibitem{Magda03}
{\sc {Arnaboldi}, M., {Freeman}, K.~C., {Okamura}, S., {Yasuda}, N., {Gerhard},
  O., {Napolitano}, N.~R., and {Pannella}, M.~E.}
\newblock {Narrowband Imaging in [O III] and H{$\alpha$} to Search for
  Intracluster Planetary Nebulae in the Virgo Cluster}.
\newblock {\em \aj 125\/} (Feb. 2003), 514--524.

\bibitem{Magda04}
{\sc {Arnaboldi}, M., {Gerhard}, O., {Aguerri}, J.~A.~L., {Freeman}, K.~C.,
  {Napolitano}, N.~R., {Okamura}, S., and {Yasuda}, N.}
\newblock {The Line-of-Sight Velocity Distributions of Intracluster Planetary
  Nebulae in the Virgo Cluster Core}.
\newblock {\em \apjl 614\/} (Oct. 2004), L33--L36.

\bibitem{1999MNRAS.307..463B}
{\sc {Balogh}, M.~L., {Babul}, A., and {Patton}, D.~R.}
\newblock {Pre-heated isentropic gas in groups of galaxies}.
\newblock {\em \mnras 307\/} (Aug. 1999), 463--479.

\bibitem{Balogh01}
{\sc {Balogh}, M.~L., {Pearce}, F.~R., {Bower}, R.~G., and {Kay}, S.~T.}
\newblock {Revisiting the cosmic cooling crisis}.
\newblock {\em \mnras 326\/} (Oct. 2001), 1228--1234.

\bibitem{1987ApJ...322..585B}
{\sc {Baron}, E., and {White}, S.~D.~M.}
\newblock {The appearance of primeval galaxies}.
\newblock {\em \apj 322\/} (Nov. 1987), 585--596.

\bibitem{begelman}
{\sc {Begelman}, M.~C.}
\newblock {AGN feedback mechanism}.
\newblock {\em ArXiv Astrophysics e-prints\/} (Mar. 2003).

\bibitem{1995AJ....110.1507B}
{\sc {Bernstein}, G.~M., {Nichol}, R.~C., {Tyson}, J.~A., {Ulmer}, M.~P., and
  {Wittman}, D.}
\newblock {The Luminosity Function of the Coma Cluster Core for $-25 < M/R <
  -9.4 $}.
\newblock {\em \aj 110\/} (Oct. 1995), 1507--+.

\bibitem{2001ApJ...555..597B}
{\sc {Bialek}, J.~J., {Evrard}, A.~E., and {Mohr}, J.~J.}
\newblock {Effects of Preheating on X-Ray Scaling Relations in Galaxy
  Clusters}.
\newblock {\em \apj 555\/} (July 2001), 597--612.

\bibitem{2002ApJ...566...93B}
{\sc {B{\"o}hringer}, H., {Collins}, C.~A., {Guzzo}, L., {Schuecker}, P.,
  {Voges}, W., {Neumann}, D.~M., {Schindler}, S., {Chincarini}, G., {De
  Grandi}, S., {Cruddace}, R.~G., {Edge}, A.~C., {Reiprich}, T.~H., and
  {Shaver}, P.}
\newblock {The ROSAT-ESO Flux-limited X-Ray (REFLEX) Galaxy Cluster Survey. IV.
  The X-Ray Luminosity Function}.
\newblock {\em \apj 566\/} (Feb. 2002), 93--102.

\bibitem{2007MNRAS.376.1588B}
{\sc {Booth}, C.~M., {Theuns}, T., and {Okamoto}, T.}
\newblock {Molecular cloud regulated star formation in galaxies}.
\newblock {\em \mnras 376\/} (Apr. 2007), 1588--1610.

\bibitem{2006astro.ph..5575B}
{\sc {Borgani}, S.}
\newblock {Cosmology with clusters of galaxies}.
\newblock {\em ArXiv Astrophysics e-prints\/} (May 2006).

\bibitem{BorgNum}
{\sc {Borgani}, S., {Dolag}, K., {Murante}, G., {Cheng}, L.-M., {Springel}, V.,
  {Diaferio}, A., {Moscardini}, L., {Tormen}, G., {Tornatore}, L., and {Tozzi},
  P.}
\newblock {Hot and cooled baryons in smoothed particle hydrodynamic simulations
  of galaxy clusters: physics and numerics}.
\newblock {\em \mnras 367\/} (Apr. 2006), 1641--1654.

\bibitem{2001ApJ...559L..71B}
{\sc {Borgani}, S., {Governato}, F., {Wadsley}, J., {Menci}, N., {Tozzi}, P.,
  {Lake}, G., {Quinn}, T., and {Stadel}, J.}
\newblock {Preheating the Intracluster Medium in High-Resolution Simulations:
  The Effect on the Gas Entropy}.
\newblock {\em \apjl 559\/} (Oct. 2001), L71--L74.

\bibitem{2001Natur.409...39B}
{\sc {Borgani}, S., and {Guzzo}, L.}
\newblock {X-ray clusters of galaxies as tracers of structure in the Universe}.
\newblock {\em \nat 409\/} (Jan. 2001), 39--45.

\bibitem{Borg}
{\sc {Borgani}, S., {Murante}, G., {Springel}, V., {Diaferio}, A., {Dolag}, K.,
  {Moscardini}, L., {Tormen}, G., {Tornatore}, L., and {Tozzi}, P.}
\newblock {X-ray properties of galaxy clusters and groups from a cosmological
  hydrodynamical simulation}.
\newblock {\em \mnras 348\/} (Mar. 2004), 1078--1096.

\bibitem{2001ApJ...561...13B}
{\sc {Borgani}, S., {Rosati}, P., {Tozzi}, P., {Stanford}, S.~A., {Eisenhardt},
  P.~R., {Lidman}, C., {Holden}, B., {Della Ceca}, R., {Norman}, C., and
  {Squires}, G.}
\newblock {Measuring $\Omega_m$ with the ROSAT Deep Cluster Survey}.
\newblock {\em \apj 561\/} (Nov. 2001), 13--21.

\bibitem{2001ApJ...553..103B}
{\sc {Brighenti}, F., and {Mathews}, W.~G.}
\newblock {Entropy Evolution in Galaxy Groups and Clusters: a Comparison of
  External and Internal Heating}.
\newblock {\em \apj 553\/} (May 2001), 103--120.

\bibitem{ByrdValt}
{\sc {Byrd}, G., and {Valtonen}, M.}
\newblock {Tidal generation of active spirals and S0 galaxies by rich
  clusters}.
\newblock {\em \apj 350\/} (Feb. 1990), 89--94.

\bibitem{1997ApJ...478..462C}
{\sc {Carlberg}, R.~G., {Yee}, H.~K.~C., and {Ellingson}, E.}
\newblock {The Average Mass and Light Profiles of Galaxy Clusters}.
\newblock {\em \apj 478\/} (Mar. 1997), 462--+.

\bibitem{Castro-Rodr}
{\sc {Castro-Rodr{\'{\i}}guez}, N., {Aguerri}, J.~A.~L., {Arnaboldi}, M.,
  {Gerhard}, O., {Freeman}, K.~C., {Napolitano}, N.~R., and {Capaccioli}, M.}
\newblock {Narrow band survey for intragroup light in the Leo HI cloud.
  Constraints on the galaxy background contamination in imaging surveys for
  intracluster planetary nebulae}.
\newblock {\em \aap 405\/} (July 2003), 803--812.

\bibitem{1976A&A....49..137C}
{\sc {Cavaliere}, A., and {Fusco-Femiano}, R.}
\newblock {X-rays from hot plasma in clusters of galaxies}.
\newblock {\em \aap 49\/} (May 1976), 137--144.

\bibitem{1998ApJ...501..493C}
{\sc {Cavaliere}, A., {Menci}, N., and {Tozzi}, P.}
\newblock {Diffuse Baryons in Groups and Clusters of Galaxies}.
\newblock {\em \apj 501\/} (July 1998), 493--+.

\bibitem{1992ApJ...393...22C}
{\sc {Cen}, R., and {Ostriker}, J.}
\newblock {A hydrodynamic treatment of the cold dark matter cosmological
  scenario}.
\newblock {\em \apj 393\/} (July 1992), 22--41.

\bibitem{2002coec.book.....C}
{\sc {Coles}, P., and {Lucchin}, F.}
\newblock {\em {Cosmology: The Origin and Evolution of Cosmic Structure, Second
  Edition}}.
\newblock Cosmology: The Origin and Evolution of Cosmic Structure, Second
  Edition, by Peter Coles, Francesco Lucchin, pp.~512.~ISBN
  0-471-48909-3.~Wiley-VCH , July 2002., July 2002.

\bibitem{Cypr06}
{\sc {Cypriano}, E.~S., {Sodr{\'e}}, L.~J., {Campusano}, L.~E., {Dale}, D.~A.,
  and {Hardy}, E.}
\newblock {Shrinking of Cluster Ellipticals: A Tidal Stripping Explanation and
  Implications for the Intracluster Light}.
\newblock {\em \aj 131\/} (May 2006), 2417--2425.

\bibitem{2008MNRAS.387.1431D}
{\sc {Dalla Vecchia}, C., and {Schaye}, J.}
\newblock {Simulating galactic outflows with kinetic supernova feedback}.
\newblock {\em \mnras 387\/} (July 2008), 1431--1444.

\bibitem{1983ApJ...267..465D}
{\sc {Davis}, M., and {Peebles}, P.~J.~E.}
\newblock {A survey of galaxy redshifts. V - The two-point position and
  velocity correlations}.
\newblock {\em \apj 267\/} (Apr. 1983), 465--482.

\bibitem{2002ApJ...567..163D}
{\sc {De Grandi}, S., and {Molendi}, S.}
\newblock {Temperature Profiles of Nearby Clusters of Galaxies}.
\newblock {\em \apj 567\/} (Mar. 2002), 163--177.

\bibitem{1960ApJ...131..585D}
{\sc {de Vaucouleurs}, G.}
\newblock {The Apparent Density of Matter in Groups and Clusters of Galaxies.}
\newblock {\em \apj 131\/} (May 1960), 585--+.

\bibitem{1969ApL.....4...17D}
{\sc {de Vaucouleurs}, G.}
\newblock {Photometry of the Outer Corona of Messier 87}.
\newblock {\em \apjl 4\/} (1969), 17--+.

\bibitem{dolag2008}
{\sc {Dolag}, K., {Borgani}, S., {Schindler}, S., {Diaferio}, A., and {Bykov},
  A.~M.}
\newblock {Simulation Techniques for Cosmological Simulations}.
\newblock {\em Space Science Reviews 134\/} (Feb. 2008), 229--268.

\bibitem{Elena}
{\sc {D'Onghia}, E., {Sommer-Larsen}, J., {Romeo}, A.~D., {Burkert}, A.,
  {Pedersen}, K., {Portinari}, L., and {Rasmussen}, J.}
\newblock {The Formation of Fossil Galaxy Groups in the Hierarchical Universe}.
\newblock {\em \apjl 630\/} (Sept. 2005), L109--L112.

\bibitem{1999ApJ...511....5E}
{\sc {Eisenstein}, D.~J., and {Hu}, W.}
\newblock {Power Spectra for Cold Dark Matter and Its Variants}.
\newblock {\em \apj 511\/} (Jan. 1999), 5--15.

\bibitem{Eke96}
{\sc {Eke}, V.~R., {Cole}, S., and {Frenk}, C.~S.}
\newblock {Cluster evolution as a diagnostic for Omega}.
\newblock {\em \mnras 282\/} (Sept. 1996), 263--280.

\bibitem{1991ApJ...383...95E}
{\sc {Evrard}, A.~E., and {Henry}, J.~P.}
\newblock {Expectations for X-ray cluster observations by the ROSAT satellite}.
\newblock {\em \apj 383\/} (Dec. 1991), 95--103.

\bibitem{1996ApJ...469..494E}
{\sc {Evrard}, A.~E., {Metzler}, C.~A., and {Navarro}, J.~F.}
\newblock {Mass Estimates of X-Ray Clusters}.
\newblock {\em \apj 469\/} (Oct. 1996), 494--+.

\bibitem{1994ARA&A..32..277F}
{\sc {Fabian}, A.~C.}
\newblock {Cooling Flows in Clusters of Galaxies}.
\newblock {\em \araa 32\/} (1994), 277--318.

\bibitem{2001MNRAS.321L..20F}
{\sc {Fabian}, A.~C., {Mushotzky}, R.~F., {Nulsen}, P.~E.~J., and {Peterson},
  J.~R.}
\newblock {On the soft X-ray spectrum of cooling flows}.
\newblock {\em \mnras 321\/} (Feb. 2001), L20--L24.

\bibitem{Feld03}
{\sc {Feldmeier}, J.~J., {Ciardullo}, R., {Jacoby}, G.~H., and {Durrell},
  P.~R.}
\newblock {Intracluster Planetary Nebulae in the Virgo Cluster. II. Imaging
  Catalog}.
\newblock {\em \apjs 145\/} (Mar. 2003), 65--81.

\bibitem{Feld04}
{\sc {Feldmeier}, J.~J., {Ciardullo}, R., {Jacoby}, G.~H., and {Durrell},
  P.~R.}
\newblock {Intracluster Planetary Nebulae in the Virgo Cluster. III. Luminosity
  of the Intracluster Light and Tests of the Spatial Distribution}.
\newblock {\em \apj 615\/} (Nov. 2004), 196--208.

\bibitem{FeldIAU03}
{\sc {Feldmeier}, J.~J., {Mihos}, J.~C., {Durrell}, P.~R., {Ciardullo}, R., and
  {Jacoby}, G.~H.}
\newblock {Kinematics of Planetary Nebulae in M51's Tidal Tail}.
\newblock In {\em IAU Symposium\/} (2003), S.~{Kwok}, M.~{Dopita}, and
  R.~{Sutherland}, Eds., pp.~633--+.

\bibitem{Feld2002}
{\sc {Feldmeier}, J.~J., {Mihos}, J.~C., {Morrison}, H.~L., {Rodney}, S.~A.,
  and {Harding}, P.}
\newblock {Deep CCD Surface Photometry of Galaxy Clusters. I. Methods and
  Initial Studies of Intracluster Starlight}.
\newblock {\em \apj 575\/} (Aug. 2002), 779--800.

\bibitem{GalYam}
{\sc {Gal-Yam}, A., {Maoz}, D., {Guhathakurta}, P., and {Filippenko}, A.~V.}
\newblock {A Population of Intergalactic Supernovae in Galaxy Clusters}.
\newblock {\em \aj 125\/} (Mar. 2003), 1087--1094.

\bibitem{1972AJ.....77..288G}
{\sc {Gallagher}, III, J.~S., and {Ostriker}, J.~P.}
\newblock {A Note on Mass Loss during Collisions between Galaxies a nd the
  Formation of Giant Systems}.
\newblock {\em \aj 77\/} (May 1972), 288--+.

\bibitem{1991ApJ...383...72G}
{\sc {Gebhardt}, K., and {Beers}, T.~C.}
\newblock {Bound populations around cD galaxies and cD velocity offsets in
  clusters of galaxies}.
\newblock {\em \apj 383\/} (Dec. 1991), 72--89.

\bibitem{Ortwin}
{\sc {Gerhard}, O., {Arnaboldi}, M., {Freeman}, K.~C., {Kashikawa}, N.,
  {Okamura}, S., and {Yasuda}, N.}
\newblock {Detection of Intracluster Planetary Nebulae in the Coma Cluster}.
\newblock {\em \apjl 621\/} (Mar. 2005), L93--L96.

\bibitem{Ger97}
{\sc {Gerritsen}, J.~P.~E., and {Icke}, V.}
\newblock {Star formation in N-body simulations. I. The impact of the stellar
  ultraviolet radiation on star formation.}
\newblock {\em \aap 325\/} (Sept. 1997), 972--986.

\bibitem{1999AJ....117.2608G}
{\sc {Gioia}, I.~M., {Henry}, J.~P., {Mullis}, C.~R., {Ebeling}, H., and
  {Wolter}, A.}
\newblock {RX J1716.6+6708: A Young Cluster at Z=0.81}.
\newblock {\em \aj 117\/} (June 1999), 2608--2616.

\bibitem{2001ApJ...548...79G}
{\sc {Girardi}, M., and {Mezzetti}, M.}
\newblock {Evolution of the Internal Dynamics of Galaxy Clusters}.
\newblock {\em \apj 548\/} (Feb. 2001), 79--96.

\bibitem{Gned03}
{\sc {Gnedin}, O.~Y.}
\newblock {Dynamical Evolution of Galaxies in Clusters}.
\newblock {\em \apj 589\/} (June 2003), 752--769.

\bibitem{Gonzales2000}
{\sc {Gonzalez}, A.~H., {Zabludoff}, A.~I., {Zaritsky}, D., and {Dalcanton},
  J.~J.}
\newblock {Measuring the Diffuse Optical Light in Abell 1651}.
\newblock {\em \apj 536\/} (June 2000), 561--570.

\bibitem{Gov07}
{\sc {Governato}, F., {Willman}, B., {Mayer}, L., {Brooks}, A., {Stinson}, G.,
  {Valenzuela}, O., {Wadsley}, J., and {Quinn}, T.}
\newblock {Forming disc galaxies in {$\Lambda$}CDM simulations}.
\newblock {\em \mnras 374\/} (Feb. 2007), 1479--1494.

\bibitem{1969BAAS....1R.191G}
{\sc {Gunn}, J.~E.}
\newblock {Visual Background Radiation in the Coma Cluster of Galaxies}.
\newblock In {\em Bulletin of the American Astronomical Society\/} (Mar. 1969),
  vol.~1 of {\em Bulletin of the American Astronomical Society}, pp.~191--+.

\bibitem{1996ApJ...461...20H}
{\sc {Haardt}, F., and {Madau}, P.}
\newblock {Radiative Transfer in a Clumpy Universe. II. The Ultraviolet
  Extragalactic Background}.
\newblock {\em \apj 461\/} (Apr. 1996), 20--+.

\bibitem{Hern}
{\sc {Hernquist}, L.}
\newblock {An analytical model for spherical galaxies and bulges}.
\newblock {\em \apj 356\/} (June 1990), 359--364.

\bibitem{1993ApJS...86..389H}
{\sc {Hernquist}, L.}
\newblock {N-body realizations of compound galaxies}.
\newblock {\em \apjs 86\/} (June 1993), 389--400.

\bibitem{1958MeLu2.136....1H}
{\sc {Holmberg}, E.}
\newblock {A photographic photometry of extragalactic nebulae.}
\newblock {\em Meddelanden fran Lunds Astronomiska Observatorium Serie II
  136\/} (1958), 1--+.

\bibitem{2000ApJ...538..543I}
{\sc {Irwin}, J.~A., and {Bregman}, J.~N.}
\newblock {Radial Temperature Profiles of 11 Clusters of Galaxies Observed with
  BEPPOSAX}.
\newblock {\em \apj 538\/} (Aug. 2000), 543--554.

\bibitem{2001ApJ...562..124J}
{\sc {Jeltema}, T.~E., {Canizares}, C.~R., {Bautz}, M.~W., {Malm}, M.~R.,
  {Donahue}, M., and {Garmire}, G.~P.}
\newblock {Chandra X-Ray Observatory Observation of the High-Redshift Cluster
  MS 1054-0321}.
\newblock {\em \apj 562\/} (Nov. 2001), 124--132.

\bibitem{1986MNRAS.222..323K}
{\sc {Kaiser}, N.}
\newblock {Evolution and clustering of rich clusters}.
\newblock {\em \mnras 222\/} (Sept. 1986), 323--345.

\bibitem{Katz92}
{\sc {Katz}, N.}
\newblock {Dissipational galaxy formation. II - Effects of star formation}.
\newblock {\em \apj 391\/} (June 1992), 502--517.

\bibitem{Katz96}
{\sc {Katz}, N., {Weinberg}, D.~H., and {Hernquist}, L.}
\newblock {Cosmological Simulations with TreeSPH}.
\newblock {\em \apjs 105\/} (July 1996), 19.

\bibitem{KauffProc}
{\sc {Kauffmann}, G.}
\newblock {Formation Histories Expected From Cosmological Simulations}.
\newblock In {\em ASP Conf. Ser. 245: Astrophysical Ages and Times Scales\/}
  (2001), T.~{von Hippel}, C.~{Simpson}, and N.~{Manset}, Eds., pp.~381--+.

\bibitem{Kay07}
{\sc {Kay}, S.~T., {da Silva}, A.~C., {Aghanim}, N., {Blanchard}, A., {Liddle},
  A.~R., {Puget}, J.-L., {Sadat}, R., and {Thomas}, P.~A.}
\newblock {The evolution of clusters in the CLEF cosmological simulation: X-ray
  structural and scaling properties}.
\newblock {\em \mnras 377\/} (May 2007), 317--334.

\bibitem{1998ApJ...498..541K}
{\sc {Kennicutt}, Jr., R.~C.}
\newblock {The Global Schmidt Law in Star-forming Galaxies}.
\newblock {\em \apj 498\/} (May 1998), 541--+.

\bibitem{2001MNRAS.327.1353K}
{\sc {Komatsu}, E., and {Seljak}, U.}
\newblock {Universal gas density and temperature profile}.
\newblock {\em \mnras 327\/} (Nov. 2001), 1353--1366.

\bibitem{Krick06}
{\sc {Krick}, J.~E., {Bernstein}, R.~A., and {Pimbblet}, K.~A.}
\newblock {Diffuse Optical Light in Galaxy Clusters. I. Abell 3888}.
\newblock {\em \aj 131\/} (Jan. 2006), 168--184.

\bibitem{1999ApJ...517..587L}
{\sc {Lewis}, A.~D., {Ellingson}, E., {Morris}, S.~L., and {Carlberg}, R.~G.}
\newblock {X-Ray Mass Estimates at Z \~{} 0.3 for the Canadian Network for
  Observational Cosmology Cluster Sample}.
\newblock {\em \apj 517\/} (June 1999), 587--608.

\bibitem{Lin03}
{\sc {Lin}, Y.-T., {Mohr}, J.~J., and {Stanford}, S.~A.}
\newblock {Near-Infrared Properties of Galaxy Clusters: Luminosity as a Binding
  Mass Predictor and the State of Cluster Baryons}.
\newblock {\em \apj 591\/} (July 2003), 749--763.

\bibitem{LinMohr04}
{\sc {Lin}, Y.-T., {Mohr}, J.~J., and {Stanford}, S.~A.}
\newblock {K-Band Properties of Galaxy Clusters and Groups: Luminosity
  Function, Radial Distribution, and Halo Occupation Number}.
\newblock {\em \apj 610\/} (Aug. 2004), 745--761.

\bibitem{2000MNRAS.315..689L}
{\sc {Lloyd-Davies}, E.~J., {Ponman}, T.~J., and {Cannon}, D.~B.}
\newblock {The entropy and energy of intergalactic gas in galaxy clusters}.
\newblock {\em \mnras 315\/} (July 2000), 689--702.

\bibitem{1992ApJ...400...65M}
{\sc {Mackie}, G.}
\newblock {The stellar content of central dominant galaxies. II - Colors of cD
  envelopes}.
\newblock {\em \apj 400\/} (Nov. 1992), 65--73.

\bibitem{1998ApJ...503...77M}
{\sc {Markevitch}, M., {Forman}, W.~R., {Sarazin}, C.~L., and {Vikhlinin}, A.}
\newblock {The Temperature Structure of 30 Nearby Clusters Observed with ASCA:
  Similarity of Temperature Profiles}.
\newblock {\em \apj 503\/} (Aug. 1998), 77--+.

\bibitem{2003MNRAS.345..561M}
{\sc {Marri}, S., and {White}, S.~D.~M.}
\newblock {Smoothed particle hydrodynamics for galaxy-formation simulations:
  improved treatments of multiphase gas, of star formation and of supernovae
  feedback}.
\newblock {\em \mnras 345\/} (Oct. 2003), 561--574.

\bibitem{2002ApJ...566..302M}
{\sc {Matzner}, C.~D.}
\newblock {On the Role of Massive Stars in the Support and Destruction of Giant
  Molecular Clouds}.
\newblock {\em \apj 566\/} (Feb. 2002), 302--314.

\bibitem{Merritt84}
{\sc {Merritt}, D.}
\newblock {Relaxation and tidal stripping in rich clusters of galaxies. II -
  Evolution of the luminosity distribution}.
\newblock {\em \apj 276\/} (Jan. 1984), 26--37.

\bibitem{Merritt95}
{\sc {Merritt}, D.}
\newblock {Dynamical Modelling of Hot Stellar Systems}.
\newblock {\em ArXiv Astrophysics e-prints\/} (Oct. 1995).

\bibitem{Mihos04}
{\sc {Mihos}, J.~C.}
\newblock {Interactions and Mergers of Cluster Galaxies}.
\newblock In {\em Clusters of Galaxies: Probes of Cosmological Structure and
  Galaxy Evolution\/} (2004), J.~S. {Mulchaey}, A.~{Dressler}, and A.~{Oemler},
  Eds., pp.~277--+.

\bibitem{MihosVirgo}
{\sc {Mihos}, J.~C., {Harding}, P., {Feldmeier}, J., and {Morrison}, H.}
\newblock {Diffuse Light in the Virgo Cluster}.
\newblock {\em \apjl 631\/} (Sept. 2005), L41--L44.

\bibitem{MO04}
{\sc {Monaco}, P.}
\newblock {Physical regimes for feedback in galaxy formation}.
\newblock {\em \mnras 352\/} (July 2004), 181--204.

\bibitem{MORGANA}
{\sc {Monaco}, P., {Fontanot}, F., and {Taffoni}, G.}
\newblock {The MORGANA model for the rise of galaxies and active nuclei}.
\newblock {\em ArXiv Astrophysics e-prints\/} (Oct. 2006).

\bibitem{PG1}
{\sc {Monaco}, P., {Murante}, G., {Borgani}, S., and {Fontanot}}.
\newblock {Diffuse Stellar Component in Galaxy Clusters and the Evolution of
  the Most Massive Galaxies at z < \~{}1}.
\newblock {\em \apjl 652\/} (Dec. 2006), L89--L92.

\bibitem{Moore96}
{\sc {Moore}, B., {Katz}, N., {Lake}, G., {Dressler}, A., and {Oemler}, A.}
\newblock {Galaxy harassment and the evolution of clusters of galaxies.}
\newblock {\em \nat 379\/} (1996), 613--616.

\bibitem{M04}
{\sc {Murante}, G., {Arnaboldi}, M., {Gerhard}, O., {Borgani}, S., {Cheng},
  L.~M., {Diaferio}, A., {Dolag}, K., {Moscardini}, L., {Tormen}, G.,
  {Tornatore}, L., and {Tozzi}, P.}
\newblock {The Diffuse Light in Simulations of Galaxy Clusters}.
\newblock {\em \apjl 607\/} (June 2004), L83--L86.

\bibitem{Nagai07}
{\sc {Nagai}, D., {Kravtsov}, A.~V., and {Vikhlinin}, A.}
\newblock {Effects of Galaxy Formation on Thermodynamics of the Intracluster
  Medium}.
\newblock {\em \apj 668\/} (Oct. 2007), 1--14.

\bibitem{Napo03}
{\sc {Napolitano}, N.~R., {Pannella}, M., {Arnaboldi}, M., {Gerhard}, O.,
  {Aguerri}, J.~A.~L., {Freeman}, K.~C., {Capaccioli}, M., {Ghigna}, S.,
  {Governato}, F., {Quinn}, T., and {Stadel}, J.}
\newblock {Intracluster Stellar Population Properties from N-Body Cosmological
  Simulations. I. Constraints at z = 0}.
\newblock {\em \apj 594\/} (Sept. 2003), 172--185.

\bibitem{NA96.1}
{\sc Navarro, J., Frenk, C., and White, S.}
\newblock The structure of cold dark matter halos.
\newblock {\em ApJ 462\/} (1996), 563.

\bibitem{NA97.1}
{\sc Navarro, J., Frenk, C., and White, S.}
\newblock A universal density profile from hierarchical clustering.
\newblock {\em ApJ 490\/} (1997), 493.

\bibitem{1995MNRAS.275..720N}
{\sc {Navarro}, J.~F., {Frenk}, C.~S., and {White}, S.~D.~M.}
\newblock {Simulations of X-ray clusters}.
\newblock {\em \mnras 275\/} (Aug. 1995), 720--740.

\bibitem{1993MNRAS.265..271N}
{\sc {Navarro}, J.~F., and {White}, S.~D.~M.}
\newblock {Simulations of Dissipative Galaxy Formation in Hierarchically
  Clustering Universes - Part One - Tests of the Code}.
\newblock {\em \mnras 265\/} (Nov. 1993), 271--+.

\bibitem{Ostr77}
{\sc {Ostriker}, J.~P., and {Hausman}, M.~A.}
\newblock {Cannibalism among the galaxies - Dynamically produced evolution of
  cluster luminosity functions}.
\newblock {\em \apjl 217\/} (Nov. 1977), L125--L129.

\bibitem{OstSte03}
{\sc {Ostriker}, J.~P., and {Steinhardt}, P.}
\newblock {New Light on Dark Matter}.
\newblock {\em Science 300\/} (June 2003), 1909--1914.

\bibitem{1999coph.book.....P}
{\sc {Peacock}, J.~A.}
\newblock {\em {Cosmological Physics}}.
\newblock Cosmological Physics, by John A.~Peacock, pp.~704.~ISBN
  052141072X.~Cambridge, UK: Cambridge University Press, January 1999., Jan.
  1999.

\bibitem{1993ppc..book.....P}
{\sc {Peebles}, P.~J.~E.}
\newblock {\em {Principles of physical cosmology}}.
\newblock Princeton Series in Physics, Princeton, NJ: Princeton University
  Press, |c1993, 1993.

\bibitem{1999PhRvL..83..670P}
{\sc {Perlmutter}, S., {Turner}, M.~S., and {White}, M.}
\newblock {Constraining Dark Energy with Type Ia Supernovae and Large-Scale
  Structure}.
\newblock {\em Physical Review Letters 83\/} (July 1999), 670--673.

\bibitem{2001A&A...365L.104P}
{\sc {Peterson}, J.~R., {Paerels}, F.~B.~S., {Kaastra}, J.~S., {Arnaud}, M.,
  {Reiprich}, T.~H., {Fabian}, A.~C., {Mushotzky}, R.~F., {Jernigan}, J.~G.,
  and {Sakelliou}, I.}
\newblock {X-ray imaging-spectroscopy of Abell 1835}.
\newblock {\em \aap 365\/} (Jan. 2001), L104--L109.

\bibitem{1999Natur.397..135P}
{\sc {Ponman}, T.~J., {Cannon}, D.~B., and {Navarro}, J.~F.}
\newblock {The thermal imprint of galaxy formation on X-ray clusters}.
\newblock {\em \nat 397\/} (Jan. 1999), 135--137.

\bibitem{Pow03}
{\sc Power, C., Navarro, J., Jenkins, A., Frenk, C., White, S., Springel, V.,
  Stadel, J., and Quinn, T.}
\newblock The inner structure of {$\Lambda$}cdm haloes - i. a numerical
  convergence study.
\newblock {\em MNRAS 338\/} (2003), 14.

\bibitem{1974ApJ...187..425P}
{\sc {Press}, W.~H., and {Schechter}, P.}
\newblock {Formation of Galaxies and Clusters of Galaxies by Self-Similar
  Gravitational Condensation}.
\newblock {\em \apj 187\/} (Feb. 1974), 425--438.

\bibitem{1992nrca.book.....P}
{\sc {Press}, W.~H., {Teukolsky}, S.~A., {Vetterling}, W.~T., and {Flannery},
  B.~P.}
\newblock {\em {Numerical recipes in C. The art of scientific computing}}.
\newblock Cambridge: University Press, |c1992, 2nd ed., 1992.

\bibitem{1991ApJ...369L...1P}
{\sc {Prestwich}, A.~H., and {Joy}, M.}
\newblock {Cooling flows and the formation of massive halos in cD galaxies}.
\newblock {\em \apjl 369\/} (Mar. 1991), L1--L4.

\bibitem{1977ApJS...35..419R}
{\sc {Raymond}, J.~C., and {Smith}, B.~W.}
\newblock {Soft X-ray spectrum of a hot plasma}.
\newblock {\em \apjs 35\/} (Dec. 1977), 419--439.

\bibitem{2002ApJ...567..716R}
{\sc {Reiprich}, T.~H., and {B{\"o}hringer}, H.}
\newblock {The Mass Function of an X-Ray Flux-limited Sample of Galaxy
  Clusters}.
\newblock {\em \apj 567\/} (Mar. 2002), 716--740.

\bibitem{2002ARA&A..40..539R}
{\sc {Rosati}, P., {Borgani}, S., and {Norman}, C.}
\newblock {The Evolution of X-ray Clusters of Galaxies}.
\newblock {\em \araa 40\/} (2002), 539--577.

\bibitem{Rudick}
{\sc {Rudick}, C.~S., {Mihos}, J.~C., and {McBride}, C.}
\newblock {The Formation and Evolution of Intracluster Light}.
\newblock {\em \apj 648\/} (Sept. 2006), 936--946.

\bibitem{2003astro.ph..9326S}
{\sc {Sakharov}, A.~S., and {Hofer}, H.}
\newblock {Development of the Universe and New Cosmology}.
\newblock {\em ArXiv Astrophysics e-prints\/} (Sept. 2003).

\bibitem{SA88.1}
{\sc Sarazin, C.~L.}
\newblock {\em X-ray emission from clusters of galaxies}.
\newblock Cambridge University Press, Cambridge, 1988.

\bibitem{2005MNRAS.364..552S}
{\sc {Scannapieco}, C., {Tissera}, P.~B., {White}, S.~D.~M., and {Springel},
  V.}
\newblock {Feedback and metal enrichment in cosmological smoothed particle
  hydrodynamics simulations - I. A model for chemical enrichment}.
\newblock {\em \mnras 364\/} (Dec. 2005), 552--564.

\bibitem{2006MNRAS.371.1125S}
{\sc {Scannapieco}, C., {Tissera}, P.~B., {White}, S.~D.~M., and {Springel},
  V.}
\newblock {Feedback and metal enrichment in cosmological SPH simulations - II.
  A multiphase model with supernova energy feedback}.
\newblock {\em \mnras 371\/} (Sept. 2006), 1125--1139.

\bibitem{1994ApJ...423..566S}
{\sc {Scheick}, X., and {Kuhn}, J.~R.}
\newblock {Diffuse Light in A2670: Smoothly Distributed?}
\newblock {\em \apj 423\/} (Mar. 1994), 566--+.

\bibitem{1988ApJ...328..475S}
{\sc {Schombert}, J.~M.}
\newblock {The structure of brightest cluster members. III - cD envelopes}.
\newblock {\em \apj 328\/} (May 1988), 475--488.

\bibitem{1959sdmm.book.....S}
{\sc {Sedov}, L.~I.}
\newblock {\em {Similarity and Dimensional Methods in Mechanics}}.
\newblock Similarity and Dimensional Methods in Mechanics, New York: Academic
  Press, 1959, 1959.

\bibitem{Shen}
{\sc {Shen}, S., {Mo}, H.~J., {White}, S.~D.~M., {Blanton}, M.~R., {Kauffmann},
  G., {Voges}, W., {Brinkmann}, J., and {Csabai}, I.}
\newblock {The size distribution of galaxies in the Sloan Digital Sky Survey}.
\newblock {\em \mnras 343\/} (Aug. 2003), 978--994.

\bibitem{2001MNRAS.323....1S}
{\sc {Sheth}, R.~K., {Mo}, H.~J., and {Tormen}, G.}
\newblock {Ellipsoidal collapse and an improved model for the number and
  spatial distribution of dark matter haloes}.
\newblock {\em \mnras 323\/} (May 2001), 1--12.

\bibitem{1936ApJ....83...23S}
{\sc {Smith}, S.}
\newblock {The Mass of the Virgo Cluster}.
\newblock {\em \apj 83\/} (Jan. 1936), 23--+.

\bibitem{1916ZPhy...17..557S}
{\sc {Smoluchowski}, M.~V.}
\newblock {Drei Vortrage uber Diffusion, Brownsche Bewegung und Koagulation von
  Kolloidteilchen}.
\newblock {\em Zeitschrift fur Physik 17\/} (1916), 557--585.

\bibitem{SommerGroups}
{\sc {Sommer-Larsen}, J.}
\newblock {Properties of intra-group stars and galaxies in galaxy groups:
  `normal' versus `fossil' groups}.
\newblock {\em \mnras\/} (June 2006), 509--+.

\bibitem{SommerLarsen}
{\sc {Sommer-Larsen}, J., {Romeo}, A.~D., and {Portinari}, L.}
\newblock {Simulating galaxy clusters - III. Properties of the intracluster
  stars}.
\newblock {\em \mnras 357\/} (Feb. 2005), 478--488.

\bibitem{GADGET2}
{\sc {Springel}, V.}
\newblock {The cosmological simulation code GADGET-2}.
\newblock {\em \mnras 364\/} (Dec. 2005), 1105--1134.

\bibitem{2005MNRAS.361..776S}
{\sc {Springel}, V., {Di Matteo}, T., and {Hernquist}, L.}
\newblock {Modelling feedback from stars and black holes in galaxy mergers}.
\newblock {\em \mnras 361\/} (Aug. 2005), 776--794.

\bibitem{2002MNRAS.333..649S}
{\sc {Springel}, V., and {Hernquist}, L.}
\newblock {Cosmological smoothed particle hydrodynamics simulations: the
  entropy equation}.
\newblock {\em \mnras 333\/} (July 2002), 649--664.

\bibitem{SpringHern03}
{\sc {Springel}, V., and {Hernquist}, L.}
\newblock {Cosmological smoothed particle hydrodynamics simulations: a hybrid
  multiphase model for star formation}.
\newblock {\em \mnras 339\/} (Feb. 2003), 289--311.

\bibitem{Springel2001MT}
{\sc {Springel}, V., {White}, S.~D.~M., {Tormen}, G., and {Kauffmann}, G.}
\newblock {Populating a cluster of galaxies - I. Results at z=0}.
\newblock {\em \mnras 328\/} (Dec. 2001), 726--750.

\bibitem{GADGET}
{\sc {Springel}, V., {Yoshida}, N., and {White}, S.~D.~M.}
\newblock {GADGET: a code for collisionless and gasdynamical cosmological
  simulations}.
\newblock {\em New Astronomy 6\/} (Apr. 2001), 79--117.

\bibitem{SK}
{\sc {Stadel}, J.~G.}
\newblock {Cosmological N-body simulations and their analysis}.
\newblock {\em Ph.D.~Thesis\/} (2001).

\bibitem{Stinson06}
{\sc {Stinson}, G., {Seth}, A., {Katz}, N., {Wadsley}, J., {Governato}, F., and
  {Quinn}, T.}
\newblock {Star formation and feedback in smoothed particle hydrodynamic
  simulations - I. Isolated galaxies}.
\newblock {\em \mnras 373\/} (Dec. 2006), 1074--1090.

\bibitem{1993ApJS...88..253S}
{\sc {Sutherland}, R.~S., and {Dopita}, M.~A.}
\newblock {Cooling functions for low-density astrophysical plasmas}.
\newblock {\em \apjs 88\/} (Sept. 1993), 253--327.

\bibitem{Suto98}
{\sc {Suto}, Y., {Sasaki}, S., and {Makino}, N.}
\newblock {Gas Density and X-Ray Surface Brightness Profiles of Clusters of
  Galaxies from Dark Matter Halo Potentials: Beyond the Isothermal beta-Model}.
\newblock {\em \apj 509\/} (Dec. 1998), 544--550.

\bibitem{2001A&A...365L..87T}
{\sc {Tamura}, T., {Kaastra}, J.~S., {Peterson}, J.~R., {Paerels}, F.~B.~S.,
  {Mittaz}, J.~P.~D., {Trudolyubov}, S.~P., {Stewart}, G., {Fabian}, A.~C.,
  {Mushotzky}, R.~F., {Lumb}, D.~H., and {Ikebe}, Y.}
\newblock {X-ray spectroscopy of the cluster of galaxies Abell 1795 with
  XMM-Newton}.
\newblock {\em \aap 365\/} (Jan. 2001), L87--L92.

\bibitem{ThacCouch00}
{\sc {Thacker}, R.~J., and {Couchman}, H.~M.~P.}
\newblock {Implementing Feedback in Simulations of Galaxy Formation: A Survey
  of Methods}.
\newblock {\em \apj 545\/} (Dec. 2000), 728--752.

\bibitem{1981ApJ...248..439T}
{\sc {Thuan}, T.~X., and {Romanishin}, W.}
\newblock {The structure of giant elliptical galaxies in poor clusters of
  galaxies}.
\newblock {\em \apj 248\/} (Sept. 1981), 439--459.

\bibitem{2003ApJ...594....1T}
{\sc {Tonry}, J.~L., {Schmidt}, B.~P., {Barris}, B., {Candia}, P., {Challis},
  P., {Clocchiatti}, A., {Coil}, A.~L., {Filippenko}, A.~V., {Garnavich}, P.,
  {Hogan}, C., {Holland}, S.~T., {Jha}, S., {Kirshner}, R.~P., {Krisciunas},
  K., {Leibundgut}, B., {Li}, W., {Matheson}, T., {Phillips}, M.~M., {Riess},
  A.~G., {Schommer}, R., {Smith}, R.~C., {Sollerman}, J., {Spyromilio}, J.,
  {Stubbs}, C.~W., and {Suntzeff}, N.~B.}
\newblock {Cosmological Results from High-z Supernovae}.
\newblock {\em \apj 594\/} (Sept. 2003), 1--24.

\bibitem{1997MNRAS.286..865T}
{\sc {Tormen}, G., {Bouchet}, F.~R., and {White}, S.~D.~M.}
\newblock {The structure and dynamical evolution of dark matter haloes}.
\newblock {\em \mnras 286\/} (Apr. 1997), 865--884.

\bibitem{2001ApJ...546...63T}
{\sc {Tozzi}, P., and {Norman}, C.}
\newblock {The Evolution of X-Ray Clusters and the Entropy of the Intracluster
  Medium}.
\newblock {\em \apj 546\/} (Jan. 2001), 63--84.

\bibitem{2000ApJ...542..106T}
{\sc {Tozzi}, P., {Scharf}, C., and {Norman}, C.}
\newblock {Detection of the Entropy of the Intergalactic Medium: Accretion
  Shocks in Clusters, Adiabatic Cores in Groups}.
\newblock {\em \apj 542\/} (Oct. 2000), 106--119.

\bibitem{Uson91}
{\sc {Uson}, J.~M., {Boughn}, S.~P., and {Kuhn}, J.~R.}
\newblock {Diffuse light in dense clusters of galaxies. I - R-band observations
  of Abell 2029}.
\newblock {\em \apj 369\/} (Mar. 1991), 46--53.

\bibitem{1983A&A...118..123V}
{\sc {Valentijn}, E.~A.}
\newblock {Calibrated B, V surface photometry of X-ray cD galaxies}.
\newblock {\em \aap 118\/} (Feb. 1983), 123--138.

\bibitem{1994A&A...283...37V}
{\sc {Vilchez-Gomez}, R., {Pello}, R., and {Sanahuja}, B.}
\newblock {Detection of intracluster light in the rich clusters of galaxies
  Abell 2390 and CL 1613+31}.
\newblock {\em \aap 283\/} (Mar. 1994), 37--50.

\bibitem{1982MNRAS.199..493V}
{\sc {Villumsen}, J.~V.}
\newblock {Simulations of galaxy mergers}.
\newblock {\em \mnras 199\/} (May 1982), 493--516.

\bibitem{1983MNRAS.204..219V}
{\sc {Villumsen}, J.~V.}
\newblock {Simulations of galaxy mergers. II}.
\newblock {\em \mnras 204\/} (July 1983), 219--236.

\bibitem{Viola}
{\sc {Viola}, M., {Monaco}, P., {Borgani}, S., {Murante}, G., and {Tornatore},
  L.}
\newblock {How does gas cool in dark matter haloes?}
\newblock {\em \mnras 383\/} (Jan. 2008), 777--790.

\bibitem{Wechsler2002}
{\sc {Wechsler}, R.~H., {Bullock}, J.~S., {Primack}, J.~R., {Kravtsov}, A.~V.,
  and {Dekel}, A.}
\newblock {Concentrations of Dark Halos from Their Assembly Histories}.
\newblock {\em \apj 568\/} (Mar. 2002), 52--70.

\bibitem{2000MNRAS.312..663W}
{\sc {White}, D.~A.}
\newblock {Deconvolution of ASCA X-ray data - II. Radial temperature and
  metallicity profiles for 106 galaxy clusters}.
\newblock {\em \mnras 312\/} (Mar. 2000), 663--688.

\bibitem{Fabio}
{\sc {Willman}, B., {Governato}, F., {Wadsley}, J., and {Quinn}, T.}
\newblock {The origin and properties of intracluster stars in a rich cluster}.
\newblock {\em \mnras 355\/} (Nov. 2004), 159--168.

\bibitem{1990ApJ...363..435W}
{\sc {Wilson}, C.~D., and {Scoville}, N.}
\newblock {The properties of individual giant molecular clouds in M33}.
\newblock {\em \apj 363\/} (Nov. 1990), 435--450.

\bibitem{2000MNRAS.318..889W}
{\sc {Wu}, K.~K.~S., {Fabian}, A.~C., and {Nulsen}, P.~E.~J.}
\newblock {Non-gravitational heating in the hierarchical formation of X-ray
  clusters}.
\newblock {\em \mnras 318\/} (Nov. 2000), 889--912.

\bibitem{1999ApJ...524...22W}
{\sc {Wu}, X.-P., {Xue}, Y.-J., and {Fang}, L.-Z.}
\newblock {The L\_X-T and L\_X-{$\sigma$} Relationships for Galaxy Clusters
  Revisited}.
\newblock {\em \apj 524\/} (Oct. 1999), 22--30.

\bibitem{1997MNRAS.284..235Y}
{\sc {Yepes}, G., {Kates}, R., {Khokhlov}, A., and {Klypin}, A.}
\newblock {Hydrodynamical simulations of galaxy formation: effects of supernova
  feedback}.
\newblock {\em \mnras 284\/} (Jan. 1997), 235--256.

\bibitem{Zibetti}
{\sc {Zibetti}, S., {White}, S.~D.~M., {Schneider}, D.~P., and {Brinkmann}, J.}
\newblock {Intergalactic stars in z\~{} 0.25 galaxy clusters: systematic
  properties from stacking of Sloan Digital Sky Survey imaging data}.
\newblock {\em \mnras 358\/} (Apr. 2005), 949--967.

\bibitem{1937ApJ....86..217Z}
{\sc {Zwicky}, F.}
\newblock {On the Masses of Nebulae and of Clusters of Nebulae}.
\newblock {\em \apj 86\/} (Oct. 1937), 217--+.

\bibitem{1951PASP...63...61Z}
{\sc {Zwicky}, F.}
\newblock {The Coma Cluster of Galaxies}.
\newblock {\em PASP 63\/} (Apr. 1951), 61--+.

\bibitem{1957moas.book.....Z}
{\sc {Zwicky}, F.}
\newblock {\em {Morphological astronomy}}.
\newblock Berlin: Springer, 1957, 1957.

\bibitem{1959HDP....53..390Z}
{\sc {Zwicky}, F.}
\newblock {Clusters of Galaxies.}
\newblock {\em Handbuch der Physik 53\/} (1959), 390--+.

\bibitem{1966cgcg.book.....Z}
{\sc {Zwicky}, F., {Herzog}, E., and {Wild}, P.}
\newblock {\em {Catalogue of galaxies and of clusters of galaxies, Vol. 3}}.
\newblock Pasadena: California Institute of Technology (CIT), |c1966, 1966.

\end{thebibliography}
\addcontentsline{toc}{chapter}{BIBLIOGRAPHY}

\end{document}